\documentclass[twoside,12pt
]{article}

\def\ss{\mbox{\boldmath $\sigma$}}

\def\AnswerYes{n}
\def\ShowLabelsVersion{y}         
\def\ShowChangesVersion{y}        
\def\ShowAnnotationsVersion{y}    

\usepackage[dvips]{thumbpdf}    
\usepackage[dvips,              
         breaklinks=true,
         colorlinks=true,           
         pdfborder={0 0 1},    
         citecolor={blue},linkcolor={blue},urlcolor={red},
         citebordercolor={1 0 0},linkbordercolor={0 0 1},urlbordercolor={1 0 0},
         pdfpagemode=UseOutlines,
         bookmarks,bookmarksopen
         ]{hyperref}         
\usepackage[numbers,sort&compress]{natbib}
\usepackage{hypernat}
\usepackage[dvips]{color}              
\usepackage[final]{graphicx}            
\graphicspath{{figures/}{./}}
\usepackage{amssymb}
\usepackage{amsmath}
\usepackage{xspace}
\usepackage{multirow}

\usepackage{slashed}

\ifx\ShowLabelsVersion\AnswerYes
   \usepackage[color]{showkeys} 
   \definecolor{refkey}{gray}{.5}   
   \definecolor{labelkey}{gray}{.5} 
   
\fi

\ifx\ShowAnnotationsVersion\AnswerYes
   \newcommand{\comment}[1]{{\scriptsize\sffamily\bfseries{#1}}}
   \newcommand{\margin}[1]{\marginpar{\scriptsize\sffamily\bfseries{#1}}}
   \usepackage{fancyhdr}
\else
   \newcommand{\comment}[1]{}
   \newcommand{\margin}[1]{}
\fi

\ifx\ShowChangesVersion\AnswerYes
   \usepackage{ulem}
    
   \newcommand{\delete}[1]{\sout{#1}}            
   \renewcommand{\emph}[1]{\textit{#1}}           
\else

   \newcommand{\sout}[1]{}
   \newcommand{\xout}[1]{}

   \newcommand{\delete}[1]{}
\fi

\newcommand{\dis}{\displaystyle}

\newcommand{\non}{\nonumber}
\newcommand{\hq}{\hspace{0.5em}}

\newcommand{\half}{\frac{1}{2}}

\newcommand{\ii}{\mathrm{i}}
\newcommand{\dd}{\mathrm{d}}
\newcommand{\rmd}{\mathrm{d}}

\newcommand{\de}{\partial}

\newcommand{\bra}{\langle}
\newcommand{\ket}{\rangle}

\newcommand{\HIGS}{HI$\gamma$S\xspace}
\newcommand{\threeHe}{${}^3$He\xspace}

\newcommand{\Q}{P}
\newcommand{\w}{\ensuremath{\omega}}
\newcommand{\wcm}{\ensuremath{\omega_\mathrm{cm}}}
\newcommand{\wlab}{\ensuremath{\omega_\mathrm{lab}}}
\newcommand{\wlabout}{\ensuremath{\omega_\mathrm{lab}^\prime}}
\newcommand{\wBreit}{\ensuremath{\omega_\mathrm{Breit}}}
\newcommand{\thetacm}{\ensuremath{\theta_\mathrm{cm}}}
\newcommand{\thetalab}{\ensuremath{\theta_\mathrm{lab}}}
\newcommand{\thetaBreit}{\ensuremath{\theta_\mathrm{Breit}}}
\newcommand{\alphae}{\ensuremath{\alpha_{E1}}}
\newcommand{\betam}{\ensuremath{\beta_{M1}}}
\newcommand{\gammaee}{\ensuremath{\gamma_{E1E1}}}
\newcommand{\gammamm}{\ensuremath{\gamma_{M1M1}}}
\newcommand{\gammaem}{\ensuremath{\gamma_{E1M2}}}
\newcommand{\gammame}{\ensuremath{\gamma_{M1E2}}}
\newcommand{\gammazero}{\ensuremath{\gamma_{0}}}
\newcommand{\gammapi}{\ensuremath{\gamma_{\pi}}}
\newcommand{\alphaep}{\ensuremath{\alpha_{E1}^{(\mathrm{p})}}}
\newcommand{\betamp}{\ensuremath{\beta_{M1}^{(\mathrm{p})}}}
\newcommand{\gammaeep}{\ensuremath{\gamma_{E1E1}^{(\mathrm{p})}}}
\newcommand{\gammammp}{\ensuremath{\gamma_{M1M1}^{(\mathrm{p})}}}
\newcommand{\gammaemp}{\ensuremath{\gamma_{E1M2}^{(\mathrm{p})}}}
\newcommand{\gammamep}{\ensuremath{\gamma_{M1E2}^{(\mathrm{p})}}}
\newcommand{\gammazerop}{\ensuremath{\gamma_{0}^{(\mathrm{p})}}}
\newcommand{\gammapip}{\ensuremath{\gamma_{\pi}^{(\mathrm{p})}}}
\newcommand{\alphaen}{\ensuremath{\alpha_{E1}^{(\mathrm{n})}}}
\newcommand{\betamn}{\ensuremath{\beta_{M1}^{(\mathrm{n})}}}
\newcommand{\gammaeen}{\ensuremath{\gamma_{E1E1}^{(\mathrm{n})}}}
\newcommand{\gammammn}{\ensuremath{\gamma_{M1M1}^{(\mathrm{n})}}}
\newcommand{\gammaemn}{\ensuremath{\gamma_{E1M2}^{(\mathrm{n})}}}
\newcommand{\gammamen}{\ensuremath{\gamma_{M1E2}^{(\mathrm{n})}}}

\newcommand{\gammapin}{\ensuremath{\gamma_{\pi}^{(\mathrm{n})}}}
\newcommand{\alphaes}{\ensuremath{\alpha_{E1}^{(\mathrm{s})}}}
\newcommand{\betams}{\ensuremath{\beta_{M1}^{(\mathrm{s})}}}

\newcommand{\alphaev}{\ensuremath{\alpha_{E1}^{(\mathrm{v})}}}
\newcommand{\betamv}{\ensuremath{\beta_{M1}^{(\mathrm{v})}}}

\newcommand{\MN}{\ensuremath{M_\mathrm{N}}} 
\newcommand{\Mp}{\ensuremath{M_\mathrm{p}}} 
\newcommand{\MDelta}{\ensuremath{M_\Delta}} 
\newcommand{\mpi}{\ensuremath{m_\pi}}     
\newcommand{\fpi}{\ensuremath{f_\pi}}
\newcommand{\wpi}{\ensuremath{\omega_\pi}}
\newcommand{\MeV}{\ensuremath{\mathrm{MeV}}}
\newcommand{\fm}{\ensuremath{\mathrm{fm}}}
\newcommand{\ChiEFT}{$\chi$EFT\xspace}
\newcommand{\EFTNoPion}{EFT($\slashed{\pi}$)\xspace}

\newcommand{\LambdaNoPion}{\ensuremath{\Lambda_\slashed{\pi}}}
\newcommand{\QNoPion}{\ensuremath{Q_\slashed{\pi}}}

\newcommand{\NXLO}[1]{N\ensuremath{{}^{#1}}LO\xspace}

\newcommand{\kv}{\vec{k}}
\newcommand{\pv}{\vec{p}}
\newcommand{\qv}{\vec{q}}
\newcommand{\wave}[3]{\ensuremath{{}^{#1}\mathrm{#2}_{#3}}\xspace}
\newcommand{\oneS}{\wave{1}{S}{0}}
\newcommand{\threeS}{\wave{3}{S}{1}}

\renewcommand{\Re}{\ensuremath{\mathrm{Re}}}
\renewcommand{\Im}{\ensuremath{\mathrm{Im}}}

\renewcommand{\deg}{\ensuremath{^\circ}}
\newcommand{\OdL}{Olmos de Le\'on}

\newcommand{\calL}{\mathcal{L}}
\newcommand{\calO}{\mathcal{O}}

\numberwithin{figure}{section}
\numberwithin{equation}{section}
\numberwithin{table}{section}

\newcommand{\ga}{g_{\scriptscriptstyle A}}
\newcommand{\gpiNN}{g_{\pi{\scriptscriptstyle\text{NN}}}}

\newcommand{\lambdachi}{\Lambda_\chi}
\newcommand{\ChPT}{\ensuremath{\chi \mathrm{PT}}}
\newcommand{\alphaEM}{\alpha_{\scriptscriptstyle\text{EM}}}

\newcommand{\GeV}{\ensuremath{\mathrm{GeV}}}
\newcommand{\NsqLO}{\ensuremath{\mathrm{N^2LO}}}
\newcommand{\gE}{\ensuremath{g_{\scriptscriptstyle E}}}
\newcommand{\gM}{\ensuremath{g_{\scriptscriptstyle M}}}
\newcommand{\be}{\begin{equation}}
\newcommand{\ee}{\end{equation}}
\newcommand{\bea}{\begin{eqnarray}}
\newcommand{\eea}{\end{eqnarray}}

\begin{document}

\title{ \vspace*{-1ex} Using effective field theory to analyse low-energy
  Compton scattering data from protons and light nuclei}
\author{H.~W.~Grie\3hammer$^1$,
  J.~A.~McGovern$^2$,\\[-0.5ex] D.~R.~Phillips$^{3}$ and G.~Feldman$^1$ \\[1ex]
  $^1$Institute for Nuclear Studies, Department of Physics,\\[-0.5ex] The
  George
  Washington University, Washington DC 20052, USA\\[-0.5ex]
  $^2$Theoretical Physics Group, School of Physics and Astronomy,\\[-0.5ex]
  The University of Manchester, Manchester, M13 9PL, UK\\[-0.5ex]
  $^3$Institute of Nuclear and Particle Physics and Department of\\[-0.5ex]
  Physics and Astronomy, Ohio University, Athens OH 45701, USA}
\maketitle

\vspace*{-4ex}

\begin{abstract} 
  Compton scattering from protons and neutrons provides important insight into
  the structure of the nucleon. For photon energies up to about $300$ MeV, the
  process can be parameterised by six dynamical dipole polarisabilities which
  characterise the response of the nucleon to a monochromatic photon of fixed
  frequency and multipolarity. Their zero-energy limit yields the well-known
  static electric and magnetic dipole polarisabilities $\alphae$ and $\betam$,
  and the four dipole spin polarisabilities. The emergence of full lattice QCD
  results and new experiments at MAMI (Mainz), \HIGS at TUNL, and MAX-Lab
  (Lund) makes this an opportune time to review nucleon Compton scattering.
  Chiral Effective Field Theory (\ChiEFT) provides an ideal analysis tool,
  since it encodes the well-established low-energy dynamics of QCD while
  maintaining an appropriately flexible form for the Compton amplitudes of the
  nucleon. The same \ChiEFT also describes deuteron and \threeHe Compton
  scattering, using consistent nuclear currents, rescattering and wave
  functions, and respects the low-energy theorems for photon-nucleus
  scattering. It can thus also be used to extract useful information on the
  neutron amplitude from Compton scattering on light nuclei.  We summarise
  past work in \ChiEFT on all of these reactions and compare with other
  theoretical approaches.
  We also discuss all proton experiments up to about $400$ MeV, as well as the
  three modern elastic deuteron data sets, paying particular attention to
  the precision and accuracy of each set.
  Constraining the $\Delta(1232)$ parameters from the resonance region, we
  then perform new fits to the proton data up to $\wlab=170$ MeV, and a new
  fit to the deuteron data.
  After checking in each case that a two-parameter fit is compatible with the
  respective Baldin sum rules, we obtain, using the sum-rule constraints in a
  one-parameter fit,
  $\alphaep=10.7\pm0.3(\text{stat})\pm0.2(\text{Baldin})\pm0.8(\text{theory})$,
  $\betamp=3.1\mp0.3(\text{stat})\pm0.2(\text{Baldin})\pm0.8(\text{theory})$,
  for the proton polarisabilities, and $ \alphaes =10.9\pm
  0.9(\text{stat})\pm0.2(\text{Baldin})\pm0.8(\text{theory})$, $\betams
  =3.6\mp 0.9(\text{stat})\pm0.2(\text{Baldin})\pm0.8(\text{theory})$, for the
  isoscalar polarisabilities, each in units of $10^{-4} \;\fm^3$.
  Finally, we discuss plans for polarised Compton scattering on the proton,
  deuteron, \threeHe and heavier targets, their promise as tools to access
  spin polarisabilities, and other future avenues for theoretical and
  experimental investigation.
\end{abstract}


\newpage
\tableofcontents
\newpage

\section{Introduction}
\label{sec:introduction}
 
Compton scattering has played a major role in the development of modern
Physics. In 1871, 
Lord Rayleigh employed the recently discovered nature of light as an
electromagnetic wave to demonstrate that the cross section for light
scattering with frequency $\w$ from neutral atoms behaves as $\sigma \propto
\w^4$, thereby explaining why the sky is blue~\cite{Rayleigh1,Rayleigh2}.
Fifty-six years later, Arthur Holly Compton won the Nobel Prize ``for his
discovery of the effect named after him'': X-rays have a longer wavelength
after they are scattered from electrons, with the difference precisely
predicted by a quantum treatment of the electromagnetic radiation and by the
relativistic kinematics of the electron~\cite{Compton}. Compton's experiment
thus provided a unified demonstration of two of the great advances in Physics
in the early 20th century: relativity and the particle-like nature of light.

In 1935, Bethe and Peierls performed the first calculation of Compton
scattering from a nucleus: the deuteron~\cite{BethePeierls:1935}. The field
gained momentum in the second half of the 20th century as it was realised that
the photon provides a clean probe for Nuclear Physics. Its interactions with
the target can be treated perturbatively, with the fine structure constant
$\alphaEM$ as a small parameter.\footnote{We use the Heaviside-Lorentz system
  of electromagnetic units and $\hbar=c=1$, so $e=-\sqrt{4\pi\alphaEM}$.  The
  factor $4\pi$ in Eq.~\eqref{eq:induceddipole} is absent in the Gaussian
  system, but present for SI units; cf.~\cite{Schumacher:2005an}.}
The leading term in the scattering of radiation from a nucleus of mass
$M_\mathrm{X}$ and atomic number $Z$ in the long-wavelength limit was first
calculated by Rayleigh's student, J.~J.~Thomson:
\begin{equation}
  \label{eq:thomsoncrosssection}
  \frac{d \sigma}{d \Omega}=\frac{Z^4 \alphaEM^2}{M_\mathrm{X}^2} 
  \;\frac{1+\cos^2\theta}{2}\;\;,
\end{equation}
where $\theta$ is the scattering angle.
Thirring, and independently Sachs and Austern, showed
that~Eq.~\eqref{eq:thomsoncrosssection} is not renormalised
non-relativisti\-cally~\cite{Thirring:1950, Sachs:1951zz}.  It is also
recovered as part of a low-energy theorem for a spin-$\half$ target due to
Low, Gell-Mann, and Goldberger~\cite{Low:1954kd,GellMann:1954kc} which invokes
only analyticity and gauge, Lorentz, parity and time-reversal invariance, and
was generalised to arbitrary spin by Friar~\cite{Friar:1975}. The theorem
states that the nuclear magnetic moment is the only additional parameter
needed to determine the leading spin dependence of the cross section.

\subsection{\it The importance of dipole electric and magnetic
  polarisabilities}
\label{sec:intropols}

For a composite system, the first spin-independent piece of the Compton
amplitude beyond the Thomson limit is parameterised by two structure
constants: the electric and magnetic (scalar dipole) polarisabilities.
Polarisabilities arise because the electric and magnetic fields of a real
monochromatic photon with frequency $\w$ displace the charged constituents of
the system and thus induce charge and current multipoles, even if the target
is overall charge neutral. The dominant contributions are typically an induced
electric dipole moment $\vec{d}_\text{ind}$, often generated by separating
positive and negative charges along the dipole component of the electric field
$\vec{E}$, and a magnetic dipole moment $\vec{m}_\text{ind}$, often generated
from currents induced by the dipole component of the magnetic field $\vec{B}$.
In addition, aligning microscopic permanent electric and magnetic dipoles in
the external fields can generate mesoscopic induced electric and magnetic
dipoles.  The linear response in frequency space demonstrates that the induced
dipoles are proportional to the incident electric and magnetic fields, i.e.,
\begin{equation}
  \label{eq:induceddipole}
  \vec{d}_\text{ind}(\w)=4\pi\alphae(\w)\;\vec{E}(\w)\;\;,\;\;
  \vec{m}_\text{ind}(\w)=4\pi\betam(\w)\;\vec{B}(\w)\;\;.
\end{equation}
Neglecting recoil corrections, these induced dipoles then re-radiate at the
same frequency $\w$ and with the angular dependence characteristic of $E1$ and
$M1$ radiation, respectively.  The proportionality constants in
Eq.~\eqref{eq:induceddipole} are the \emph{electric dipole polarisability}
$\alphae(\w)$ and the \emph{magnetic dipole polarisability} $\betam(\w)$. They
characterise the strength of the dipole radiation relative to the intensity of
the incoming fields and vanish for objects with no internal structure.  Since
they lead to different angular dependences, they can be disentangled in the
differential Compton scattering cross section at fixed frequency $\w$.

Polarisabilities encode the temporal response of the target to a real photon
of energy $\w$ and thus provide detailed information on the masses, charges,
interactions, etc.~of its active internal degrees of freedom. As an example,
in the long-wavelength approximation, the Lorentz-Drude model assumes that the
electric field of the photon displaces point-like, charged constituents $q_n$
of mass $m_n$, bound in classical harmonic oscillators with resonance
energies 
$E_{n}$ and damping factors $\Gamma_n$:
\begin{equation}
  \label{eq:lorentzmodel}
  \alphae(\w)=\sum\limits_{n}\frac{q_n^2}{4\pi
    m_n}\;\frac{1}{
    E_n^2-\w^2-\ii\Gamma_n\w}\;\;.  
\end{equation}
One can estimate $\alphae(\w)$ in the nucleon semiclassically by representing
the constituents of its charged pion cloud as objects which are harmonically
bound to the nucleon core with an eigenfrequency such that the pion cloud has
the same root-mean-square radius as the nucleon, namely about $0.7\;\fm$.
Eq.~\eqref{eq:lorentzmodel} then leads to a value of
$\alphae(0)\approx10\times10^{-4}\;\fm^3$. This model is unrealistic, of
course, but the estimate is surprisingly close to the experimental value.
Throughout this review, we therefore quote values for $\alphae(\w)$ and
$\betam(\w)$ in the ``canonical'' units of $10^{-4}\;\fm^3$, and for the
dipole spin polarisabilities discussed below, in units of $10^{-4}\;\fm^4$.

We also take the term ``polarisabilities'' to be the ``Compton
polarisabilities'', defined in parallel to the above intuitive description by
a multipole expansion of the Compton amplitudes, as elaborated in
Section~\ref{sec:multipoles}. In nonrelativistic Quantum Mechanics, they can
be related to the spectrum of nucleon excited states $|n \rangle$ with
energies $E_n$, e.g.:
\begin{equation}
  \label{eq:excitations}
  \alphae(\w)=2\sum_n \frac{\langle \mathrm{N}|D_z|n \rangle\langle n|D_z|\mathrm{N}
    \rangle}{E_n-\w}+\dots\;\;,
\end{equation}
where $D_z$ is the electric dipole operator. (In Eq.~\eqref{eq:excitations} we
have not explicitly written subtle but important corrections beyond
non-relativistic second-order perturbation theory, which were emphasised by
L'vov~\cite{Lvov:1993fp}, and Bawin and Coon~\cite{Bawin:1996nz}, and
summarised by Schumacher~\cite{Schumacher:2005an}.) Therefore, $\alphae(\w)$
and $\betam(\w)$ are strongly influenced by the lowest state with the quantum
numbers of an electric or magnetic dipole excitation of the nucleon.  For
$\alphae$, this is indeed the $\pi \mathrm{N}$ state. The $M1$ excitation of
the $\Delta(1232)$ would appear to provide a sizable paramagnetic contribution
to $\betam$ of order 10, but since the experimentally measured value is about
an order of magnitude smaller, a diamagnetic contribution of similar magnitude
but opposite sign exists whose precise nature is not yet determined.

Such excitations also set the energies at which the polarisabilities are
manifest in the single-nucleon cross section: $\w \gtrsim 50\;\MeV$.  In
addition, Eq.~\eqref{eq:excitations} implies that the dynamical
polarisabilities become complex once the first inelastic channel opens at the
$\pi$N threshold.  Finally, polarisabilities are in general related to the
dielectric function $\varepsilon(\w)$ and magnetic permeability function
$\mu(\w)$ of a macroscopic system. Since these, in turn, characterise optical
properties, nucleon polarisabilities are related to the index of refraction
and absorption coefficient of a bulk system of nucleons at a given frequency.

Though polarisabilities are naturally defined as functions of the photon
energy $\w$, historically much of the emphasis in the context of the proton
and neutron has been on the \emph{static polarisabilities}
$\alphae\equiv\alphae(\w=0)$ and $\betam\equiv\betam(\w=0)$, which are often
simply termed ``the polarisabilities''.  For clarity, we therefore refer to
the functions as \emph{energy-dependent} or \emph{dynamical}
polarisabilities~\cite{Griesshammer:2001uw,Hildebrandt:2003fm}\footnote{For
  completeness, we note that the \emph{generalised} polarisabilities of
  virtual Compton scattering are explored by an incoming photon of non-zero
  virtuality and can provide complementary information about the spatial
  charge and current distribution, see e.g.~\cite{Drechsel:2002ar} for a
  review.}.  The static polarisabilities are formally and uniquely defined via
the Compton scattering amplitudes, as detailed in Section~\ref{sec:overview}.
Along with the anomalous magnetic moment, they parameterise the deviation of
the proton Compton cross section from that of a point-like particle in a
low-energy expansion which is valid up to photon energies of roughly 80~MeV.
Static polarisabilities can be conceptualised, again up to subtle
corrections~\cite{Lvov:1993fp,Bawin:1996nz,Schumacher:2005an}, as the
proportionality constants between induced dipoles and external fields which
would be ``measured'' were the nucleon placed into a parallel-plate capacitor
or a pure N-S magnet.  They also enter in other processes with sensitivity to
nucleon structure; in particular, the relation of $\betam$ to doubly-virtual
forward Compton scattering has attracted recent interest in connection with
the two-photon-exchange contribution to the Lamb shift in muonic hydrogen
\cite{Pachucki} (see also Ref.~\cite{Carlson:2011dz} and references therein),
and with the nucleon electromagnetic mass shift, see most
recently~\cite{WalkerLoud:2012bg}.

The Effective Field Theory (EFT) methods which are discussed in this review
predict both static and dynamical polarisabilities. They can be used to
extract the values of static polarisabilities from experimental data taken at
energies too high for the low-energy expansion to be valid. We present a new
EFT extraction of polarisabilities from world data in Sections
\ref{sec:gammap} and \ref{sec:2+3N}.

\subsection{\it Compton scattering from nucleons: data, structure and analysis
  tools}
\label{sec:introdata}

Therefore, in this review, we examine real Compton scattering from the
simplest stable strongly-interacting systems, namely protons and light nuclei,
in order to obtain information on the photon-nucleon scattering amplitude.
Compton scattering from larger nuclei is reviewed in Ref.~\cite{Hutt:1999pz}.
In Section~\ref{sec:multipoles}, we provide the detailed relation of
$\alphae(\w)$, $\betam(\w)$ and the spin polarisabilities to the
single-nucleon Compton amplitude.  The spin polarisabilities have received
much recent attention in both theoretical and experimental studies, since they
are a low-energy manifestation of the spin structure of the nucleon,
parameterising its spin-dependent response to external electric and magnetic
fields.

How are polarisabilities explored? Until the last decades of the 20th century,
all experiments investigating Compton scattering from nucleons and nuclei
employed bremsstrahlung beams. This created difficulties in accurately
measuring the small (nb/sr) cross sections at energies where the
polarisabilities are particularly relevant. The advent of photon tagging in
the 1980s facilitated a clean separation of elastic and inelastic processes,
enabling measurements of the proton cross section with good energy resolution
at $\w \lesssim 200$ MeV. At the turn of the Millennium, this led to a wealth
of data for the proton from Illinois~\cite{Federspiel:1991},
Saskatoon~\cite{Hallin:1993,MacGibbon:1995} and
MAMI~\cite{Zieger:1992,Olmos:2001}, and for the deuteron from
Illinois~\cite{Lucas:1994}, Saskatoon~\cite{Hornidge:2000} and
Lund~\cite{Lundin:2003}. Since most of these experiments were reviewed by
Schumacher~\cite{Schumacher:2005an}, we only summarise the data in
Section~\ref{sec:expoverview}, with particular attention to statistical and
systematic errors.

In parallel with these developments, $\alphae$ and $\betam$ were calculated in
various theoretical models of nucleon
structure~\cite{Dattoli:1977wg,Schafer:1984hw,Weiner:1985ih,DeSanctis:1990ym,Capstick:1992tx,Liebl:1994qc,Nyman:1984ys,Chemtob:1987ut,Scoccola:1989px,Scoccola:1990yh,Saito:1993uf}.
Comparing these predictions with Compton-scattering data indicates how
accurately these models describe electromagnetic excitations of the nucleon.
Quark models which do not incorporate explicit pionic degrees of freedom tend
to underpredict $\alphae$ and overpredict $\betam$; see
e.g.~Ref.~\cite{Liebl:1994qc}.  Computations of $\alphae$ and $\betam$ in
chiral quark models incorporate both long-distance ($\pi \mathrm{N}$) and
short-distance (other excitations)
physics~\cite{Weiner:1985ih,Dong:2005kt,Dong:2006ym}.
Direct determinations of nucleon polarisabilities from lattice simulations of
the QCD path integral now appear imminent~\cite{Detmold:2006vu}, with results
reported in quenched~\cite{Christensen:2004ca,Lee:2005dq,Lujan:2011ue},
partially
quenched~\cite{Detmold:2009dx,Detmold:2010ts,Alexandru:2010dx,Lee:2011gz} and
even full QCD~\cite{Engelhardt:2010tm,Engelhardt:2011qq}.

New plans for Compton-scattering experiments on protons and light nuclei at
MAMI, the HI$\gamma$S facility at TUNL and MAX-Lab in Lund make this an
opportune time to re-examine our knowledge of nucleon Compton scattering at
energies up to a few hundred MeV. We therefore delineate in this review what
is known about the Compton amplitudes of the proton and neutron.
Equation~\eqref{eq:excitations} implies that the polarisabilities are
dominated by the lowest nucleonic states, namely by $\pi$N and $\Delta(1232)$
dynamics, while sensitivity to higher excitations is suppressed. This means
that Compton scattering at low energies, $\w \lesssim 300$ MeV, is dominated
by long-distance properties of the nucleon. In particular, we note that the
particles detected in the experiments are photons, nucleons and pions, not
quarks themselves.  Analysing Compton scattering at these energies in terms of
quark degrees of freedom, such as in lattice QCD or models of nucleon
structure, is thus not really profitable.  Instead, such calculations can be
tested against the constraints extracted from data using a theoretical
approach that includes the pertinent low-energy dynamics, provided that it is
sufficiently general to encompass the data without undue prejudice.  Effective
Field Theory (EFT) fits these requirements.

\subsection{\it The role of Effective Field Theory}
\label{sec:introEFT}

The basic principle of an EFT is that many phenomena can be
economically---i.e.~effectively---described in terms of entities which are not
elementary. The fact that the details of nucleon structure are not probed at
low energies suggests that a low-energy EFT which includes nucleons, pions and
photons should be a useful tool to extract information on nucleon
polarisabilities.

In general, for an EFT approach to be successful, a \emph{separation of
  scales} must exist between the energies involved and those required to
excite the particular degrees of freedom of the system which are not treated
dynamically.  A famous example of an EFT is the Fermi theory of weak
interactions, in which $\beta$ decay is described by a contact interaction
between the neutron and its decay products $\text{e}^-$, $\overline
\nu_{\text{e}}$ and $\text{p}$.  At energy scales of the order of a few MeV,
the threshold for production of W and Z bosons is far off, and they can be
``integrated out'' to leave a simple energy-independent four-fermion
interaction together with a series of further interactions, each of which is
suppressed by powers of the small quantity $p_{\text{typ}}/M_\text{W}$, where
$p_{\text{typ}}$ is the typical momentum of a decay product.
Similarly, at low enough energies---energies below those where pion degrees of
freedom become relevant---a theory of heavy, point-like nucleons should
suffice to describe the interactions of nucleons with one another and with
photons. This theory has come to be known as the ``Pionless Effective Field
Theory'' (\EFTNoPion) of Nuclear Physics, and for NN scattering, it is
equivalent to Bethe's low-energy Effective Range Expansion \cite{Blatt:1949zz,
  Bethe:1949yr, Schwinger, ChewGoldberger, BarkerPeierls, Kaplan:1998we,
  Gegelia:1998gn, Birse:1998dk, vanKolck:1998bw}.  As in all EFTs,
``point-like'' and ``pionless'' does not imply the absence of effects such as
a non-zero anomalous magnetic moment 
which are due to nucleon structure, pions and heavier particles. Instead, they
are taken into account through Lagrangian parameters---so-called low-energy
constants (LECs)---but are not explained within the EFT itself.

The guiding principle in constructing an EFT Lagrangian is that \emph{all}
terms compatible with the symmetries of the underlying theory must be
included, each proportional to an a priori unknown LEC.  This infinite string
of terms and couplings is organised according to the power of
$p_\mathrm{typ}/\Lambda$ that each operator contributes to amplitudes, with
$p_\mathrm{typ}$ the typical momentum of the process, and the breakdown scale
$\Lambda$ set by the mass of the lightest omitted degree of freedom ($\mpi$,
for example, in the case of \EFTNoPion).  Since all terms are included, the
counterterms needed to renormalise the divergent loops at a given order in
$p_\mathrm{typ}/\Lambda$ are automatically present.  Unless the theory has a
low-energy bound state, the renormalised loop contributions are suppressed,
typically by $p_\mathrm{typ}^2/\Lambda^2$ for each loop.  Therefore, only a
finite number of terms in the Lagrangian need to be considered when working to
a particular order in the small, dimensionless parameter
$p_\mathrm{typ}/\Lambda$. While the theory is thus not renormalisable in the
conventional sense, only a finite and usually small number of terms are needed
for renormalisability to a given order in $p_\mathrm{typ}/\Lambda$.  In
addition, a rigorous assessment of residual theoretical uncertainties can be
made for any process by estimating the accuracy of its momentum expansion.

Though the number of possible terms (and hence the number of LECs) in the
Lagrangian grows rapidly with the order, any given process usually involves
only a few of them. Once determined by one piece of data, an LEC enters in the
prediction of other observables. Some LECs are related to familiar properties
of the particles involved, like charge, mass, anomalous magnetic moment, decay
constant, etc., but others are less easy to fix and interpret. Some LECs which
govern pion interactions are now being computed by direct lattice simulations
of QCD~\cite{arXiv:1108.1380,Colangelo:2010et,arXiv:1111.3729}.  But even when
LECs can be derived from the underlying theory, calculations of more complex
processes are often more tractable in the EFT framework, as we shall see here
for Compton scattering from protons and light nuclei.

Compton scattering in \EFTNoPion is discussed in Section~\ref{sec:pionless}.
For the proton, it is of limited use, as it amounts to an expansion of the
amplitude in powers of $\w/\mpi\ll1$. Since only a small fraction of the data
is in the regime where this expansion is valid, the resulting errors on the
polarisabilities are inevitably large.  In the two-nucleon sector, though, the
situation is different.  The low-energy NN scattering amplitude is given as an
expansion in powers of momenta, with the LECs determined by the scattering
length, effective range, etc.  Nuclear binding effects and photon-nuclear
interactions are then fixed and the leading deuteron Compton amplitudes are
predicted with no free parameters, as demonstrated in 1935 in a calculation by
Bethe and Peierls~\cite{BethePeierls:1935}.

The \EFTNoPion expansion breaks down for typical momenta of order $\mpi$, and
above this point, the pion itself must be included. While an EFT including the
pion must exist because the next excitation, the $\Delta(1232)$, is far enough
away to provide a separation of scales, there is no guarantee that it is
viable.  The size of the pion-nucleon coupling constant, $g_{\pi \mathrm{NN}}
\approx 13$, suggests that multi-loop diagrams including dynamical pions may
not be suppressed. However, two crucial aspects make the theory manageable.
First, the fact that the pion is much lighter than other hadrons is now
understood as a consequence of the spontaneously broken (hidden) chiral
symmetry of QCD. The up and down quarks are nearly massless on the scale of
typical QCD energies.  If they were actually massless, the Lagrangian would be
invariant under independent isospin rotations of the left- and right-handed
quarks, $\text{SU}(2)_{\text{L}}\times \text{SU}(2)_{\text{R}}$. However, only
the vector (isospin) subgroup, $\text{SU}(2)_{\text{V}}$, is manifest in the
hadron spectrum and the full symmetry is hidden in the physical QCD vacuum.
The pions are then identified as the three Nambu-Goldstone bosons
corresponding to the three axial rotations which are symmetry operations on
the QCD Lagrangian, but not on the QCD vacuum. The small non-zero quark masses
lead to pion masses which are again much smaller than typical QCD scales,
$\mpi \ll \Lambda_\chi$.  The fact that the pion is a (pseudo-)Nambu-Goldstone
boson leads to the second key point: in the ``chiral'' or zero-quark-mass
limit, soft pions decouple so that their interactions with one another and
with other hadrons vanish linearly with their momentum.

This provides a perturbative expansion of amplitudes in powers of $P/
\lambdachi \equiv (p_{\text{typ}},m_\pi)/\lambdachi\ll1$, with the light EFT
scales being the pion momenta and mass, and $\lambdachi$ being the scale
associated with hadrons which are not explicitly included in the EFT,
$\lambdachi\approx m_\rho$.  This EFT is known as ``chiral EFT'' (\ChiEFT).
The version without explicit $\Delta(1232)$ degrees of freedom is often
referred to as ``Baryon Chiral Perturbation Theory'' (B$\chi$PT), and the
purely pionic one as Chiral Perturbation Theory ($\chi$PT). Details pertaining
to Compton scattering are discussed in Section~\ref{sec:singleN}, including
the special role of the $\Delta$.  There we also show that photons are
included in \ChiEFT\ partly by invoking minimal substitution and partly by
including the field strength tensor $F_{\mu\nu}$ as a building block in the
Lagrangian.  Since the latter generates photon couplings which are not
constrained by gauge invariance, new LECs enter, including ones which can (at
successively higher orders) be related to the anomalous magnetic moment and
the charge radius of the nucleon, and to non-chiral contributions to
polarisabilities.

The resulting EFT framework provides predictions for the low-energy
interaction of pions, photons and nucleons through calculations consistent
with the known pattern of QCD symmetries and their breaking.  Furthermore,
because it is a quantum field theory, \ChiEFT incorporates the requisite
consequences of unitarity and Lorentz invariance at low energies. It is this
framework which we use to analyse proton Compton scattering in
Sections~\ref{sec:singleN} and~\ref{sec:protonanalysis}.  In
Sections~\ref{sec:drs}, \ref{sec:comp} and \ref{sec:other}, we also compare
$\chi$EFT to other approaches for analysing low-energy Compton scattering,
most notably dispersion relations (DRs). (Full details of DRs are given in the
review of Drechsel et al.~\cite{Drechsel:2002ar}.)  The main differences
between $\chi$EFT and DRs lie in the careful incorporation of chiral
constraints in the former, its stringent agnosticism regarding high-energy
details, and the fact that the presence of a small parameter allows an a
priori estimate of residual theoretical uncertainties.

\subsection{\it The elusive neutron: Compton scattering from light nuclei}
\label{sec:introneutron}

Experiments on the proton reveal only half of the information in Compton
scattering on the nucleon. The 14.7 minute lifetime of the neutron, coupled
with the relative weakness of its electromagnetic interactions, means that the
neutron Compton amplitude must be inferred indirectly.  While some results
exist for scattering neutrons directly in the Coulomb field of heavy nuclei,
more accurate data are available for Compton scattering in few-nucleon
systems, where nuclear effects can be precisely calculated and taken into
account in the analysis. However, it is clearly advantageous to use the same
theoretical framework for both the nuclear and photon-nucleon dynamics. The
\ChiEFT low-momentum expansion again provides a controlled, model-independent
framework for subtracting the nuclear binding effects and analysing the
available elastic deuteron scattering data. Both the number and quality of
these data, reviewed in Section~\ref{sec:expdeutlow}, are appreciably inferior
to the proton case, since the experiments are markedly harder. The substantial
progress made to determine the neutron polarisabilities via this route is
reviewed in Section~\ref{sec:2+3N}. The key to extracting neutron
polarisabilities from few-nucleon targets is that the coherent nature of
deuteron Compton scattering allows us to observe the photon-neutron amplitude
through its interference with the proton and meson-exchange amplitudes, while
the Thomson limit imposes a stringent constraint on the few-nucleon amplitude
that can be used to check these non-trivial calculations.

\ChiEFT work on the deuteron originates from Weinberg's seminal
papers~\cite{Weinberg:1992yk}, where the nucleon-nucleon potential is computed
up to a fixed order in the momentum expansion and then iterated using the
Schr\"odinger equation to obtain the scattering amplitude.  The resulting wave
functions are combined with operators for Compton scattering derived in
\ChiEFT. The photon-nucleon operators are given by the \ChiEFT photon-nucleon
amplitudes described above.  However, \ChiEFT also expands the nuclear current
operators in powers of $P$.  This has the crucial advantage that the chiral
dynamics which drives low-energy Compton scattering is treated consistently in
both the single- and few-nucleon operators, so that \ChiEFT results for
deuteron Compton scattering can be assessed for order-by-order convergence,
too. Meanwhile, Compton scattering from the deuteron in quasi-free kinematics
for the neutron was measured in
Refs.~\cite{Rose:1990a,Rose:1990b,Kolb:2000,Kossert:2002,Kossert:2002ws}, and
values for $\alphaen$ were extracted. These experiments are discussed in
Section~\ref{sec:expdeutqf}, but \ChiEFT has not yet been extended to $\gamma
\text{d} \rightarrow \gamma \text{np}$. Model calculations of elastic and
inelastic scattering on the deuteron are briefly reviewed in
Section~\ref{sec:gammadmodels}.
\ChiEFT is also used to produce the first calculations, in any framework, of
elastic scattering on
\threeHe~\cite{Choudhury:2007bh,Shukla:2008zc,ShuklaThesis}; see
Section~\ref{sec:gammaHe3chiral}. Since \threeHe is doubly charged, its cross
section is larger by a factor of up to $4$ compared to the proton or deuteron.
In addition, polarised \threeHe is interesting since it serves as an effective
polarised neutron target.

\subsection{\it Results and the future}
\label{sec:introfuture}

In Sections~\ref{sec:protonanalysis} and \ref{sec:deuteronanalysis}, we apply
the \ChiEFT methodology to the proton and deuteron databases, respectively,
and present new central values and uncertainties for the proton and neutron
dipole polarisabilities, as well as a detailed comparison of the \ChiEFT
predictions with data. Here we preview our best values of the proton and
isoscalar (i.e.~average nucleon) scalar dipole polarisabilities in a
two-parameter fit:
\begin{equation}
  \begin{split}
    \alphaep=10.5\pm0.5(\text{stat})\pm0.8(\text{theory})&\;\;,\;\;
    \betamp =2.7\pm0.5(\text{stat})\pm0.8(\text{theory})\\
    \alphaes=10.5\pm 2.0(\text{stat})\pm0.8(\text{theory}) &\;\;,\;\; \betams=3.6\pm 1.0(\text{stat})\pm0.8(\text{theory})\;\;.
  \end{split}
\end{equation}
In a one-parameter fit employing the Baldin sum rules for the proton and
isoscalar nucleon:
\begin{equation}
  \begin{split}
    \alphaep&=10.7\pm0.3(\text{stat})\pm0.2(\text{Baldin})\pm0.8(\text{theory})\\
    \betamp
    &=\phantom{0}3.1\mp0.3(\text{stat})\pm0.2(\text{Baldin})\pm0.8(\text{theory})
    \\
    \alphaes
    &=10.9\pm 0.9(\text{stat})\pm0.2(\text{Baldin})\pm0.8(\text{theory})
    \\
    \betams
    &=\phantom{0}3.6\mp
    0.9(\text{stat})\pm0.2(\text{Baldin})\pm0.8(\text{theory}) \; \; .
  \end{split}
\end{equation}
In Section~\ref{sec:future}, we close this review with a discussion of future
experiments which hold the promise of improving the Compton database and the
determination of polarisabilities.  We also describe upcoming experimental
efforts to extract the still relatively unexplored spin polarisabilities.
Finally, we outline the anticipated \ChiEFT developments that will help refine
the analysis of these forthcoming data.



\section{Foundations}
\label{sec:foundations}

\subsection{\it Overview}
\label{sec:overview}

We begin by parameterising the $T$-matrix for Compton scattering of a photon
of incoming energy $\w$ from a nucleon with spin $\vec{\sigma}/2$ by six
independent invariant amplitudes~\cite{Prange:1958}, which read in the
operator basis first appearing in~Ref.~\cite{Jacob:1960} and usually employed
in \ChiEFT:
\begin{equation}
  \label{eq:Tmatrix}
  \begin{array}{rcl}
    T(\w,z)&=& A_1(\w,z)\;(\vec{\epsilon}\,'^*\cdot \vec{\epsilon}) +
    A_2(\w,z)\;(\vec{\epsilon}\,'^*\cdot\hat{\vec{k}})\;(\vec{\epsilon}
    \cdot\hat{\vec{k}}')
    \\&&
    +\ii\,A_3(\w,z)\;\vec{\sigma}\cdot\left(\vec{\epsilon}\,'^*\times\vec{\epsilon}\,\right)
    +\ii\,A_4(\w,z)\;\vec{\sigma}\cdot\left(\hat{\vec{k}}'\times\hat{\vec{k}}\right)
    (\vec{\epsilon}\,'^*\cdot\vec{\epsilon}) \\&&
    +\ii\,A_5(\w,z)\;\vec{\sigma}\cdot
    \left[\left(\vec{\epsilon}\,'^*\times\hat{\vec{k}}
      \right)\,(\vec{\epsilon}\cdot\hat{\vec{k}}')
      -\left(\vec{\epsilon}\times\hat{\vec{k}}'\right)\,
      (\vec{\epsilon}\,'^*\cdot\hat{\vec{k}})\right]
    \\&& +\ii\,A_6(\w,z)\;\vec{\sigma}\cdot
    \left[\left(\vec{\epsilon}\,'^*\times\hat{\vec{k}}'\right)\,
      (\vec{\epsilon}\cdot\hat{\vec{k}}') -\left(\vec{\epsilon}
        \times\hat{\vec{k}} \right)\,
      (\vec{\epsilon}\,'^*\cdot\hat{\vec{k}})\right] \; \;,
  \end{array}
\end{equation}
where $\hat{\vec{k}}$ ($\hat{\vec{k}}'$) is the unit vector in the momentum
direction of the incoming (outgoing) photon with polarisation $\vec{\epsilon}$
($\vec{\epsilon}\,'^*$), $\theta$ is the scattering angle, and $z=\cos\theta$.
This form holds in the Breit and centre-of-mass (cm) frames.  The first two
amplitudes are spin-independent, while the other four parameterise
interactions with the nucleon spin.  The amplitude is related to the
differential cross section by
\begin{equation}
  \label{eq:diffxsecfermion}
  \left.\frac{\dd\sigma}{\dd\Omega}\right|_\text{frame}= \Phi^2_\text{frame}\;
  \left| T\right|^2\;\;,
\end{equation}
where $\Phi_\text{frame}$ is a frame-dependent flux factor which tends to
$1/(4\pi)$ at low energies; see also Section~\ref{sec:kinematics}.

For forward scattering, $\vec{k}=\vec{k}'$, and only the structures of $A_1$
and $A_3$ survive.  The behaviour of these amplitudes as $\omega\to0$ is
determined by low-energy theorems (LETs) which rely only on analyticity and
gauge, Lorentz, parity and time-reversal invariance~\cite{Low:1954kd,
  GellMann:1954kc}
\begin{equation}
  \label{eq:LET}
  \lim\limits_{\w\to0}T(\w,1)=-\frac{e^2Q^2}{\MN}\;
  \vec{\epsilon}\,'^*\cdot \vec{\epsilon}-\ii \frac{e^2\kappa^2\w}{2\MN} \vec{\sigma}\cdot\left(\vec{\epsilon}\,'^*\times\vec{\epsilon}\,\right)+\calO(\w^2)\;\;,
\end{equation}
where $-eQ$ is the nucleon charge, $\kappa$ is the anomalous magnetic moment
(in nuclear magnetons) and $\MN$ is the mass of the target.  At $\omega=0$, we
recover the spin- and energy-independent Thomson term, and hence the
corresponding cross section of Eq.~\eqref{eq:thomsoncrosssection}.

\begin{figure}[!htb]
  \begin{center}
    \includegraphics*[width=0.7\linewidth]{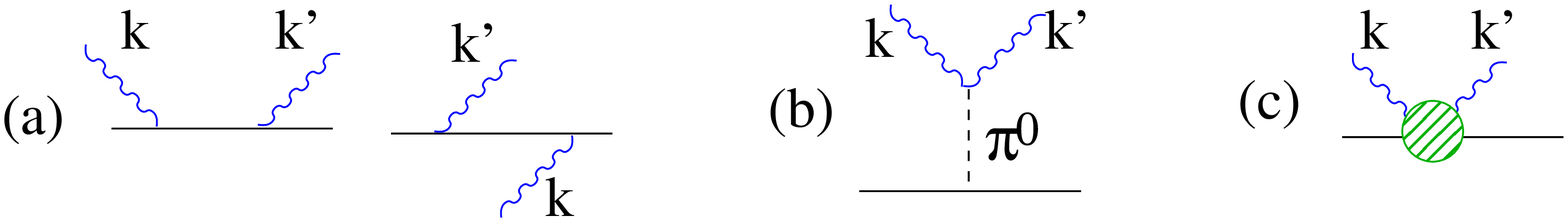}\qquad
    \caption {(a) Nucleon Born graphs for Dirac nucleons; (b) pion Born graph;
      (c) structure contribution.}
    \label{fig:structure}
  \end{center}
\end{figure}

Eq.~\eqref{eq:LET} may also be obtained in the limit $\w\to 0$ from the pole
diagrams in a calculation of Compton scattering using photons coupled to a
Dirac nucleon via its charge and anomalous magnetic moment (with the spinors
normalised to $\bar u u=1$), as shown in diagram (a) of
Fig.~\ref{fig:structure}.

As explained in the Introduction, our primary interest 
is to obtain information on the structure of the two-photon response of the
nucleon, which requires going beyond the single-photon processes of the Born
terms.  It is therefore useful to separate the amplitudes into
``non-structure'' and ``structure'' parts
\begin{equation}
  A_i(\w,z)=A_i^\text{Born}(\w,z)+\bar{A}_i(\w,z)\;\;. 
\end{equation}
Identifying the former with the pole (Born) contribution mentioned above,
while not a unique choice, ensures that the LETs are satisfied by these terms
alone.  It also agrees with usage in dispersion relation (DR) calculations
(see Section \ref{sec:drs}).  There is another known contribution to
spin-dependent scattering which comes from the $t$-channel exchange of a
neutral pion, as shown in Fig.~\ref{fig:structure}(b), which first contributes
at $\calO(\w^3)$; we choose to include this term in the Born or
``non-structure'' part as well, but the choice is not universal.  The other
term, $\bar{A}_i$, represented by Fig.~\ref{fig:structure}(c), parameterises
deviations from the ``known'' part of the amplitude and describes the nucleon
as a polarisable object.

Expanding both nucleon Born and structure amplitudes in the Breit frame to
order $\omega^3$, we have {\setlength\arraycolsep{0.1em}\begin{eqnarray}
    A_1(\omega,z)&=&-\frac{Q^2e^2}{\MN}+\frac{e^2\,\omega^2}{ 4\MN^3}
    \Bigl((Q+\kappa)^2(1+z)-Q^2\Bigr)(1-z)+
    4\pi\omega^2(\alphae+z\,\betam)\,+\calO(\omega^4)\nonumber\\
    A_2(\omega,z)&=&\frac{e^2\,\omega^2}{ 4\MN^3}\kappa(2Q+\kappa)z
    -4\pi\omega^2\betam+\calO(\omega^4)\nonumber\\
    A_3(\omega,z)&=& \frac{e^2 \omega}{ 2\MN^2}\Bigl(Q(Q+2\kappa)-(Q+\kappa)^2
    z\Bigr) +A_3^{\pi^0}
    +4\pi\omega^3(\gamma_1-(\gamma_2+2\gamma_4)z)+\calO(\omega^5)\nonumber\\
    A_4(\omega,z)&=& -\frac{e^2\omega }{ 2\MN^2 }(Q+\kappa)^2
    +4\pi\omega^3 \gamma_2 +\calO(\omega^5)\nonumber\\
    A_5(\omega,z)&=& \frac{e^2 \omega}{ 2\MN^2 }(Q+\kappa)^2
    +A_5^{\pi^0}+4\pi\omega^3\gamma_4 +\calO(\omega^5)\nonumber\\
    A_6(\omega,z)&=& -\frac{e^2 \omega}{ 2\MN^2 }Q(Q+\kappa)
    +A_6^{\pi^0}+4\pi\omega^3\gamma_3 +\calO(\omega^5) \;\;,
    \label{eq:low-en-amps}
  \end{eqnarray}}%
where the $\pi^0$ pole amplitudes are, with $t=2\w^2(z-1)$ and $\tau_3$ the
third Pauli matrix in isospin space:
\begin{equation}
  A_3^{\pi^0}=
  \tau_3\;\frac{e^2\gpiNN\omega^3(z-1)}{4\pi^2\fpi\MN(m_{\pi^0}^2-t)}\;\;,
  \;\; A_6^{\pi^0}=
  -A_5^{\pi^0}=\tau_3\;\frac{e^2\gpiNN\omega^3}{8\pi^2\fpi\MN(m_{\pi^0}^2-t)}
  \;\;. 
  \label{eq:pi-pole}
\end{equation}
To this order, the only difference in the cm frame is that $A_2$ has an
additional Born term $e^2 Q^2 \w/ \MN$. The omitted terms start one order
lower in the cm frame (e.g.\ $\calO(\omega^3)$ in $A_1$), because the
amplitude does not have manifest crossing symmetry, whereas in the Breit frame
it does.  The ``structure'' parts of the amplitude are represented at this
order by $\alphae$ and $\betam$, namely the static spin-independent
polarisabilities discussed above, and by the analogous spin polarisabilities
$\gamma_i$ \cite{Ragusa:1993rm}; see Section~\ref{sec:multipoles}. The
canonical units are $10^{-4}\;\fm^3$ for the former and $10^{-4}\;\fm^4$ for
the latter.  We denote proton and neutron polarisabilities by a superscript
(e.g.\ $\alphaep$) and define isoscalar and isovector polarisabilities as
averages and differences, such that $\alphaes=\half(\alphaep+\alphaen)$ and
$\alphaev=\alphaep-\alphaen$.

For forward scattering, only $\alphae+\betam$ and
$\gamma_0=\gamma_1-\gamma_2-2\gamma_4$ augment the LETs of Eq.~\eqref{eq:LET},
while the $\pi^0$ pole terms do not contribute. Since forward scattering
amplitudes can be related to inelastic cross sections via the optical theorem,
two sum rules can be constructed for these quantities:
\begin{equation}
  \alphae + \betam=\frac{1}{2 \pi^2} \int\limits_{\wpi}^{\infty} \dd \omega'\;
  \frac{\sigma_T(\omega')}{\omega'^2}\;\;,\;\; \gamma_0=\frac{1}{4
    \pi^2} \int\limits_{\wpi}^\infty \dd \omega'\;
  \frac{\sigma_{1/2}(\omega')-\sigma_{3/2}(\omega')}{\omega'^3} \;\;,
  \label{eq:bothSR}
\end{equation}
where $\sigma_T$ is the total cross section for an unpolarised target and
$\sigma_{1/2}$ ($\sigma_{3/2}$) is the cross section for a photon-nucleon
system of total helicity 1/2 (3/2).  The integrals are over lab energies,
starting at the pion photoproduction threshold.  The first of these is known
as the Baldin Sum Rule~\cite{Baldin:1960,Lapidus:1963}.  We adopt the
evaluation of \OdL et al.~\cite{Olmos:2001} for the proton and that of Levchuk
and L'vov~\cite{Levchuk:1999zy} for the neutron:
\begin{equation}
  \alphaep + \betamp=13.8 \pm 0.4 \;\;,\;\; \alphaen + \betamn=15.2 \pm 0.4 \;\;,
  \label{eq:ourBaldin}
\end{equation}
which combine to give the isoscalar sum-rule value, with errors added in
quadrature:
\begin{equation}
  \label{eq:baldinisoscalar}
  \alphaes + \betams=14.5 \pm 0.3\;\;.
\end{equation}
Using information on pion photoproduction multipoles, together with a
parameterisation of data at intermediate energies and a Regge form at higher
energies, Babusci et al.~\cite{Babusci:1997ij} obtained $\alphaep +
\betamp=13.69 \pm 0.14$. However, using different parameterisations for
$\sigma_T$, Levchuk and L'vov~\cite{Levchuk:1999zy} obtained a central value
of 14.0. The evaluation quoted above is more recent and more
conservative~\cite{Olmos:2001}.  Given the uncertainties associated with
extracting neutron multipoles from deuterium data, perhaps it is not
surprising that evaluations for the neutron are more variable. We regard the
direct use of deuterium photodisintegration data above pion threshold as
ill-advised (c.f. Ref.~\cite{Babusci:1997ij}). The value we quote uses neutron
photoproduction multipoles obtained from proton ones via isospin
considerations in a more reliable approach, although significant model
dependence is still present~\cite{Levchuk:1999zy}. 

Only the combinations $\alphae-\betam$ and
$\gamma_\pi=\gamma_1+\gamma_2+2\gamma_4$ enter for backward scattering.
Because we have separated out the pion-pole contribution, our values for
$\gammapip$ differ from those obtained when it is included. The difference is
$-e^2\gpiNN/(8\pi^3\fpi\MN m_{\pi^0}^2)=-46.4$, using the proton value
$\gpiNN^2/4\pi=13.64$~\cite{Rentmeester:1999}.

Various low-energy cross sections have been constructed in the literature. All
of these can be found by using various approximations to
Eq.~\eqref{eq:low-en-amps} in Eqs.~\eqref{eq:Tmatrix} and
\eqref{eq:diffxsecfermion}. The simplest is the Klein-Nishina formula of a
point-like Dirac particle, with $\kappa=0$~\cite{Klein:1929}. If $\kappa\ne0$
is included but the target is still structureless and terms of order
$\omega^3$ are discarded, one obtains the Powell cross
section~\cite{Powell:1949}.  The Petrun'kin cross section additionally allows
the inclusion of terms of order $\omega^2$ arising from the interference of
the leading structure contributions $\alphae$ and $\betam$ with the Thomson
term in $|A_1|^2$ and hence is complete at
$\calO(\w^2)$~\cite{Petrun'kin:1961,Petrun'kin:1981}.  None of these includes
spin polarisabilities, the $\pi^0$ pole, or the energy dependence of the
polarisabilities. We shall say more about their impact in the next section.

\subsection{\it Polarisabilities from a multipole expansion}
\label{sec:multipoles}

Many multipoles are induced inside an object that interacts with an
electromagnetic field of frequency $\w$.  Each oscillates with that same
frequency and thus emits radiation with a characteristic angular distribution.
The proportionality constant between each photon field and the corresponding
induced multipole moment is called a polarisability; each is an
energy-dependent function which parameterises the stiffness of the internal
degrees of freedom with particular quantum numbers against deformations of a
given electric or magnetic multipolarity and energy. In this section, we
generalise the picture presented in the Introduction to consider
polarisabilities beyond the scalar dipole ones.

Hildebrandt et al.~\cite{Griesshammer:2001uw,Hildebrandt:2003fm} used the
formalism of an energy-dependent multipole analysis established by
Ritus~\cite{Ritus:1957,Contogouris:1962,Nagashima:1965} and summarised in
Ref.~\cite{Pfeil:1974ib} to define energy-dependent polarisabilities.
Here we proceed differently, constructing the first few multipoles via the
most general field-theoretical Lagrangian which describes the interactions
between a nucleon field $N$ with spin $\vec{\sigma}/2$ and two photons of
fixed, non-zero energy $\w$ and definite multipolarities. This includes the
structure effects, i.e.~the local coupling of the two photons to the nucleon.
Taking into account gauge and Lorentz invariance, as well as invariance under
parity and time-reversal, the interactions with the lowest photon
multipolarities are
\begin{eqnarray}
  \label{eq:polsfromints}
  \calL_\text{pol}&=&2\pi\;N^\dagger \big[\alphae(\w)\;\vec{E}^2\;+
  \;\betam(\w)\;\vec{B}^2\;+\;\gammaee(\w)
  \;\vec{\sigma}\cdot(\vec{E}\times\dot{\vec{E}})\; 
  \\&&
  +\;\gammamm(\w)\;\vec{\sigma}\cdot(\vec{B}\times\dot{\vec{B}}) 
  -\;2\gammame(\w)\;\sigma_i\;B_j\;E_{ij}\;+
  \;2\gammaem(\w)\;\sigma_i\;E_j\;B_{ij} \;+\;\dots
  \big]\;N\;\;,\non
\end{eqnarray}
with $T_{ij}\equiv\half (\de_iT_j + \de_jT_i)$, $\vec{T}=\vec{E},\vec{B}$.
These terms are straightforward extensions to the effective Lagrangian of
zero-energy scattering in Refs.~\cite{Babusci:1998ww,Holstein:1999uu}. The
photons couple electrically or magnetically ($X,Y=E,M$) and undergo
transitions $Xl\to Yl^\prime$ of definite multipolarities $l$ and
$l^\prime=l\pm\{0,1\}$.  The interactions are unique up to field redefinitions
using the equations of motion. Dipole couplings are proportional to the
electric and magnetic field directly, or to their time derivatives. Quadrupole
interactions couple to the irreducible second-rank tensors $E_{ij}$ and
$B_{ij}$.
Eq.~\eqref{eq:polsfromints} lists all contributions with coupling to at least
one dipole field.
Polarisabilities of higher multipolarity, e.g.~the electric and magnetic
quadrupole polarisabilities~\cite{Babusci:1998ww,Holstein:1999uu}, are denoted
by ellipses.  Thus far, such terms are not relevant for Compton scattering
below $300\;\MeV$~\cite{Hildebrandt:2003fm,Hildebrandt:2003md,
  Hildebrandt:2005iw,Hildebrandt:2005ix,Griesshammer:2010pz}.

The two-photon response of the nucleon in the dipole approximation is
therefore characterised by the six linearly independent, energy-dependent
polarisabilities of Eq.~\eqref{eq:polsfromints}.
The spin-independent terms are parameterised by the two scalar functions
already encountered in Section~\ref{sec:intropols}: the electric dipole
polarisability $\alphae(\w)$, and the magnetic dipole polarisability
$\betam(\w)$.
The four spin polarisabilities parameterise the response of the nucleon spin
to an external field. The two corresponding to dipole-dipole transitions,
$\gammaee(\w)$ and $\gammamm(\w)$, are analogous to the classical Faraday
effect related to birefringence inside the nucleon~\cite{Holstein:2000yj}.
They describe how an incoming photon causes a dipole deformation in the
nucleon spin, which in turn leads to dipole radiation.
The two mixed spin polarisabilities, $\gammaem(\w)$ and $\gammame(\w)$, encode
scattering where the angular momenta of the incident and outgoing photons
differ by one unit.  In principle, the polarisabilities can be defined in any
coordinate system in which the initial and final photon energies are
identical. In practice, the centre-of-mass frame is usually used.\footnote{The
  dynamical polarisabilities introduced here via Eq.~(\ref{eq:polsfromints})
  differ from those given in
  Refs.~\cite{Griesshammer:2001uw,Hildebrandt:2003fm} by a factor of $\sqrt
  s/\MN$.  The polarisabilities are linear combinations of the multipole
  moments $f_{XY}^{L\pm}$ of the Compton amplitudes;
  the details, including the relevant projection formulae, are given in
  Ref.~\cite{Hildebrandt:2003fm}.}

We can now translate the interactions~\eqref{eq:polsfromints} into
contributions to the structure amplitudes $\bar{A}_i(\w,z)$:
\begin{align}
  \bar{A}_1(\w,\,z) &\dis=4\pi\,
  \left[\alphae(\w)+z\,\betam(\w)\right]\,\w^2+\dots
  \nonumber \\
  \bar{A}_2(\w,\,z) &\dis =-4\pi\,
  \betam(\w)\,\w^2 +\dots
  \nonumber\\
  \bar{A}_3(\w,\,z) &\dis=-4\pi\,
  \left[\gammaee(\w)+z\,\gammamm(\w)+\gammaem(\w)
    +z\,\gammame(\w)\right]\,\w^3+\dots
  \nonumber \\
  \bar{A}_4(\w,\,z) &\dis=4\pi\,
  \left[-\gammamm(\w) +\gammame(\w)\right]\,\w^3+\dots
  \label{eq:strucamp}\\
  \bar{A}_5(\w,\,z) &\dis=4\pi\,
  \gammamm(\w)\,\w^3 +\dots
  \;\; \nonumber \\
  \bar{A}_6(\w,\,z) &\dis=4\pi\,
  \gammaem(\w)\,\w^3+\dots
  \nonumber \;\;,
\end{align}
where the dots refer to omitted higher multipoles.  The relations between the
static polarisabilities of the multipole expansion and those of Ragusa
\cite{Ragusa:1993rm} defined in Eq.~\eqref{eq:low-en-amps} are
\begin{equation}
  \label{eq:poltranslate}
  \begin{split}
    &\gamma_1=-\gammaee-\gammaem\;\;,\;\;\gamma_2=\gammame-\gammamm\;\;,\;\;
    \gamma_3=\gammaem\;\;,\;\; \gamma_4=\gammamm\\
    &\gammazero=-\gammame-\gammamm-\gammaee-\gammaem\;\;,\;\;
    \gammapi=\gammame+\gammamm -\gammaee-\gammaem\;\;.
  \end{split}
\end{equation}
The first of the photon multipolarities in the subscripts of the spin
polarisabilities is sometimes dropped in the literature (so
$\gamma_{M2}\equiv\gammaem$ etc).

We reiterate that polarisabilities are identified by a multipole analysis
\emph{at fixed energy}, i.e.~only by the angular and spin dependence of the
amplitudes. Eq.~\eqref{eq:strucamp} emphasises that the complete set of
energy-dependent polarisabilities does not contain more or less information
than the untruncated Compton amplitudes $\bar{A}_i(\w,z)$.  However, the
information is more accessible, since any hadronic mechanism and interaction
leaves a characteristic signature in a particular multipole polarisability, as
discussed in the Introduction. Thus, when the multipole expansion is truncated
after the dipole terms, determining the six energy-dependent dipole
polarisabilities is reduced, in principle, to an energy-dependent multipole
analysis of the Compton scattering database. Such proof-of-principle results
were reported in~\cite{Miskimentalk,Griesshammer:2004yn,Hildebrandt:2005ix}
but suffer at present from rather large error bars.

At very low energies, the functions encoding the dynamical polarisabilities
can be approximated by their zero-energy values and therefore some experiments
(most recently Ref.~\cite{Federspiel:1991}) have used the Petrun'kin formula
to extract the static polarisabilities $\alphaep$ and $\betamp$ directly from
their data.  Indeed, the low-energy expansion of the cross section could
conceivably be extended to higher powers in $\w^2$. At fourth order, not only
do the spin polarisabilities enter, but also the next terms (slope parameters)
in the expansion of $\alphae(\w)$ and $\betam(\w)$:
\begin{equation}
  \label{eq:alphaexpanded}
  \lim\limits_{\w\to0}\alphae(\w)=\alphae+
  \w^2\alpha_{E\nu}+\calO(\omega^4)\;\;,\;\;
  \lim\limits_{\w\to0}\betam(\w)=\betam+\w^2\beta_{M\nu}+\calO(\omega^4)\;\;.
\end{equation}
Static values of the next multipoles, the scalar quadrupole polarisabilities,
also enter at $\calO(\omega^4)$. (Contributions to $T$ at $\w^4$ and $\w^5$
were discussed by Babusci et al.~\cite{Babusci:1998ww} and Holstein et
al.~\cite{Holstein:1999uu}.)
However, the convergence is governed by the pion-production threshold (the
first non-analyticity) and the expansion is thus in powers of $\w/\mpi$.  With
the slope correction to the scalar polarisabilities of size $(\w/\mpi)^2$,
this leads to a correction of about $10\%$ for photon energies as low as
$50\;\MeV$. The expansion is therefore useless where most high-accuracy data
are taken.

Consequently, modern extractions of the static polarisabilities choose a
different route.  It is assumed that the energy dependence of the
polarisabilities, and hence that of the amplitudes $\bar{A}_i$, is adequately
captured in some framework (e.g. dispersion relations, \ChiEFT). With the
long-range part of the interactions thus fixed, one fits up to six low-energy
constants which encode the short-distance dynamics and thereby determines the
static polarisabilities from data.  This is the approach taken in this review.

\subsection{\it Kinematics}
\label{sec:kinematics}

Consider Compton scattering on a nucleus $\mathrm{X}$ with mass
$M_\mathrm{X}$, $\gamma(\w,\vec{k})\;\mathrm{X}(E,\vec{p})\to
\gamma(\w^\prime,\vec{k^\prime})\;\mathrm{X}(E^\prime,\vec{p}^\prime)$, with
$E$ and $\vec{p}$ ($E^\prime$ and $\vec{p}^\prime$) the kinetic energy and
momentum of the target before (after) the reaction. So far, $\w$ denoted a
generic photon momentum and $\theta$ a generic scattering angle.  We now add
subscripts to differentiate between the centre-of-mass (cm), laboratory (lab)
and Breit frames, using relativistic kinematics throughout. The coordinate
axes are specified as follows: the incident photon beam direction defines the
$z$-axis, and the $xz$-plane is the scattering plane, with the $y$-axis
perpendicular to it.

In the centre-of-mass (cm) frame, the total energy is the square-root of the
Mandelstam variable $s$,
\begin{equation}
  \label{eq:s}
  \sqrt{s}=\wcm+\sqrt{M_\mathrm{X}^2+\wcm^2}\;\;.
\end{equation}
On the other hand, experiments are performed in the lab frame, where the
incident and outgoing photon energy are related to one another by a recoil
correction, and to $\wcm$, as follows:
\begin{equation}
  \label{eq:wcmtowlab}\wlab=\wcm\frac{\sqrt s}{M_\mathrm{X}} \quad , \quad
  \wlabout=\frac{M_\mathrm{X}\, \wlab}{M_\mathrm{X} + \wlab (1 - \cos
    \thetalab)}\;\;.
\end{equation}
The scattering angle transforms as
\begin{equation}
  \label{eq:trafoscattangles}
  \cos\thetacm = \frac{\cos\thetalab-\beta} {1-\beta\cos\thetalab}\;\;,
\end{equation}
where $\beta=\wlab/(\wlab+M_\mathrm{X})$ is the relative velocity between the
cm and lab frames.  The frame-dependent flux (phase-space) factor for cross
sections (see Eq.~\eqref{eq:diffxsecfermion}) is
\begin{equation}
  \label{eq:flux}
  \Phi_{\mathrm{cm}} = \frac{M_\mathrm{X}}{4 \pi \sqrt{s}} \quad ,\quad 
  \Phi_{\mathrm{lab}} = \frac{\wlabout}{4\pi \wlab}  \;\;.
\end{equation}
It is also useful to introduce the Mandelstam variables $t$ and
$\nu=(s-u)/(4M_\mathrm{X})$:
\begin{equation}
  t=2\wlab\wlabout(\cos\thetalab-1)=2\wcm^2(\cos\thetacm-1)
  =2\wBreit^2(\cos\thetaBreit-1) \quad , \quad \nu={\textstyle \frac 1 2}
  (\wlab+\wlabout) \;\;,
\end{equation}
where we have also listed variables in the Breit frame, with
\begin{equation}
  \label{eq:Breit}
  \wBreit=\frac{2M_\mathrm{X}\nu}{\sqrt{4M_\mathrm{X}^2-t}}\;\;,\;\;
  \cos\thetaBreit=1-\frac{\wcm^2}{\wBreit^2}\left(1-\cos\thetacm\right)\;\;.
\end{equation}  
For $\omega/M_\mathrm{X}\to0$, i.e.~small photon energy or large target mass,
the three coordinate frames coincide. The Breit and lab frames coincide for
$\theta=0\deg$, and the Breit and cm frames for $\theta=180\deg$.

In the Breit or ``brick-wall'' frame, the photon transfers no energy and the
target recoils with the magnitude of its momentum unchanged but its direction
exactly reversed, $\vec{p}_\text{Breit}^\prime=-\vec{p}_\text{Breit}$. This
has the advantage that the Compton amplitude is manifestly crossing-symmetric,
i.e.~invariant under the interchange of initial and final states:
$\wBreit\leftrightarrow-\wBreit$,
$\vec{k}_\text{Breit}\leftrightarrow-\vec{k}_\text{Breit}^\prime$
$\vec{\epsilon}\leftrightarrow\vec{\epsilon}^{\prime*}$.
By inspection of Eq.~\eqref{eq:Tmatrix}, the spin-independent amplitudes
$A_{1,2}$ are even in $\wBreit$, and the spin-dependent ones, $A_{3-6}$, are
odd.

\subsection{\it Observables}
\label{sec:observables}

We can now relate the six independent amplitudes of Eq.~\eqref{eq:Tmatrix} to
scattering observables.  A complete classification of unpolarised, single- and
double-polarisation observables for a spin-$\half$ target (nucleon or
\threeHe) by Babusci et al.~\cite{Babusci:1998ww} demonstrated that
experiments in which up to two polarisations are fixed can be described by
four independent observables below the first threshold and four more
observables above it. Here, we list some of the combinations which either have
been or will soon be explored experimentally or theoretically: unpolarised,
linearly or circularly polarised beams and unpolarised or polarised targets,
without the detection of final-state polarisations.  We define them as they
are measured, and---with the exception of the differential cross
section---refer to the literature for formulae which relate them to the
Compton amplitudes $A_i$. Observables with a vector-polarised nucleus can be
understood by replacing the spin-polarised nucleon by the polarised nucleus.
These will be discussed in Section~\ref{sec:spinpols}.

In order to take into account nuclear binding for the deuteron and \threeHe,
it is convenient to use a helicity basis for these targets:
\begin{equation}
  \label{eq:ampsd}
  A(M_f,\lambda_f;M_i,\lambda_i)= \bra M_f,\lambda_f|T| M_i,\lambda_i \ket\;\;,
\end{equation}
where $\lambda_{i/f}=\pm$ is the circular polarisation of the initial/final
photon, and $M_{i/f}$ is the magnetic quantum number of the initial/final
target spin, i.e.~$M_{i/f}\in\{0;\pm1\}$ for the deuteron and
$M_{i/f}\in\{\pm\half\}$ for \threeHe. A target of spin $S_\mathrm{X}$ has
$[2\text{(photon helicities)}\times(2S_\mathrm{X}+1)\text{(target spins)}]^2$
(in and out state) amplitudes, but parity and time reversal leave only
$2(S_\mathrm{X}+1)(2S_\mathrm{X}+1)$ independent components: $4$ spin-$0$
helicity amplitudes become $2$ independent ones, $A_{1,2}$; $16$ helicity
amplitudes for spin-$\half$ reduce to the $6$ of Eq.~\eqref{eq:Tmatrix}; and
$36$ amplitudes for spin-$1$ yield $12$ independent structures, constructed
e.g.~by Chen et al.~\cite{Chen:2004wwa}.

\paragraph{Unpolarised beam and target:}
In this case, the only observable is the differential cross section.  For the
nucleon, it is obtained from Eq.~\eqref{eq:diffxsecfermion} after averaging
over the initial target spins and photon polarisations and summing over the
final states; one finds in the basis of Eq.~\eqref{eq:Tmatrix} (see
e.g.~Refs.~\cite[Chapter~IV.2]{Bernard:1995dp}
and~\cite[Chapter~4.1]{Hildebrandt:2005ix}):
\begin{eqnarray}
 \label{eq:Tsquared}
  \left|T\right|^2&\!\!\!=\!\!\!
  &\;\textstyle{\frac{1}{2}}\,|A_1|^2\,\left(1+z^2\right)+\textstyle{\frac{1}{2}}\,|A_3|^2\, 
  \left(3-z^2\right)
  +\left(1-z^2\right)\Bigl[4\,\Re[A^*_3A_6]+
    \Re[A_3^*A_4+2A_3^*A_5-A_1^*A_2
    ]\,z\Bigr]\non\\
  &&+\left(1-z^2\right)\Bigl[\textstyle{\frac{1}{2}}\,|A_2|^2\,\left(1-z^2\right)+
 \textstyle{\frac{1}{2}}\,|A_4|^2\,\left(1+z^2\right)\\
  &&\phantom{+\left(1-z^2\right)\biggl[}+|A_5|^2\left(1+2z^2\right)+3\,|A_6|^2+
  2\,\Re[A_6^*\left(A_4+3A_5\right)]z+2\,\Re[A_4^* A_5]z^2
  \Bigr]\;\;.\non
\end{eqnarray}
In the helicity basis, the unpolarised differential cross section is built
with the flux factors of Eq.~\eqref{eq:flux} by summing over all combinations
of incident and outgoing quantum numbers $(M_f,\lambda_f;M_i,\lambda_i)$ and
including a symmetry factor from averaging over the initial target and photon
polarisations:
\begin{equation}
  \label{eq:diffxsecdeuteron}
  \left.\frac{\dd\sigma}{\dd\Omega}(\w,\theta)\right|_\text{frame}= 
  \frac{1}{2(2S_\mathrm{X}+1)}\;\Phi^2_{\text{frame},\mathrm{X}}
  \sum \limits_{M_f, M_i; \lambda_f,\lambda_i}
  \left|A(M_f,\lambda_f;M_i,\lambda_i) \right|^2\;\;.
\end{equation}

\begin{figure}[!htb]
  \begin{center}
    \includegraphics*[width=0.65\linewidth]{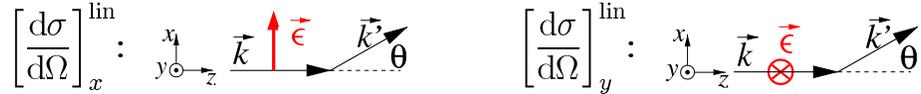}
    \caption {(Colour online) Observables for linearly polarised photon
      incident on unpolarised target.}
\label{fig:lin-nospin}
\end{center}
\end{figure}

\paragraph{Polarised beam, unpolarised target:}
The cross section with a circularly polarised photon beam of arbitrary helicity
is half of the unpolarised one and thus provides no additional information.
With a linearly polarised beam, two observables can be constructed, see
Fig.~\ref{fig:lin-nospin}, with corresponding experiments approved at
\HIGS~\cite{Weller:2009zza,Weller}.
The cross section for an incoming photon is
denoted by $\left[ \dfrac{\dd\sigma}{\dd\Omega}\right]^\text{lin}_{x(y)}$
for an incoming photon polarised parallel (perpendicular) to the scattering
plane. Their sum gives the unpolarised cross section, and the difference,
\begin{equation}
  \Sigma_3(\w,\theta)=\left(
    \left[ \dfrac{\dd\sigma}{\dd\Omega}\right]^\text{lin}_x-
    \left[ \dfrac{\dd\sigma}{\dd\Omega}\right]^\text{lin}_y\right)
  /\dfrac{\dd\sigma}{\dd\Omega}\;\;,
  \label{eq:polasym}
\end{equation}
is the beam asymmetry\footnote{The subscript is omitted in
  Ref.~\cite{Choudhury:2004yz}; the symbol used in
  Ref.~\cite{Griesshammer:2010pz} is $\Pi^\text{lin}$.}.  Its relation to the
deuteron helicity basis is given in
Refs.~\cite{ShuklaThesis,Choudhury:2004yz,Griesshammer:2010pz}.

\begin{figure}[!htb]
\begin{center}
\includegraphics*[width=0.18\linewidth]{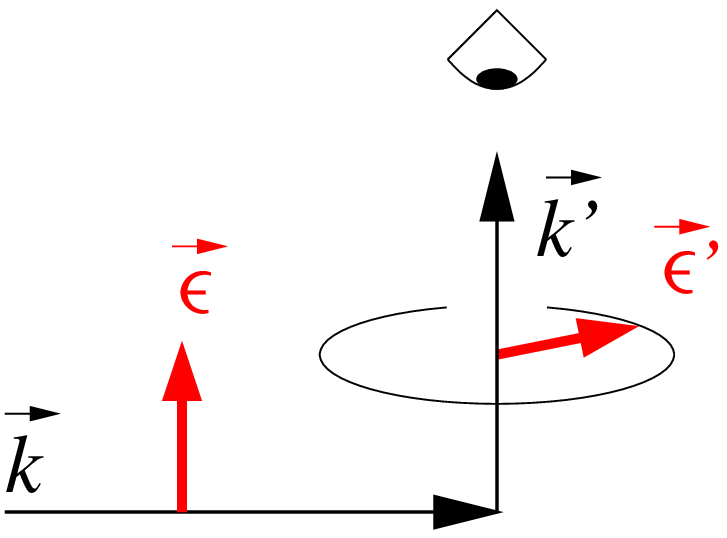}\hq\hq\hq\hq\hq\hq\hq\hq
\includegraphics*[width=0.18\linewidth]{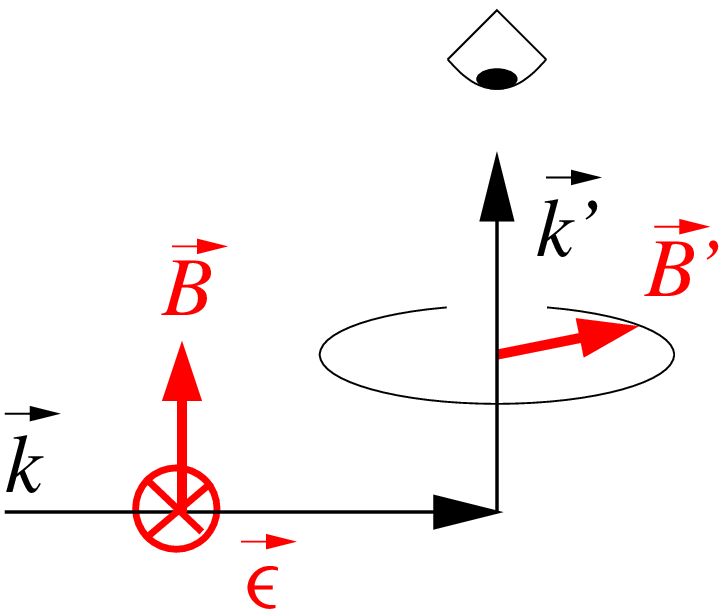}
\caption{\label{fig:switchingoffpols} (Colour online) Left: Configuration for
  which an induced electric dipole cannot radiate an $E1$ photon to an
  observer (``eye'') at $\thetaBreit=90^\circ$; right: same for radiating an
  $M1$ photon. }
\end{center}
\end{figure}

For the forward and backward spin polarisabilities, $\gamma_{0}$ and $\gamma_{\pi}$, 
we already saw that the multipole expansion is a convenient tool to identify
configurations in which a specific polarisability is isolated or suppressed.
Such configurations were found for the proton by
Maximon~\cite{Maximon:1989zz}.  The interaction in Eq.~\eqref{eq:polsfromints}
parameterised by $\alphae(\w)$ vanishes when the polarisations of the incoming
and outgoing photons are orthogonal, e.g.~when a photon which is linearly
polarised in the scattering plane scatters at $90\deg$ as in
$\left[\frac{\dd\sigma}{\dd\Omega}\right]^\text{lin}_{x}
(\thetaBreit=90\deg)$, see Fig.~\ref{fig:switchingoffpols}. 
In that case, since the incoming magnetic field is orthogonal to the scattering plane,
the induced magnetic dipole radiates most strongly at $90\deg$, providing
maximal sensitivity to $\betam(\w)$.
Similarly,
$\left[\frac{\dd\sigma}{\dd\Omega}\right]^\text{lin}_{y}(\thetaBreit=90^\circ)$
is independent of $\beta_{M1}(\w)$ but maximally sensitive to $\alphae(\w)$.
When the nucleon is embedded in a nucleus, the relative motion of the
$\gamma$N system may complicate this analysis.

\begin{figure}[!htb]
  \begin{center}
    \includegraphics*[width=0.97
    \linewidth]{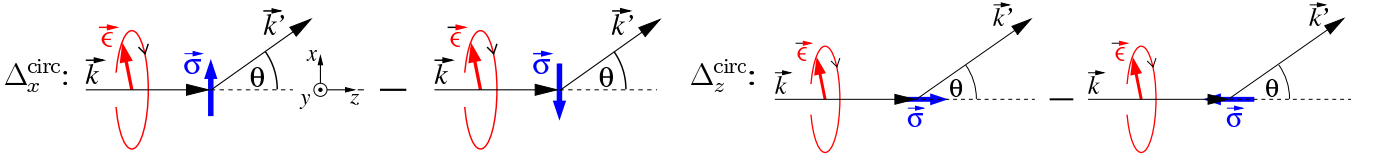}
    \caption {(Colour online) Observables for circularly polarised photon
      incident on polarised target.}
    \label{fig:circ-spin}
  \end{center}
\end{figure}

\paragraph{Polarised beam, polarised target:}
Figure~\ref{fig:circ-spin} depicts double-polarisation observables with
circularly polarised photons, as an example of observables that will be
explored in the future; see Section~\ref{sec:futureexp}.
The target can 
be polarised in the scattering plane along $\hat{x}$ or along the beam
direction, $\hat{z}$. Cross-section differences can be defined by flipping the
target polarisation:
\begin{equation}
  \Delta_{x}^\text{circ}(\w,\theta)=
  \left(\frac{\dd\sigma}{\dd\Omega}\right)_{\uparrow
    \rightarrow} - \left(\frac{\dd\sigma}{\dd\Omega}\right)_{\uparrow
    \leftarrow}\;\;,\;\;
  \Delta_{z}^\text{circ}(\w,\theta)
  =\left(\frac{\dd\sigma}{\dd\Omega}\right)_{\uparrow \uparrow} -
  \left(\frac{\dd\sigma}{\dd\Omega}\right)_{\uparrow \downarrow} \;\;.
  \label{eq:deltaxz} 
\end{equation}
The first arrow of the subscript denotes a positive beam helicity, the second
the target polarisation.  $\Delta_x^\text{circ}$ compares a target polarised
along $+\hat{x}$ vs.~$-\hat{x}$, i.e.~perpendicular to the beam direction but
in the scattering plane.  Similarly, $\Delta_z^\text{circ}$ is the difference
with the target polarised parallel vs.~anti-parallel to the beam
helicity\footnote{In Ref.~\cite{Chen:2004wwa}, $\Delta_z^\text{circ}$ is
  denoted by $2\Delta_{1\sigma}$.}. Both observables change sign for
left-circularly polarised photons (negative beam helicity).

Polarisation asymmetries are defined as the ratio of the cross-section
differences $\Delta$ to their sums\footnote{$\Sigma_{x,z}^\text{circ}$ is
  denoted by $\Sigma_{2x,2z}$ in Ref.~\cite{Babusci:1998ww} and $\Sigma_{x,z}$
  in Refs.~\cite{Choudhury:2004yz,Chen:2004wwa}.}:
\begin{eqnarray}
  \Sigma_{x}^\text{circ}=
  \frac{\left(\frac{\dd\sigma}{\dd\Omega}\right)_{\uparrow
      \rightarrow} - \left(\frac{\dd\sigma}{\dd\Omega}\right)_{\uparrow
      \leftarrow}}{\left(\frac{\dd\sigma}{\dd\Omega}\right)_{\uparrow
      \rightarrow} + \left(\frac{\dd\sigma}{\dd\Omega}\right)_{\uparrow
      \leftarrow}}
  \label{eq:sigmax} 
  &,&
  \Sigma_{z}^\text{circ}=
  \frac{\left(\frac{\dd\sigma}{\dd\Omega}\right)_{\uparrow \uparrow} -
    \left(\frac{\dd\sigma}{\dd\Omega}\right)_{\uparrow
      \downarrow}}{\left(\frac{\dd\sigma}{\dd\Omega}\right)_{\uparrow
      \uparrow} + \left(\frac{\dd\sigma}{\dd\Omega}\right)_{\uparrow
      \downarrow}}\;\;.
\end{eqnarray} 
Except for $\Sigma_{x,z}^\text{circ}$ for a spin-$\half$ target, the
denominators are not the unpolarised cross
sections~\cite{ShuklaThesis,Choudhury:2004yz,Hildebrandt:2005ix}.  Normalising
to sums of cross sections removes many experimental systematic uncertainties
and the frame dependence associated with the flux factors $\Phi_\text{frame}$.
However, a small spin-averaged cross section in the denominator may enhance
theoretical uncertainties or hide unfeasibly small count rates: cross-section
differences $\Delta$ set the scale for the beamtime necessary to perform these
experiments.  The asymmetries $\Sigma_{x,z}^\text{circ}$ vanish in the static
limit only for a proton target, but are nonzero for the neutron.

The relations of the double-polarisation observables (with arbitrary angle
$\phi$ between polarisation and scattering planes) to the amplitudes $A_i$ are
compiled in Ref.~\cite[Chapter~4.1]{Hildebrandt:2005ix} for complex
amplitudes.  For real amplitudes, i.e.~below threshold, the relations for
$\Delta^\text{circ}_{x,z}$ and $\Sigma^\text{circ}_{x,z}$ were first reported
in Ref.~\cite[Chapter~IV.2]{Bernard:1995dp}. Those for the helicity amplitudes
are compiled for the deuteron in~\cite{Choudhury:2004yz,Griesshammer:2010pz}
and for \threeHe in~\cite{Shukla:2008zc,ShuklaThesis}.
Similarly, double-polarisation observables with linearly polarised photons can
also be defined--- see
e.g.~\cite{Babusci:1998ww,Hildebrandt:2005ix,Griesshammer:2010pz} for
definitions and figures analogous to the ones above.




\section{Experimental overview} 
\label{sec:expoverview}

In this section, we review the experimental efforts on Compton scattering
using proton and deuteron targets, spanning the past half-century.  For the
proton, we have divided the discussion into low-energy measurements (below
pion threshold) and high-energy measurements (above pion threshold).  There is
a relatively clear distinction between experiments in these two energy
regions, although some cases do overlap with both regions.  There are also
some polarised measurements on the proton in the modern era.  For the
deuteron, there are again two categories---elastic Compton-scattering
experiments and quasi-free measurements in which deuteron breakup is exploited
to study the neutron explicitly in quasi-free kinematics (with the proton
acting as a spectator).

Special emphasis will be placed on enumerating the statistical and systematic
uncertainties of the proton and deuteron experiments, since these issues
figure prominently in the fitting of the various data sets using the \ChiEFT
formalism.  Somewhat cursory details are given about the experiments: many are
discussed at greater length in the review by
Schumacher~\cite{Schumacher:2005an}.

\subsection{\it Low-energy proton Compton scattering}
\label{sec:expprotlow}

The earliest low-energy Compton-scattering experiments on the proton (up to
about $\wlab\sim100$ MeV) were reported in the mid-to-late 1950s by Pugh et
al.~\cite{Pugh:1957}, Oxley~\cite{Oxley:1958}, Hyman et al.~\cite{Hyman:1959},
Bernardini et al.~\cite{Bernardini:1960} and Goldansky et
al.~\cite{Goldansky:1960}.  These early experiments were not aimed at
measuring the electromagnetic polarisabilities of the proton, as we think of
them today.  In fact, these experiments were more motivated by their ability
to test recently developed dispersion-theory calculations of Gell-Mann and
Goldberger and others.  Nevertheless, it is noteworthy that in some of these
early papers (most notably Goldansky et al.~\cite{Goldansky:1960}), attempts
were made to extract the proton polarisability.

These early experiments were pioneering efforts, given the difficulty of
working with continuous bremsstrahlung photon beams and detector systems with
very poor energy resolution.  Large NaI photon detectors with good energy
resolution were not yet available, and photon-tagging facilities were still
decades in the future.  Normalising the photon flux to obtain an absolute
cross section is notoriously difficult with bremsstrahlung beams, not to
mention that the incident continuous bremsstrahlung beam itself has no
well-defined energy resolution.  For example, the experiment of
Oxley~\cite{Oxley:1958} had a central photon energy of 60 MeV, with a full
width of 55 MeV---so it is remarkable that this experiment produced results
for the cross section between 10.6 and 14.7 nb/sr from 70\deg\ to 150\deg,
which is generally consistent with modern results.

Directed efforts were made in the experiment of Goldansky et
al.~\cite{Goldansky:1960} in 1960 to determine the proton polarisability.  The
same group continued these efforts many years later in the mid-1970s in the
experiment of Baranov et al.~\cite{Baranov:1974,Baranov:1975}.  While
uncertainties in the extracted polarisabilities were reduced by a factor of
two in the later experiment, the experimental techniques during that period
were still relatively crude compared to today.  For this reason, it is
reasonable to consider the data prior to 1980 to be exploratory in nature, and
to not give very much credence to the absolute scale of the cross sections
from those experiments.

Almost 20 years passed before the next proton Compton-scattering experiment
was attempted.  Two major developments in experimental techniques emerged in
that period to make the new generation of Compton measurements considerably
more reliable than their predecessors.  First, the method of photon tagging
was introduced, providing both a mono-energetic beam of photons and a means of
normalising the photon flux by direct counting of the post-bremsstrahlung
electrons.  This revolutionised many photonuclear experiments.  Second, new
large-volume high-resolution NaI detectors (25.4~cm diameter $\times$ 25.4~cm
long, and even larger) were available with a resolution $\Delta E/E \sim3\%$
for photons of 50--100 MeV.  These two improvements in the beam and the
detectors changed the game significantly, paving the way for a new era of
Compton-scattering experiments designed specifically to pin down the proton
polarisability.

The major experiments that have contributed in the modern era to the
determination of the proton polarisability are those of Federspiel et
al.~\cite{Federspiel:1991}, Zieger et al.~\cite{Zieger:1992}, Hallin et
al.~\cite{Hallin:1993}, MacGibbon et al.~\cite{MacGibbon:1995}, and \OdL\ et
al.~\cite{Olmos:2001} covering the 10-year period between 1991 and 2001.  All
but two of them used tagged-photon beams of energy $\wlab \leq$ 165 MeV---the
exceptions are Hallin, who used a continuous bremsstrahlung beam with energies
up to 289 MeV, and Zieger, who used a bremsstrahlung beam with a proton-recoil
detection technique.  It is worth examining all of these low-energy
experiments in more detail to elucidate their relative merits and potential
weaknesses.

The first of the modern experiments was conducted by Federspiel et
al.~\cite{Federspiel:1991} at the tagged-photon facility at Illinois.  Tagged
photons in the energy range $\wlab$ = 32--72 MeV impinged on a liquid hydrogen
target; scattered photons were detected at fixed lab angles of 60\deg\ and
135\deg\ by two large 25.4~cm $\times$ 25.4~cm NaI detectors.  This
experimental configuration can be considered a ``standard'' tagged-photon
experiment, in the sense that the well-characterised beam and the single-arm
detector conditions are conceptually extremely simple.  The data were divided
into 8 bins over this energy range, each 4 MeV wide.  The resulting cross
sections had statistical uncertainties that were roughly $\pm$10\%.
Systematic uncertainties due to photon flux ($\pm$1\%), target thickness
($\pm$1\%), detector acceptance ($\pm$1.4\%) and simulation uncertainties $\pm$1\%)
were correlated among all data points and contributed in quadrature to 
give a total systematic error of $\pm$2.2\% for this experiment.  The data were
predominantly limited by statistical accuracy, due to the low counting
statistics in each tagger energy bin, as well as error propagation due to
subtraction of random coincidences and empty-target background in the data
analysis.

A follow-up experiment was performed by MacGibbon et al.~\cite{MacGibbon:1995}
(including the Illinois group) at the tagged-photon facility at the
Saskatchewan Accelerator Laboratory (SAL).  This experiment was essentially
identical to the lower-energy version at Illinois and used the same large NaI
detectors, which had been transported to Saskatoon for this purpose.  The
tagged-photon energy range was $\wlab$ = 70--100 MeV, so it overlapped with
the previous measurement, and the detector lab angles were $90\deg$ and
$135\deg$.  One clever twist to this experiment, however, was the ability to
extend the usable energy range into the untagged region ($\wlab$ = 100--148
MeV).  The normalisation of the untagged data could be linked directly to the
well-known normalisation of the data in the tagged region using the
bremsstrahlung spectrum shape, so a reliable extrapolation of the cross
section up to the endpoint energy of 148 MeV could be achieved.  The tagged
data were divided into 4 energy bins, each 8~MeV wide, and the untagged data
were divided into 5 bins, each 10~MeV wide.  Here again, the statistical
accuracy of the data was the limiting factor---the lowest statistical
uncertainty was about $\pm$10\% and some points had errors close to $\pm$20\%.
Systematic uncertainties came from similar sources as the previous experiment,
but were slightly larger. Correlated errors such as photon flux ($\pm$1--2\%), 
target thickness ($\pm$1.2\%) and detector acceptance ($\pm$1.2\%), as well as additional
rate-dependent corrections ($\pm$1--3\%), added up to an overall systematic
error of $\pm$2.9\% for the untagged data and a range of $\pm$3--4\% for the
tagged data.  Only a small part of that overall error ($\pm$0.4\% for the tagged 
data and $\pm$1.0--1.4\% for the untagged data) was uncorrelated among the data points. 
In the overlap region near 70 MeV for the 135\deg\ data, the MacGibbon results are 
in excellent agreement with those of Federspiel et al.~\cite{Federspiel:1991}.

The experiment of Hallin et al.~\cite{Hallin:1993} was also performed at SAL
and included participants from the two experiments discussed above.  The
Hallin experiment had three major differences: (1) the photon beam was a
continuous bremsstrahlung beam, (2) the beam energies were mostly above pion
threshold (endpoint energies of $\wlab$ = 170--298 MeV), and (3) a single,
large-volume NaI detector from Boston University was used (the BUNI detector,
50~cm diameter $\times$ 56~cm long).  Since only a single photon detector was
used, data for each lab angle in the 25\deg\ to 135\deg\ range of this
experiment had to be measured sequentially (as opposed to simultaneously).  As
a bremsstrahlung experiment, the statistical uncertainties of the data can be
relatively low, depending on how large a region of the bremsstrahlung
distribution is utilised to generate the cross section.  In this case,
different regions were used for the angular distributions (15 MeV wide) as
compared to the excitation functions (5 MeV wide)---hence, the statistical
errors in the angular distributions tended to be lower, ranging from
$\pm$3.5--5.0\% at backward angles to $\pm$10--18\% at forward angles.
The major sources of (correlated) systematic uncertainties arose from determining the photon
flux ($\pm$3\%) and the target thickness ($\pm$2\%), giving $\pm$3.6\%.  Additional 
contributions from, for example, detector efficiency ($\pm$0.3\%) and solid angle
($\pm$0.3\%), as well as other factors, led to point-to-point systematic errors in the
range of $\pm$1--2\%.  Taken together, these systematic errors (correlated + uncorrelated) 
amounted to about $\pm$3.7--4.2\% for the total systematic uncertainty 
in the final cross sections, which is quite good considering the inherent 
difficulties of normalising a bremsstrahlung experiment.

One of the more novel experiments was a 180\deg~scattering measurement by
Zieger et al.~at MAMI~\cite{Zieger:1992}.  In this case, the forward proton
(recoiling at 0\deg) was detected in a magnetic spectrometer, yielding the
scattered photon cross section in the backward direction (180\deg), where the
result is exclusively sensitive to the difference of the electric and magnetic
polarisabilities ($\alphaep-\betamp$).  The incident photon energies were
$\wlab$ = 98 and 132 MeV, and each of the two data points constituted an
energy bin of width 16 MeV.  The forward proton cross sections were determined
by comparing the Compton proton and electron yields---this enabled the
absolute normalisation to be deduced without detailed knowledge of the shape
or intensity of the incident bremsstrahlung spectrum.  The statistical
uncertainties for the two data points varied widely ($\pm$18.4\% and
$\pm$5.4\%), but the systematic uncertainties due to the subtraction of target
backgrounds and the normalisation procedure were more consistent between the 
two data points ($\pm$6.4\% and $\pm$4.3\%, respectively).  This experiment is 
the only measurement of the 180\deg\ (backward) Compton-scattering cross section for the proton.

The most comprehensive experiment to date was the one by \OdL\ et
al.~\cite{Olmos:2001} at MAMI, covering the tagged-photon energy range $\wlab$
= 59--164 MeV and the lab angular range 59\deg--155\deg\ simultaneously.  This
was possible by using the large TAPS array consisting of six separate
assemblies of hexagonal BaF$_2$ cells.  Statistical errors varied from
$\pm$5\% at the lowest energies to about $\pm$10\% at the higher energies.
Systematic errors from photon flux ($\pm$2\%) and target thickness ($\pm$2\%)
were added in quadrature to give an overall systematic error of $\pm$3\% for
the entire data set.  In addition, random systematic uncertainties
(point-to-point) due to the geometry of the individual array elements and the
detector simulation contributed another $\pm$5\% to each data point
separately.  These data overlapped with (and extended higher than) the energy
regions of the previous data from Federspiel~\cite{Federspiel:1991} and
MacGibbon~\cite{MacGibbon:1995}, and all three data sets are in excellent
agreement in the overlap region.  Regarding the extraction of proton
polarisabilities, it is worth mentioning that the \OdL\ data set tends to
dominate most global analyses due to the large number of data points and their
relatively low statistical errors.

A summary of the low-energy proton experiments, from the 1950s up to the
present day, is shown in the upper portion of Table~\ref{table-proton-low}.
Plots of sample cross sections are also shown in Fig.~\ref{fig:sect5-fig1}.

\begin{table}[!htb]
  \begin{center}
    \begin{tabular}{|l|l|c|c|c|c|c|}
      \hline
      Legend & First Author & $\wlab$ (MeV) & $\thetalab$ (deg) & Ref. & $N_{\rm data}$ & Symbol\\
      \hline
      Chicago 58 & Oxley & 25--87 & 70--150 & \cite{Oxley:1958} & 4 &\includegraphics*[width=7pt]{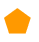}\\
      MIT 59 & Hyman & 50--140 & 50, 90 & \cite{Hyman:1959} & 12 &\includegraphics*[width=7pt]{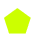} \\
      Moscow 60 & Goldansky & 40--70 & 45--150 & \cite{Goldansky:1960} & 5 &\includegraphics*[width=7pt]{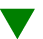}\\
      Illinois 60 & Bernardini & 100--290 & 90, 129, 139 & \cite{Bernardini:1960} & 17 &\includegraphics*[width=7pt]{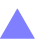}\\
      MIT 67 &Pugh & 50--130 & 45, 90, 135 & \cite{Pugh:1957} & 16 &\includegraphics*[width=7pt]{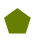}\\
      Moscow 74 & Baranov & 82--111 & 90, 150 & \cite{Baranov:1974,Baranov:1975} & 7 &\includegraphics*[width=7pt]{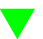} \\
      Illinois 91 & Federspiel & 32--72 & 60, 135 & \cite{Federspiel:1991} & 16 &\includegraphics*[width=7pt]{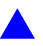} \\
      MAMI 92 & Zieger & 98, 132 & 180 & \cite{Zieger:1992} & 2 &\includegraphics*[width=7pt]{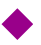}\\
      SAL 93 & Hallin  & 135--291 & 25--135 & \cite{Hallin:1993} & 77 &\includegraphics*[width=7pt]{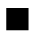}  \\
      SAL 95 & MacGibbon & 70--148 & 90, 135 & \cite{MacGibbon:1995} & 18 &\includegraphics*[width=7pt]{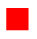} \\
      MAMI 01 & Olmos de Le\'on & 59--164 & 59--155 & \cite{Olmos:2001} & 65 &\includegraphics*[width=7pt]{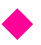}\\
      \hline
      Cornell 61& DeWire & 275--425 & 75, 90, 120  & \cite{DeWire:1961}& 5 &\includegraphics*[width=7pt]{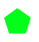}\\
      Tokyo 64& Nagashima & 310--420 & 90  & \cite{Gray:1967,Genzel:1976}& 2 &\includegraphics*[width=7pt]{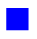}\\
      Moscow 66& Baranov & 214,237,249 & 55--150  & \cite{Baranov:1966a,Baranov:1966b}& 12 &\includegraphics*[width=7pt]{symbol6}\\
      Illinois 67& Gray & 185--335 & 90, 135  & \cite{Gray:1967}& 5 &\includegraphics*[width=7pt]{symbol4}\\
      Bonn 76& Genzel & 237--430 & 50--130  & \cite{Genzel:1976}& 19 &\includegraphics*[width=7pt]{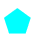}\\
      MAMI 96 & Peise & 200--410 & 75  & \cite{Peise:1996,Hunger:1997}& 17 &\includegraphics*[width=7pt]{symbol9}\\
      MAMI 96a & Molinari & 250-500 & 90  & \cite{Molinari:1996,Hunger:1997}& 13 &\includegraphics*[width=7pt]{symbol9}\\
      MAMI 99 & Wissmann & 199--410 & 131  & \cite{Wissmann:1999}& 11 &\includegraphics*[width=7pt]{symbol9}\\
      MAMI 01a & Wolf & 250--800 & 30--150  & \cite{Galler:2001,Wolf:2001}& 294 &\includegraphics*[width=7pt]{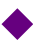}\\
      LEGS 01& Blanpied & 213--334 & 65--135  & \cite{Blanpied:2001}& 77 &\includegraphics*[width=7pt]{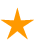}\\
      MAMI 02 & Camen & 210--470 & 136  & \cite{Camen:2002}& 10 &\includegraphics*[width=7pt]{symbol9}\\

      \hline
    \end{tabular}
  \end{center}
  \caption{\label{table-proton-low} Compton-scattering experiments on the proton.  
    The column for $N_{\rm data}$ shows \emph{only} the number of points for each data set 
    up to a cutoff of 400 MeV.  
    The last column indicates the corresponding symbols used in the figures.}
\end{table}

\subsection{\it High-energy proton Compton scattering}
\label{sec:expprothigh}

Compton-scattering experiments on the proton have also been explored above
pion threshold, dating back to the 1960s.  In the early experiments of DeWire
et al.~\cite{DeWire:1961}, Baranov et al.~\cite{Baranov:1966a,Baranov:1966b},
Gray and Hanson~\cite{Gray:1967} and Genzel et al.~\cite{Genzel:1976}, the
main intent was to investigate the region of the $\Delta(1232)$ resonance.  In
the 1990s, the emphasis of these experiments evolved to become more
specific---studying the E2/M1 ratio and determining the backward spin
polarisability, $\gammapip$, of the proton.  These experiments had been
interpreted in a dispersion-relation framework, but the recent evolution of
EFT has made these experiments (below about 400 MeV) accessible to this
theoretical treatment.

Most of the experiments have been conducted at MAMI or LEGS---in the main,
they have been competing efforts and less complementary in nature.  There have
been some disagreements in the published cross sections, and as a result,
there have been diverging interpretations, particularly with regard to
$\gammapip$.  We now review the experiments from each laboratory in turn.

The MAMI experiments covered the energy region of the $\Delta$ resonance and
higher, roughly $\wlab$ = 200--500 MeV, using the photon-tagging facility with
a typical tagger-channel energy width of about 2.0--2.5 MeV.  For most of the
cases, the scattered photon was detected in coincidence with the recoil
proton---this kinematic redundancy helps tremendously in the reduction of
background due to photons from the decay of neutral pions which are copiously
produced above threshold.  In some cases below about 300 MeV, however, the
CATS large-volume NaI detector (48~cm diameter $\times$ 64~cm long) was
sufficient to cleanly separate the Compton events from the pion-production
events due to its excellent intrinsic energy resolution

In the paper by H\"unger et al.~\cite{Hunger:1997}, two independent
experiments at forward angles were reported---this constituted a summary of
work already published.  In one case, also reported earlier by Peise et
al.~\cite{Peise:1996}, the CATS detector was located at $\thetalab$ = 60\deg\
($\thetacm$ = 75\deg) and a multi-element array of small NaI detectors was
located at the corresponding angle of $\thetalab$ = 47\deg\ on the other side
of the beamline to intercept the forward recoil proton.  For additional
background suppression, a $2\pi$ array of BaF$_2$ crystals was positioned
close to the target, directly opposite CATS, serving as a supplemental veto
for $\pi^0$ production events.  In the reported cross sections, the
statistical uncertainties varied from as large as $\pm$40\% for the very
lowest energies, to $\pm$10--15\% below the $\Delta$ resonance, and then
closer to $\pm$5\% at and above the resonance peak.  Systematic errors have
been estimated to be $\pm$4.4\% below 350 MeV, $\pm$6\% for 350--380 MeV, and
$\pm$8\% above 380 MeV.  Within these systematic errors, common sources of
uncertainty for all data points arose from the photon flux ($\pm$2\%) and the
target thickness ($\pm$2\%).

In the other case, also reported earlier by Molinari et
al.~\cite{Molinari:1996}, sets of 12 photon/proton detector pairs were
arranged azimuthally around the target, with all photon detectors at a polar
angle of $\thetalab$ = 76\deg\ ($\thetacm$ = 90\deg) and the corresponding
proton detectors at $\thetalab$ = 44\deg\ on the opposite side.  The photon
detectors were Pb-glass blocks and the proton detectors were $\Delta E/E$
plastic scintillator telescopes.  The good energy resolution of the proton arm
and the small acceptance of each proton telescope enabled a clean separation
of Compton and pion events.  Moreover, the Pb-glass blocks are insensitive to
neutrons, so the background from charged-pion production was reduced.
Statistical uncertainties in the data are $\pm$4--10\% below 400 MeV,
$\pm$10--25\% up to 450 MeV, and then rather large ($\pm$50\% or more) up to
490 MeV.  Systematic errors have been estimated to be $\pm$4\% below 360 MeV
and $\pm$5\% above that energy due to background subtraction of $\pi^0$
events.  This included the common uncertainties due to the photon flux
($\pm$2\%), target thickness ($\pm$2\%), and detector efficiency ($\pm$3\%).

A broader survey of energies and angles was presented by Wolf et
al.~\cite{Wolf:2001}, which was also identically reported in Galler et
al.~\cite{Galler:2001}.  This experiment used a large-acceptance detector
array covering lab angles of 30\deg--150\deg\ and photon energies of $\wlab$ =
250--800 MeV.  Much of this data set (the upper half, generally) is beyond the
applicable scale of \ChiEFT.  The photon arm of the detector configuration
consisted of 10 large Pb-glass segments, each containing 15 individual
Pb-glass blocks.  The proton arm had two wire chambers and an array of tall
plastic scintillator bars used for proton time-of-flight measurements.  Below
400 MeV, relatively good separation between Compton and pion events could be
achieved.  However, above that energy, considerable overlap required a
separate subtraction procedure based on comparing in-plane (Compton-scattering
plane) and out-of-plane (pion production with a $2\gamma$ decay) events to
isolate the Compton events.  This method could add a considerable uncertainty
to the extracted cross sections for the data above 400 MeV.  The random errors
associated with each data point (which varied widely in magnitude across the 
extensive data set) included both the
purely statistical error due to counting statistics as well as individual
systematic errors due to detector efficiency, geometrical acceptance and
background subtraction.  Beyond that, there were overall scaling systematic
errors related to photon flux ($\pm$2\%) and target thickness ($\pm$2\%) that 
affected the entire data set as a whole.

A separate experiment by Wissmann et al.~\cite{Wissmann:1999} focused on a
single backward angle ($\thetalab$ = 131\deg) in the energy range $\wlab$ =
200--410 MeV.  The motivation was to compare free proton Compton scattering
with quasi-free scattering, as a precursor to a similar quasi-free Compton
experiment on the neutron.  The experimental setup was simple, consisting of a
single photon arm using the large-volume CATS detector, which could achieve
good separation between Compton and pion-production events below 300 MeV.
Above that point, there is some overlap which can be subtracted using a
simulation of the pion events.  The resulting data set had statistical
uncertainties as large as $\pm$20--40\% at the lowest energies and improved
to $\pm$8--13\% above about 275 MeV (approaching the $\Delta$ peak and
beyond).  Systematic uncertainties due to photon flux, target thickness and
detector acceptance were common to all data points and were combined in
quadrature to be $\pm$3\% overall.

Finally, in a related experiment, Camen et al.~\cite{Camen:2002} obtained data
at a similar backward angle ($\thetalab$ = 136\deg) at energies of $\wlab$ =
200--470 MeV in an effort to extract \gammapip.  This was related to a
controversy based on an earlier extraction by
LEGS~\cite{Tonnison:1998,Blanpied:2001} which disagreed with standard
dispersion theory and $\chi$PT.  As expected, the disagreement can be traced
to discrepancies in the measured cross sections at the two laboratories near
the $\Delta(1232)$ peak.  The results from LEGS are discussed below.  In the
MAMI experiment, the CATS detector was used in conjunction with a recoil
proton detector array consisting of 30 liquid scintillator cells located at
$\thetalab$ = 18\deg\ on the other side of the beamline.  This experiment was
similar to that of Wissmann~\cite{Wissmann:1999}, but the redundancy of
detecting the coincident recoil proton helped discriminate Compton events from
pion-production events without relying on a simulation for the latter.  The
statistical errors were typically in the range of $\pm7\%$ to $\pm15\%$ for
all data points, but there was no mention in the published report of any
information regarding systematic uncertainties.  It is likely, however, that
they would be similar to the ones stated by Wissmann~\cite{Wissmann:1999}.

Although LEGS covers a more limited energy range than MAMI, it offers the
unique capability of delivering polarised photon beams.  This is accomplished
by backscattering polarised ultraviolet laser photons from 2.6 GeV electrons
in the Brookhaven National Laboratory synchrotron ring.  Energies of $\wlab$ =
210--333 MeV can be covered using the photon-tagging facility at LEGS---the
photon energy resolution is about 5 MeV, which is twice the width of the MAMI
tagger channels.  At LEGS (as at MAMI), these higher-energy experiments were
performed by detecting the recoil proton in coincidence with the scattered
photon, allowing a clean separation of Compton-scattering and pion-production
events.

The LEGS polarised experiments were performed in a series of three groups of
runs~\cite{Blanpied:1996,Blanpied:1997,Blanpied:2001} covering scattering
angles of $\thetacm$ = 65\deg--135\deg.  Scattered photons were detected in a
large-volume NaI detector (48~cm diameter $\times$ 48~cm long); a large array
of plastic scintillator bars on the opposite side of the beamline was used to
detect the recoil proton in coincidence.  The proton arm also had wire
chambers located near the target to reconstruct the proton angle more
precisely.  The kinematic overdetermination for Compton events not only served
to isolate them from the larger background due to $\pi^0$ decay photons, but
it also enabled detector efficiencies to be determined directly from the data
(instead of relying on simulations).  This helped considerably in reducing
systematic uncertainties.

The statistical precision of the LEGS cross sections is excellent in the
vicinity of the $\Delta(1232)$ peak, and the systematics also seem to be very
much under control.  The reported cross sections list the combination (in
quadrature) of statistical and point-to-point systematic errors.  Even with
these combined errors, the uncertainties are only about $\pm2\%$ to $\pm5\%$
for data points between 275 and 325 MeV.  Outside that energy range (the
highest energy, 333 MeV, and 213--265 MeV), the error bars are in the
$\pm$(5--10)\% range.  The overall systematic error in the cross-section scale
was evaluated to be $\pm$2\% based on uncertainties in the target thickness,
photon flux, and geometric solid angles.

With the exception of a much earlier measurement from
Frascati~\cite{Barbiellini:1968} at 318 MeV, these are the first polarisation
asymmetry measurements in this energy region for proton Compton scattering.
Near the $\Delta$ peak, the asymmetries are not small (0.1--0.5) and the
combined statistical plus point-to-point systematic errors (arising solely
from uncertainties in the photon beam polarisation for the two different
polarisation states) are typically $\pm$(7--25)\% for most data points.  Owing
to the nature of the asymmetry measurement, there are no overall systematic
errors that do not cancel out completely.

Regarding the cross sections, there are clearly discrepancies between the LEGS
and the MAMI data.  This is most apparent in Fig.~11 of Ref.~\cite{Wolf:2001}
at backward angles close to the $\Delta$ peak.  It is this discrepancy which
gives rise to a disagreement in the values extracted for the E2/M1 ratio and
$\gammapip$.  At the backward angles, both of these quantities display the
greatest sensitivity, especially the spin polarisability.  The same
discrepancy is shown in Fig. 2 of Ref.~\cite{Camen:2002}, which is in
agreement with~\cite{Wolf:2001}.  Data at the highest SAL energy of 286
MeV~\cite{Hallin:1993} appear to be intermediate between the two sets.  It is
worth recalling that both the LEGS
experiments~\cite{Blanpied:1996,Blanpied:1997,Blanpied:2001} and the MAMI
experiments~\cite{Wolf:2001,Camen:2002} used coincidence methods to detect the
scattered photon and the recoil proton, greatly reducing their susceptibility
to backgrounds from pion production.  By contrast, the experiment of
Ref.~\cite{Hallin:1993} used bremsstrahlung (untagged photons) and had a
single photon arm using the large-volume BUNI detector.  It is difficult to
know the source of the discrepancy between the MAMI and LEGS data sets.

A summary of the high-energy proton experiments is shown in the lower portion
of Table~\ref{table-proton-low}.  Plots of sample cross sections are also
shown in Fig.~\ref{fig:sect5-fig2}.

\subsection{\it A critical look at the proton data}
\label{sec:expdatacritical}

\begin{figure}[!htbp]
  \begin{center}
    \includegraphics*[width=\linewidth]{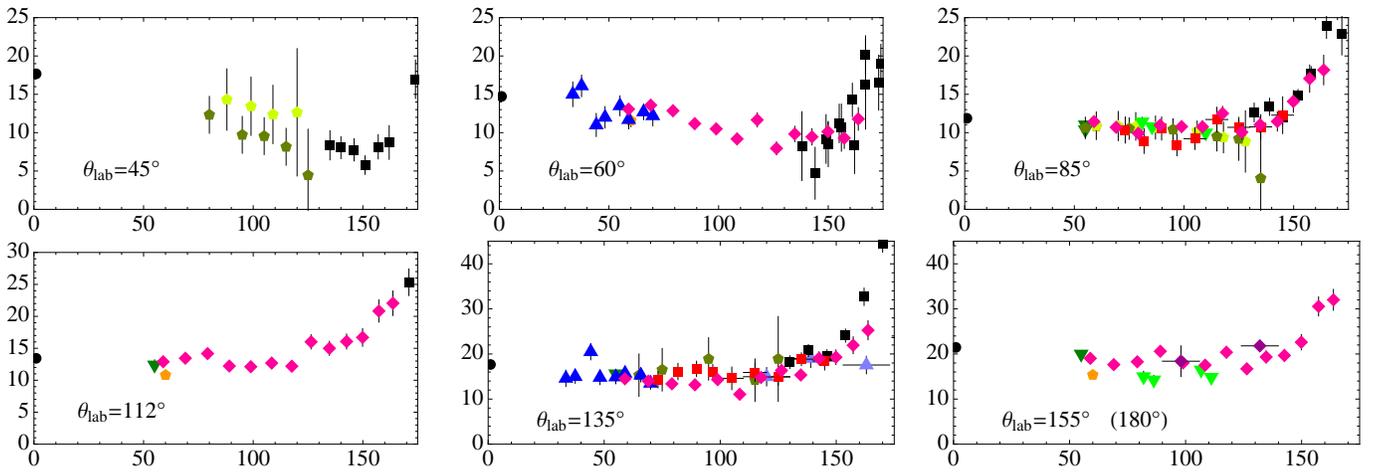}
    \caption {(Colour online) Data for Compton scattering on the proton below
      $\wlab=175$~MeV.  Lab cross sections in nb/sr are plotted in bins of
      10\deg\ lab angle as a function of lab photon energy in MeV.  For
      greater comparability of data, a phenomenological parameterisation of
      the angular dependence has been used to shift data points to the nominal
      bin angle, but the effect is only perceptible for the Oxley and
      Goldansky points at 70\deg\ and 120\deg.  The black dot at $\wlab=0$
      indicates the cross section corresponding to the Thomson limit.  See
      Table~\ref{table-proton-low} for the key to the data symbols.}
    \label{fig:sect5-fig1}
  \end{center}
\end{figure}

\begin{figure}[!htbp]
  \begin{center}
    \includegraphics*[width=\linewidth]{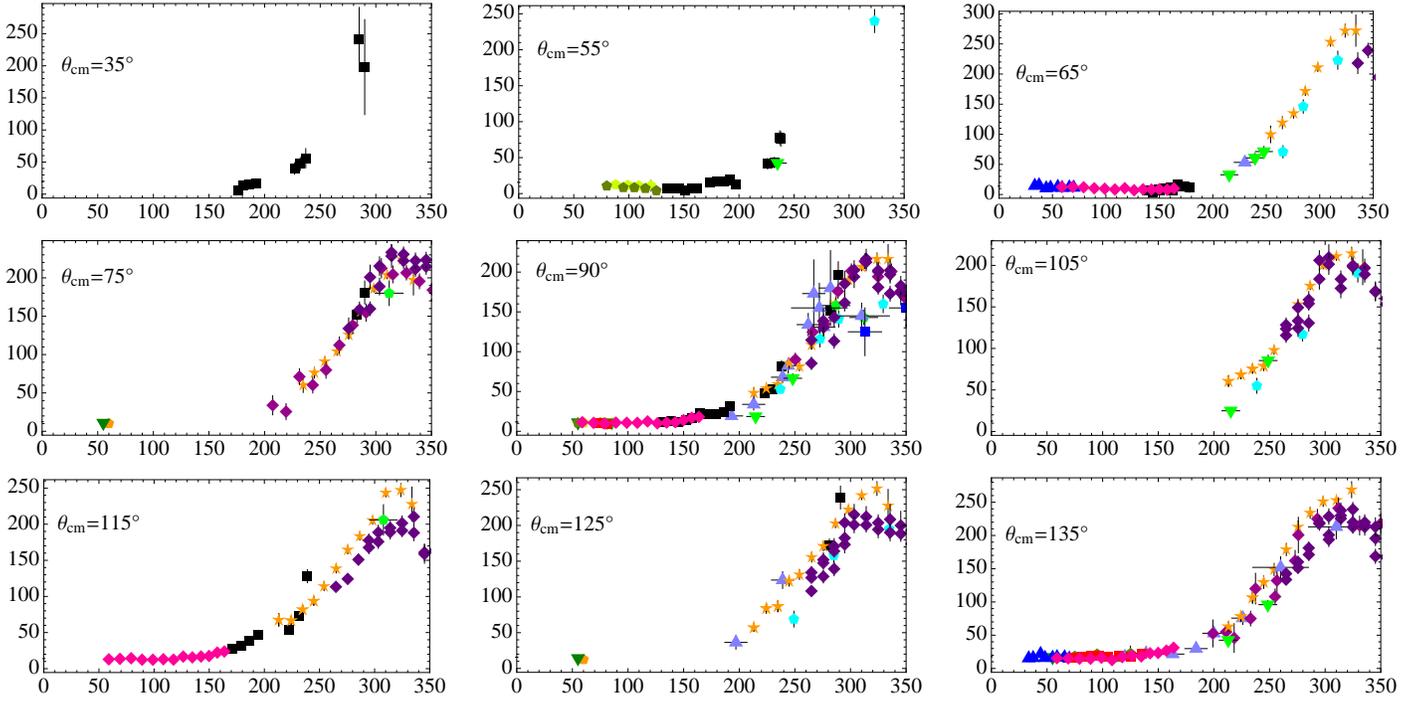}
    \caption {(Colour online) Selected data for Compton scattering on the
      proton below 350~MeV lab energy.  Centre-of-mass cross sections in nb/sr
      are plotted in bins of 10\deg\ cm angle as a function of lab photon
      energy in MeV.  See Table~\ref{table-proton-low} for the key to the data
      symbols.}
    \label{fig:sect5-fig2}
  \end{center}
\end{figure}

As a precursor to the EFT fits presented in Section~\ref{sec:protonanalysis},
we briefly provide a comparison of the various data sets that have been used
in the proton fitting.  Unless otherwise stated, energies and angles will
always refer to the lab frame.  We will use ``low energy'' to mean the region
up to 170~MeV (ending just at the upper limit of the \OdL\ data),
``intermediate energy'' to mean 170--250~MeV (ending where the Wolf data set
starts) and ``high energy'' to mean 250--400~MeV (``high'' is to be understood
in a low-energy EFT context).  We have attempted to judge the consistency of
the world data sets by plotting them as a function of energy for a wide range
of angles, as seen in Figs.~\ref{fig:sect5-fig1} and \ref{fig:sect5-fig2} .
Initially we do not add any theory curves.  There is not space here to present
all the data in sufficient detail to appreciate all the points that can be
made; more plots can be found in the supplemental material available online.

In the low-energy region (see Fig.~\ref{fig:sect5-fig1}), we note that there
are no data below 45\deg\ and none above 155\deg\ (with the exception of the
two Zieger 180\deg\ points).  The only data below 55~MeV are from Federspiel.
From 55~MeV to around 120~MeV, multiple data sets contribute; in addition to
modern experiments (Federspiel, \OdL\ and MacGibbon), there are also a number
of older data sets which look compatible with the modern data.  From 120 to
170~MeV, the only data are from MacGibbon, \OdL\ and Hallin (with the
exception of some points from Pugh with large error bars).  Figure
\ref{fig:sect5-fig2} shows a selection of data over the whole energy range up
to 350~MeV (chosen to highlight kinematics where data exist from more than one
experiment).  From 170--210~MeV, Hallin is essentially the only data set, with
Bernardini contributing three points and two separate MAMI experiments one
point each (both with very large error bars).  Above 210~MeV, much more data
are available, both old and new, but the agreement is very problematic.  Above
250~MeV, the Wolf data set dominates statistically.

It has been recognised that the extensive data set of \OdL\ has larger
fluctuations than are compatible with the quoted statistical errors, and we
follow Wissmann \cite{Wissmann:2004} in adding a point-to-point systematic
error of $\pm5\%$ in quadrature with the statistical error when fitting this
set (not shown in plots)\footnote{For the other proton data sets, we include
  all systematic uncertainties in a single floating normalisation for each
  respective set.}. The \OdL\ 133\deg\ data tend to be low at energies around
100~MeV compared to other data, especially MacGibbon.  There is an apparent
discrepancy between the Baranov 150\deg\ and the \OdL\ 155\deg\ data, with the
latter lying high.

Three points from the modern low-energy data are clearly outliers: Federspiel
($135\deg$, 44~MeV), \OdL\ ($133\deg$, 108~MeV) and Hallin ($141\deg$ (cm),
170~MeV).  The Bernardini data set consists of 7 points in the range
120--200~MeV (in wide energy bins), and sufficiently many of these are low
compared with the trend of the rest of the data that we do not include this
set in our fits.  The Oxley data set is broadly consistent with other sets,
but the statistical errors are so tiny as to give it undue weight in any fit.
Though we continue to plot all of these, we do not use them in fits.

In the intermediate-energy region above about 200~MeV, the problems with
consistency of the data are most significant.  The lowest-energy Blanpied data
are noticeably high compared to any plausible smooth interpolation between the
low- and high-energy data; the Hallin data partially reinforce this trend. The
older Gray and Baranov data lie much lower in this region, while agreeing well
a little higher, and the few MAMI data points in this region are also lower,
though with large error bars.  That said, the 215~MeV Baranov data points at
91.9\deg\ and 107.2\deg\ (cm) are too low to be plausible.

In the high-energy region, the data are dominated by the Wolf data, and the
other MAMI data look compatible with it (although two of the Wissmann points
beyond the peak lie decidedly high).  A number of points are available from
older experiments and most are in good agreement, though as is well known, the
Genzel, Nagashima and Gray data sets all have one or more points which are
implausibly low.  The major problem, however, is the lack of agreement between
the Wolf and Blanpied data for angles above 100\deg, where the Blanpied data
sit substantially above the Wolf data.  At the worst point, the $\Delta(1232)$
peak for $\thetacm$ = 115\deg, the disagreement is almost 25\%.  But agreement
at more forward angles such as $\thetacm$ = 90\deg\ precludes an explanation
in terms of an overall normalisation error.  Viewed as a function of angle for
constant energy, both data sets show more structure than one might expect, and
it is hard to draw a conclusion about which is more internally consistent
based on the data alone.  The older data sets favour Wolf, while the few
Hallin points in this region favour or even exceed Blanpied.  This issue will
be revisited later when discussing the proton EFT fits in
Section~\ref{sec:protonanalysis}.

\subsection{\it Low-energy deuteron elastic Compton scattering}
\label{sec:expdeutlow}

Compton-scattering experiments that have sought to investigate the
electromagnetic polarisabilities of the neutron are considerably fewer in
number than those on the proton, primarily due to the lack of free neutron
targets.  Moreover, for the proton, the electric ($\alphaep$) and magnetic
($\betamp$) polarisabilities enter at order $\omega^2$ in the cross section
(where $\omega$ is the photon energy) due to an interference with the leading
Thomson amplitude.  For a ``free" neutron, there is no Thomson term (the
neutron is uncharged), so the polarisabilities enter at order $\omega^4$ and
are much harder to determine.

It is worthwhile to get a historical perspective on the subject of the neutron
polarisabilities before proceeding with the Compton discussion.  The first
experiments designed to investigate this subject used a totally different
technique.  Neutron scattering at very low energies ($E_n <$ 600 keV) in the
Coulomb field of a high-$Z$ target (such as $^{208}$Pb) is sensitive to the
electric polarisability of the projectile neutron.  The initial
experiments~\cite{Alexandrov:1986,Koester:1986} had very large error bars, but
the later experiments of Schmiedmayer et
al.~\cite{Schmiedmayer:1988,Schmiedmayer:1991} reported a rather precise
extraction of the neutron electric polarisability: $\alphaen = 12.0
\pm1.5(\mathrm{stat})\pm2.0(\mathrm{syst})$.  These results were challenged by
Koester et al.~\cite{Koester:1988,Koester:1995}, who obtained $\alphaen=0 \pm
5$, and by Enik et al.~\cite{Enik:1997}, who argued that
Schmiedmayer~\cite{Schmiedmayer:1991} had underreported the systematic errors
of the experiment and extracted $\alphaen=7$--$19$ using the Schmiedmayer data
set.  A more recent measurement by Laptev et al.~\cite{Laptev:2002} also
raises questions about the result of Ref.~\cite{Schmiedmayer:1991}. This
debate has not been resolved, and here we can only say that the
neutron-scattering technique, although successful in determining
polarisabilities in more weakly bound systems, appears to have led to a weak
constraint on $\alphaen$, at best.

The failure of these attempts to measure the neutron electric polarisability
in a reliable way motivated alternate efforts to access the neutron
polarisabilities (both $\alphaen$ and $\betamn$). This led to the use of
Compton scattering on deuterium, the simplest nuclear target containing a
neutron, via either the quasi-free $\mathrm{d}(\gamma,\gamma'
\mathrm{n})\mathrm{p}$ reaction or the elastic-scattering
$\mathrm{d}(\gamma,\gamma)\mathrm{d}$ reaction.  To date, there have been
three quasi-free experiments---a pioneering experiment at
MAMI~\cite{Rose:1990a,Rose:1990b}, a later experiment at SAL~\cite{Kolb:2000}
and the most recent one at MAMI~\cite{Kossert:2002ws}.  The last is the most
precise, but, even so, the value of $\alphaen$ was obtained to no better than
30\% when the statistical, systematic and model errors are combined linearly.
This emphasises the difficulty of the quasi-free measurements.  By contrast,
for elastic scattering on deuterium, the Thomson term is recovered, so the
polarisability extraction is similar (in principle) to the proton case, except
for the fact that only the \emph{sum} of the proton and neutron
polarisabilities ($\alphaep + \alphaen$ and $\betamp + \betamn$) can be
unambiguously deduced from the data.  However, a major experimental challenge
must be confronted due to the fact that the deuteron breakup channel is only
separated from the elastic channel by $2.225\;\MeV$ (plus recoil).  As a
result of this stringent requirement, only four measurements of the
$\mathrm{d}(\gamma,\gamma)\mathrm{d}$ reaction have been performed to date.
These experiments are reviewed below, followed by the quasi-free experiments
in Section~\ref{sec:expdeutqf}.

The first elastic-scattering experiment was performed at Illinois by
Lucas~\cite{Lucas:1994}, using an identical setup to the experiment of
Federspiel~\cite{Federspiel:1991} on the proton.  Two large 25.4~cm $\times$
25.4~cm NaI detectors were used to detect scattered photons at four lab angles
(50\deg, 75\deg, 110\deg, 140\deg) at an incident photon energy of $\wlab$ =
49 MeV, as well as at two scattering angles (60\deg\ and 135\deg) at a photon
energy of $\wlab$ = 69 MeV.  To extract the elastic-scattering cross sections,
the inelastic contributions to the measured scattering spectra were generated
using the Impulse Approximation, and then the sum of elastic and inelastic
contributions was fitted to the individual spectra.  The statistical
uncertainties for the deduced elastic cross sections were quite good
($\pm$4--13\%).  However, one drawback is that these statistics were obtained
at the cost of summing all of the tagger channels in the entire tagger focal
plane for each energy point, so that each data point constitutes an average
photon-energy bin width of 6.5 or 7.7 MeV, respectively. This is a rather wide
energy bin, although if the cross section is linear with energy, there are no
serious complications when comparing the data to theory.  Systematic
uncertainties were in the range $\pm$3.6--4.0\% for the data, primarily due to
\textit{correlated} factors such as photon flux ($\pm$1\%), detector
acceptance ($\pm$1--2\%), target thickness ($\pm$2\%), possible contamination
of the yield due to inelastic contributions ($\pm$1--2\%), and simulation
uncertainties ($\pm$1\%), leading to a \textit{correlated} systematic
uncertainty of $\pm$3\%.  The remaining systematic uncertainties, at the level
of $\pm$2.0--2.5\%, were due to subtraction of randoms and empty-target
contributions, giving an overall total of $\pm$3.6--4.0\%.

The experiment of Lundin et al.~\cite{Lundin:2003} was conducted at Lund and
was very similar in energy and angle to the Lucas
experiment~\cite{Lucas:1994}.  Three large NaI detectors were used at two
``nominal" photon energies ($\wlab$ = 55 and 66 MeV) and at three ``nominal"
lab angles (45\deg, 125\deg, 135\deg).  In this case, however, many separate
runs were taken over an extended period of time, resulting in multiple
independent data points which more or less overlap.  There were 3 runs at
54.6--55.9 MeV and 3 runs at 65.3--67.0 MeV, where the actual photon energy
varied slightly over those ranges.  In each of the 6 runs, the detector angles
also varied slightly from the nominal values, but the measured angles
basically form two clusters---a forward cluster near 45\deg\ and a backward
cluster near 130\deg.  In the end, the total data set comprised 18 data
points, although the overall kinematic coverage is actually rather limited due
to the overlap of the various points.  For this experiment, the inelastic
contribution from deuteron breakup was not simulated, but rather it was
excluded (to a large extent) from the elastic strength by setting a tight
summing window in the scattered $\gamma$-ray spectra.  The statistical
uncertainties of the extracted cross sections are in the range $\pm$(8--24)\%,
which are roughly double those of Lucas~\cite{Lucas:1994}.  In addition, the
energy width of each point was fairly broad (10 MeV wide) due to summing all
tagger channels in the entire tagger focal plane to increase the statistics.
Systematic uncertainties due to photon flux ($\pm$5\%), detector acceptance
($\pm$4\%), target thickness ($\pm$2\%), and possible contamination of the
yield due to inelastic contributions ($\pm$3\%) were \textit{correlated} among
all the data points, for a total of $\pm$7.5\%.  In addition, background
subtraction gave an uncorrelated systematic error of at most $\pm$4\%, leading
to an overall reported systematic uncertainty of $\pm$7--14\% for the measured
data.  We note that the lower limit of this range ($\pm$7\%) is slightly
inconsistent with the overall correlated systematic uncertainty stated above
($\pm$7.5\%); we cannot explain this discrepancy, although there are indeed
several data points in Table I of Ref.~\cite{Lundin:2003} which have
systematic errors below $\pm$7.5\%.

Both of the previous experiments spanned an energy range which is only
moderately sensitive to the polarisabilities.  The experiment of Hornidge et
al.~\cite{Hornidge:2000} pushed the energy higher, where the sensitivity
increases.  This experiment was performed at SAL using the large-volume BUNI
detector.  In many respects, this experiment was similar to the Hallin
experiment on the proton~\cite{Hallin:1993}, except that Hornidge used tagged
photons at energies $\wlab$ = 84--105 MeV and measured 5 angles sequentially
in the lab angular range 35\deg--150\deg.  The BUNI detector was crucial for
separating the elastic-scattering peak from the inelastic contributions which
are only 2.2 MeV apart.  With a scattered photon energy of $\sim$100 MeV, a
large-volume NaI detector with an energy resolution better than $\Delta E/E
\sim$ 2\% was required to achieve this separation.  With the use of a tight
summing region in the scattering spectra, the BUNI resolution ensured that the
elastic-scattering contributions could be cleanly determined.  The resulting
statistical uncertainties for the 5 data points are fairly good
($\pm$5--10\%).  One issue, however, as seen before, is that the full focal
plane was added together to improve the statistics, and, as such, the data
points each constitute a broad energy bin with a width of 21 MeV.  Systematics
due to photon flux ($\pm$1\%), solid angle ($\pm$1.6\%), detection efficiency
($\pm$3.6\%), and target thickness ($\pm$2.5\%) were \textit{correlated} among
the data points for a total of $\pm$5\%.  In addition, uncertainties due to
energy calibration ($\pm$1--5\%) give a range of $\pm$(5--7)\% for the overall
systematic uncertainty of the data.

A potential issue in the analysis of the deuteron elastic-scattering data, as
mentioned above, is that the widths of the energy bins themselves are often
non-negligible, e.g. the data of Ref.~\cite{Hornidge:2000} are combined into a
single photon-energy bin that is 21 MeV wide.  One might be concerned about
how the energy dependence of experimental acceptances and systematic errors
affect these data.  However, some reassurance is provided by the analysis of
Ref.~\cite{Hildebrandt:2004hh}, which showed that comparing the average cross
section over a wide experimental bin with the point theoretical cross section
at the centre of the bin, as done in theoretical analyses thus far, is
entirely reasonable.  In fact, the difference between carefully averaging over
the energy bin and simply taking the cross section at the central energy is
less than $1$\%, even for the SAL bin widths of 21 MeV, and nearly an order of
magnitude smaller for the other experiments; see also~\cite{Griesshammer:2012}
for more details.

As in the proton case, the experimental systematic errors have been separated
into two pieces: (1) point-to-point contributions which have been added in
quadrature to the statistical error, and (2) overall scaling contributions for
the whole data set which have been subsumed into a floating normalisation, see
Eq.~\eqref{eq:chisqare} in Section~\ref{sec:protonanalysis}.  We have checked
that these separate contributions to the systematic error combine to give the
overall stated systematic error of the data points.  In cases where a
discrepancy exists, as for some of the Lund data~\cite{Lundin:2003}, the
point-to-point contributions have been taken to be zero.

A summary of all of the deuteron experiments is included in
Table~\ref{table-deuteron-low}.  Overall, the three published
experiments~\cite{Lucas:1994,Lundin:2003,Hornidge:2000} provide a total of 29
data points in essentially 4 energy bins, centred at lab energies between $49$
and $95\;\MeV$.  The angular coverage is rather limited and overlaps between
experiments, with only $2$ to $5$ angles per energy bin. To a large extent,
this is attributed to the difficulties in clearly differentiating between
elastic and inelastic (deuteron breakup) Compton events, which are only
separated by the small deuteron binding energy.
\begin{table}[!htbp]
  \begin{center}
    \begin{tabular}{|l|l|c|c|c|c|c|c|}
      \hline
      Legend & First Author & $\wlab$ (MeV) & $\thetalab$ (deg) & Ref. & Reaction & $N_{\rm data}$ & Symbol\\
      \hline
      Illinois 94 & Lucas & 49, 69 & 50--140 & \cite{Lucas:1994} & elastic & 6 & \includegraphics*[width=7pt]{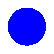}\\
      SAL 00 & Hornidge & 94 & 35--150 & \cite{Hornidge:2000} & elastic & 5 & \includegraphics*[width=7pt]{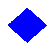}\\
      Lund 03 & Lundin & 55, 66 & 45, 125, 135 & \cite{Lundin:2003} & elastic & 18 & \includegraphics*[width=7pt]{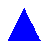}\\
      Lund 10 & Myers & 67--116 & 60, 120, 150 & \cite{Myers:2010} & elastic & 8 &\\
      Lund 12 & Shoniyozov & 81--116 & 60,90,120,150 & \cite{Shoniyozov:2011} & elastic &  & \\
      \hline
      MAMI 90 & Rose & 80--130 & 90, 135 & \cite{Rose:1990a,Rose:1990b} & quasi-free & 2& \\
      SAL 00 QF & Kolb & 247 & 135 & \cite{Kolb:2000} & quasi-free & 1 &\\
      MAMI 02 & Kossert  & 211--377 & 136 & \cite{Kossert:2002,Kossert:2002ws} & quasi-free & 9& \\
      \hline
    \end{tabular}
  \end{center}
  \caption{\label{table-deuteron-low} Compton-scattering experiments on the deuteron, including elastic and quasi-free scattering. 
    The last column indicates the corresponding symbols used in the figures for elastic scattering.  
    The most recent Lund data (2010 and 2012) were not yet available to be included in these figures.}
\end{table} 
Experimental statistical and systematic errors range over $\pm5$ to $\pm24\%$
and $\pm4$ to $\pm14\%$, respectively.  At the present time, these are the
data that have been included in global EFT fits aimed at extracting the
neutron polarisabilities. We note that this limited deuteron database contains
roughly 10\% of the volume of data available for the proton, where the latter
extends far into the $\Delta$(1232) resonance region and generally has
substantially better energy resolution and angular coverage, as well as lower
statistical and systematic uncertainties.

There is some promise, however, that the database will be expanded.  A more
recent experimental programme has been conducted at
Lund~\cite{Myers:2010,Shoniyozov:2011} in an energy range higher than the
earlier Lundin data and comparable to that of Hornidge.  The tagged energy
range was broader than any previous experiment ($\wlab$ = 67--116 MeV) and
covered four lab angles (60\deg, 90\deg, 120\deg\ and 150\deg).  A key
improvement, moreover, is that the measurements were performed using three
large-volume high-resolution NaI detectors simultaneously in a single
configuration---the BUNI detector already described, the CATS detector which
had been transported from MAMI to Lund for this purpose, and the DIANA
detector (60~cm diameter $\times$ 50~cm long) from the University of Kentucky.
This enabled three angles to be measured simultaneously in a single run
period.  It also facilitated a cross-check on systematics by enabling the
measurement of repeat angles with different NaI detectors over the various run
periods.  These data are still being finalised, and so there are no results to
be shown at the time of this review.

\subsection{\it Deuteron quasi-free Compton scattering}
\label{sec:expdeutqf}

Compton scattering on a ``free" neutron is the objective of the quasi-free
reaction $\mathrm{d}(\gamma,\gamma' \mathrm{n})\mathrm{p}$ in which a
kinematic configuration is chosen such that the proton is predominantly a
spectator, and the scattered photon (backward direction) and recoil neutron
(forward direction) are detected in coincidence.  Since this is a breakup
process, it has not yet been treated in Effective Field Theory, although it is
now feasible to do so and work is beginning along these lines.  Nevertheless,
a brief review of quasi-free experiments to date is appropriate, since it so
closely relates to the objective of obtaining precise information about the
neutron polarisability.

The pioneering experiment was performed by Rose et
al.~\cite{Rose:1990a,Rose:1990b} at MAMI using a 130 MeV brems\-strahlung
photon beam.  Two 25.4~cm $\times$ 35.6~cm NaI detectors were located at
135\deg\ and 90\deg\ on one side of the beamline and four plastic
scintillators were located at 22\deg, 31\deg, 39\deg\ and 49\deg\ on the other
side.  The data were summed over a wide energy range, from 80 MeV up to the
endpoint.  Since the endpoint was below pion threshold, all coincidence events
were unambiguously from Compton scattering (i.e. no contamination from $\pi^0$
production).  Due to poor statistics and the low energy range of the
measurement, only an upper limit could be established for the neutron electric
polarisability $\alphaen$.  Using a dispersion theory analysis, this
experiment gave results of $\alphaen = 10.7^{+3.3}_{-10.7}$ and $\betamn =
5.3^{+10.7}_{-3.3}$ for the neutron, where a sum-rule value of $\alphaen +
\betamn$ = 16 has been assumed.  We note that this was the first experimental
determination (1990) of the neutron polarisability via Compton scattering,
since it precedes the initial elastic-scattering experiment of Lucas (1994) by
4 years.

At the low energies of the above experiment, there is a sizable model
dependence in the polarisability values extracted from the dispersion
analysis.  Levchuk et al.~\cite{Levchuk:1994,Levchuk:2000} determined that the
model dependence could be minimised (and the sensitivity to polarisability
maximised) by moving to higher energies like 200--300 MeV.  At SAL, this was
undertaken by Kolb et al.~\cite{Kolb:2000} using tagged photons at an average
energy of 247 MeV (ranging over 236--260 MeV).  Scattered photons were
measured at 135\deg\ in the large-volume BUNI detector on one side of the
beamline and the recoil neutrons were detected in an 85-cell
liquid-scintillator array centred at 20\deg\ on the other side (note that this
is much more forward than Rose~\cite{Rose:1990b}).  An important consistency
check was performed in this experiment by simultaneously measuring the
quasi-free scattering reaction on the proton using the same setup and
comparing those results to the known case of the free proton.  For the
extraction of the neutron quasi-free cross section at these energies, a
simulation of the significant $\pi^0$ background is absolutely necessary.
This is rendered more difficult by the smearing of the Compton and pion events
due to Fermi motion of the bound nucleon in deuterium.  In the end, this
experiment yielded an independent lower limit for the neutron electric
polarisability $\alphaen$.  Combining this lower limit with the upper limit
from Rose, the final values obtained from a dispersion theory analysis gave
$\alphaen = 13.6^{+0.4}_{-6.0}$ and $\betamn = 1.6^{+6.0}_{-0.4}$ for the
neutron, based on a sum-rule value of $\alphaen + \betamn =
15.2$~\cite{Levchuk:2000}.

The most definitive quasi-free scattering experiment was performed at MAMI by
Kossert et al.~\cite{Kossert:2002,Kossert:2002ws} using the large-volume CATS
detector at 136\deg\ and a 30-cell liquid-scintillator array at 21\deg\ on the
opposite side of the beamline.  The tagged-photon energy range was 200--400
MeV.  This experiment was very similar to the SAL experiment~\cite{Kolb:2000}
in design, but it produced 9 data points over this energy range, as compared
to a single SAL point at 247 MeV.  This data set allowed much tighter
constraints to be placed on the extracted polarisability values for the
neutron.  The results from the Kossert experiment were $\alphaen$ = 12.5
$\pm$1.8(stat)$^{+1.1}_{-0.6}$(syst)$\pm$1.1(theory) and $\betamn$ = 2.7
$\mp$1.8(stat)$^{+0.6}_{-1.1}$(syst)$\mp$1.1(theory) with the inclusion of the
sum-rule condition $\alphaen + \betamn = 15.2$.
The model used to obtain these numbers will be discussed in
Section~\ref{sec:gammadmodels}, where a critical assessment of the theory
error bar will be provided as well. In fact, these results for neutron
polarisabilities depended on assuming the model value of $\gammapin$ obtained
by L'vov et al.~\cite{L'vov:1996xd}. As pointed out by Levchuk et
al.~\cite{Levchuk:1994}, the quasi-free cross section in these kinematics is
rather sensitive to the backward spin polarisability of the neutron,
$\gammapin$. It is therefore reassuring that the experiment also afforded an
independent determination of this quantity. By allowing both
$\alphaen-\betamn$ and $\gammapin$ to vary as free parameters, a fit to the
quasi-free cross section yielded a value of $\gammapin= 58.6 \pm 4.0$
(including the $\pi^0$ pole contribution; cf.~Section~\ref{sec:overview}).
This is entirely consistent with the model value employed to obtain the
numbers quoted above.  The addition of $\gammapin$ as a free parameter in
fitting the cross section did not alter the extracted values of $\alphaen$ and
$\betamn$ at all, except to slightly increase the statistical error bar. Up to
the present, this remains the only experimental determination of the backward
spin polarisability of the neutron.



\section{Compton scattering from the nucleon}
\label{sec:gammap}

\subsection{\it Dispersion relations} 
\label{sec:drs}

In this section, we review dispersion-relation (DR) calculations which have
been used to extract proton polarisabilities from Compton-scattering data. For
a thorough review, the reader is referred to Ref.~\cite{Drechsel:2002ar}. A
short description of the DR approach can also be found in
Ref.~\cite{Pasquini:2011ek}.

To write down DRs, we must construct a complete set of amplitudes for Compton
scattering in accordance with the dictates of Lorentz covariance, parity and
time-reversal invariance. These amplitudes must be free of kinematical
singularities. The basis of
L'vov~\cite{Akhmedov:1981ct,Lvov:1980wp,L'vov:1996xd} has been employed in
Refs.~\cite{Drechsel:1999rf,Babusci:1998ww}. These amplitudes are linear
combinations of the $A_i$'s defined in Eq.~\eqref{eq:Tmatrix}, with
coefficients which vary with angle. The relationship can be derived from
Appendix A of Ref.~\cite{Babusci:1998ww}.  The L'vov amplitudes, here denoted
by $B_i$, are written as functions of Mandelstam $t$ and the variable $\nu$,
defined as $\nu=(s-u)/(4 \MN)=\wlab + t/(4\MN)$.  Under crossing symmetry,
they obey $B_i(-\nu,t)=B_i(\nu,t)$. Within the DR approach, the $B_i$'s are
constructed by adding the nucleon and pion-pole contributions to an integral
over the spectrum of intermediate excitations. Applying Cauchy's theorem (with
a suitable contour) to the function $B_i(\nu',t)/(\nu'-\nu-\ii \epsilon)$
yields
\begin{equation}
  \Re \; B_i(\nu,t)=B_i^\mathrm{Born}(\nu,t)+ \frac{2}{\pi} P \int\limits_{\nu_0}^\infty \rmd \nu' \nu' \frac{\Im_s B_i(\nu',t)}{\nu'^2-\nu^2}\;\;,
  \label{eq:unsubAi}
\end{equation}
with $B_i^\mathrm{Born}$ the contributions from the Born (nucleon- and
pion-pole) graphs and $P$ indicating a principal-value integral. This is an
unsubtracted dispersion relation which applies at the chosen value of $t$ {\it
  if} we may neglect the contribution from the semi-circle at infinity. $B_i$
can then be reconstructed from the Born terms and its imaginary part, $\Im_s
B_i$, which itself is $1/(2\ii)$ times the discontinuity across the s-channel
cut of the Compton process.  This imaginary part can thus be obtained from
nucleon-pion photoproduction amplitudes, and the integral starts at
$\nu_0=\mpi + (\mpi^2 + t/2)/(2\MN)$, which is the $\pi $N threshold in the
lab frame---the frame in which we choose to work throughout this sub-section.
Pion photoproduction is the lowest channel that contributes to the imaginary
part, since we neglect the process $\gamma \mathrm{p} \rightarrow \mathrm{e}^+
\mathrm{e}^- \gamma$p (Delbr\"uck scattering) and the effect of $\gamma$p
intermediate states, as these contributions to $\sigma_T$ are suppressed by
powers of $\alphaEM$ compared to the hadronic ones.

The dispersion relations for $t=0$ (i.e.\ $\theta=0$) are of particular
interest because the optical theorem states that the imaginary part of the
forward amplitude is equal to a kinematic factor times the total cross section
for photoabsorption. As discussed in Section~\ref{sec:overview}, only two
amplitudes ($A_1$ and $A_3$) survive for forward Compton scattering, and
$\sigma_T(\omega')$ does {\it not} fall off fast enough as $\omega'
\rightarrow \infty$ for $A_1$ to obey an unsubtracted dispersion relation such
as Eq.~\eqref{eq:unsubAi}. To deal with this, we replace $A_1(\wlab,z=1)$ with
the function $A_1(\wlab,1)-A_1(0,1)$ in the argument that led to
Eq.~\eqref{eq:unsubAi} and obtain the once-subtracted dispersion relation:
\begin{equation}
  \Re \; A_1(\wlab,1)=A_1(0,1) + \frac{2 \wlab^2}{\pi} P \int\limits_{\wpi}^\infty \rmd\omega' \frac{\sigma_T(\omega')}{\omega'^2 - \wlab^2}\;\;,
  \label{eq:fsub}
\end{equation}
with $\wpi=\mpi + \mpi^2/(2\MN)$.  In the case of $A_3$, however, Regge
arguments allow us to anticipate an unsubtracted dispersion relation,
\begin{equation}
  \Re \; A_3(\wlab,1)=\frac{\wlab}{\pi} P \int\limits_{\wpi}^\infty \rmd
  \omega' \omega' \frac{\sigma_{1/2}(\omega')-\sigma_{3/2}(\omega')}{\omega'^2
    - \wlab^2} \;\;, 
  \label{eq:gunsub}
\end{equation}
where $\sigma_{1/2}$ ($\sigma_{3/2}$) is the total cross section for a
photon-nucleon system of helicity 1/2 (3/2).

If the integrals in Eqs.~(\ref{eq:fsub}, \ref{eq:gunsub}) converge, then each
side may be expanded as a Taylor series about $\wlab=0$. Inserting the
low-energy expansions of Eq.~\eqref{eq:low-en-amps} on the left-hand side then
yields a variety of sum rules for the structure coefficients in those
expansions. For example, from
Eq.~\eqref{eq:fsub} we obtain the Baldin sum rule quoted in
Eq.~\eqref{eq:bothSR}. Meanwhile, equating terms of $\calO(\wlab)$ in
Eq.~\eqref{eq:gunsub} yields the famous sum rule of Gerasimov, Drell and Hearn
(GDH)~\cite{Gerasimov:1965et,Drell:1966jv} relating the square of
$\kappa^{(\mathrm{p})}$ to an integral over $\sigma_{1/2}-\sigma_{3/2}$. The
terms of $\calO(\wlab^3)$ yield the Forward Spin Polarisability (or
$\gamma_0$) sum rule~\cite{GellMann:1954kc,GellMann:1954db}, which was also
listed in Eq.~\eqref{eq:bothSR}.  Recently, Pasquini, Drechsel and Pedroni
took such arguments one order further, obtaining and evaluating the sum rule
for the ``higher-order forward spin polarisability",
$\bar{\gamma}_0$~\cite{Pasquini:2010zr}:
\begin{equation}
  \bar{\gamma}_0=\frac{1}{4 \pi^2} \int\limits_{\wpi}^\infty \rmd \omega'
  \frac{\sigma_{1/2}(\omega')-\sigma_{3/2}(\omega')}{\omega'^5}\;\;, 
  \label{eq:DPP}
\end{equation}
which is defined as the second term in the Taylor expansion of the dynamical
forward spin polarisability $\gamma_0(\omega)$ about $\omega=0$. These three
sum rules demonstrate that higher-order terms in the Taylor expansion of $A_1$
and $A_3$ about $\omega=0$ correspond to integrands with additional powers of
the excitation energy in the denominator. Consequently, higher-order
coefficients are increasingly dominated by near-threshold physics, and so
their dependence on assumptions regarding high-energy pieces of the integral
decreases markedly. In particular, the GDH sum rule receives approximately
10\% of its total value from contributions above $\sqrt{s} - \Mp=2$ GeV, while
less than 4\% of the Forward Spin Polarisability sum rule comes from $\sqrt{s}
- \Mp > 800$ MeV~\cite{Drechsel:2002ar}. One corollary is that
dispersion-relation results for $\gamma_0$ and (even more so) for
$\bar{\gamma_0}$ are quite stable against assumptions about the high-energy
behaviour. So, for instance, Ref.~\cite{Drechsel:2002ar} quoted
$\gammazerop=-1.01 \pm 0.08 {\rm (stat)} \pm 0.10 {\rm (syst)}$, which was
updated in Ref.~\cite{Pasquini:2010zr} to $\gammazerop=-0.90 \pm 0.08 {\rm
  (stat)} \pm 0.11 {\rm (syst)}$.  Ref.~\cite{Schumacher:2011gs} makes
additional model assumptions and found a marginally consistent result:
$\gammazerop=-0.58 \pm 0.20$.  Another corollary is that dispersion relations
for $A_1$ and $A_3$, upon which multiple subtractions have been made, predict
that the energy dependence of these functions is driven entirely by chiral
(i.e.  near-threshold) physics.

From such analyses we learn that unsubtracted dispersion relations do not hold
for the L'vov amplitudes $B_1$ and $B_2$ and that the unsubtracted dispersion
relation for the amplitude $B_3$ converges only slowly. It is therefore more
appropriate to consider once-subtracted dispersion relations at fixed
$t$~\cite{Drechsel:1999rf}:
\begin{equation}
  \Re \; \bar{B}_i(\nu,t)=\bar{B}_i(0,t) + \frac{2}{\pi} \nu^2 P
  \int\limits_{\nu_0}^\infty \rmd \nu' \nu' \frac{\Im_s
    B_i(\nu',t)}{\nu'(\nu'^2-\nu^2)}\;\;, 
  \label{eq:subAi}
\end{equation}
at least in the cases $i=1,2,3$. (Here the $\bar{B}_i$'s are the structure
parts, in analogy to the $\bar{A}_i$'s of Section~\ref{sec:foundations}.) In
order to obtain predictions from Eq.~\eqref{eq:subAi}, it is necessary to
provide the function $B_i(0,t)$ as input. A dispersion relation can also be
written for $\bar{B}_i(0,t)$:
\begin{equation}
  \bar{B}_i(0,t)=\bar{B}_i(0,0)+B_i^{\pi^0}(0,t)-B_i^{\pi^0}(0,0)+\frac{t}{\pi} \int\limits_{4 \mpi^2}^\infty \rmd t' \frac{\Im_t B_i(0,t')}{t'(t'-t)} + \frac{t}{\pi} \!\!\!\!\int\limits_{-\infty}^{-2 \mpi^2 - 4 M \mpi} \!\!\! \rmd t' \frac{\Im_t B_i(0,t')}{t'(t'-t)}\;\;. 
  \label{eq:tchannDR}
\end{equation}
The possible intermediate states in the $t$-channel yield cuts along the
positive $t$-axis, but as long as $t < 4 m_\mathrm{K}^2$, this is saturated by
$\pi \pi$ intermediate states. The main contribution which needs to be
calculated to evaluate the function $B_i(0,t)$ is thus $\gamma \gamma
\rightarrow \pi \pi \rightarrow \mathrm{N} \bar{\mathrm{N}}$. In
Ref.~\cite{Drechsel:1999rf}, the $\pi \pi \rightarrow \mathrm{N}
\bar{\mathrm{N}}$ amplitudes of Ref.~\cite{Hoehler} were employed, and a model
was constructed for $\gamma \gamma \rightarrow \pi \pi$.

Meanwhile, the $s$-channel integral in Eq.~\eqref{eq:subAi} is largely
saturated by $\pi $N intermediate states, which mitigates the need for
modelling of higher-energy physics in photoproduction. The $\pi $N
contribution is evaluated using a contemporary parameterisation of the $\gamma
\mathrm{p} \rightarrow \pi $N data.  Multi-pion intermediate states in the
$s$-channel integral are evaluated from data on the inelastic decay channels
of $\pi $N resonances, and their helicity structure is assumed to be the same
as that of the $\gamma \mathrm{p} \rightarrow \pi $N amplitudes. This
introduces some model dependence in the DR evaluation, and there is also model
dependence in the computation of the second, $u$-channel, integral in
Eq.~\eqref{eq:tchannDR}, since it requires the evaluation of amplitudes well
into the unphysical region. However, the overall uncertainty in the evaluation
of the DRs is very small since ``...subtracted dispersion relations are
essentially saturated at $\nu=0.4$ GeV.''~\cite{Drechsel:2002ar}. Indeed,
neglecting both the $u$-channel integral and the two-pion contributions
changes the cross sections by only 3-5\%~\cite{Drechsel:2002ar}.

The six subtraction constants $b_i \equiv \bar{B}_i(0,0)$ then remain as the
only free parameters in this approach. These are equal to linear combinations
of the six static dipole polarisabilities. In the results quoted below,
$b_4$--$b_6$ are assumed to be derivable from unsubtracted forward dispersion
relations, i.e. to obey:
\begin{equation}
  b_i=\frac{2}{\pi} \int\limits_{\wpi}^\infty \rmd\omega' \frac{\Im_s
    B_i(\omega',0)}{\omega'}\;\;\mbox{ with } \;\; i=4, 5, 6\;\;.
  \label{eq:biunsub}
\end{equation}
where $b_3$ is often fixed from the Baldin sum rule.  Therefore, at least
three of the six static polarisabilities are assumed to be calculable from
analyticity arguments, and only $\alphae$, $\betam$ and one spin
polarisability (typically chosen to be $\gamma_\pi$) are taken as free
parameters in the fit. The formalism does not, however, demand that
Eq.~\eqref{eq:biunsub} be invoked. All six $b_i$'s could be taken as fit
parameters, if sufficient data were available to determine them.

If one does not wish to take these parameters as input, one can continue to
work with an unsubtracted dispersion relation instead, provided the
Cauchy-theorem contour used in their construction is closed at a finite
radius, rather than at $\nu'=\infty$~\cite{L'vov:1996xd}.  This produces a
third contribution on the right-hand side of Eq.~\eqref{eq:unsubAi}, $B_i^{\rm
  as}$, which is a piece resulting from integrating along a finite semi-circle
of radius $\nu_{\rm max}$ in the complex plane. In Ref.~\cite{L'vov:1996xd}
and subsequent works (see Ref.~\cite{Schumacher:2005} for a review and
Refs.~\cite{Schumacher:2006cy,Schumacher:2007xr,Schumacher:2009xt,Schumacher:2011gs}
for more recent developments), this contribution is parameterised by the
exchange of a single particle in the $t$-channel, e.g. the $\sigma$ meson in
the case of $B_1$. Proceeding in this manner means that one is modelling the
high-energy behaviour of the Compton amplitudes, and so a significant set of
additional dynamical assumptions is introduced. The advantage of
once-subtracted fixed-$t$ dispersion relations in which the subtraction
function is fixed via Eq.~(\ref{eq:tchannDR}) is that they contain minimal
dynamical assumptions and are sensitive mainly to the well-constrained
low-energy dynamics of the $\pi $N system.

Such dispersion relations do, however, require extrapolation of the imaginary
parts into an unphysical region. In practice, this extrapolation is difficult
to control for large values of $t$. Hyperbolic DRs provide a related approach,
which, however, does not necessitate such an extrapolation.
In hyperbolic DRs, the integration paths are hyperbolae in the Mandelstam
plane which correspond to a fixed value of cm (or lab) scattering angle. Thus
they are often referred to as ``fixed-angle dispersion relations".  The
integrals in these calculations have a form similar to that found in
subtracted fixed-$t$ DRs, but additional kinematical factors are now present
in the integrand. These factors cure the difficulties with the extrapolation
of the integrand at large $t$, but lead to some problems with convergence at
small angles. The two types of DRs are complementary: fixed-$t$ DRs are good
at small angles, while fixed-angle dispersion relations work best at
$\theta_{\rm lab}=180^\circ$. We note that such backward DRs have been used in
their unsubtracted form to directly obtain sum rules for $\alphaep -
\betamp$~\cite{Holstein:1994tw} and $\gammapip$~\cite{L'vov:1998ez}. Results
for these quantities from other analogous evaluations at $\theta_{\rm
  lab}=180^\circ$ are given in Table~\ref{table-polcomp} below. The values
found for other angles within the domain of validity of the hyperbolic
dispersion relations differ by $\approx 10\%$ for $\alphaep - \betamp$ and
$\gammapip$ and by a few percent for other spin polarisabilities.

\begin{figure}[!htbp]
  \begin{center}
    \includegraphics*[width=0.8\linewidth]{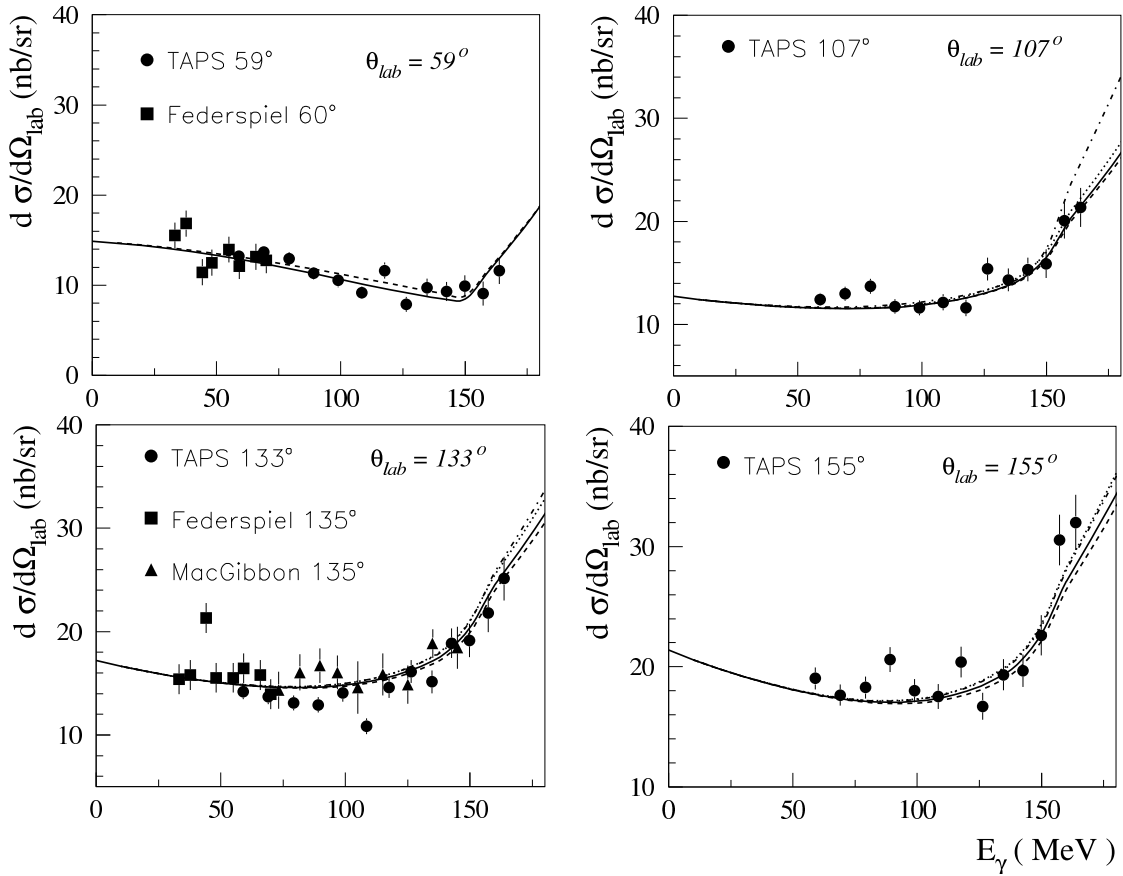}
  \end{center}
  \caption{\label{fig:DRs} Differential cross section for $\gamma$p scattering
    as a function of $\wlab$ at fixed lab angles. Experimental data are
    represented by the symbols, as indicated. Curves are for fixed-$t$
    subtracted/unsubtracted DRs (solid/dashed) and for fixed-angle
    subtracted/unsubtracted DRs (dotted/dash-dotted). All results are shown
    for $\alphaep-\betamp=13.8$, $\alphaep-\betamp=10.0$, $\gammapip=7$.
    Figure modified from Fig.~20 in Ref.~\cite{Drechsel:2002ar}. Used with
    permission.}
\end{figure}

Reassuringly, if the same photoproduction data are used as input for both the
hyperbolic and fixed-$t$ dispersion relations, then very good agreement is
obtained for the shape of the low-energy $\gamma$p data (see
Fig.~\ref{fig:DRs}). (The predictions for hyperbolic dispersion relations are
only shown for $\theta_{\rm lab} \geq 107^\circ$, as formally they do not
converge forward of this point.) Within their expected range of validity, it
is gratifying that all approaches agree at a level which is in accord with the
size of the experimental error bars. However, we caution that when these
curves are continued into the resonance region, more significant differences
are seen~\cite{Drechsel:2002ar,Pasquini:2007hf}.

Regardless, the good agreement in the vicinity of the pion threshold lends
strong support to the extraction of proton polarisabilities using dispersion
relations. Using fixed-$t$, subtracted dispersion relations, the results of a
fit to the data of
Refs.~\cite{MacGibbon:1995,Federspiel:1991,Olmos:2001,Zieger:1992} for
$\alphaep + \betamp$ and $\alphaep - \betamp$ are shown in
Table~\ref{table-polcomp} in Section~\ref{sec:comp}. The fit also determines
$\gammapip=10.4 \pm 1.8 \pm 3.2$ and has a $\chi^2/{\rm d.o.f.}=1.2$. In this
global analysis, the normalisation of each data set was allowed to float (see
Eq.~\eqref{eq:chisqare} below) within an assigned uncertainty of 3\%. The
first error bar is obtained by fitting with fixed values of the normalisation
constants, while the second represents the systematic uncertainty due to these
normalisation effects.


\subsection{\it Chiral EFT for $\gamma \mathrm{p}$ and $\gamma \mathrm{n}$
  scattering} 
\label{sec:singleN}

\subsubsection{Principles of \ChiEFT}
\label{sec:principlesEFT}

There are many excellent reviews of chiral EFTs, and we will give only a
summary here. For those interested in learning more, the most comprehensive
introduction to the subject are the lectures by Scherer and his subsequent
book with Schindler~\cite{Scherer:2002tk,Scherer:2012}. Recent reviews include
Bijnens \cite{Bijnens:2006zp} for the meson sector and Bernard
\cite{Bernard:2007zu} for the baryon sector; see also references therein.
The classic review of
Bernard et~al.~\cite{Bernard:1995dp} is still quite extensively used.  Reviews
covering the few-nucleon sector include those of Beane
et~al.~\cite{seattle_review}, Bedaque and van Kolck \cite{Bedaque:2002mn},
Epelbaum et~al.~\cite{Epelbaum:2008ga}, and Machleidt and Entem
\cite{Machleidt:2011zz}. Phillips' review specifically covers the applications to electromagnetic
reactions on nucleons and nuclei \cite{Phillips:2009}. 

We start by briefly reviewing \ChPT, in which the dominant dynamics is
directly linked to the chiral symmetry of QCD.  Subsequently, we will
introduce the $\Delta(1232)$ as a dynamical degree of freedom in a more
general chiral EFT.

Pionic \ChPT{} is a mature subject with a voluminous literature, but most of
the details are scarcely relevant here.  The Lagrangian initially includes all
allowed terms with either two derivatives or one factor of $\mpi^2$,
reflecting the fact that soft pions are non-interacting in the chiral limit.
The resulting ${\cal L}_\pi^{(2)}$---see Eq.~\eqref{eq:Lpipi2}---contains one
term of each type and generates the free-pion propagator away from the chiral
limit.  Writing ${\cal L}_\pi^{(2)}$ in a chirally symmetric way generates
non-vanishing amplitudes for 4, 6, $\ldots$ pions interacting at a point,
which are predictive at this order.  These reproduce Weinberg's low-energy
theorem for $\pi\pi$ scattering~\cite{Wei66} and give lowest-order predictions
for the scattering lengths which are within 25\% of currently accepted values.
At next-to-leading order (NLO, $\calO(P^4)$) there are contributions both from
terms in the fourth-order Lagrangian $\calL^{(4)}$ (with, e.g., four
derivatives) and also from one-loop diagrams such as a single rescattering; at
next-to-next-to leading order (\NsqLO) there are contributions from
$\calL^{(6)}$, from two-loop diagrams and also from one-loop diagrams in which
one of the vertices is taken from $\calL^{(4)}$ \cite{Gasser:1983yg}. Of
course, beyond ${\cal L}^{(2)}$ unknown LECs enter, and various strategies
exist.  One can fit to data and excellent agreement with the scattering
amplitudes near threshold is
obtained~\cite{Colangelo:2001df,Caprini:2005an,Bijnens:2006zp}.
Alternatively, lattice QCD can now constrain the fourth-order
LECs~\cite{arXiv:1108.1380,Colangelo:2010et,arXiv:1111.3729}.  Whichever
approach is taken, $\pi\pi$ scattering is a textbook example of a
systematically improvable EFT calculation.

A chiral Lagrangian with nucleon fields coupled to pions and photons was first
investigated beyond tree level by Gasser et~al.~\cite{Gasser:1987rb}.  Since
the Dirac equation is linear in derivatives, terms with both odd and even
numbers of derivatives appear. It was immediately recognised that the simplest
possible loop contribution to the self-energy, from one $\pi\mathrm{N}$ loop
using dimensional regularisation, does not obey the power counting found in
the pionic case; rather than being suppressed by some specific power of
$\mpi/\lambdachi$ relative to the leading nucleon mass, it generates an entire
series of powers of $\mpi/\MN$, including a contribution $\sim M_N$.  This
effectively precludes the possibility of a systematic calculation in which all
omitted diagrams contribute at a higher order in $\mpi/\lambdachi$ than the
desired order.

In subsequent years, methods to tame this problem and restore power counting
to the theory with Dirac nucleons were developed.  The first solution which
was proposed, and the one which has been used in most work on Compton
scattering, is called Heavy-Baryon \ChPT{} (HB\ChPT)
\cite{Jenkins:1990jv,Jenkins:1991ne,Bernard:1995dp}. Other approaches will be
mentioned in Section~\ref{sec:includeDelta2}.  In the framework of HB\ChPT,
the nucleon mass is recognised to be of the same magnitude as $\lambdachi$,
and so the free Dirac Lagrangian contains terms which contribute at different
chiral orders in low-energy processes: specifically the mass and energy are
zeroth order while the momentum is first order (i.e.~of order $\mpi$).  By
decomposing the nucleon field into ``large" and ``small" components (upper and
lower components in the nucleon rest frame) and by integrating out the
``small" component, one is left with a theory of Pauli spinors.  As guaranteed
by the relativistic starting point, Lorentz invariance is maintained in the
new Lagrangian, but perturbatively, with corrections of order $(P/\MN)^{n+1}$
when working to $n$th order.  The mass and its contribution to the energy
cancel and disappear from the leading-order Lagrangian $\calL_{\pi
  \mathrm{N}}^{(1)}$---see Eq.~\eqref{eq:LpiN1}.  From
$\calL_{\pi\mathrm{N}}^{(2)}$ onward, terms appear which have been generated
by the heavy-baryon reduction, and so terms with a fixed coefficient
proportional to $1/\MN^{n-1}$ at $n$th order occur, in addition to those with
an undetermined LEC.  In this theory, corrections to the bare nucleon mass
arise as expected at second order, with a contribution proportional to
$\mpi^2$ (or equivalently $m_q$) with the LEC $c_1$ as its coefficient (this
also gives the leading contribution to the pion-nucleon sigma term---see
Eq.~\eqref{eq:LpiN2}). Then, at third order, the loop diagram gives a finite
contribution proportional to $\mpi^3$ only.  (The contribution needs to be
finite, because terms in the Lagrangian can only involve integer powers of
$m_q$ or equivalently, even powers of $\mpi$.  Thus there can be no LEC to
cancel a divergence at $\mpi^3$.)  HB\ChPT{} does have a transparent power
counting which matches that in the mesonic theory.  The Lagrangian to third
order was developed in Refs.~\cite{kra90,eck94,em96,fms98}, and then to fourth
order in Refs.~\cite{mms00,fmms00}. 

One comment on power counting and electromagnetic processes: since a single
Lagrangian term with a gauged derivative can contribute either a power of
momentum or a factor of $|e|=\sqrt{4\pi\alphaEM}$, and since the magnitude of
$|e|$ and of $\mpi/\lambdachi$ are comparable, they will be considered
interchangeable in the perturbative expansion. This is just a matter of
bookkeeping provided the only photons in the process are external ones:
Compton scattering from the nucleon, for instance, will start with the Thomson
term at $\calO(P^2)$, and working to \NsqLO{} will involve terms from the
fourth-order Lagrangian.  Photon loops can also be included; their effects
will generally be of the same size as strong isospin-splitting effects at the
appropriate order, and both have so far been ignored in Compton scattering.

\subsubsection{The Lagrangian for Compton scattering in \ChiEFT}
\label{sec:comptonEFT}

\newcommand{\piN}{\pi\mathrm{N}} Here we discuss the relevant terms in the
Lagrangian for the construction of the Compton scattering amplitude to fourth
order in HB\ChPT.  The full Lagrangian needs to be written in terms of
building blocks with appropriate chiral properties, and hence every term can
give rise to interactions with multiple pions.  The usual notation is compact
but far from transparent. Below we retain only the relevant structures for our
purposes:
\begin{align} {\cal L}_\pi^{(2)}=&\textstyle{\frac{1}{2}}\,\partial_\mu
  {\boldsymbol\phi} \cdot \partial^\mu {\boldsymbol\phi}
  +eA_\mu\,\epsilon_{3ij}\,\phi_i\partial^\mu\phi_j
  +\textstyle{\frac{1}{2}}\,e^2A_\mu A^\mu(\phi_1^2+\phi_2^2)-\textstyle{\frac{1}{2}}\mpi^2{\boldsymbol\phi}^2+\ldots \label{eq:Lpipi2}\\
  {\cal L}_{\piN}^{(1)}=&\psi^\dagger( \ii v\cdot D+\ga u\cdot S )\psi \label{eq:LpiN1} \\
  {\cal L}_{\piN}^{(2)}=&\psi^\dagger\left\{\frac{1}{2\MN }\Bigl((v\cdot
    D)^2-D^2
    -\ii g\{S\cdot D,v\cdot u\}\Bigr)+4 c_1\mpi^2 \left(1-\frac 1 {2\fpi^2} \phi^2\right)\right. \notag \\
  &\left.+\left(c_2-\frac{\ga^2}{8\MN }\right)(v\cdot u)^2 +c_3 u\cdot u-\frac
    \ii {4\MN }[S^\mu,S^\nu]e F_{\mu\nu}\bigl((1+\kappa^{(\mathrm{s})})
    +(1+\kappa^{(\mathrm{v})})\tau_3\bigr)\right\}\psi+\ldots \label{eq:LpiN2} \\
  {\cal L}_{\piN}^{(4)}=&2 \pi e^2\psi^\dagger \Bigl\{ {\textstyle
    \frac{1}{2}}\left(\delta\beta^{(\mathrm{s})}
    +\delta\beta^{(\mathrm{v})}\tau_3\right)
  g_{\mu\nu}-\left((\delta\alpha^{(\mathrm{s})} +
    \delta\beta^{(\mathrm{s})})+(\delta\alpha^{(\mathrm{v})} +
    \delta\beta^{(\mathrm{v})})\tau_3\right) v_\mu v_\nu\Bigr\}F^{\mu\rho}
  F^{\nu}_{\; \; \rho}\psi +\ldots\label{eq:LpiN4} .
\end{align}
where $\psi$ is the nucleon field, $\phi_a$ are the pion fields, $F^{\mu\nu}$
is the electromagnetic field tensor and $D^\mu\equiv\partial^\mu-\ii e Q
A^\mu$ is the gauged derivative; $v^\mu=g^{0\mu}$ and $S^\mu=(0,\vec\sigma/2)$
in the rest frame of the nucleon. Since this is meant to be an expansion
around the chiral limit of QCD, the LECs written here as $\ga$, $\fpi$,
$\kappa$ etc.\ are actually ``bare" quantities which will differ from the
physical quantities by loop corrections which only vanish in the chiral limit.
This distinction is relevant at $\calO(P^4)$. The object $u_\mu$ is given by
\begin{equation}
  u_\mu=-\frac 1 \fpi (\tau_a\partial_\mu\phi_a+e\epsilon^{a 3b}\tau_a\phi_b
  A_\mu   +\ldots)\;\;.
\end{equation}
For the physical values, we use $\mpi^\pm= 139.6$~\MeV, $\fpi= 92.42$~\MeV,
$\MN=\Mp= 938.3$~\MeV, $\ga= 1.267$, $\kappa^{(\mathrm{s})}= -0.22$ and
$\kappa^{(\mathrm{v})}= 3.71$.  The neutral pion mass,
$m_{\pi^0}=134.98$~\MeV, is typically used in the $\pi^0$ pole diagram but not
elsewhere; other isospin-breaking effects are neglected.  The second-order
LECs are less well known.  They have been determined by fits to both $\pi$N
and NN scattering (since two-pion exchange is a significant part of the NN
force); we take the values of Bernard \cite{Bernard:2007zu}
$c_1=-0.9^{+0.2}_{-0.5}$, $c_2=3.3 \pm 0.2$, $ c_3=-4.7^{+1.2}_{-1.0}$, all in
$\GeV^{-1}$.

\subsubsection{Born terms}
\label{sec:EFT-Born}

\begin{figure}[!htb]
  \begin{center}
    \raisebox{0.25 cm}{(a)}
    \includegraphics*[width=0.25\linewidth]{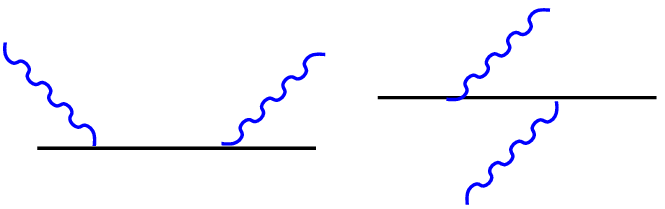}\qquad \raisebox{0.25
      cm}{(b)} \includegraphics*[width=0.1\linewidth]{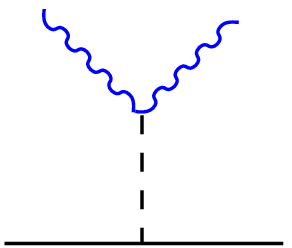}\qquad
    \raisebox{0.25 cm}{(c)}
    \includegraphics*[width=0.45\linewidth]{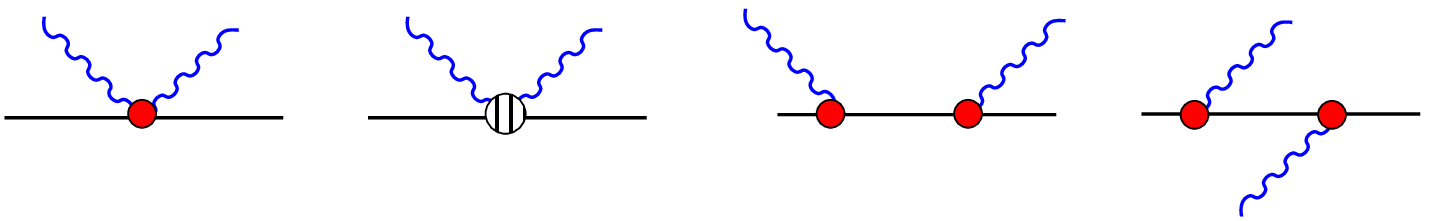}\qquad
    \caption {(Colour online) (a) Nucleon Born diagrams for Dirac nucleons;
      (b) Pion Born diagram; (c) Born diagrams in HB\ChPT up to $\calO(P^3)$:
      solid (red) dots indicate vertices from $\calL_{\piN}^{(2)}$ and open
      sliced dots from $\calL_{\piN}^{(3)}$.  In the counting schemes to be
      introduced in Sections \ref{sec:includeDelta} and
      \ref{sec:includeDelta2}, the first diagram of (c) counts as $\epsilon^2$
      and $e^2$, and (b) and the rest of (c) count as $\epsilon^3$ and
      $e^2\delta^2$.}
    \label{fig:Born}
  \end{center}
\end{figure}

We now discuss the HB$\chi$PT nucleon Compton amplitude in $\chi$PT. First, we
focus on the Born terms.  The low-energy theorems of Compton scattering
provide a simple example of how the same physics plays out in the Dirac and
heavy-baryon pictures; we refer to Eqs.~\eqref{eq:LET} and
\eqref{eq:low-en-amps}. In the Dirac picture, the Born (non-structure) terms
come from the two diagrams with an intermediate nucleon, with direct and
crossed photons (Fig.~\ref{fig:Born}(a)).  If the resulting amplitude is
expanded in powers of $1/\MN$, the Thomson term is reproduced in the amplitude
$A_1$; at one order higher, when the anomalous magnetic moment is included in
the photon coupling, the Born terms in $A_{3-6}$ appear.

In HB\ChPT, however, the leading (electric) photon-nucleon coupling is
sufficiently simple that the direct and crossed diagrams just cancel.  Indeed,
for real photons it is possible to work in a gauge in which this vertex is
actually absent (transverse or radiation gauge, corresponding to a purely
space-like photon polarisation vector $\vec k\cdot\vec \epsilon=0$ in the
nucleon's rest frame).  The HB$\chi$PT diagrams which do contribute to the
Born terms of Eq.~\eqref{eq:low-en-amps} up to $\calO(P^3)$ are shown in
Fig.~\ref{fig:Born}(c).  In transverse gauge, the leading $\gamma$NN vertex
comes from $\calL^{(2)}_{\pi\text{N}}$ and includes a magnetic coupling
proportional to the sum of the Dirac and anomalous magnetic moments. Thus, the
two right-most diagrams in Fig.~\ref{fig:Born}(c) give a third-order
contribution proportional to $1/\MN^2$, accounting for the $(Q+\kappa)^2$ Born
terms in $A_3$ to $A_5$ and also for the Born term in $A_6$.  However, the
missing terms in $A_1$ and $A_3$ come from photon-nucleon seagulls at second
and third order, respectively. In particular, the Thomson term comes from
gauging the leading kinetic term in the Lagrangian; as discussed above, it
counts as second order even though it has no powers of chiral momenta because
it is proportional to $\alphaEM$.  This demonstrates that the distinction
between Born and non-Born in HB\ChPT{} cannot be equated with (apparent)
one-particle reducibility or the lack thereof. It is always safest to define
Born terms via Dirac nucleons.

Of course, strictly speaking, in all these Born diagrams, it is the
chiral-limit magnetic moment, $\kappa^{(0)}$, which appears, not the
experimental value. The difference between $\kappa$ and $\kappa^{(0)}$ is
$\calO(m_\pi)$, and so this affects the amplitude only at fourth order. This
issue will, however, recur below. The $\pi^0$-pole diagram
Fig.~\ref{fig:Born}(b) also contributes to amplitudes $A_3$ to $A_6$ at third
order; its contribution is denoted by $A_i^{\pi^0}$ in Eq.~\eqref{eq:pi-pole}.
 
\subsubsection{Leading-order structure contributions and polarisabilities}
\label{sec:EFT-structure}

\begin{figure}[!htb]
  \begin{center}
    \includegraphics*[width=0.5\linewidth]{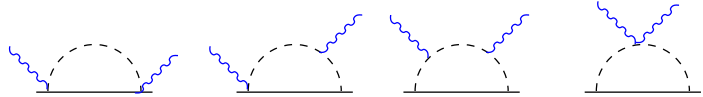}
    \caption {(Colour online) $\calO(P^3)$ loop diagrams in HB\ChPT; all
      orderings of vertices and crossed as well as direct photons are implied.
      Vertices (shown without dots) are all from the LO Lagrangian, that is,
      $\calL_{\piN}^{(1)}$ for the nucleonic coupling and $\calL_\pi^{(2)}$
      for the $\gamma\pi$ couplings.  These also count as $\epsilon^3$ and
      $e^2\delta^2$. }
    \label{fig:loops-3rd}
  \end{center}
\end{figure}

It is gratifying that the basic low-energy theorems of Eq.~\eqref{eq:LET} are
reproduced in this EFT, but our interest is in the predictions made by the
theory for the structure-dependent amplitudes, including the static
polarisabilities $\alphae$, $\betam$ and the $\gamma$'s. As just described,
the leading-order HB\ChPT\ Compton-scattering amplitude is simply the Thomson
term.  At NLO---$\calO(P^3)$---there are the spin-dependent Born contributions
described above, but there are also contributions from pion loops
\cite{bkkm92}, specifically the diagrams depicted in Fig.~\ref{fig:loops-3rd}.
Individually these diagrams are divergent and violate the LETs, but the sum is
finite and leaves the Born contributions intact.  Thus the sum of the loop
diagrams contributes only to the structure parts of the six amplitudes and
hence vanishes quadratically for $A_1$ and $A_2$ as $\omega\to 0$ and as the
third power of $\omega$ for $A_{3-6}$.  The coefficients of these terms are
the polarisabilities, and at this order they are the same for both the proton
and neutron. The results, first calculated by Bernard
et~al.~\cite{bkkm92,Bernard:1995dp}, are
\begin{equation}
  \alphae=10\betam=\frac{10\alphaEM\ga^2}{192\pi\mpi\fpi^2}=12.5\;\;,\;\;
  \gammaee=5\gammamm=-5\gammame=-5\gammaem=-\frac{5\alphaEM\ga^2}{96\pi^2\mpi^2\fpi^2}=-5.6.
  \label{eq:op3preds}
\end{equation}

It should be stressed that up to third order the full amplitudes, as well as
the polarisabilities, are entirely predicted in terms of the well-known
quantities $\mpi$, $\fpi$ and $\ga$; there are no free parameters.  Of course,
the best method to analyse experiments for extracting even $\alphae$ and
$\betam$ is the subject of this review, but nonetheless, the many attempts
made in the past to measure these quantities all come out close to these
values for both the proton and neutron; in particular, the order-of-magnitude
difference between $\alphae$ and $\betam$ and their nearly isoscalar nature is
not easily understood in most models.  This has long been lauded as a stunning
early success of HB\ChPT.  (As the spin polarisabilities are less well known,
it is harder to judge these predictions; see Section~\ref{sec:comp}.)

There are a number of caveats, however.  Even strictly within HB\ChPT, one
would expect higher-order corrections to be of order $P/\lambdachi$---around
20\% if the scale of the expansion were $\lambdachi\sim m_\rho$.  There is
also good reason to expect that for $\betam$ (as well as $\gammamm$), the
scale is actually set by the much smaller $\Delta$-nucleon mass difference
$\MDelta-\MN$.  Furthermore, in a relativistic framework, the predictions from
the diagrams in Fig.~\ref{fig:loops-3rd} are substantially smaller: $\alphaep=6.8$, $\betamp=-1.8$
\cite{Bernard:1991rq,Lensky:2009uv}.  But, before dismissing the success of
third-order HB$\chi$PT as a fluke, we should step back and remember that the
calculation gives us full amplitudes as a function of $\omega$, not merely the
static polarisabilities.  As will be shown in more detail subsequently, the
full third-order cross section extends the region in which data can be well
described substantially beyond that where the Petrun{}'kin cross section (Born
plus static scalar polarisabilities) is valid. In particular, it reproduces
the pronounced cusp at the photopion threshold which is seen at forward
scattering angles (see Fig.~\ref{fig:sect5-fig1}).  Beyond that point, the
data show a huge rise in the cross section which is obviously due to the
$\Delta(1232)$ (see Fig.~\ref{fig:sect5-fig2}), and one could not expect a
theory without the $\Delta(1232)$ to work in that region.

For completeness, we should mention that a handful of calculations of
polarisabilities have been done in the framework of SU(3)$\times$SU(3) chiral
perturbation theory, involving kaons as well as pions and all the octet
baryons.  Bernard et al.\ calculated the spin-independent static
polarisabilities in HB\ChPT\ \cite{Bernard:1992xi} and showed that for
nucleons the effect of kaon loops was small (see also Butler and Savage
\cite{Butler:1992ci}); Vijaya Kumar et al.\ found a similar result for
$\gammazero$ \cite{VijayaKumar:2011uw}.  Dynamical polarisabilities
$\alphae(\w)$ and $\betam(\w)$ have also been calculated at NLO in a covariant
framework by Aleksejevs and Barkanova \cite{Aleksejevs:2010zw}.

\subsubsection{Structure beyond leading order}
\label{sec:EFT-fourth}

\begin{figure}[!htb]
  \begin{center}
    \includegraphics*[width=0.95\linewidth]{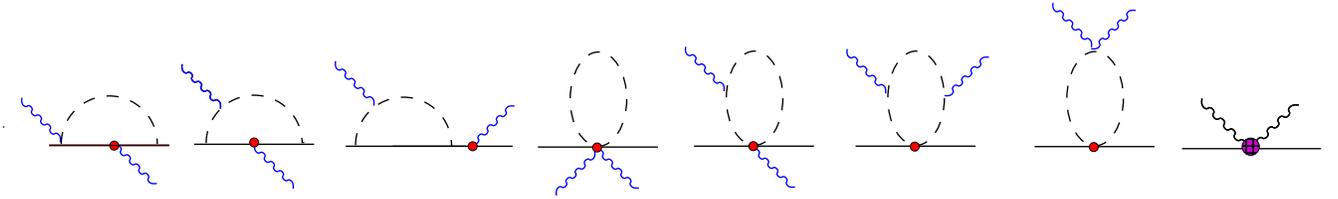}
    \caption{(Colour online) $\calO(P^4)$ diagrams in HB\ChPT; vertices
      labelled as in Figs~\ref{fig:Born} and \ref{fig:loops-3rd} with the
      addition of a (magenta) diced dot for the fourth-order counterterms
      $\delta\alphaep$ etc.\ of $\calL^{(4)}_{\piN}$.  All orderings of
      vertices and crossed as well as direct photons are implied.  Omitted are
      all diagrams obtained from those in Fig.~\ref{fig:loops-3rd} by
      substituting an NLO vertex or propagator for an LO one.  These also
      count as $\epsilon^4$ and $e^2\delta^4$, though the final diagram is
      included at one order lower if polarisabilities are fit.}
    \label{fig:loops-4th}
  \end{center}
\end{figure}

Although the ability of third-order HB\ChPT{} to qualitatively describe
low-energy data is encouraging, the lack of any free parameters limits its use
as a tool to extract more information from those data.  This situation changes
at fourth order, because at that order we can construct Lagrangian terms like
$\psi^\dagger F^{\mu\nu}F_{\mu\nu}\psi $ which are multiplied by new,
undetermined LECs.  Such terms give rise to photon-nucleon seagull diagrams
which contribute terms proportional to $\omega^2$ to the amplitudes $A_1$ and
$A_2$ \cite{bksm93}.  In the enumeration of Ref.~\cite{fmms00}, there are
actually six such terms (numbers 89-94) but in the photon-nucleon sector only
four independent combinations of LECs enter, which we can call
$\delta\alphaep$, $\delta\alphaen$, $\delta\betamp$ and $\delta\betamn$ (see
$\calL^{(4)}_{\piN}$, Eq.~\eqref{eq:LpiN4}, and Fig.~\ref{fig:loops-4th}).
These are contributions to the spin-independent polarisabilities of the proton
and neutron which come from non-chiral physics---for example, quark
substructure, or resonances, according to perspective, and they obviously
encode the leading effects of a $\Delta(1232)$ pole.  In addition, at fourth
order a new set of $\pi$N diagrams has to be included.  Finally, all the
\NsqLO{} terms in the expansion of the relativistic Born contributions to
$A_1$ and $A_2$ are also generated via fourth-order seagulls and diagrams like
those of Fig.~\ref{fig:loops-4th} with either one vertex taken from
$\calL^{(2)}_{\pi\text{N}}$ or with an NLO nucleon propagator.

Of the loop diagrams, many are $1/\MN$ corrections to the diagrams of
Fig.~\ref{fig:loops-3rd} (no new LECs enter in these).  However, there are
also two new types of diagrams---those with magnetic-moment couplings as well
as a pion loop, and those with a pion-nucleon seagull, as shown in
Fig.~\ref{fig:loops-4th}.  In the former, the only new LECs are the well-known
proton and neutron anomalous magnetic moments.  In the latter, however, the
three $\pi$N LECs $c_1$, $c_2$ and $c_3$ enter.

This time, the sum of all loop diagrams does make a $\calO(\w)$ contribution
to the Born terms.  The contributions are exactly those which are needed to
replace the chiral-limit $\kappa^{(0)}$ with the correction that shifts
$\kappa$ to its experimental value at this order.  In the expansion of the
$\gamma$N vertex, this shift comes from a diagram in which the photon couples
to a pion loop, as in the third diagram of Fig.~\ref{fig:loops-4th}, but in
Compton scattering this is not the only diagram which gives $\delta\kappa$
corrections to the Born term, nor is such a correction the only contribution
from this diagram~\cite{jko00a,kmb99,VijayaKumar:2000pv}.  The
$\calO(\omega^2)$ piece of the sum of all fourth-order loop diagrams produces
a logarithmically divergent result for the spin-independent polarisabilities.
These divergences are cancelled by the divergent parts of $\delta\alphaep$
etc. to leave a finite but undetermined total fourth-order contribution to the
spin-independent polarisabilities \cite{bksm93}.  By contrast, the
$\calO(\omega^3)$ parts of all these loop diagrams yield a finite shift for
the spin polarisabilities~ \cite{jko00a,kmb99,ghm00,VijayaKumar:2000pv,bjm01};
in particular the contribution of the one-nucleon-reducible diagrams in
Fig.~\ref{fig:loops-4th} may not be omitted~\cite{bjm01,Gellas:2001yv}.

There are actually a number of subtleties in the fourth-order calculation
which are not present at third order, one of which is the frame dependence.
Since the kinetic energy $p^2/(2\MN)$ of a nucleon is one chiral order higher
than that of a photon (which is the same as its three momentum and assumed to
be of order $\mpi$), at lowest order the incoming and outgoing photon momenta
are the same, and the same in any frame in which the heavy-baryon reduction is
valid.  That includes the lab, Breit and centre-of-mass (cm) frames.  Thus, in
the lowest-order ($\calO(P^3)$) loop calculation, we do not need to specify in
which frame we are working.  At $\calO(P^4)$, however, that is no longer the
case.  The Breit frame is the simplest in which to calculate and has the
additional merit over the commonly used cm frame that the amplitudes are
crossing-symmetric.  (Already at third order the cm frame violates this: the
Born terms, but not the loops, are different in the two frames.)  Furthermore,
in identifying the terms $\alphae^{(\textrm p)}\omega^2$ etc.\ in the
amplitudes, it is desirable that this is as close as possible to what is used
in other determinations of the polarisabilities. In the Petrun{}'kin cross
section, it is $\wlab\wlabout$ which multiplies the polarisabilities; in DR
analyses, it is $\nu^2$.  These differ from one another and from $\wBreit^2$
by terms of order $\w^2/\MN^2$ (which matters only at $\calO(P^5)$), whereas
$\wcm^2$ differs at $\calO(w/\MN)$.  Having obtained the Breit-frame
amplitudes, those in other frames can be obtained from a boost, expanded to
the appropriate order in $1/\MN$.  Whereas at $\calO(P^3)$ the cross section
differs noticeably depending on which frame is chosen, at $\calO(P^4)$ this
dependence is negligible.  (This frame dependence at $\calO(P^3)$ can be
confusing, given the lack of frame dependence in the \emph{form} of the
amplitudes.  But the variables in the two frames, ($\wcm$, $\thetacm$) and
($\wBreit$, $\thetaBreit$), are \emph{not} the same as functions of the lab
variables, see Section~\ref{sec:kinematics}.)

A more physical issue than the choice of frame, passed over in the discussion
of third order above, is the fact that any amplitude calculated to finite
order will put the photoproduction threshold at $\omega=\mpi$ (irrespective of
the frame in which $\omega$ is defined).  This is because the incoming nucleon
kinetic energy is included perturbatively and is not present in the nucleon
propagator in the $\pi$N loop.  Of course, within the heavy-baryon amplitude
there will be a string of terms which would be generated by the $1/\MN$
expansion of a loop in which the nucleon is allowed to recoil, and, as noted by
Bernard et~al.~\cite{bkms93}, the solution is to resum these terms to put the
threshold at the correct place.  In practice, that involves replacing $\omega$
within loop integrals by a new variable $\omega_s(\omega)$, which differs from
$\omega$ by terms of $\calO(\omega/\MN)$ and which equals $\mpi$ when
$\omega=\omega_\text{th}$, and then writing a Taylor expansion in powers of
$\omega_s/\MN$ and retaining only those terms required at the order to which
one is working.  At third order, no expansion is required, and there is simply
a substitution of variable; that has always been done before comparison with
experiment.  At fourth order, the third-order loop pieces do need to be
expanded, and indeed only with such a prescription are the amplitudes finite
at threshold \cite{bkms93}.  $\omega_s$ is not uniquely determined by this
prescription, but it is commonly taken to be $\omega_s=\sqrt s-\MN$.  A
different choice was made in Refs.~\cite{bkms93,mcg01,
  Beane:2002wn,Beane:2004ra}.

In Ref.~\cite{mcg01}, McGovern took the values of $\alphaep$ and $\betamp$
from the Particle Data Group (which were very close to the third-order values)
and made a comparison with all extant data below 180~\MeV.  The resulting
conclusion was that there was a modest improvement over the third order up to
around the photoproduction threshold, but no improvement beyond that point,
with particularly poor agreement at backward angles.  Of course, there is
still no $\Delta(1232)$, but the tadpole diagrams of Fig.~\ref{fig:loops-4th}
would be obtained from a theory in which an explicit $\Delta(1232)$ had been
integrated out to leave the pion seagull terms proportional to the LECs $c_2$
and $c_3$, so it was somewhat surprising that they did not extend the fit a
little further.  Of course, the biggest contribution from an explicit
$\Delta(1232)$ would be the pole diagram, and it is clear from comparison to
the data that by 200~MeV, replacing this with one term of a polynomial
expansion ($\delta\betamp\omega^2$ etc.) is quite inadequate (see
Fig.~\ref{fig:oldfits}).  In Ref.~\cite{Beane:2002wn,Beane:2004ra}, Beane et
al.\ instead fitted $\alphaep$ and $\betamp$ to the data in the same region
(with a cut on $|t|$ as well as $\omega$).  As will be discussed in
Section~\ref{sec:protonanalysis}, there are problems with the data set which
precluded a fit with an acceptable $\chi^2$.  Nevertheless, a fairly robust
conclusion emerged in which a higher value of $\betamp$ than that generally
accepted was required by the data---the central value quoted was $3.4$.  If
the Baldin sum rule was imposed, this value was decreased to $2.8$ (see
Table~\ref{table-polcomp} in Section~\ref{sec:comp} for detailed results).

\subsubsection{Including the $\Delta(1232)$ in the small-scale expansion}
\label{sec:includeDelta}
\newcommand{\DeltaM}{\Delta_M}

The $\Delta(1232)$ resonance has long been recognised as hugely important in
the physics of nucleons.  In \ChPT{} it is not explicitly present, but its
influence is felt through LECs such as $c_{2,3}$ and $\delta\betamp$.  But, as
the radius of convergence of an EFT is set by the scale of the lowest degree
of freedom which has not been included, the $\Delta(1232)$ can be expected to
severely restrict the applicability of \ChPT\, at least in those processes in
which it contributes, with the convergence governed by the scale
$\DeltaM\equiv\MDelta-\MN$.  And any glance at Compton-scattering data above
200~MeV, as in Fig.~\ref{fig:sect5-fig2}, confirms that this is such a
process.

In fact, since $\DeltaM\approx 2\mpi$, it could be argued that these two
scales are similar and should be included on the same footing.  This is the
basis of the so-called ``small-scale" or ``$\epsilon$" expansion developed by
Hemmert et~al.~\cite{Hemmert:1996xg,Hemmert:1997ye} (see also the earlier work
of Butler et al.~\cite{Butler:1992ci,Butler:1992pn}). Explicit $\Delta(1232)$
fields are included in the Lagrangian, as first shown by Manohar and Jenkins
\cite{Jenkins:1991ne}, and $\DeltaM/\lambdachi$ is counted like
$p_{\text{typ}}/\lambdachi$ and $\mpi/\lambdachi$ in determining the order of
a diagram.  The heavy-baryon expansion is used, ensuring that only $\DeltaM$
and not $\MDelta$ appears.

The relevant terms in the Lagrangian are the following:
\begin{align}
  {\cal L}_{\Delta}^{(1)}=&(\Delta^i_\nu)^\dagger\left(-\ii v\cdot D+\DeltaM\right){\Delta^{i\nu}}\\
  {\cal L}_{\pi \mathrm{N}\Delta}^{(1)}=&
  -\frac{g_{\pi \mathrm{N}\Delta}}{\fpi} (\psi^\dagger \partial^\nu \phi^i\Delta^i_\nu +(\Delta^i_\nu)^\dagger \partial^\nu \phi^i \psi+\ldots)\\
  {\cal L}_{\gamma \text{N}\Delta}^{(2)}&=\frac{-\ii e
    b_1}{\MN}\Big(\psi^\dagger S_\rho F^{\mu\rho}\Delta^3_\mu -
  (\Delta^3_\mu)^\dagger S_\rho F^{\mu\rho} \psi\Big)
  \label{eq:Delta-lagrangian}
\end{align}
where $\Delta^i_\nu$ is the heavy-baryon reduction of an $I=\frac 3 2$,
$S=\frac 3 2$ Rarita-Schwinger field $\Psi^i_\nu$, with $i$ and $\mu$ being
the indices on the (iso)spin-1 vector coupled to the (iso)spin-$\frac 1 2$
spinor.  The coupling constant $g_{\pi\mathrm{N}\Delta}$ is fit to the
$\Delta(1232)$ width and varies significantly depending on whether
relativistic or non-relativistic kinematics are used.  The transition magnetic
moment $b_1$ also has no single widely accepted value.  Both will be specified
when we use them later.\footnote{There is some confusion in the literature by
  what is meant by the magnetic coupling $b_1$.  The early papers of Hemmert,
  Holstein et~al.~\cite{Hemmert:1996xg,Hemmert:1997ye,hhk97,hhkk98} were not
  always consistent with one another and did not always specify which
  Lagrangian was being used, so that signs and factors of 2 come and go.
  Although it is not stated in the paper, Hildebrandt
  et~al.~\cite{Hildebrandt:2005ix} use the Lagrangian displayed above, which
  gives a contribution to $\betam$ of $2 \alphaEM b_1^2/(9\DeltaM\MN^2)$.}  It
is also worth displaying here the alternative form of ${\cal L}_{\gamma
  \text{N}\Delta}^{(2+3)}$ used in the $\delta$ expansion, see later (here
$\psi$ is the nucleon Dirac spinor and other notation is that of
Ref.~\cite{Pascalutsa:2002pi}):
\begin{equation} {\cal L}=\frac{3e}{2\MN(\MN+\MDelta)}\Big(\bar\psi (\ii\gM
  \tilde F^{\mu\nu}-\gE \gamma_5 F^{\mu\nu})\partial_\mu\Psi^3_\nu -
  \bar\Psi^3_\nu \overleftarrow{\partial}_\mu(\ii\gM \tilde
  F^{\mu\nu}-\gE\gamma_5 F^{\mu\nu})\psi\Big)\;\;.
  \label{eq:PP-Lagrangian}
\end{equation}
The leading (magnetic) term in the heavy-baryon reduction of this Lagrangian
is equivalent to the one above with the identification
\begin{equation}\label{eq:gm-b1}
  g_M= b_1(1+\MDelta/\MN)/3
\end{equation} 
(though it is important to note that if the full vertex is used, there are
substantial sub-leading terms), but there is also a sub-leading electric
coupling \gE.  The ratio of these two couplings (at the $\Delta(1232)$ pole)
can be obtained from the E2/M1 ratio to be $-0.34$
\cite{Pascalutsa:2005vq,Pascalutsa:2006up}.  Being third order, the electric
contribution to the amplitudes is suppressed by a power of $\omega/\MN$
relative to the magnetic one.  In the following, we will sometimes use $b_2$
for the electric coupling, defined via $g_E= b_2(1+\MDelta/\MN)/3$.

\begin{figure}[!htb]
  \begin{center}
    \includegraphics*[width=0.7\linewidth]{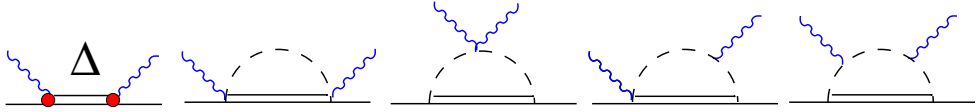}
    \caption {(Colour online) $\Delta(1232)$ contributions to Compton
      scattering at $\epsilon^3$ and $e^2\delta^3$ (see text for explanation).
      All orderings of vertices and crossed as well as direct photons are
      implied.}
    \label{fig:delta-contrib}
  \end{center}
\end{figure}

In the small-scale expansion, $\Delta(1232)$ propagators $(v\cdot
p-\Delta_M+\ii\eta)^{-1}\sim\epsilon^{-1}\sim \mpi^{-1}$ have the same
counting as nucleon ones $(v\cdot p+\ii\eta)^{-1}$, and so the diagrams of
Fig.~\ref{fig:delta-contrib} are also third order (denoted now
$\calO(\epsilon^3)$ rather than $\calO(P^3)$).  Those vertices of the theory
without external $\Delta$s are unchanged, except that the LECs $c_2$ and $c_3$
need to be retuned to exclude the $\Delta(1232)$ contribution; this is done by
subtracting the $\Delta$-pole contribution to soft $\pi$N scattering.  There
are significant complications in including spin-$\frac 3 2$ fields, not least
of which is the need to ensure that unphysical spin-$\frac 1 2$ degrees of
freedom do not propagate, but at leading order this is not an issue.  The
contribution of the $\Delta(1232)$ to the polarisabilities in this framework
was calculated in Refs.~\cite{hhk97,hhkk98} (see also
\cite{Butler:1992ci,Butler:1992pn}) and to the full Compton amplitudes in
Refs.~\cite{hhk97,Hildebrandt:2005ix,Hildebrandt:2003md}.  Of course, these
are dependent on new and rather poorly known coupling constants, namely the
$\pi$N$\Delta$ coupling constant and the $\gamma$N$\Delta$ transition magnetic
moment.

Long before Hemmert's work, though, it was known that the $\Delta$-pole
diagrams give a huge paramagnetic (isoscalar) contribution to $\betam$.  The
exact value depends on the parameter values, but Mukhopadhyay et~al.\
estimated $\delta\betam^\Delta=7.0$ \cite{Mukhopadhyay:1993zx}, and Hemmert
et~al.\ obtained values of 7.2--12~\cite{hhk97,hhkk98}.  In the latter work
and in Ref.~\cite{Butler:1992ci}, it was also shown that the $\pi\Delta$ loop
diagrams made a substantial contribution to $\alphae$.  With no counterterm to
adjust at third order, there was no possibility of agreement with data.  The
pragmatic solution, adopted by Hildebrandt
et~al. in Refs.~\cite{Hildebrandt:2003fm,Hildebrandt:2003md,Hildebrandt:2005ix}, was to promote the fourth-order isoscalar counterterms
$\delta\alphae$ and $\delta\betam$ to third order,
so that these unwanted large contributions to the polarisabilities could be
cancelled.  The prediction for the rest of the amplitudes remained intact, of
course.

In these works, the
third-order theory with an explicit $\Delta(1232)$ was fit to the \OdL\ and
Hallin data up to 240~MeV ($\wcm=200$~MeV).  In view of the uncertainty in the
$\Delta$-sector parameters in this formulation, the transition magnetic-moment
strength $b_1$ was also fit along with $\alphae$ and $\betam$.
($\Delta_M=271.3$~MeV was fit to the real part of the $\Delta$ pole position,
and $g_{\pi \text{N}\Delta}=1.12$ to the imaginary part using the
non-relativistic formula for the width.) A good description of the data was
obtained. The value of $b_1$ = 4.7 may seem high compared to the value of 3.7
obtained from a recent fit to photoproduction data
\cite{Pascalutsa:2005vq,Pascalutsa:2006up}, but since the latter used the
Lagrangian of Eq.~\eqref{eq:PP-Lagrangian}, no direct comparison can be made.
As expected, the problem at backward angles that was evident without the
$\Delta(1232)$ was cured and no restriction on $|t|$ was required.  Perhaps
surprisingly, the values of $\alphae$ and $\betam$ obtained were close to
those of Beane et~al., although with less tension with the Baldin sum rule.
The value of $\betamp$ = 3.4, in particular, is again higher than the
traditional value and appears to be a feature of chiral EFT fits.  It should
be noted that though the theory now includes the $\Delta(1232)$, its reach is
still limited to $\wlab \lesssim 240$~MeV, since the power counting suppresses
the loop diagrams which need to be resummed to give the $\Delta$ a width, and
so the amplitudes diverge as $\omega\to\DeltaM$.

\subsubsection{The $\delta$ expansion}
\label{sec:includeDelta2}

As important as the $\Delta(1232)$ is above the photoproduction threshold, its
influence on the cross section diminishes rapidly as the energy is reduced.
Arguably, with the particular values of $\DeltaM(=\MDelta-\MN)$ and $\mpi$
obtaining in the real world, counting the two as the same scale gives undue
prominence to the $\Delta(1232)$ in the region where static polarisabilities
are important.  An alternative counting was proposed by Pascalutsa and
Phillips \cite{Pascalutsa:2002pi} in which $\mpi/\DeltaM$ and
$\DeltaM/\lambdachi$ are counted as proportional to the same expansion
parameter $\delta$ (the so-called ``$\delta$ expansion"). In this counting,
the $\Delta$-less theory is an expansion in powers of $\delta^2$, and, for low
energies, the first contributions from the $\Delta$ (the pole diagrams and the
$\pi\Delta$ loops) intercalate between the third and fourth orders of \ChPT{}.
The main advantage of this expansion, however, is that it allows for two
separate energy regions, $\omega\sim\mpi$ and $\omega\sim\DeltaM$, and, in the
latter regime, there is no suppression of $\pi$N loop contributions to the
$\Delta$ propagator.  Thus, these must be resummed, and the $\Delta$ becomes
an unstable particle with a width $\Gamma$, with the new propagator going as
$({p\llap/}-\MDelta +\ii\Gamma(p^2)/2)^{-1}$.  For $\omega\sim\DeltaM$, the
one-$\Delta$ reducible diagram dominates (that is, the direct $\Delta$-pole
diagram), with all other contributions being sub-leading.  In order to have
amplitudes which could be used over the entire energy region, Pascalutsa and
Phillips added just the resummed, direct $\Delta$-pole diagram to the standard
third-order \ChPT{} amplitudes.  This includes all contributions to
$\calO(e^2\delta^2)$ in each region, at the expense of having an incomplete
set of higher-order contributions in each.\footnote{In the literature on
  $\delta$ counting, contrary to common usage in \ChPT, a factor of $e^2$ is
  pulled out before the counting starts, so $\calO(e^2\delta^2)
  \sim\calO(P^3)$.} This meant that the sub-leading electric $\gamma$N$\Delta$
coupling was also included.  (Two further technical details relevant to
Ref.~\cite{Pascalutsa:2002pi} are that: (a) the $\Delta$ Lagrangian used was
one developed by Pascalutsa \cite{pasc98,pt99} which ensures freedom from
spin-$\frac 1 2$ degrees of freedom via the imposition of an extra local gauge
symmetry; and (b) the direct $\Delta$-pole contribution was included with
relativistic kinematics and the full Dirac vertex rather than being expanded
in powers of $1/\MN$.)  Importantly, in this approach there is no promotion of
the $\delta\alphae$ etc.\ terms of Eq.~\eqref{eq:LpiN4}, and the
polarisabilities are \emph{predicted} in terms of the two $\gamma$N$\Delta$
couplings $\gM$ and $\gE$, which were fit to Compton data.  Because
$\pi\Delta$ loops are absent, $\alphae$ is close to its HB\ChPT{} value (there
is a small negative contribution $\propto\gE^2$).  Since the $u$-channel
$\Delta$ diagram is missing and the magnetic coupling is somewhat smaller than
previously assumed, the usual huge enhancement of $\betam$ is significantly
reduced, and the net value is $\betamp=3.9$.  The fit in the low-energy region
is not dissimilar to that of Hildebrandt, but a good description of the data
is obtained even in the $\Delta$-resonance region.  From the point of view of
deducing $\alphae$ and $\betam$ from the low-energy data, the main message
again is that a good description is obtained with a larger value of $\betam$
than the previously accepted one.  Unfortunately, the amplitude used by
Pascalutsa and Phillips for $\pi\mathrm{N}$ loops in this work did not have
the correct analytical continuation above the $\pi\mathrm{N}$ threshold,
thereby rendering both the real and imaginary parts of those loops incorrect.
The description of data between $\wlab \approx 150$ MeV and the $\Delta(1232)$
peak published in Ref.~\cite{Pascalutsa:2002pi} is influenced by these loops,
and thus cannot be regarded as definitive.  This may also account for their
finding that the best results were obtained with a value of $\gE\sim-2.3\gM$
which is much larger than that usually deduced from the E2/M1
ratio.\footnote{Note also the following misprints in that paper: in the
  expression for the $\Delta$ width in Eq.~(42), $s+\MN^2-\mpi^2$ should be
  replaced by $\bigl((\sqrt{s}+\MN)^2-\mpi^2\bigr)/2$; Eq.~(51) should contain
  $\calO_5^{\mu\nu}={q}^\mu{q'}_\alpha\gamma^{\alpha\nu}+\gamma^{\mu\alpha}{q}_\alpha{q'}^\nu$,
  $\calO_6^{\mu\nu}={q}^\mu{q}_\alpha\gamma^{\alpha\nu}+\gamma^{\mu\alpha}{q'}_\alpha{q'}^\nu$
  and $\calO_8^{\mu\nu}=\ii\epsilon^{\mu\nu\alpha\beta} {q}_\alpha{q'}_\beta$.
  Eq.~(54) should have $-\calO_7$ on the left-hand side. The unwritten
  convention is $\epsilon_{0123}=1$ so the penultimate equation of footnote
  (2) should be
  $\gamma^{\mu\nu\alpha}=-\ii\epsilon^{\mu\nu\alpha\beta}\gamma_\beta\gamma_5$.}

More recently, Lensky and Pascalutsa \cite{lp09,Lensky:2009uv} extended the
previous calculation to include all contributions that enter at
$\calO(e^2\delta^3)$, which includes both $s$- and $u$-channel $\Delta$-pole
diagrams, as well as $\pi\Delta$ loops.  However, unlike previous work on
Compton scattering in EFT, they did not use the heavy-baryon framework, but
worked in terms of Dirac nucleons.  It is beyond the scope of this review to
detail modern developments in baryon \ChPT; these avoid expanding the
Lagrangian in powers of $1/\MN$, while ensuring that positive powers of $\MN$
which would spoil the power counting do not enter (see Bernard
\cite{Bernard:2007zu} for a summary).  The various schemes in use all yield
results which have several features in common: they aim at a Lorentz-covariant
amplitude; when expanded in powers of $1/\MN$, they agree with one another and
with HB\ChPT{} up to the order of validity of the calculation; they also agree
for any term which is non-analytic in $\mpi^2$; but they may differ with
regard to effects that are higher-order in $1/\MN$.  Ultimately, of course,
LECs (which are different in the different schemes) will absorb these
differences.  For low-order calculations, however, the difference between
schemes can produce significant effects in observables.
The calculations of Refs.~\cite{lp09,Lensky:2009uv} were performed in one such
version of Baryon $\chi$PT. Although the nucleons obey relativistic
kinematics, antinucleons are not---and should not be---included in the
effective theory.

While some of the $1/\MN$ terms which are automatically included in the
covariant case produce effects relevant to low-energy dynamics, e.g.  terms
which must be resummed to put the threshold in the right place in the
HB\ChPT{} approach, others act as a regulator on loop integrals. They
therefore have no meaning when separated from the higher-order LECs, except
perhaps as an estimate of the likely size of uncertainties due to omitted
higher-order physics. Nevertheless, the polarisabilities constitute one case
in which such higher-order effects make a sizable difference in a low-order
calculation. The covariant B\ChPT{} diagrams to be evaluated at third order
include all those in Fig.~\ref{fig:loops-3rd}, as well as those from
Fig.~\ref{fig:loops-4th} where the $\gamma$NN vertex is generated by the
$1/\MN$ expansion of the relativistic Lagrangian. An example is the Dirac
magnetic moment of the Lagrangian; diagrams where both photons couple to a
nucleon line also enter the relativistic calculation of polarisabilities at
LO.  The integrals differ from the HB\ChPT{} case in that they depended on
the nucleon mass as well as on the pion mass, and the net effect of the
sub-leading $\mpi/\MN$ terms is to substantially reduce $\alphae$ and
$\betam$.  
Once the $\Delta$(1232) is added in (with the parameters of
Ref.~\cite{lp09,Lensky:2009uv}, which are taken from photoproduction studies),
we return to familiar values of $\alphae= 10.8$ and $\betam=4.0$---but the
difference from the fourth-order HB\ChPT{} case is that these are predictions
rather than fits.  Since this calculation includes all the dynamics relevant
for $\wlab \leq 170$ MeV, it is not surprising that the data description in
this domain is good (see Fig.~\ref{fig:oldfits}).

\begin{table}[!htbp]
  \begin{center}
\begin{tabular}{|l|c|cc|c|}
\hline
&  \multirow{2}{*}{$\pi\mathrm{N}$ loops} & \multirow{2}{*}{$\Delta$ pole} & \multirow{2}{*}{$\pi \Delta$ loops} & Higher resonances, \\
& & & &$\sigma$ pole\\
\hline
$\calO(P^3)$ & LO & $\times$& $\times$ & $\times$ \\
$\calO(\epsilon^3)$& LO & $s$- and $u$-channel & $\checkmark$ & $\times$  \\
$\calO(\epsilon^3)$(mod.)~\cite{Hildebrandt:2003fm} & LO & $s$- and $u$-channel & $\checkmark$ & In 2 CTs \\
$\calO(P^4)$ \cite{Beane:2004ra} & NLO & In 2 CTs & In 2 CTs & In 2 CTs\\
$\calO(e^2\delta^2+)$~\cite{Pascalutsa:2002pi} & LO & $s$-channel, width, M1+ E2& $\times$ &$\times$ \\
$\calO(e^2\delta^3+)$~\cite{Lensky:2009uv}& LO + pNLO &   $s$- and $u$-channel, width, M1 + E2\hspace*{-0.5cm} & $\checkmark$ & $\times$ \\
\hline
\end{tabular}
  \end{center}
  \caption{\label{table-theorycomp} Comparison of the different mechanisms
    included in different EFT variants. All calculations include the nucleon
    Born and pion-pole term.   $\calO(X^n+)$ indicates some contributions
    beyond the stated order;  pNLO indicates some effects which would be NLO
    in HB\ChPT. CT denotes the $\gamma$N contact interactions of
    Eq.~\eqref{eq:LpiN4}; at $\calO(P^4)$ the $\pi$N contact interactions
    $\propto c_i$ also encode resonance physics.} 
\end{table}

\subsubsection{EFT fits in relation to the data}
\label{sec:earlyfit}

At this point, it will be useful to summarise the different counting schemes
in use.  In HB\ChPT\ without the $\Delta$(1232), the counting is the usual
one, in powers of $P/\lambdachi$ with as $P$ the pion mass, a typical momentum
or the photon energy.  For the multipoles to which the $\Delta$ is known to
contribute, though, $\lambdachi\sim M_\rho$ is replaced by
$\DeltaM(=\MDelta-\MN)$.  The lowest-order contribution to Compton scattering
is the Thomson term which is $\calO(P^2)$ (since $|e|\sim P/\lambdachi$).
When the $\Delta$ is included in the small-scale expansion
\cite{Hemmert:1996xg,Hemmert:1997ye}, we treat $\DeltaM\sim P$ and call the
expansion parameter $\epsilon\equiv(\DeltaM,\,P)/\lambdachi$.  In the
low-energy regime, both $\Delta$ and nucleon propagators scale as
$\epsilon^{-1}$ and so $\pi\Delta$ loops and $\Delta$-pole diagrams enter at
$\calO(\epsilon^3)$ along with the corresponding nucleon diagrams.
Short-range contributions to $\alphaep$ etc.\ only enter at fourth order, but
have often been promoted to third order---partly to provide diamagnetic strength to
counteract the $\Delta$; we refer to this as modified $\calO(\epsilon^3)$.  In
the region of the $\Delta$ resonance, diagrams which give the $\Delta$ a
finite width must be included, but this has not been done in the $\epsilon$
expansion.  An alternative counting is given by the $\delta$ expansion
\cite{Pascalutsa:2002pi}, in which away from the resonance region $\Delta$
propagators are suppressed relative to nucleon ones by one power of
$\delta\approx\mpi/\DeltaM$; here the Thomson term is $\calO(e^2)$, LO N-pole
and $\pi$N are $\calO(e^2\delta^2)$, LO $\pi\Delta$ and $\Delta$-pole are
$\calO(e^2\delta^3)$, and NLO N-pole and $\pi$N diagrams are
$\calO(e^2\delta^4)$ (see Figs.~\ref{fig:Born}--\ref{fig:loops-4th}).  The
ingredients of several calculations that use different \ChiEFT expansions are
summarised in Table~\ref{table-theorycomp}.

An important difference not noted in the table is that, in order to facilitate
the treatment of the pole region where finite-width $s$-channel $\Delta$-pole
diagrams are leading order, all work in the $\delta$ expansion has used a
relativistic form for the $\Delta$ propagator with the $\gamma$N$\Delta$
vertex from Eq.~\eqref{eq:PP-Lagrangian}, even for $\w \ll \DeltaM$.  At
sufficiently low energies, the two magnetic $\gamma$N$\Delta$ coupling
constants, $b_1$ from Eq.~\eqref{eq:Delta-lagrangian} and $\gM$ from
Eq.~\eqref{eq:PP-Lagrangian}, are related by Eq.~\eqref{eq:gm-b1}, which
translates $\gM=2.9$ \cite{Pascalutsa:2005vq,Pascalutsa:2006up} to $b_1=3.8$;
with this identification, the two give exactly the same contribution to
$\betam$.  However, the two vertices are not equivalent at higher energies.
If the full $\betam$ is fit to data and hence does not directly depend on
$b_1$ ($\gM$), significant $\w/M_{\rm N}$ effects from
Eq.~\eqref{eq:PP-Lagrangian} mean that a substantially higher value of $b_1$
is required to give broadly the same cross section.  Numerical studies show
that values of $\gM=2.9$ and $b_1=4.8$--5.0 give similar cross sections up to
150~MeV, if the parameters and other ingredients of the modified $\epsilon^3$
and modified $e^2\delta^3$ are the same. (The closeness of the dynamical
polarisabilities can be seen in Fig.~\ref{fig:dynpolas}.) Hildebrandt
et~al.~\cite{Hildebrandt:2003fm,Hildebrandt:2003md,Hildebrandt:2005ix} found
that a slightly lower value of $b_1\approx 4.7$ gave the best results, but
they use pole-position rather than $\Delta$-resonance parameters.

\begin{figure}[!htb]
  \begin{center}
    \includegraphics*[width=\linewidth]{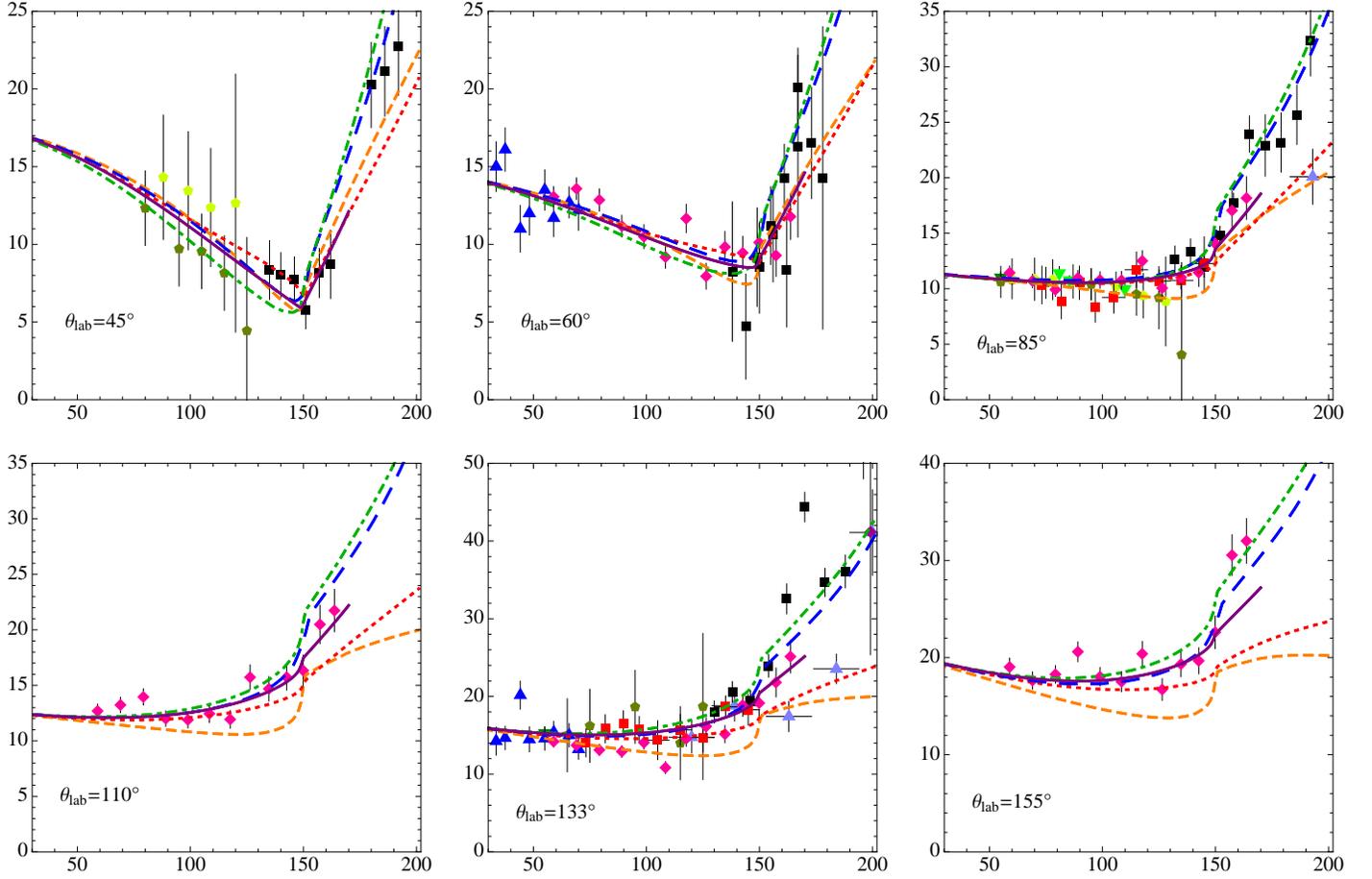}
    \caption {(Colour online) Previous results. Working from the lowest cross
      section up in the last frame, the lines are as follows: Medium dash
      (orange): $\calO(P^3)$ Babusci et~al.~\cite{Babusci:1997ij}; short dash
      (red): $\calO(P^4)$ Beane et. al.~\cite{Beane:2004ra}; solid (purple):
      partial $\calO(e^2\delta^3+)$ Lensky and Pascalutsa~\cite{Lensky:2009uv};
      long dash (blue): $\calO(\epsilon^3)$ Hildebrandt
      et~al.~\cite{Hildebrandt:2003fm}; dash-dot (green): $\calO(e^2\delta^2+)$
      Pascalutsa and Phillips.~\cite{Pascalutsa:2002pi}.  For
      Refs.~\cite{Beane:2004ra,Hildebrandt:2003fm}, the fits without
      imposition of the Baldin sum rule are shown.  For
      Ref.~\cite{Pascalutsa:2002pi}, the coding error mentioned in the text
      has been corrected, and photoproduction values are used for $b_1$ and
      $b_2$ \cite{Pascalutsa:2005vq,Pascalutsa:2006up}.  The data key is in
      Table~\ref {table-proton-low}.  The 180\deg\ curves are all extremely
      close to the 155\deg\ ones and hence are not shown.}
    \label{fig:oldfits}.
  \end{center}
\end{figure}

Figure~\ref{fig:oldfits} shows the results of the various calculations and
fits described in the previous section; see Table~\ref{table-theorycomp} for
ingredients and Table~\ref{table-polcomp} for a summary of results for the
static polarisabilities.  In attempting to extract polarisabilities from the
low- and possibly intermediate-energy data, there are competing pressures.  On
the one hand, EFTs are clearly most reliable at low energy.  On the other
hand, the effects of the polarisabilities grow with energy, and the more data
that can be included in a fit, the better the resulting statistical errors are
expected to be.  Using HB\ChPT\ to $\calO(P^4)$ but without the
$\Delta$(1232), Beane et~al. applied a cut of 180~MeV in both $\wlab$ and
$\sqrt{-t}$, the latter meaning that the maximum energy was in fact only
reached for angles up to 60\deg, dropping to 90~MeV at 180\deg.  The resulting
extraction was therefore free from any of the issues discussed above
concerning the data at intermediate energies, but the statistical errors of
$\pm 1.1$ on $\alphaep$ and $\betamp$ compared unfavourably with those
obtained, for instance, by Baranov et~al.\ \cite{Baranov:2001} using all data
up to 150~MeV prior to \OdL\ ($\pm 0.8$ on $\alphaep$ and $\pm0.9$ on
$\betamp$).  Hildebrandt et~al.~did include the $\Delta$ in the small-scale
expansion and fit to the \OdL\ and Hallin data up to 240~MeV.  Without a
finite width, their $\Delta$-pole amplitude starts to deviate from a
Breit-Wigner-like shape which could continue into the resonance region above
200~MeV, more strongly at backward angles.  Some of the nice agreement with
the Hallin data in the intermediate-energy region shown in Hildebrandt's
thesis may therefore be fortuitous.  The relatively high errors (of $\pm 1.4$
on $\alphaep$ and $\betamp$) may be due to the fact that they did not allow
the normalisation of the data sets to float or that they did not include the
older data.

Pascalutsa and Phillips resum $\Delta$ self-energy diagrams, however,
generating a width and a Breit-Wigner-like propagator, and the resulting form
gives at least a qualitatively good description of the data up to the pole
region.  All the ingredients for an extraction that also incorporates the
effects of $\pi\mathrm{N}$ loops, $\pi \Delta$ loops and the $\Delta$-pole
diagrams---with a finite width where necessary---now exist.  While Hildebrandt
et~al.\ could not access the resonance region and hence had to fit the crucial
$\gamma$N$\Delta$ magnetic coupling constant to the low- and
intermediate-energy data, in the $\delta$ expansion it should be possible to
fit purely $\Delta$ parameters in the resonance region, leaving only
$\alphaep$ and $\betamp$ to be fit to low-energy data.  Such a strategy should
also be able to bypass the problems of the intermediate region, although the
issue of Blanpied (LEGS) versus Wolf (MAMI) cannot be avoided.  Details of
such a calculation will be presented in Section~\ref{sec:protonanalysis}.

%


\subsection{\it Comparing \ChiEFT with Dispersion Relations}
\label{sec:comp}

We compare the values for various polarisabilities extracted from various DR
and $\chi$EFT calculations in Table~\ref{table-polcomp}. Overall, the
agreement is very good. We note that the integral for $\alphaep - \betamp$ in
the hyperbolic dispersion relation is somewhat sensitive to assumptions about
high-energy physics, but the value found there is within the error bar of the
various EFT fits.  The result for $\alphaep + \betamp$ is even more sensitive
to high-energy physics, which makes it remarkable that the agreement with the
$\calO(P^3)$ $\chi$PT prediction is so good. The apparent variation of the
static values of the spin polarisabilities may be settled in the near future
by a series of ongoing and planned experiments with polarised targets and
beams (see Section~\ref{sec:spinpols}).

\begin{table}[!htbp]
  \begin{center}
    \begin{tabular}{|l|cc|c|cc|}
      \hline
      & $\calO(P^3)$&$\calO(P^4)$~\cite{Beane:2002wn,VijayaKumar:2000pv}& Mod. $\calO(\epsilon^3)$~\cite{Hildebrandt:2003fm} &Fixed-t~\cite{Drechsel:2002ar,Holstein:1999uu,Babusci:1998ww}&
      Fixed-$\theta$=180\deg~\cite{Drechsel:2002ar}\\
      \hline
      \alphaep + \betamp & 13.8 &  15.4 $\pm$ 1.4$^*$  &  13.8 $\pm$ 0.4$^\dagger$ & 13.2 $\pm$ 0.9 $\pm$ 0.7$^*$ & N/A\\
      \alphaep - \betamp & 11.3 & 8.8 $\pm$ 1.6$^*$ &  8.3 $\pm$ 1.9$^*$  & 11.1 $\pm$ 1.1 $\pm$ 0.8$^*$ & 10.9\\
      \alphaep & 12.5 & 12.1 $\pm$ 1.1 $\pm$ 0.5$^*$& 11.0 $\pm$ 1.4$^*$ & 12.2 $\pm$ 0.7 $\pm$ 0.5$^*$&N/A \\ 
      \betamp & 1.25 & 3.4 $\pm$ 1.1 $\pm$ 0.1$^*$  & 2.8 $\mp$ 1.4$^*$  & 1.1 $\pm$ 0.7 $\pm$ 0.5$^*$& N/A\\ \hline
      \gammaeep & -5.7 & -1.3 & -5.7 & -3.85 $\pm$ 0.45 & -3.8\\
      \gammammp & -1.1 & 3.3 & 3.1 & 2.8 $\pm$ 0.1 & 2.9\\
      \gammaemp & 1.1 & 0.2 & 1.0  & -0.15 $\pm$ 0.15 & 0.5\\
      \gammamep & 1.1 & 1.7 & 1.0  & 2.0 $\pm$ 0.1  & 1.6\\
      \hline
    \end{tabular}
  \end{center}
  \caption{\label{table-polcomp} Previous calculations of proton
    polarisabilities. $\dagger$ indicates that the Baldin sum rule was used. $*$
    indicates that the result was obtained by fitting to $\gamma \mathrm{p}$
    data. In the 
    $\calO(P^4)$ column, the second errors are theory errors; the errors of the
    modified $\calO(\epsilon^3)$ calculation are statistical only. In the
    fixed-t DR  column, the second errors are from floating normalisations (see
    Section~\ref{sec:drs}) and the numbers for spin polarisabilities are
    obtained by averaging the calculations of
    Refs.~\cite{Holstein:1999uu,Babusci:1998ww} with uncertainties reflecting
    the range. The DR calculations presented here predict  {\it all} spin
    polarisabilities, in contrast to the fit of Ref.~\cite{Drechsel:2002ar}
    discussed in Sec.~\ref{sec:drs}.} 
\end{table} 

\begin{figure}[!htb]
  \begin{center}
    \includegraphics*[width=\linewidth]{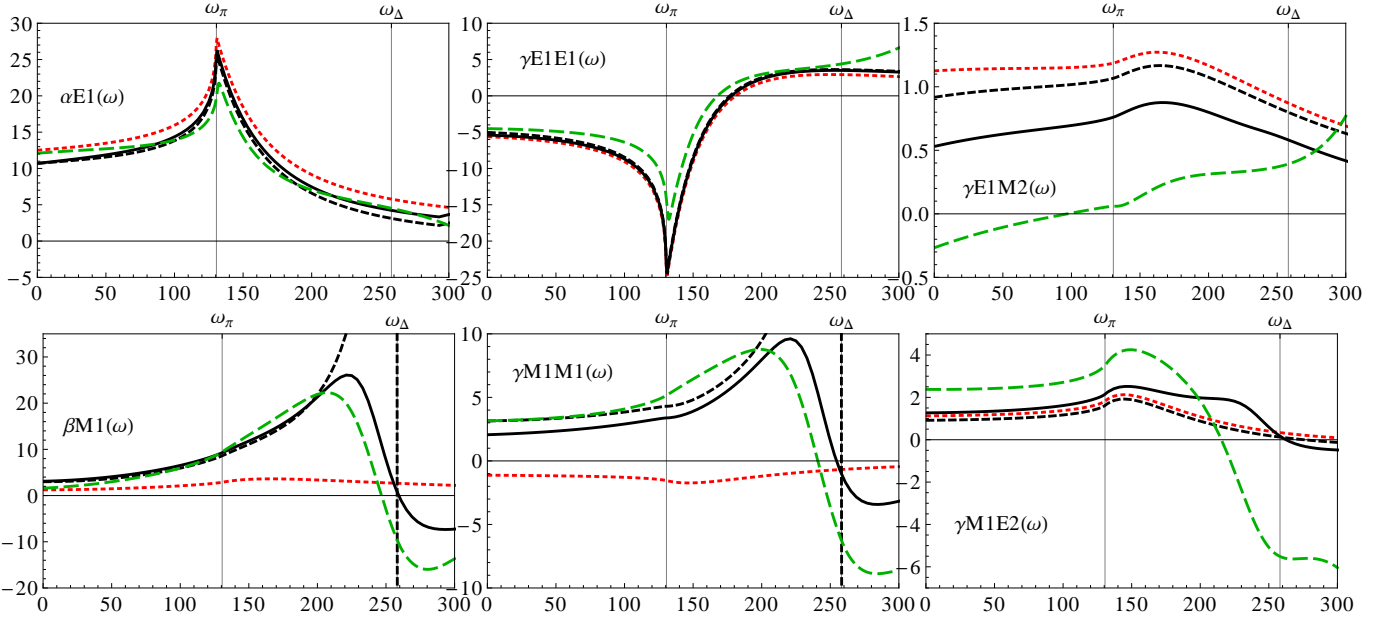}
    \caption {(Colour online) Comparison of DR results \cite{Drechsel:2002ar}
      (green long-dashed line) and $\calO(P^3)$, modified $\calO(\epsilon^3)$
      and modified $\calO(e^2\delta^3)$ EFT results (red dotted, and black
      dashed and solid lines respectively) for the real parts of the six
      dipole proton polarisabilities, as a function of cm energy.  The
      $\calO(e^2\delta^3)$ EFT results use the parameters $\alphae=10.7$,
      $\betam=3.1$ and $b_1=3.66$ as determined in the Baldin-constrained fit
      of Section~\ref{sec:protonanalysis}, whereas the $\calO(\epsilon^3)$
      result uses the same polarisabilities but a non-relativistic
      $\Delta(1232)$ propagator and hence $b_1=5$ as discussed in
      Section~\ref{sec:earlyfit}. The threshold for pion production and the
      peak of the $\Delta$(1232) Breit-Wigner cross section are marked by
      $\omega_\pi$ and $\omega_\Delta$ respectively. Note the difference in
      scales and the inherent theoretical uncertainties of each approach.}
    \label{fig:dynpolas}
  \end{center}
\end{figure}

Far more informative than the static ($\w=0$) values of the polarisabilities
are their full functional forms.  These were calculated in (modified)
$\calO(\epsilon^3)$ by Hildebrandt et al.\ \cite{Hildebrandt:2003fm} and
compared with the results of the DR analysis of Drechsel et al.\
\cite{Drechsel:2002ar}, as updated in Ref.~\cite{Hildebrandt:2003fm}. In
Fig.~\ref{fig:dynpolas} we show results for EFT fits with the parameters which
will be obtained from the proton Compton data in
Section~\ref{sec:protonanalysis}. Keeping in mind the inherent theoretical
uncertainties of the DR and EFT approaches, the shapes of all six dynamical
dipole polarisabilities agree remarkably well up to photon energies of around
200~MeV (centre-of-mass).  For the modified $\calO(e^2\delta^3)$ EFT which
includes a $\Delta$(1232) width, the shape agreement continues above 200~MeV
for all but the two mixed spin polarisabilities. However, for most energies,
these are numerically small so that higher-order effects may be more
prominent.  The similarity emphasises the point that the energy dependence of
$B_i(\nu,t)$ is driven by near-threshold dynamics, even if the value of
$B_i(0,t)$ is sensitive to higher-energy physics. Its origin is the fact that
HB$\chi$PT captures the large $E_{0+}$ photoproduction multipole that
generates a significant $E1$ excitation of the pion cloud of the nucleon. This
is clearly seen in the cusps at the pion-production threshold, most prominent
in the polarisabilities of $E1E1$ multipoles. We see in the table that the
static $\calO(P^3)$ value for $\alphaep$ agrees
exceptionally well with the DR result, while that for $\gammammp$ is in reasonable
accord with DR.  

However, the $M1$ excitation is not properly described in $\chi$EFT until the
$\Delta(1232)$ is included.  This is reflected in the improved agreement
between DRs and the $\chi$EFT calculation with dynamical $\Delta(1232)$ in the
second block of the table, compared to the $\Delta$-less calculation of the
first block, and even more so by the prominence of the $\Delta(1232)$
resonance in the polarisabilities containing an $M1$ multipole in
Fig.~\ref{fig:dynpolas}. The poor agreement around the $\Delta$ peak for
$\gammame$ is rather surprising in this regard, since the $E2$ electric
transition vertex is included in the EFT Lagrangian (\ref{eq:PP-Lagrangian}).
The pronounced structure in the DR description in the 160--250~MeV region
is markedly different from what is seen in the $\chi$EFT calculation; this
discrepancy needs further study.

Finally, even a $\calO(e^2\delta^4)$ $\chi$EFT calculation does not have the
strong $\pi \pi$ interactions in the $t$-channel which generate a substantial
contribution to $\alphaep - \betamp$ that offsets the effect of the
$\Delta(1232)$ in that quantity.  Whether this contribution is modelled as a
$\sigma$ or constructed through the $t$-channel DR (\ref{eq:tchannDR}), it is
present in the numbers given in the last two columns. It is incorporated in
current $\chi$EFT calculations only as a contribution to a $\alphaep-\betamp$
$\gamma \mathrm{N}$ contact operator. Therefore, the fact that the static and
dynamical scalar polarisabilities in \ChiEFT and DR agree suggests that the
energy dependence of this process is negligible below about $300$~MeV, which
is consistent with the phenomenology of the correlated $2\pi$ state.


\subsection{\it A new fit to the  proton data}
\label{sec:protonanalysis}

As outlined in Section \ref{sec:earlyfit}, in the $\delta$ expansion, it
should be possible to fit purely $\Delta$(1232) parameters in the resonance
region, leaving only $\alphaep$ and $\betamp$ to be fit to low-energy data.
This section presents the results of a fit using this strategy.

All the experiments have an overall normalisation error.  We incorporate this
by adding a piece to the usual $\chi^2$ function:
\begin{eqnarray}
  \chi^2&=&\chi^2_{\text{stat.}}+\chi^2_{\rm syst.}\nonumber\\
  &=&\sum_{j=1}^{N_{\text{sets}}}\sum_{i=1}^{N_j}
  \left(\frac{ {\cal N}_j (\text{d}\sigma_{ij}/\text{d}\Omega)_{\rm expt}
      -(\text{d}\sigma_{ij}/\text{d}\Omega)_{\rm theory}} {{\cal N}_j
      \Delta_{ij}}\right)^2  +\sum_{j=1}^{N_{\rm sets}} 
  \left(\frac{{\cal N}_j - 1}{{\cal N}_j \delta_j}\right)^2\;\;,
  \label{eq:chisqare}
\end{eqnarray}
where $(\text{d}\sigma_{ij}/\text{d}\Omega)_{\rm expt}$ and $\Delta_{ij}$ are
the value and statistical error of the $i$th observation from the $j$th
experimental set, $\delta_j$ is the fractional systematic error of set $j$,
and ${\cal N}_j$ is an overall normalisation for set $j$.  The additional
parameters ${\cal N}_j$ are to be optimised by minimising the combined
$\chi^2$.  This can be done analytically, leaving a $\chi^2$ that is a
function of $\alphaep$ and $\betamp$ alone.  The number of degrees of freedom
in the final minimisation is reduced by the number of independent data sets
used.  For more details on this formalism, see Ref.~\cite{Baranov:2001}.  The
best justification we have found in the modern literature for the precise form
of this expression (specifically, the inclusion of ${\cal N}_j$ in the
denominators) is from d'Agostini \cite{D'Agostini:1994}.

We perform a fit at modified third order, $\calO(e^2\delta^3)$ plus the
counterterms of Eq.~\eqref{eq:LpiN4}, then one at fourth order,
$\calO(e^2\delta^4)$. For reasons which will become clear below, we regard the
latter only as a consistency check on the former.  The basic ingredients of
the modified third-order calculation were described in Section
\ref{sec:singleN} and shown in Figs. \ref{fig:Born}--\ref{fig:delta-contrib}.
These are nucleon Born diagrams and pion- and $\Delta$-pole diagrams, $\pi$N
loops, $\pi\Delta$ loops and the photon-nucleon seagull terms from
$\calL^{(4)}$ which give LEC contributions $\delta\alphaep$ and
$\delta\betamp$ to the polarisabilities.  The nucleon Born diagrams are
calculated to fourth order as given by McGovern~\cite{mcg01}, and they are
indistinguishable from the full relativistic form given by Babusci et~al.\
\cite{Babusci:1998ww} for the energies of interest.  The pion-pole
contributions are given by Bernard et~al.\ \cite{Bernard:1995dp}; this form
does not change at fourth order and is indeed relativistically invariant,
although at higher orders, form factors would still enter at the vertices.
The $\pi$N loops are given at third order by Bernard
et~al.~\cite{Bernard:1995dp} and at fourth order by McGovern~\cite{mcg01}.  At
both third and fourth order, we use $\omega_s=\sqrt{s}-\MN$ to shift the
threshold to the correct place.  The $\pi\Delta$ loops are given by
Hildebrandt et~al.~\cite{Hildebrandt:2003fm} in his Appendix B, except that we
use $\wBreit$ throughout in place of $\omega_s$ and $\omega_u$ to preserve
crossing symmetry.  The $s$- and $u$-channel $\Delta$-pole diagrams are
calculated using the Lagrangian of Pascalutsa and Phillips
\cite{Pascalutsa:2002pi}; the expressions given in Appendix A3 of that paper
refer to a redundant, covariant set of operators which we reduce to our usual
six in the Breit frame. Strict consistency in the $\delta$ counting would
require us to use the third-order nucleon Born contribution at this order.
However, since the main contribution at fourth order consists of terms of
order $\omega^2$ in $A_1$ and $A_2$, polarisabilities extracted at different
orders will be more comparable if these terms are always included.  With
regard to the $\Delta$ pole, Pascalutsa and Phillips have detailed which terms
are leading or sub-leading in the low-energy and resonance regions; following
their scheme, we include all terms in both regions to avoid discontinuities.
At third and fourth order, the $\gamma$N$\Delta$ vertices and $\Delta$
propagator should strictly be expanded in powers of $\omega/\MN$, $\Delta/\MN$
and $\omega/\Delta$, but again, to avoid artificial shifts in polarisabilities
in going from one order to the next, we keep the full Dirac structures.  The
$\Delta$ parameters we use are $\Delta_M= 293\;\MeV$ and
$g_{\pi\mathrm{N}\Delta} =1.425$, which are fit to the Breit-Wigner peak
position and width, the latter via the relativistic formula.  These, and not
the pole position, must be used if the resonance is to be reproduced.

\begin{figure}[!htbp]
  \begin{center}
    \includegraphics*[width=\linewidth]{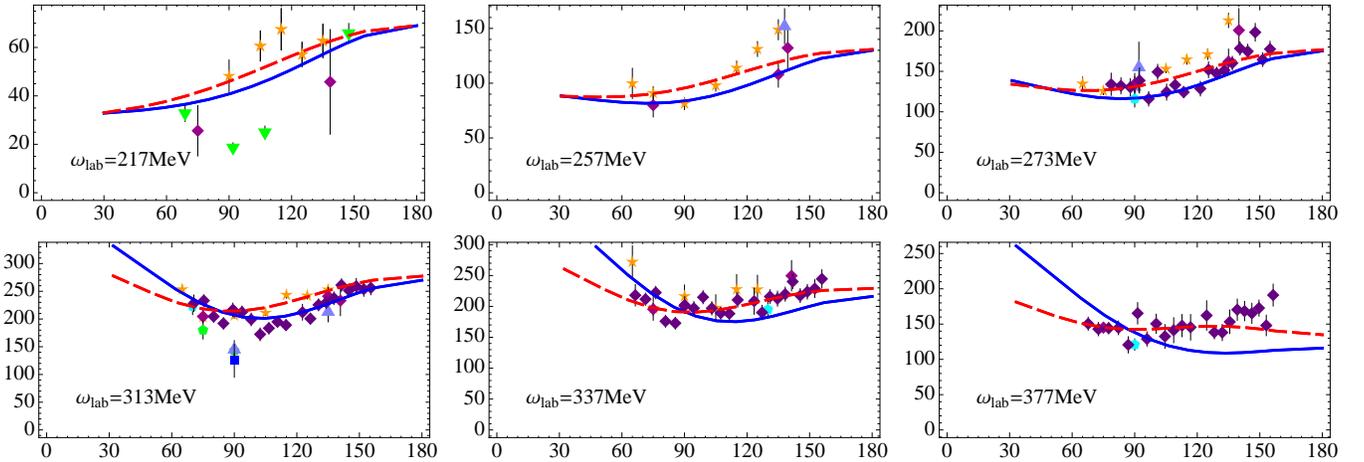}
    \caption {(Colour online) Comparison of third-order (blue, solid) and
      fourth-order (red, dashed) cross sections with Compton-scattering data
      in the intermediate- and high-energy regions.  Centre-of-mass (cm) cross
      section in nb/sr plotted in bins of 8~MeV as a function of cm photon
      angle. In both cases, $\alphaep=10.5$ and $\betamp=2.8$; $b_1=3.66$ and
      $3.47$, respectively.  See Table~\ref {table-proton-low} for key.}
    \label{fig:sect5-fig4}
  \end{center}
\end{figure}

First we present our main result, from a modified $\calO(e^2\delta^3)$
calculation.  Our strategy is to first determine the $\gamma$N$\Delta$
couplings by considering the resonance region.  By eye it is apparent that
with typical values of $\alphaep=11$ and $\betamp=3.4$ and with
$b_2/b_1=-0.34$, a reasonably convincing overall reproduction of the MAMI data
up to the peak around 325~MeV can be obtained with $b_1=3.7$ (close to the
Pascalutsa and Vanderhaeghen value of $b_1=3.76$ \cite{Pascalutsa:2005vq}).
Above the peak, the cross section falls off too slowly at forward angles and
too fast at backward angles, but we hardly expect to fit well at these high
energies.  (See Fig.~\ref{fig:sect5-fig4}, solid blue curve.)  For that
reason, unlike Pascalutsa and Phillips, we do not adjust $b_2$.  Below the
peak, the fit is always better to the MAMI data than the LEGS data (where they
disagree) even if the peak height is raised.  In view of this problem, and
considering the fact that the former will dominate statistically, we choose to
fit only to the MAMI data.

Thus we fit $b_1$ to the MAMI data from 200~MeV up to 325~MeV, then we fit
$\alphaep$ and $\betamp$ to the low-energy data up to 170~MeV, iterating until
convergence is reached.  If we exclude the Hallin data, we obtain a solution
with a good $\chi^2$.  Looking at individual experiments, however, we see that
the Baranov data at 150~MeV (which we treat as an independent data set,
following Ref.~\cite{Baranov:2001}) have a $\chi^2$ per degree of freedom
(d.o.f.) of over 3 and should therefore be excluded from the fits.  The \OdL\
data, which form nearly half the total, are fit with an acceptable $\chi^2$ of
73 for 65 d.o.f.  If we include the 27 Hallin points in this energy range,
though, the $\chi^2$ of 41 for 27 d.o.f.\ is hard to accept, and we prefer to
quote our best results without it.  (The specific $\chi^2$ above is taken from
the fit constrained by the Baldin sum rule but is hardly changed if that
constraint is lifted.)

\begin{figure}[!t]
  \begin{center}
    \includegraphics*[width=\linewidth]{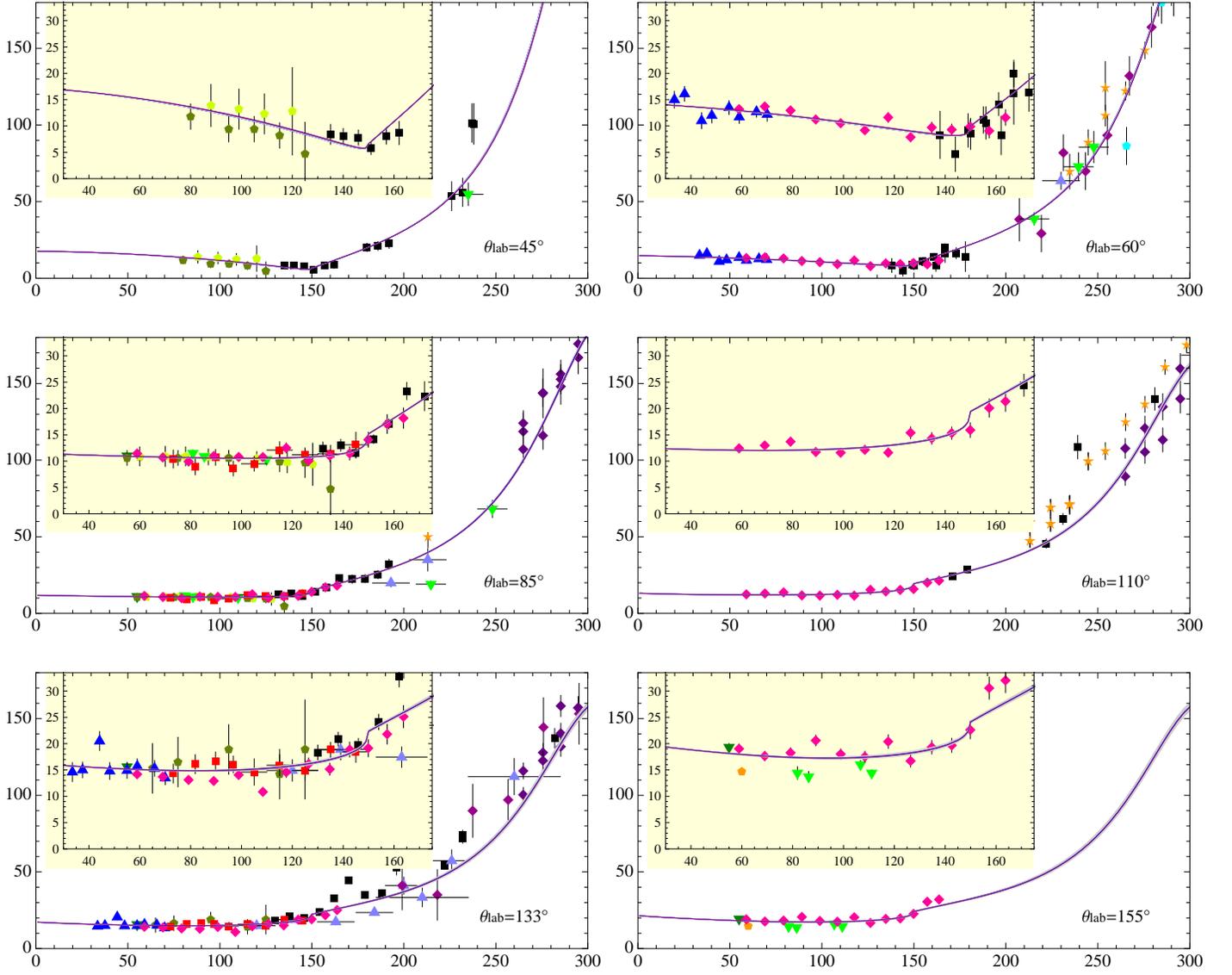}
    \caption {(Colour online) Comparison of third-order result with
      Compton-scattering data.  Lab cross section in nb/sr plotted in bins of
      10\deg\ lab angle as a function of lab photon energy in MeV.  The insets
      show the fit region.  The fits with and without the Hallin data are both
      plotted, but are indistinguishable. The barely visible narrow grey band
      shows the variation within the statistical error of the one-parameter
      fit. See Table~\ref {table-proton-low} for key.}
    \label{fig:sect5-fig3}
  \end{center}
\end{figure}

The central result for the modified $\calO(e^2\delta^3)$ fit without using the Baldin
sum rule is then (rounded to one decimal place)
\begin{equation}
  \alphaep=10.5\pm0.5(\text{stat})\pm0.8(\text{theory})\;\;,\;\;
  \betamp = 2.7\pm0.5(\text{stat})\pm0.8(\text{theory}) 
  \label{eq:p2parameterfinalfit}
\end{equation}
with $\chi^2=106.1$ for 124 d.o.f.\ (or $\alphaep+\betamp = 13.2\pm0.9$ and
$\alphaep-\betamp = 7.8\pm0.6$, since the principal axes of the
$\chi_{\text{min}}^2+1$ ellipses align fairly closely with these axes).  These
are obtained with $b_1=3.66\pm 0.03$.  With the Baldin constraint of
$\alphaep+\betamp = 13.8\pm0.4$, the result is $\alphaep-\betamp=7.7\pm 0.6$
with $\chi^2=106.5$ for 125 d.o.f.\ and the same $b_1$ as before, which gives
\begin{equation}
  \alphaep=10.7\pm0.3(\text{stat})\pm0.2(\text{Baldin})\pm0.8(\text{theory})
  \;,\;
  \betamp = 3.1\mp0.3(\text{stat})\pm0.2(\text{Baldin})\pm0.8(\text{theory}).
  \label{eq:p1parameterfinalfit}
\end{equation}
If the low-energy fits are repeated with $b_1$ varied within its limits,
$\alphaep-\betamp$ changes by less than 0.1.  Varying the upper cutoff between
300 and 350~MeV scarcely changes $\alphaep-\betamp$.  Restoring the two
omitted data sets (Hallin and Baranov 150\deg) also results in an upward shift
in $\alphaep-\betamp$ of 0.1.  We therefore conclude that our fit is stable.
Visually, the agreement with the bulk of the data continues far above the
low-energy region, without any obvious systematic problems, as can be seen in Fig.~\ref{fig:sect5-fig3}.

To arrive at the quoted theory error on our results, we note that we perform
an $\calO(e^2\delta^3)$ fit in a framework in which the polarisabilities first
enter at $\calO(e^2\delta^2)$.  We would expect corrections to be of order
$\delta^2\sim16\%$ of the lowest-order result, or $\delta\sim 40\%$ of the
shift between the LO and NLO results; taking $(\alphaep+\betamp)/2\approx 7$
to set the scale for the first approach gives 1.1, while taking the shifts 
in the values of $\alphaep$ and $\betamp$ from third order to 
fourth order to be $\approx 2$ gives 0.8 in the second approach.  In view
of the similarity between our third- and fourth-order results (see later), the
stability under inclusion or exclusion of data sets, and the values obtained
in the $\calO(P^4)$ and $\calO(\epsilon^3)$ fits
\cite{Beane:2002wn,Hildebrandt:2003fm}, we consider the latter to be already
conservative and so we use it.

Next we look at the $\calO(e^2\delta^4)$ fit, which differs from the previous one by
the inclusion of the NLO pieces of $\pi$N loops (in the \ChiEFT expansion).
The first observation is that, with $\alphaep$ and $\betamp$ as above but with
a somewhat reduced $b_1$ of 3.48, the general trend of the data in the
resonance region and above is improved; the exaggerated angular dependence of
the third-order cross section is tamed (see Fig.~\ref{fig:sect5-fig4}).
However, at low energy, the fit is systematically worse, with the fourth-order
curve lying above the third-order one---and above the data---at all angles.
If we repeat the fit procedure detailed above, the value $b_1= 3.48$ is
confirmed, but the best fit constrained by the Baldin sum rule has a $\chi^2$
of 170; relaxing the Baldin constraint drops the $\chi^2$ to 137 but with
$\alphaep+\betamp = 18.7$.  In fact, this is not a surprise: McGovern and
Hildebrandt et~al.\ independently found that adding either NLO pion loops or
the $\Delta$ separately raised the cross section around the photoproduction
threshold and improved the fit compared with third-order pions alone; adding
both is too great a correction.

With both $\alphaep$ and $\betamp$ as fit parameters, the primary contribution
of the $\Delta$(1232) at low energies is via the spin polarisability
$\gammammp$.  Just as inclusion of the $\Delta$ at third order requires
promotion of the strictly fourth-order counterterms for $\alphae$ and
$\betam$, it would appear that, at fourth order, promotion of one or more
fifth-order counterterms for the spin polarisabilities is required.  Various
strategies are possible: we could promote all four and fit them all, or we
could find out which is the most important and fit just that one, or we could
take $\gammamm$ as the one with the largest $\Delta$ contribution.  With any
of these strategies, we can get an acceptable low-energy fit.  However, there
are very flat directions in parameter space if we promote all four, and in the
absence of further constraints, it is not clear what we learn from the
procedure.  In fact, the best low-energy results from promoting a single
polarisability are obtained with $\gammamm$, so we choose this; the
Baldin-constrained results are then $\alphaep-\betamp=7.6\pm 0.8$ and
$\gammammp=2.6\pm0.5$ with $\chi^2=116.7$; this is obtained with $b_1=3.59$.

It is reassuring to note that the fitted value of $\gammammp$ is ``sensible"
when compared to DR estimates (see Table.~\ref{table-polcomp}), whereas
unfitted it is distinctly large at 6.4.  However, in view of the choices we
have had to make in obtaining $\gammammp$, we would caution against treating
our result as a chiral-EFT extraction of this parameter.  Only low-energy
polarisation measurements will have real power to constrain spin
polarisabilities.  It should also be noted that the fit at higher energies is
degraded in the forward direction.  For this reason, we prefer to consider the
third-order results for $\alphaep$ and $\betamp$ of
Eqs.~\eqref{eq:p2parameterfinalfit} and \eqref{eq:p1parameterfinalfit} as the
most reliable, using the fourth-order fit simply for reassurance that these
results are stable against the inclusion of higher orders.

Our results agree within errors with the results of Beane et~al.\ and
Hildebrandt et~al.\ (see Table.~\ref{table-polcomp}).
The errors quoted on the former were taken from the limits of the 1$\sigma$
curve ($\chi^2_{\text{min}}+2.3$) as a function of $\alphaep$ and $\betamp$
and hence are large compared with those obtained from the more conventional
$\chi^2_{\text{min}}+1$ measure used here (equivalent to marginalising over
the other parameter) which explains our tighter errors.  In general, fits
which include data above 170~MeV prefer higher values of both $\alphaep$ and
$\betamp$, which may account for their higher central values.  Like Baranov
et~al.\ \cite{Baranov:2001}, we find that most low-energy data sets, old and
new, are compatible with one another (including, with the caveats of
Section~\ref{sec:expdatacritical}, the \OdL\ data set which was not available
to Baranov at the time), and our results are again compatible with his but
with a distinctly smaller $\alphaep-\betamp$.

Further details of the calculation described in this section are the subject of a forthcoming paper~\cite{McGovern:2012}.


\subsection{\it Other methods} 
\label{sec:other}

We close this section by briefly discussing other approaches which incorporate
the $\pi $N and $\Delta(1232)$ dynamics that are key to describing $\gamma$p
data in this energy domain.

The approach of Gasparyan and Lutz~\cite{Gasparyan:2010xz,Gasparyan:2011yw}
emphasises causality and unitarity. In these works, Compton scattering is
analysed using the $\chi$PT Lagrangian. Partial-wave amplitudes are obtained
by an analytic continuation of amplitudes computed in ($\Delta$-less) $\chi$PT
in the sub-threshold region. An integral equation that implements $\pi
$N-analyticity and unitarity of the six Compton amplitudes is used to
extrapolate to the kinematic region of interest. The difference from the work
presented in Section~\ref{sec:drs} is that, in
Refs.~\cite{Gasparyan:2010xz,Gasparyan:2011yw}, the input for pion
photoproduction that saturates the integral equation is from ${\cal O}(P^3)$
$\chi$PT, not from data. Hence, multi-pion states have not, as yet, been
included in the $s$-channel integral. Moreover, the absence of an explicit
$\Delta(1232)$ field in the Lagrangian is an additional dynamical assumption
in this approach.

Nevertheless, this assumption is somewhat vindicated by comparison with the
data---at least for the specific regulator choice and $\calO(P^3)$ calculation
discussed in Ref.~\cite{Gasparyan:2010xz}. Gasparyan and Lutz can explain the
$\gamma$p data up to energies of roughly $\wlab\approx450$ MeV with a quality
comparable to that obtained in the $\calO(e^2 \delta^3)$ computation
(c.f.~Fig.~\ref{fig:sect5-fig3}). Broadly speaking, this happens because
photoproduction multipoles in this approach are in good agreement with extant
data, and the constraints of gauge invariance, analyticity, and unitarity are
all implemented.  In contrast to the DR results displayed in
Section~\ref{sec:drs} or the $\chi$EFT results discussed in
Section~\ref{sec:protonanalysis}, the $\gamma$p cross sections displayed in
Ref.~\cite{Gasparyan:2010xz} are not a fit, since there are no additional free
parameters in the Compton amplitude at this order. Because the calculation of
sub-threshold amplitudes is matched to $\chi$PT, the $\calO(P^3)$ results for
$\alphaep$ and $\betamp$ of Eq.~\eqref{eq:op3preds} are obtained.  The
$\chi$PT results for $\gammaemp$ and $\gammamep$ shown in
Eq.~\eqref{eq:op3preds} are also recovered, but the values found for
$\gammaeep$ and $\gammammp$ are closer to those obtained in DRs.

Finally, we consider the dressed $K$-matrix model of Kondratyuk and
Scholten~\cite{Kondratyuk:2001qu,Kondratyuk:2001ys}, a relativistic approach
that respects crossing symmetry. Two-body $\pi $N-unitarity is maintained by
computing the scattering amplitude $T$ in a coupled-channels ($\pi $N,
photoproduction and Compton scattering) approach. Analyticity constraints are
fully respected for one-particle reducible diagrams, but violated by the
one-particle irreducible diagrams.  The degrees of freedom used are the
nucleon, the $\Delta(1232)$, and the $\rho$ and $\sigma$ mesons, together with
higher nucleon resonances.  Most parameters are determined by a combined fit
to $\pi $N, $\gamma \mathrm{p} \rightarrow \pi $N and $\gamma$p data. For
Compton scattering, the two key parameters are in the $\gamma \gamma \sigma$
vertex, and they are adjusted to reproduce the backward $\gamma$p cross
section at moderate energies. In order to reproduce the differential cross
section around pion-production threshold, and in particular the cusp seen
there, the one-particle irreducible diagrams shown in Fig.~\ref{fig:loops-3rd}
are added to the calculation in an ad hoc and approximate way.  As in
Ref.~\cite{Gasparyan:2010xz}, the good description of $\pi $N data and
photoproduction up to $\wlab \approx 450$ MeV, together with the
implementation of analyticity and unitarity, helps ensure a fairly successful
result for $\gamma$p cross sections.  After the fit to the data was performed,
polarisabilities were extracted from the model, with results similar to the DR
and $\chi$PT evaluations: $\alphaep=12.1$, $\betamp=2.4$.  The small value of
$\betam$ again results from the $\sigma$ meson cancelling most of the
$\Delta(1232)$ contribution. The diagrams of Fig.~\ref{fig:loops-3rd} are
crucial to describing $\alphaep$ and $\gammaeep$, but are only a small
perturbation in $\betamp$. Together with the subsequent results of
Ref.~\cite{Kondratyuk:2001ys}, this implies that reducibility and analyticity
are not the best guides for constructing a consistent description of
Compton-scattering data for $\wlab < 350$ MeV. It also reminds us that,
especially for structure constants at higher orders in the $\omega$ expansion,
$\pi $N and $\Delta(1232)$ physics play a key role and ultimately determine
many aspects of the shape of Compton observables.



\section{Compton scattering from two- and three-nucleon systems}
\label{sec:2+3N}

Does the neutron have a similar Compton response as the proton? We saw in
Section~\ref{sec:singleN} that \ChiEFT predicts the isoscalar polarisabilities
to be one order in $P/\Lambda_\chi\approx\frac{1}{5}$ larger than the
isovector ones, indicating that neutron and proton polarisabilities should
agree at the 20\% level. This is not unexpected, since the largest
contributions to nucleon polarisabilities of all kinds come from charged-pion
dynamics and excitation of the $\Delta(1232)$, and both of these mechanisms
are predominantly isoscalar.

As already mentioned, the structure functions $\bar{A}_i$~\eqref{eq:strucamp}
of the neutron can be probed indirectly by embedding it into a light, stable
nucleus.  In \ChiEFT, elastic Compton scattering has indeed been explored for
the deuteron in Refs.~\cite{Griesshammer:2010pz, Choudhury:2004yz,
  Hildebrandt:2005ix, Hildebrandt:2005iw, Beane:2004ra,
  Hildebrandt:2004hh,Chen:1998rz,Chen:1998vi,Chen:1998ie,Beane:1999uq} and for
\threeHe in Refs.~\cite{Choudhury:2007bh,Shukla:2008zc,ShuklaThesis}.  Since
the deuteron is isoscalar, its Compton amplitude is sensitive only to the
isoscalar polarisabilities. Neutron polarisabilities are then obtained by
combination with the proton polarisabilities found in the preceding section,
Eqs.~\eqref{eq:p2parameterfinalfit} and \eqref{eq:p1parameterfinalfit}.  We
first concentrate on the deuteron, since this is the only case for which data
are presently available, and we compare these data to non-\ChiEFT calculations
in Section~\ref{sec:gammadmodels}.  The case of \threeHe is discussed in
Sections~\ref{sec:gammaHe3chiral} and~\ref{sec:spinpols} below.  In that case,
both isoscalar and isovector polarisabilities affect the cross section, but
\threeHe behaves approximately as a free-neutron target for polarisation
observables.

In nuclei, nuclear effects such as meson-exchange currents and nuclear binding
act in concert with the single-nucleon Compton amplitude. EFT allows the
rigorous and systematic calculation of such effects in a model-independent way
and with a well-defined theoretical uncertainty. Analysing the proton,
deuteron and \threeHe in one common EFT framework can thus lead to
high-accuracy determinations of the proton and neutron polarisabilities.
However, the corresponding coherent Compton-scattering experiments are
challenging, as described in Section~\ref{sec:expdeutlow}, and the accuracy
claimed in an EFT calculation must be checked by comparing to data. To
overdetermine polarisabilities by multiple extractions from experiments on
different systems thus provides important cross-checks on both experiment and
theory.

\subsection{{\it \ChiEFT for few-nucleon systems}}

The effective theory of few-nucleon systems is significantly more invilved
than that discussed in Section~\ref{sec:singleN}, as the shallow binding of
light nuclei complicates the picture. The S-wave NN scattering lengths
$a(\threeS ) \approx 5\;\fm$ and $a(\oneS) \approx-24\;\fm$ are much larger
than typical large-distance scales of the one-nucleon sector, such as the pion
Compton wavelength. Concomitantly, the deuteron binding energy $B_\mathrm{d}
\approx 2.225\;\MeV$ is small compared to the typical QCD energy scale, and
the corresponding ``binding momentum'' (inverse size) $\gamma \equiv \sqrt{\MN
  B_\mathrm{d}} \approx 46\;\MeV \approx 1/a(\threeS)$ is appreciably less
than even the typical ``chiral'' \ChiEFT scale $m_\pi$.  Therefore, low-energy
S-wave NN rescattering is nonperturbative at low energies. Iterates of the NN
potential are not suppressed by chiral symmetry and must at least partially be
resummed.

In practice this is achieved by defining a leading-order NN potential, which
is then iterated via the Schr\"odinger equation, in order to generate the LO
wave function of the nuclear bound state.  In \ChiEFT, at LO, the long-range
part of the NN potential is given by one-pion exchange, constructed from the
chiral Lagrangian, Eqs.~\eqref{eq:Lpipi2} to \eqref{eq:LpiN2}. Weinberg
proposed~\cite{Weinberg:1990rz,Weinberg:1991um} that this long-range potential
should be supplemented at LO by two S-wave contact interactions, whose
strengths are determined from the above-mentioned fine-tuned physical scales.
This potential yields a Hamiltonian which is unbounded from below, and so a
cutoff must be placed on the Schr\"odinger equation in order to obtain
physically sensible predictions.
Refs.~\cite{Eiras:2001hu,Nogga:2005hy,PVRA06,Birse,Yang:2009kx} have pointed
out that additional contact interactions (e.g.~in attractive P waves) must be
added to this LO potential in order to make the NN phase shifts independent of
the cutoff used in solving the Schr\"odinger equation. A number of proposals
for dealing with orders beyond LO also exist (see,
e.g.,~Refs.~\cite{Beane:2001bc,BKV,Birse:2010fj,Valderrama:2011mv,Long:2011xw}).
The question of how to build a \ChiEFT for NN scattering that is valid over a
large range of cutoffs must be regarded as presently unresolved.
    
However, Epelbaum and Mei\ss ner argue in Ref.~\cite{EM06} that the \ChiEFT
proposed by Weinberg for few-nucleon systems is consistent if the cutoff is
kept at $\lesssim 800 \;\MeV$, namely roughly the mass of the $\rho$ meson.
Weinberg's proposal is that all diagrams contributing to the nuclear potential
are classified according to their $\chi$PT order.  For example, the NLO
potential includes two-pion exchange (constructed from the first-order
vertices in Eq.~\eqref{eq:LpiN2}), together with additional contact
interactions.  The entire potential at a fixed order is then inserted into the
Schr\"odinger equation to obtain the nuclear wave function. For cutoffs in the
range $500$ to $800$~MeV, NN potentials based on this power counting have been
constructed to \NXLO{3}, with very little cutoff dependence observed in this
range~\cite{Entem:2003ft,Epelbaum:2004fk}. Furthermore, this N$^3$LO potential
produces a $\chi^2$ per degree of freedom with respect to the NN data which is
comparable to that of ``high-precision'' NN potential models. Consistent
three-nucleon forces have also been obtained to
N$^3$LO~\cite{vK94,Bernard:2007sp,Bernard:2011zr}. This method has been
successfully applied in a number of reactions on deuterium: $\pi$d,
$\mathrm{e}^-$d, \ldots; see e.g.~\cite{Phillips:2009} for a recent review.
It is this scheme that we shall use to construct \ChiEFT deuteron wave
functions.  Additional details regarding the pros and cons of this procedure
can be found e.g.~in~\cite{seattle_review,Bedaque:2002mn, Epelbaum:2008ga,
  Machleidt:2011zz}.  But, as far as Compton scattering is concerned, we will
demonstrate in Section~\ref{sec:dunify} that {\it any} NN potential can be
used for deuteron Compton scattering at the energies and level of accuracy of
the present data---as long as that potential captures the correct
long-distance physics of one-pion exchange and reproduces NN scattering data
reasonably well. The advantage of Weinberg's power counting for the NN
potential and operators of nuclear Compton scattering is that the consequences
of chiral-symmetry breaking are included in the calculation in a
straightforward manner, and a potential of the minimal complexity needed to
obtain results at the desired level of accuracy is employed.

\subsection{\it Compton scattering from the deuteron in \ChiEFT}

\subsubsection{Scales and regimes}

We now turn to the presentation of \ChiEFT calculations of $\gamma$d
scattering.  As before, we begin by examining the \ChiEFT variant without
dynamical $\Delta(1232)$ degrees of freedom and postpone including the
$\Delta$ to Section~\ref{sec:RegimeII}. Thus we first count in powers of the
generic scale $\Q$ and still define the leading-order nuclear Thomson
term\footnote{Different power-counting notations exist in the literature,
  including renaming the parameter to $Q$. In some, the deuteron Compton
  amplitudes start at order $e^2Q^{-1}$~\cite{Griesshammer:2010pz,
    Griesshammer:2008, Griesshammer:2007aq, Griesshammer:2004yn}, but our
  choice allows an intuitive translation between the one- and two-nucleon
  sectors; see also the note below Eq.~\eqref{eq:1Nseagull}.} as
$e^2\sim\Q^2$.  

The deuteron binding momentum is not the only new scale in few-nucleon
systems. In the one-nucleon sector of Section~\ref{sec:singleN}, recoil
effects are suppressed because the nucleon mass is much larger than the energy
of any pions or photons participating in the reactions of interest.  Recoil
between two nucleons cannot be neglected, however, because they have the same
mass. The nonrelativistic kinetic energy of a nucleon with momentum $\pv$
propagating close to its mass shell in the few-nucleon system is thus
$E\approx{\pv^2}{/(2\MN)} \sim \Q^2 \ll |\pv| \sim \Q$. After interaction with
a photon, its propagator is given by
\begin{equation}
  \label{eq:simplepropagator} 
  \frac{\ii\MN}{\MN E \pm \MN \w- (\pv \pm\kv)^2}\;\;, 
\end{equation}
which converts the photon energy into a new NN momentum scale $\sqrt{\MN \w}
\gg \w$ which is ``hard'' relative to $\w$. The $(+)$ sign applies to photon
absorption by the nucleon and the $(-)$ sign to photon emission.
 This scale only appears when the photon's energy-momentum
flow must be routed through a nucleon and not when two photons couple
instantaneously to the same nucleon; cf.~discussion of
Fig.~\ref{fig:deuteronLOThomson} below.

While a fourfold expansion in the low \ChiEFT scales $\w$, $\sqrt{\MN \w}$,
$\gamma$ and $\mpi$ is possible, it is more convenient to approximately
identify scales until only one is left. We set $\gamma \sim \mpi$ and define
two different \textbf{regimes} of photon energy, relative to energy scales
built out of the chiral scale $\Q
$~\cite{Chen:1998ie, Chen:1998vi, Beane:1999uq, Beane:2004ra}.

\textbf{Regime I} has comparable chiral and hard photon scales, $\sqrt{\MN \w
}\sim \mpi \sim \Q$, i.e.~$\w\sim20\;\MeV$, so that the intermediate-state
propagator~\eqref{eq:simplepropagator} scales as $\Q^{-2}$.  This Regime is
discussed in Section~\ref{sec:2Npionless}.

\textbf{Regime II}, on the other hand, treats the photon energy as a soft
scale which is close to the chiral one, $\w \sim 140\;\MeV \sim \Q$, so that
the hard scale is $\sqrt{\MN \w} \sim \Q^{\half}$ and the intermediate-state
propagator~\eqref{eq:simplepropagator} counts as $\Q^{-1}$, i.e.~one order
less.  Numerically, this implies $\sqrt{\MN \w} \sim 360\;\MeV$, which is
still small compared to the breakdown scale of \ChiEFT when the $\Delta(1232)$
is included dynamically. But convergence for these energies is in powers of
$\sqrt{\MN \w}/\Lambda_\chi$ and hence not as rapid as in Regime I. This will
be discussed in Section~\ref{sec:2Nhigherenergies}, where we also include a
dynamical $\Delta(1232)$ as we did in the single-nucleon sector of
Section~\ref{sec:includeDelta2}.

We will see that the mechanisms probed in each Regime are quite different.
The transition from one to the other is obviously not abrupt but rather
gradual, and the central values defined here provide only a priori estimates.
For example, Regime II involves an expansion in $
{(\gamma^2,m_\pi^2)}{/(\MN \omega)}$ which breaks down completely when $\w
\sim 
{\mpi^2}{/\MN}\lesssim20\;\MeV$. The discussion in
Section~\ref{sec:2Nhigherenergies} will even show that its higher-order terms
already have a marked effect for $\w \approx 50$ MeV.  Since the available
deuteron data lie in that intermediate region, $49\;\MeV\le\wlab\le95\;\MeV$,
we will analyse them in Section~\ref{sec:deuteronanalysis} by a formulation
which is applicable in both Regimes, introduced in Section~\ref{sec:dunify}.

\subsubsection{Very low energies and the Thomson limit} 
\label{sec:2Npionless}

\label{sec:2Nmotivation}

The Thomson limit of vanishing photon energy (see Eqs.
\eqref{eq:thomsoncrosssection} and \eqref{eq:LET}) imposes a stringent
constraint on the few-nucleon Compton amplitudes in Regime I since the photon
cannot resolve the target structure or spin:
\begin{equation}
  \label{eq:thomson}
  \lim\limits_{\w\to0}T(\w,\theta)=-\frac{Z^2e^2}{M_X}\;
  \vec{\epsilon}\,'^*\cdot \vec{\epsilon}+\calO(\w)\;\;.
\end{equation}
Since $e^2\sim\Q^2$, the amplitude indeed scales as $\Q^2$.  While such a
low-energy expansion breaks down as $\w$ approaches the first intrinsic
low-energy scale of the object, it not only dominates the deuteron amplitude
for $\w \lesssim B_\mathrm{d}$, but constrains the cross section throughout
Regime I; see Section~\ref{sec:dunify}.

The significance of the Thomson limit comes from the fact that it is based on
gauge invariance. This, in turn, implies that the total matrix element of all
electromagnetic current operators is conserved\footnote{Gauge invariance and
  current conservation are identical in the absence of on-shell photons in
  loops, as in all Compton calculations to date.}, including photon couplings
to charged mesons like the pion which provide nuclear binding. Thus, Friar
used the Thomson constraint in conjunction with the generalised Siegert
theorem to avoid specifying the exact form of meson-exchange current
operators~\cite{Friar:1975,Friar:1977zq}.  In later work, Arenh\"ovel and
Weyrauch constructed explicit expressions for these operators and showed that
the theorem enforces cancellations between different
contributions~\cite{Arenhovel:1980jx,Weyrauch:1984tf}. These significant
simplifications and stringent numerical tests form the foundation of the
modern theoretical description of deuteron Compton scattering in Regime
I~\cite{Griesshammer:2010pz, Hildebrandt:2005ix, Hildebrandt:2005iw,
  Weyrauch:1988zz, Weyrauch:1990zz, Wilbois:1995, Levchuk:1995,
  Levchuk:1999zy, Karakowski:1999pt, Karakowski:1999eb}. Ultimately, the
practical importance of the Thomson limit is not only the \emph{fact} that it
is fulfilled, but also \emph{how} it is fulfilled. Here, we motivate the
results of Refs.~\cite{Friar:1975,Friar:1977zq,
  Arenhovel:1980jx,Weyrauch:1984tf, Sachs:1951zz} from the EFT perspective,
cf.~\cite{Griesshammer:2010pz, Griesshammer:2008, Griesshammer:2007aq,
  Griesshammer:2004yn}.

\begin{figure}[!htbp]
  \centerline{\includegraphics*[width=0.8\linewidth]{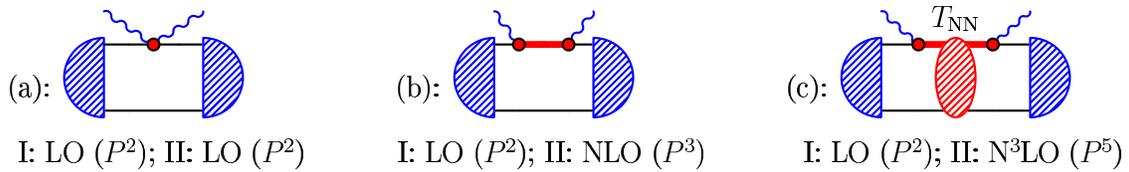}}
  \caption{\label{fig:deuteronLOThomson} (Colour online) Deuteron Compton
    scattering in Regime I at LO, $\calO(\Q^{2})$: one-nucleon seagull term
    (a); intermediate-nucleon propagation without, (b), and with rescattering
    (c; hatched ellipse: $T_\mathrm{NN}$).  Thick (red) line: nucleon carrying
    photon energy as in Eq.~\eqref{eq:comptonpropagator}; dot: coupling from
    $\calL^{(2)}_{\pi\text{N}}$ via minimal substitution and the magnetic
    moment.  Crossed and permuted diagrams not displayed. Also indicated is
    the order at which each graph contributes in Regimes I and II,
    respectively.}
\end{figure}

An exact low-energy theorem must be observed at each individual order in the
generic expansion parameter of any EFT. In the one-nucleon sector of
Section~\ref{sec:singleN}, the Thomson limit is automatically fulfilled; but
NN rescattering complicates the picture in few-nucleon systems. Figure
\ref{fig:deuteronLOThomson} shows all leading-order, $\calO(\Q^{2})$,
contributions in Regime I.
In diagram (a), the one-nucleon seagull term is convoluted with the deuteron
wave function, and no momentum is transferred at zero energy:
\begin{equation}
  \label{eq:1Nseagull}
  T_\text{seagull}(\w\to0)=\langle \Psi_\mathrm{d}|
  \left(-\frac{e^2}{\MN}\;
    \vec{\epsilon}\,'^*\cdot \vec{\epsilon}\right)|\Psi_\mathrm{d}\rangle=
  -\frac{e^2}{\MN}\;  \vec{\epsilon}\,'^*\cdot \vec{\epsilon}\;\;.
\end{equation}
This is the Thomson limit for an individual nucleon, i.e.~twice the deuteron
result, and would therefore overpredict the deuteron cross section by a factor
of $4$.  Sachs and Austern showed that contributions (b) and (c) yield an
amplitude $+
{e^2}\; (\vec{\epsilon}\,'^*\cdot \vec{\epsilon}){/(2 \MN)}$ at zero energy,
and so cancel half of the seagull diagram for the correct Thomson
limit\footnote{A 0.1\% correction from relativistic effects at higher orders
  replaces $2\MN$ with the deuteron mass,
  $M_\mathrm{d}=2\MN-B_\mathrm{d}$.}~\cite{Sachs:1951zz}. This occurs because,
in contrast to (a), the photon energy in (b) and (c) flows through the
two-nucleon state between emission and absorption, with the (highlighted)
intermediate-state propagators of the form in Eq.~\eqref{eq:simplepropagator}:
\begin{equation}
  \label{eq:comptonpropagator}
  \frac{\ii \MN}{\w^2\pm\MN\w-\gamma^2 -\qv^2}\;\;.
\end{equation} 
The $(+)$ sign applies to the diagrams in Fig.~\ref{fig:deuteronLOThomson}(b)
and (c), while the $(-)$ sign applies to the ``crossed'' graphs.  In Regime I,
$\MN \w \sim \gamma^2 \sim \Q^2$, the propagator scales as $\Q^{-2}$, and loop
momenta $q \sim \Q$ dominate. Intuitively, the initial coherent two-nucleon
state is not perturbed very much by photon absorption and propagates
coherently on a typical length scale $1/q\gtrsim1/\mpi$, which is larger than
the anomalously large NN scattering lengths. Multiple NN interactions are
therefore not parametrically suppressed before another photon is emitted to
produce the final deuteron--photon state, and rescattering must be re-summed.
Power counting bears this picture out. Diagram (c) contributes at LO, just
like (a) and (b) in this Regime, since the scattering amplitude of two
nucleons close to their mass shell scales as $T_\mathrm{NN} \sim \Q^{-1}$, as
in any EFT with fine-tuned NN scales~\cite{Griesshammer:2010pz,
  Griesshammer:2008, Griesshammer:2007aq,
  Griesshammer:2004yn,seattle_review,Bedaque:2002mn}.

Clearly, the three contributions are of very different computational
complexity: (a) is analytically known in the Thomson
limit~\eqref{eq:1Nseagull}, irrespective of the deuteron wave function used;
(b) contains a convolution of the propagator~\eqref{eq:comptonpropagator} with
$|\Psi_\mathrm{d}\rangle$, which is usually not known in closed form; (c)
involves an off-shell rescattering matrix $T_\mathrm{NN}$ and thus depends not
only on $|\Psi_\mathrm{d}\rangle$, but also directly on all NN partial waves
and the potential used to generate them. In addition, only the sum of diagrams
(b) and (c) is independent of the UV regulator.  Since the Thomson limit
requires that the sum of the last two must give $-\half$ of the seagull, it
can be used as a nontrivial check of numerical evaluations.  This condition
must hold irrespective of the particular potential or wave function chosen and
does not even depend on the particle content of the theory.  After recent
numerical improvements in the \ChiEFT implementation of Hildebrandt et
al.~\cite{Griesshammer:2010pz, Hildebrandt:2005ix, Hildebrandt:2005iw}, the
cancellation is fulfilled for each helicity amplitude with a relative
numerical accuracy of $
\le 0.01\%$. 

\begin{figure}[!htbp]
  \centerline{\includegraphics*[width=0.8\linewidth]{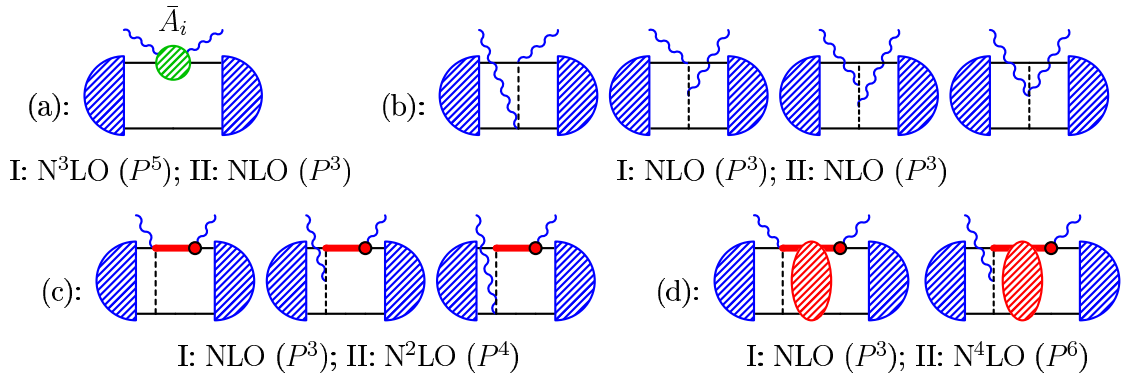}}
  \caption{\label{fig:deuteronNLOThomson} (Colour online) Deuteron Compton
    scattering in Regime I at NLO, $\calO(\Q^{3})$: one-nucleon structure
    terms (a); photons coupling to the same pion-exchange currents (b); or to
    exchange current and nucleon without, (c), and with, (d), rescattering.
    Crossed and permuted diagrams not displayed.}
\end{figure}

All contributions at NLO ($\Q^3$) in Regime I are listed in
Fig.~\ref{fig:deuteronNLOThomson}, providing a projected relative accuracy of
$(\Q/\Lambda_\chi)^2 \approx 1/5^2\approx4\%$.  For them, the constraint is
even more intricate.
Since the Thomson limit is already fulfilled at LO, all NLO contributions must
sum to zero as $\w \to 0$.  Equation~\eqref{eq:strucamp} implies that the
one-nucleon structure amplitudes $\bar{A}_i$ of (a) trivially obey the
constraint since they vanish as $\w\to 0$.  In (b), wave functions are
convoluted by six-dimensional integrals with an instantaneous meson-exchange
current to which the photons couple directly. The other graphs involve
convolutions of intermediate-state nucleon propagators without, (c), and with,
(d), rescattering amplitudes $T_\mathrm{NN}$ which have to be constructed
separately.  Fortunately, Arenh\"ovel and Weyrauch demonstrated three decades
ago that, while (b), (c) and (d) are individually large and cutoff-dependent,
they indeed cancel at $\w=0$ in the sum for any potential and wave
function~\cite{Arenhovel:1980jx, Weyrauch:1984tf}.  Since rather different
techniques are employed to calculate the different contributions, this
provides another nontrivial cross-check. In the \ChiEFT implementation of
Hildebrandt et al.~\cite{Griesshammer:2010pz, Hildebrandt:2005ix,
  Hildebrandt:2005iw}, it holds in each helicity amplitude to a relative
accuracy of $\le 0.9\%$.  The Thomson limit therefore provides a stringent
check on the numerical implementation and the consistent treatment of the
potential, the wave function and the NN current operator. Subsequent numerical
improvements result in the Thomson limit now being restored to better than
$0.2$\%.

\subsubsection{Beyond the Thomson limit: reducing rescattering contributions,
  adding the $\Delta(1232)$}
\label{sec:2Nhigherenergies}
\label{sec:RegimeII}

At higher energies, $\w \sim \Q \approx 140 \;\MeV$ and $\sqrt{\MN \w} \sim
\Q^\half$, the Thomson limit is not directly
significant~\cite{Griesshammer:2008, Beane:1999uq, Beane:2004ra,
  Griesshammer:2007aq, Griesshammer:2004yn}.  The struck nucleon is far
off-shell, $E \sim \sqrt{\MN \w} \gg q^2,\,\gamma^2 ,\,\w^2$, and at LO the
intermediate-state propagator~\eqref{eq:comptonpropagator} becomes $\ii/\w$.
The nucleon behaves as if static, and the standard \ChiEFT counting of the
one-nucleon sector prevails.  Physically, each nucleon propagates
incoherently, i.e.~as if the other were absent, because its wavelength $1/q
\sim 1/\sqrt{\MN \w} \approx 1/(360 \;\MeV)$ is much shorter than the
rescattering scale $1/\gamma\approx1/(50\;\MeV)$. The struck nucleon has
little time to scatter with its partner before the second photon is radiated
to restore the coherent final state. Thus, rescattering is suppressed and can
be treated perturbatively. This leads to significant computational
simplifications and changes the relative importance of many contributions.

The formal power counting, indicated in Figs.~\ref{fig:deuteronLOThomson}
and~\ref{fig:deuteronNLOThomson}, comes to the same conclusion.  The seagull
in Fig.~\ref{fig:deuteronLOThomson}(a) has no two-nucleon intermediate state
and thus still scales as $\Q^{2}$, constituting LO.  Since an
intermediate-state propagator (Eq.~\eqref{eq:comptonpropagator} and thick
lines in Figs.~\ref{fig:deuteronLOThomson} and~\ref{fig:deuteronNLOThomson})
now scales only as $\ii/\w\sim \Q^{-1}$, each of its occurrences moves a
diagram to one order higher. A diagram with one intermediate NN state (but
without rescattering, Fig.~\ref{fig:deuteronLOThomson}(b)) is demoted from LO
to NLO ($\Q^{3 }$) and the diagrams of Fig.~\ref{fig:deuteronNLOThomson}(c)
from NLO to \NXLO{2} ($\Q^4 $). Since $T_\mathrm{NN}\sim\Q^0$ now, each
rescattering in Figs.~\ref{fig:deuteronLOThomson}(c)
and~\ref{fig:deuteronNLOThomson}(d) costs one additional power of $\Q$
(c.f.~Regime I). In addition, there is one more power of $P$ than in the
Regime I counting of this diagram from each intermediate-state NN propagator
that is present. Thus, these two graphs are each demoted by at least $P^3$, to
order $\Q^5 $ (\NXLO{3}) and beyond. On the other hand, the nucleon-structure
contributions $\bar{A}_i$ of Fig.~\ref{fig:deuteronNLOThomson}(a) are enhanced
from order $\Q^5$ (\NXLO{3}) in Regime I to order $\Q^3$ (NLO) since they
scale with $ {\w^2}{/\mpi}$, see (\ref{eq:strucamp}) and
Section~\ref{sec:comptonEFT}.  The meson-exchange diagrams of
Fig.~\ref{fig:deuteronNLOThomson}(b) still count as NLO, $\Q^{3}$.
Contributions from loop momenta $q \sim \Q^\half$ make the counting proceed in
half-integer powers of $\Q$ which however only appear beyond NLO.

This implies that, up to \NXLO{3} corrections, contributions from the
off-shell matrices $T_\mathrm{NN}$ of NN rescattering are absent in Regime II.
This simplification was first observed by Beane et al.~\cite{Beane:1999uq} and
is crucial to the success of the \ChiEFT deuteron Compton-scattering
calculations in Refs.~\cite{Beane:1999uq, Beane:2004ra, Hildebrandt:2004hh}.
The Compton amplitude for the deuteron can then be computed as
\begin{equation}
  T_{\gamma \mathrm{d}}=\langle \Psi_\mathrm{d}|\left[T_{\gamma \mathrm{N}} +
    T_{\gamma \mathrm{NN}}\right]|\Psi_{\mathrm{d}} \rangle\;\;.
  \label{eq:wein}
\end{equation}
Both contributions are direct convolutions of irreducible photonuclear kernels
with deuteron wave functions.  $T_{\gamma \mathrm{N}}$ is the single-nucleon
Compton amplitude, and $T_{\gamma \mathrm{NN}}$ the irreducible amplitude for
$\gamma \mathrm{NN} \rightarrow \gamma \mathrm{NN}$, with all NN interactions
truncated after (before) the departure (arrival) of the outgoing (incoming)
photon.
A complete calculation for $T_{\gamma\mathrm{N}}$ at \NXLO{2} ($\Q^4$)
therefore consists of only the diagrams of
Figs.~\ref{fig:deuteronLOThomson}(a) and (b) and
Fig.~\ref{fig:deuteronNLOThomson}(a), and for $T_{\gamma\mathrm{NN}}$ consists
of the ``exchange current'' diagrams of Figs.~\ref{fig:deuteronNLOThomson}(b)
and (c).  For other light nuclei, the only change is to replace the wave
function in Eq.~\eqref{eq:wein} with that of the target nucleus.

The pioneering work of Beane et al.~\cite{Beane:1999uq} used all NLO ($\Q^3$)
contributions in Regime II, namely the isoscalar exchange currents of
Fig.~\ref{fig:deuteronNLOThomson}(b) and the one-body mechanisms of
Figs.~\ref{fig:deuteronLOThomson}(a) and (b),
and~\ref{fig:deuteronNLOThomson}(a) with the single-nucleon
$T_{\gamma\mathrm{N}}$ amplitude computed in \ChiEFT without explicit
$\Delta(1232)$ at the matching order, $\calO(P^3)$.  The amplitude is thus
complete up to corrections $\sim {(\gamma^2,\,\mpi^2)}{/(\MN \w)}$, which
indicates effects in the amplitude of a few percent at $\omega=100$ MeV,
increasing to $\approx 10$\% at the lower end of the data range. Indeed,
agreement with higher-energy data is good, but the $49$~MeV data of
Lucas~\cite{Lucas:1994} are significantly overestimated; see
Fig.~\ref{fig:d-rescattdep}.  The two-body currents due to Compton
scattering from one-pion exchange turned out to be sizable at all energies and
are crucial to reasonably describe all of the data.
 
Subsequently, Beane et al.~fitted the scalar nucleon polarisabilities from
deuteron data~\cite{Beane:2002wn,Beane:2004ra}. They extended the calculation
to \NXLO{2} by incorporating the $P^4$ mechanisms for single-nucleon
scattering discussed in Section~\ref{sec:singleN} and the two-body mechanisms
of Fig.~\ref{fig:deuteronNLOThomson}(c). At this order, these two-body
diagrams are indeed \emph{all} contributions to $T_{\gamma\mathrm{NN}}$ with
one vertex from ${\cal L}_{\pi\mathrm{N}}^{(2)}$ and one from ${\cal
  L}_{\pi\mathrm{N}}^{(1)}$.


Neither implementation included the effects of an explicit $\Delta(1232)$
degree of freedom. As discussed in Section~\ref{sec:includeDelta}, the
$\Delta$ contributes substantially in the one-nucleon sector, in particular at
backward angles for $\omega\gtrsim100$ MeV due to the rise of $\betam$ to
about four times its static value. It would therefore be expected to have a
similarly large effect in the deuteron, in particular for Hornidge's SAL data
at $\wlab=94.5\;\MeV$~\cite{Hornidge:2000}.
Except in the one-nucleon amplitude itself, no new contributions arise with a
dynamical $\Delta$. Since the deuteron is isoscalar, there is no $\Delta$N
component in its wave function. The dynamical $\Delta\Delta$ component is
strongly suppressed below $\w\sim2\Delta_M$. No explicit $\Delta$ appears in
intermediate-nucleon graphs analogous to
Figs.~\ref{fig:deuteronLOThomson}(b,c) and \ref{fig:deuteronNLOThomson}(c,d),
or in two-body currents similar to Fig.~\ref{fig:deuteronNLOThomson}(b).  The
two-nucleon part at order $P^3$ (NLO) in the $\Delta$-less theory is thus
identical to the one at orders $\epsilon^3$ and $e^2\delta^3$ with explicit
$\Delta$. We continue to use $\Q$ to parameterise contributions of the
two-nucleon sector for the variants with or without an explicit $\Delta$, and
to indicate the counting of the single-nucleon amplitudes embedded in
Fig.~\ref{fig:deuteronNLOThomson}(a) separately.  Recall that for diagrams
with only pions and nucleons, the three power countings translate as
$P^n\sim\epsilon^n\sim e^2\delta^{2n-4}$, see Sections~\ref{sec:includeDelta}
to \ref{sec:earlyfit}.


To summarise, an explicit $\Delta(1232)$ does not appear in the two-nucleon
contributions until high orders, and one can therefore insert any of the three
versions in Section~\ref{sec:singleN} for the single-nucleon contributions:
without a dynamical $\Delta(1232)$ at $\calO(P^4)$; with explicit $\Delta$ at
$\calO(\epsilon^3)$, modified by adding the LECs
$\delta\alphaes,\,\delta\betams$; and with even more pion-nucleon loops at
$\calO(e^2\delta^4)$.  Since the two-nucleon sector must be treated to the
same order, the latter case warrants including the two-nucleon contributions
of Fig.~\ref{fig:deuteronNLOThomson}(c). As discussed in
Section~\ref{sec:singleN}, a single-nucleon calculation at order $\epsilon^3$
is actually equivalent to one at order $e^2\delta^3$ for $\omega\sim\mpi$,
i.e.~over the whole energy range of Regimes I and II. The proton fit above
used the resummed, covariant $\Delta$ propagator, but that is not necessary
here.  All published deuteron data lie below the pion-production threshold, so
we can treat the nonzero $\Delta$ width and relativistic effects
perturbatively and consider corrections for the pion-threshold position to be
negligible~\cite{Hildebrandt:2005ix,Hildebrandt:2004hh}, in contrast to the
single-nucleon case in Section~\ref{sec:singleN}.

Hildebrandt et al.~performed such a NLO calculation in Regime II, with a
dynamical $\Delta(1232)$ at modified order
$\epsilon^3$~\cite{Hildebrandt:2004hh}. The total accuracy is $\epsilon^2$
relative to LO, i.e.~\NXLO{2}, as the first order in which not all
contributions are consistently retained in both the one- and two-nucleon
sectors.  The formulation does not improve the accuracy found by Beane et
al.~\cite{Beane:1999uq} at lower energies, but it markedly improves the
agreement with Hornidge's SAL data~\cite{Hornidge:2000}; cf.~discussion of
Fig.~\ref{fig:d-rescattdep} below.

\subsubsection{A unified description of Regimes I and II}
\label{sec:dunify}

The deuteron database covers $49$ to $95\;\MeV$, i.e.~energies that overlap
both Regimes I and II.  An accuracy better than 10\% necessary for a
meaningful extraction of nucleon polarisabilities can therefore only be
achieved if one incorporates \emph{all} the effects which are LO or NLO in
either Regime I or Regime II, as well as the energy dependence of the
$\Delta(1232)$ at $e^2\delta^3\sim\epsilon^3$(modified) that so prominently
affects $\betam(\omega)$. This guarantees both the correct Thomson limit and
an accurate description at higher energy.

This was achieved by Hildebrandt et
al.~\cite{Hildebrandt:2005ix,Hildebrandt:2005iw,Griesshammer:2010pz}.
Combining the discussions of Regime I and II above, we see that when all
diagrams of Figs.~\ref{fig:deuteronLOThomson} and \ref{fig:deuteronNLOThomson}
are included, one actually performs a calculation which is complete for the
two-nucleon sector up to \NXLO{2 } ($\Q^4$) corrections in both Regimes
simultaneously and can move smoothly between them. In Regime I, the only
higher-order diagram is the nucleon-structure contribution of
Fig.~\ref{fig:deuteronNLOThomson}(a). In Regime II, one includes all
corrections at \NXLO{2} (Fig.~\ref{fig:deuteronNLOThomson}(c)) as well as
several higher-order contributions, as in Figs.~\ref{fig:deuteronLOThomson}(c)
and~\ref{fig:deuteronNLOThomson}(d). In each Regime, the accuracy of the
two-nucleon diagrams is still set by the last order for which all terms are
included, namely $\Q^3$ in Regime I and $\Q^4$ in Regime II. The accuracy of
the single-nucleon sector must match and thus can at most include all diagrams
at orders $P^4$ without explicit $\Delta(1232)$, or, with it, $\epsilon^3$ and
$e^2\delta^4$, respectively.

One may wonder about the predicted decrease in accuracy at lower energies,
where an EFT description is supposed to be better. However, since the Thomson
limit is already fulfilled at LO, higher-order corrections approach zero
order-by-order as $\w$ is decreased. The relative size of the total NLO
correction is thus not given by $\Q$ at low energies, but rather is
proportional to $\Q$ multiplied by powers of $\w$. As the importance of the
Thomson limit decreases in Regime II, the relative size approaches $\Q$.

It is this formulation which we will use to extract the isoscalar electric and
magnetic dipole polarisabilities from all deuteron data in
Section~\ref{sec:deuteronanalysis}.  Since isovector polarisabilities enter
only at higher orders in \ChiEFT, agreement with deuteron data should be good
when the isoscalar amplitudes obtained in a proton fit are inserted into the
deuteron calculation, which is then parameter-free. This is confirmed in the
following discussion.

\begin{figure}[!htbp]
  \centerline{
    \includegraphics*[width=0.4\linewidth]{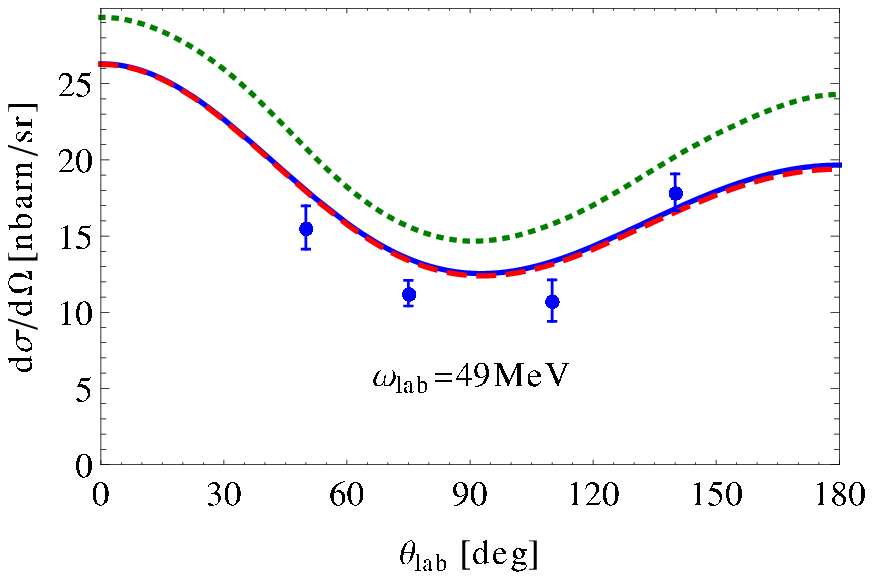}
    \hq\hq\hq\hq
    \includegraphics*[width=0.4\linewidth]{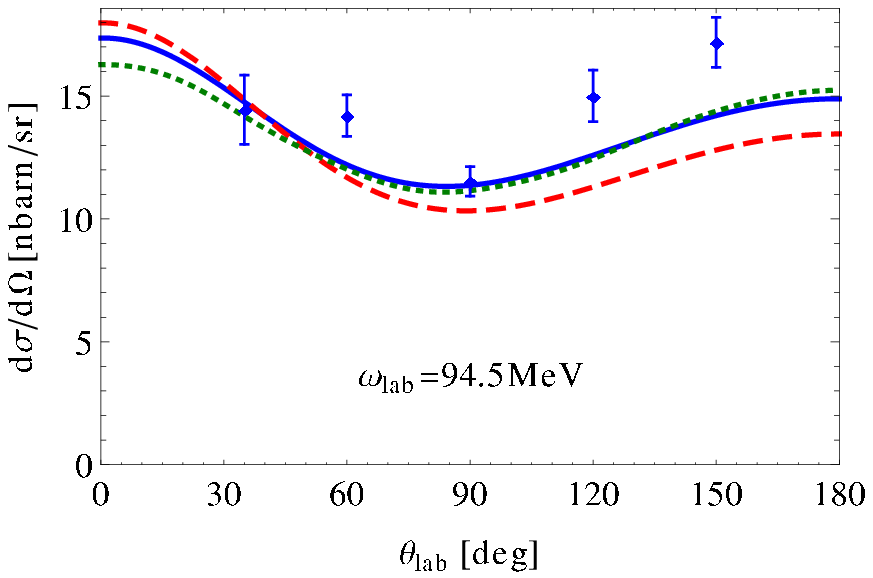}
  }
  \caption{\label{fig:d-rescattdep}\label{fig:ddeltadep} (Colour online)
    Comparison of the deuteron data at $49$ and $94.5\;\MeV$ with \ChiEFT
    predictions (data key in Table~\ref{table-deuteron-low}).  Solid (blue):
    with both dynamical $\Delta(1232)$ and rescattering (``unified approach'',
    modified order $\epsilon^3\sim e^2\delta^3$); dotted (green): with
    $\Delta(1232)$, but without rescattering (strict Regime II counting);
    dashed (red): without dynamical $\Delta(1232)$, but with rescattering
    (``unified'' $\calO(P^3)$).  Static scalar polarisabilities are always set
    at the $\calO(P^3)$ values: $\alphaes=10\betams=12.5$.}
\end{figure}

\begin{figure}[!htbp]
  \centerline{\includegraphics*[width=0.4\linewidth]{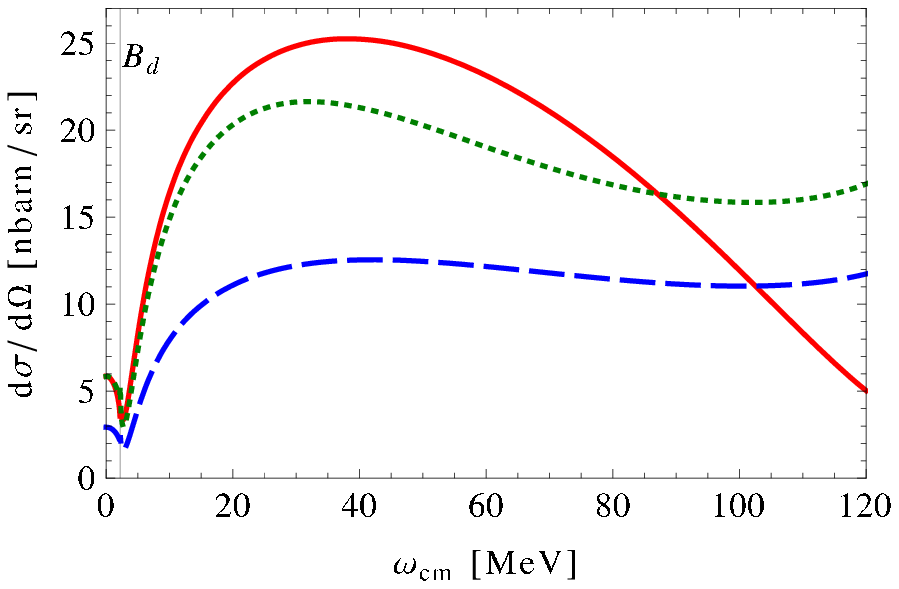}
    \hq\hq\hq\hq
    \includegraphics*[width=0.4\linewidth]{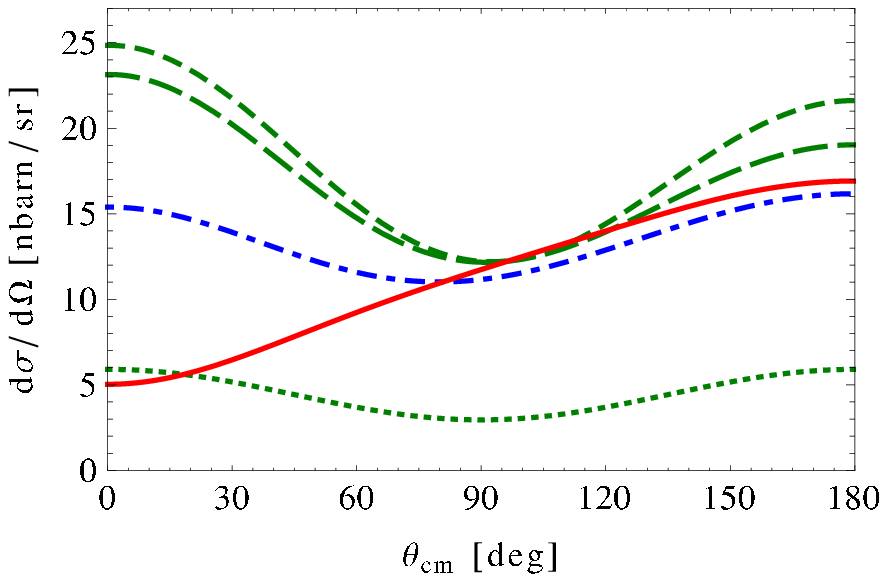}}
  \caption{\label{fig:d-omegadependence} (Colour online) Energy and angle
    dependence of unpolarised deuteron Compton scattering in the
    centre-of-mass frame in the unified approach of \ChiEFT with explicit
    $\Delta(1232)$ and the static scalar polarisabilities set to the
    $\calO(P^3)$ values $\alphaes=10\betams=12.5$; see
    Eq.~\eqref{eq:op3preds}.  Left: at $\thetacm=0^\circ$ (solid red line),
    $90^\circ$ (dashed blue), $180^\circ$ (dotted green). Right: at the
    Thomson limit (dotted green), $\wcm=30\;\MeV$ (short-dashed green),
    $60\;\MeV$ (long-dashed green), $90\;\MeV$ (dash-dotted blue), $120\;\MeV$
    (solid red).}
\end{figure}

First, we assess the transition between Regimes I and II.
Figure~\ref{fig:d-rescattdep} compares two otherwise identical calculations in
\ChiEFT with a dynamical $\Delta(1232)$ at order $e^2\delta^3\sim\epsilon^3$
(both modified)~\cite{Hildebrandt:2005ix, Hildebrandt:2005iw} with the same
static scalar polarisabilities of Eq.~\ref{eq:op3preds}.
The first takes the unified approach; the other follows
Refs.~\cite{Beane:1999uq,Hildebrandt:2004hh} for strict Regime II counting at
NLO, using only \eqref{eq:wein} and no rescattering.  The results indeed
converge to each other with increasing $\w$ as claimed by Beane et
al.~\cite{Beane:1999uq}, but the calculation with the correct Thomson limit
clearly provides a better description of the data at $\wlab=49$ MeV.
In the strict Regime II calculation, corrections from higher orders in $
{(\gamma^2,\mpi^2)}{/(M_N \omega)}$ (including rescattering) increase
dramatically as the deuteron breakup at $\wcm=B_\mathrm{d}$ is approached from
above, as also seen in Fig.~\ref{fig:d-omegadependence}.
Comparison with Fig.~\ref{fig:sect5-fig3} shows that the deuteron and proton
results are of similar size for $\gtrsim70\;\MeV$.  The angular distribution
becomes quickly skewed: forward scattering becomes weaker than backward
scattering for $\wcm\gtrsim80\;\MeV$, and weaker than even the $90^\circ$
cross section above $100\;\MeV$.

Of course, this is exactly the effect of the $\Delta(1232)$ discussed above.
In Fig.~\ref{fig:ddeltadep}, we compare the deuteron data to the
parameter-free predictions of the unified calculation with two variants of the
single-nucleon amplitudes which produce the static polarisabilities of
Eq.~\eqref{eq:op3preds}: without a dynamical $\Delta$ at order $P^3$, and with
an explicit $\Delta$ at order $e^2\delta^3\sim\epsilon^3$(modified).  The
results are depicted in Fig.~\ref{fig:ddeltadep} for the lowest- and
highest-energy deuteron data.
At low energies, including the $\Delta$ as a dynamical degree of freedom has
no significant effect, and either variant agrees well with the data, as
required by the decoupling theorem. With increasing energy, however, effects
of the $\Delta$ are particularly important to match the angular dependence of
Hornidge's SAL backward-angle data at $94.5\;\MeV$~\cite{Hornidge:2000}. As
mentioned in Section~\ref{sec:RegimeII}, this was first demonstrated in the
variant without rescattering contributions by Hildebrandt et
al.~\cite{Hildebrandt:2005ix, Hildebrandt:2004hh}.  Indeed, the strong energy
dependence from the $\Delta$ excitation helps resolve the discrepancy between
these data and earlier calculations in both \ChiEFT
(Section~\ref{sec:RegimeII}) or models (Section~\ref{sec:gammadmodels} below).
An alternative remedy is to exclude the two backward-angle points since they
correspond to a momentum transfer beyond \ChiEFT without explicit
$\Delta(1232)$ even at $\calO(P^4)$, as described in
Section~\ref{sec:earlyfit}~\cite{Beane:2002wn,Beane:2004ra}.

For the rest of this section, we address some more technical issues of this
approach~\cite{Hildebrandt:2005ix, Hildebrandt:2005iw, Griesshammer:2010pz}.
Quite different numerical implementations were merged.  The Compton amplitudes
involving the one- and two-nucleon kernels $T_{\gamma \mathrm{N}}$ and
$T_{\gamma \mathrm{NN}}$ of Eq.~\eqref{eq:wein} are numerically evaluated as
three- and six-dimensional integrals, respectively, against the deuteron wave
function, following Beane et al.~\cite{Beane:1999uq}. The
two-nucleon-reducible contributions may be represented in the form:
\begin{equation}
  \epsilon_i \;
  \langle\Psi_{\mathrm{d}}|\left[{\bf J}_i G(-B_\text{d} + \omega) {\bf
      J}_j^\dagger +{\bf J}_j^\dagger G(-B_\text{d} - \omega) {\bf J}_i
  \right]|\Psi_{\mathrm{d}} \rangle  \;\epsilon_j\;\;. 
  \label{eq:Mrescatt} 
\end{equation}
The second term represents the crossed diagrams not shown in
Figs.~\ref{fig:deuteronLOThomson}(b) and (c) and
\ref{fig:deuteronNLOThomson}(c) and (d).  ${\bf J}$ denotes the NN current
operator including mesonic currents and $G$ is the (interacting) NN Green's
function, i.e.~the off-shell $S$-matrix of the NN system.  Therefore,
Eq.~\eqref{eq:Mrescatt} includes both free propagation of the NN pair and
rescattering via $T_\mathrm{NN}$. It is constructed in coordinate space
following Arenh\"ovel et
al.~\cite{Arenhovel:1980jx,Weyrauch:1984tf,Arenhovel:1986}; cf.~\cite[Chapter
3.7.1]{ericsonweise} and the particularly clear presentation by
Karakowski~\cite{Karakowski:1999pt}.  The fact that the cross section at the
Thomson limit is now obtained to better than $ 0.2\%$
accuracy~\cite{Griesshammer:2010pz} when the three contributions are combined
provides a nontrivial check.

\begin{figure}[!htbp]
  \centerline{\includegraphics*[width=0.4\linewidth]{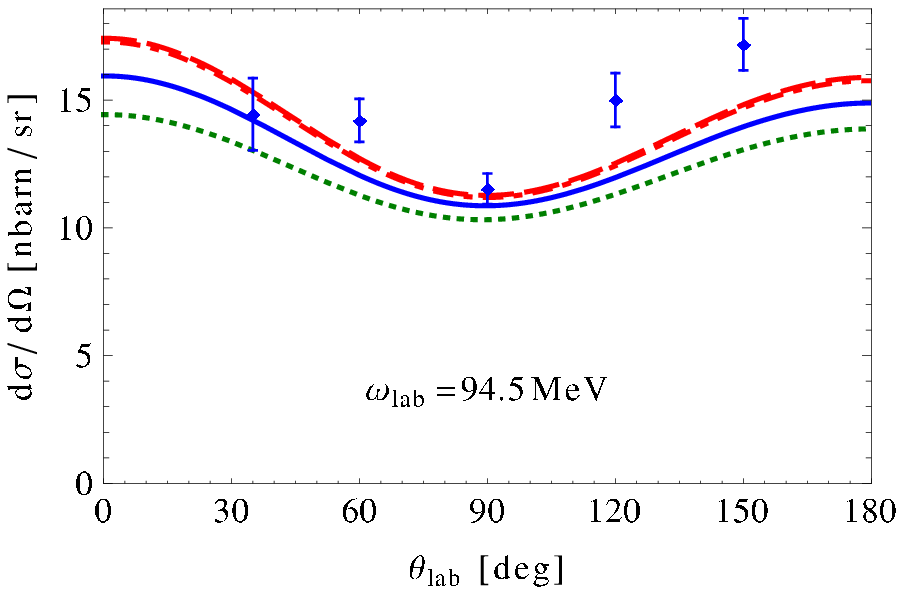}
    \hq\hq\hq\hq
    \includegraphics*[width=0.4\linewidth]{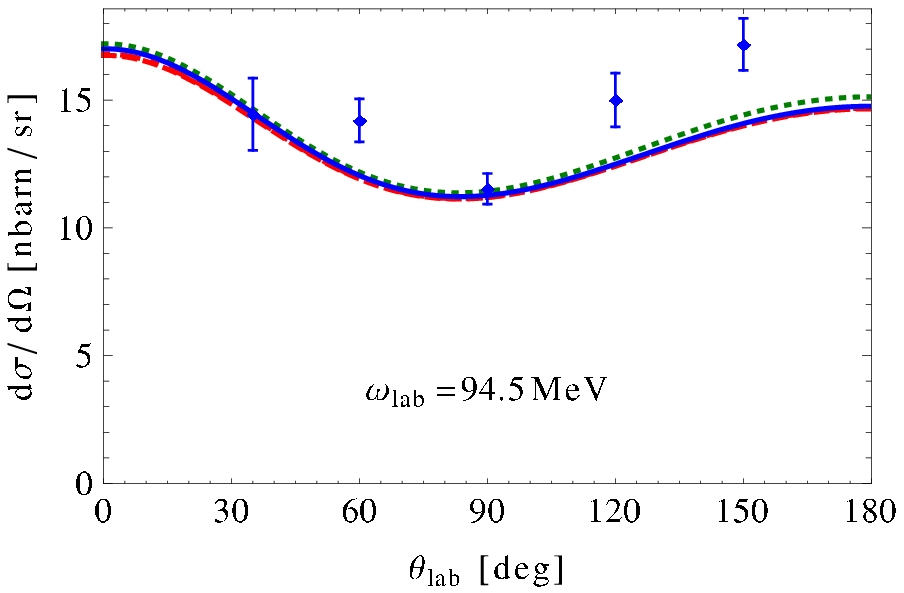}
  }
  \caption{\label{fig:d-wfdep} (Colour online) Unpolarised deuteron Compton
    cross section and data~\cite{Hornidge:2000} at high energies without
    (left) and with (right) rescattering effects in NLO \ChiEFT with explicit
    $\Delta(1232)$, using the static scalar values from
    Eq.~\eqref{eq:op3preds}. Deuteron wave functions: \ChiEFT at \NXLO{2}
    (cutoff $650\;\MeV$, identical to Fig.~\ref{fig:d-rescattdep}; blue solid)
    and NLO (cutoff $600\;\MeV$ with $B_\mathrm{d}=2.175\;\MeV$, green
    dotted)~\cite{Epelbaum:1999dj}; AV18 (red dot-dashed)~\cite{av18};
    Nijmegen 93 (red dashed)~\cite{Nijm}.}
\end{figure}

Besides providing a unified framework for energies from the Thomson limit to
the pion-production threshold, a crucial benefit of keeping higher-order
rescattering diagrams even in Regime II is that they also considerably reduce
the dependence on the deuteron wave function; see Fig.~\ref{fig:d-wfdep}.
For the counting which is valid only in Regime II, Beane et al.~had
established that the dependence of the cross-section predictions on the NN
wave function amounts to about $10$\% at $94.5$~MeV at
NLO~\cite{Beane:1999uq}.  The effect is not diminished at \NXLO{2} ($\Q^4$),
as they subsequently demonstrated in Ref.~\cite{Beane:2004ra} using wave
functions calculated from a variety of NN potentials based on
$\chi$EFT~\cite{Phillips:1999am,Epelbaum:1999dj}. Since all deuteron wave
functions share the same long-distance physics, the variability is associated
with differences in the short-distance physics of the photon--deuteron
interaction.  This is largely independent of angle and represents an
irreducible theoretical uncertainty of calculations that strictly employ the
Regime II power counting, severely compromising the accuracy with which
$\alphaes$ could be extracted from the deuteron data~\cite{Beane:2004ra}.
However, as discussed in Section~\ref{sec:2Nmotivation}, the Thomson limit
dictates wave-function independence as $\omega \rightarrow 0$.
Figure~\ref{fig:d-wfdep} shows that with the zero-energy point thus fixed, the
dependence is reduced even at $100$~MeV from about $\pm5\%$ to
$\lesssim\pm0.5\%$, and is therefore virtually eliminated in the
experimentally relevant energy range~\cite{Hildebrandt:2005ix,
  Hildebrandt:2005iw}. Therefore, a wave function derived from any modern,
``high-precision'' potential can be used.

The dependence on the choice of NN potential could still be of importance when
the partial waves which enter rescattering in the NN intermediate state are
computed. However, Hildebrandt et al.~\cite{Hildebrandt:2005ix,
  Hildebrandt:2005iw} compared rescattering contributions derived from the
AV18 potential~\cite{av18} with those from a rather crude LO \ChiEFT
potential~\cite{Rho}.  They found that nearly perfect agreement at the Thomson
limit still resulted, and there was only a $\lesssim4\%$ deviation at
$90\;\MeV$.

Hildebrandt et al.~\cite{Hildebrandt:2005ix, Hildebrandt:2005iw,
  Griesshammer:2010pz} also combined one ``high-order chiral'' or
``high-precision traditional'' potential for rescattering with a deuteron
wave function generated by another potential. These apparent mismatches did
not affect the degree to which the Thomson limit is restored, nor did they
show significant differences in a variety of unpolarised and polarised
observables at $\lesssim100\;\MeV$.  This is an excellent test for
independence from details of different short-distance physics and implies that
issues of matching electromagnetic currents with both the wave function and
the potential can be neglected.  Indeed, such effects related to gauge
invariance only appear two orders higher than considered here.

All of this suggests that cross sections and polarisabilities in the present
approach will be essentially unchanged once a fully systematic \ChiEFT is
established which is cutoff-independent over a wide range of cutoffs and has a
strict perturbative expansion. The use of Weinberg's power counting for the
NN potential and NN operators yields results which should differ from the
result in this full theory by only a small amount, thanks largely to the
stringent constraint imposed by the Thomson limit.


\subsection{\it A new fit to the deuteron data } 
\label{sec:deuteronanalysis}

We now use the \ChiEFT variant from Section~\ref{sec:dunify}, which is valid
from the Thomson limit to the pion production threshold, for a new extraction
of the isoscalar electric and magnetic dipole polarisabilities from the
available deuteron Compton scattering data: Lucas
(Illinois)~\cite{Lucas:1994}, Hornidge (SAL)~\cite{Hornidge:2000} and Lundin
(MAX-Lab)~\cite{Lundin:2003}.  Comparisons to the static values for the proton
will then allow us to assess whether isovector (scalar dipole)
polarisabilities are indeed small, as predicted by \ChiEFT.  However, first we
recall the conclusion of Section~\ref{sec:expdeutlow} that the deuteron data
are not as refined as the proton data, cf.~Table~\ref{table-deuteron-low}.
Limited angle and energy coverage, as well as difficulties in cleanly
differentiating elastic and inelastic events, gives rise to large statistical
and systematic errors. It therefore comes as no surprise that the experimental
errors are at least as large as the residual theoretical uncertainties in
extractions of the
polarisabilities~\cite{Beane:2002wn,Beane:2004ra,Hildebrandt:2005ix,
  Hildebrandt:2005iw,Hildebrandt:2004hh}. This is confirmed by our
re-analysis.

We present an update of the deuteron results by Hildebrandt et
al.~\cite{Hildebrandt:2005ix,Hildebrandt:2005iw}. As discussed in
Sections~\ref{sec:RegimeII} and~\ref{sec:dunify}, it actually represents a
complete (modified) $\calO(e^2\delta^3)$ calculation in both the two- and
one-nucleon sectors, valid from the Thomson limit up to $\w\sim\mpi$.
All two-nucleon contributions are included which enter at this order in either
Regime I or II, Figs.~\ref{fig:deuteronLOThomson} and
\ref{fig:deuteronNLOThomson}. The single-nucleon sector diagrams are listed in
Figs.~\ref{fig:Born}, \ref{fig:loops-3rd} and~\ref{fig:delta-contrib}. They
contain the two short-distance coefficients $\delta\alphaes$ and
$\delta\betams$ (last graph of Fig.~\ref{fig:loops-4th}), which, strictly
speaking, only enter at order $e^2\delta^4$. They also include the
$\Delta(1232)$, but in the deuteron calculation it is treated
nonrelativistically and without a width, as described in
Section~\ref{sec:includeDelta}. With one exception all parameters are the same
as for the proton fit, including the $\Delta$ parameters $\Delta_M=293\;\MeV$,
$g_{\pi\mathrm{N}\Delta}=1.425$. The exception is the $\gamma\mathrm{N}\Delta$
coupling for which we use the translation to the nonrelativistic value
$b_1=5$, as explained in Section~\ref{sec:earlyfit}.
This treatment of the $\Delta(1232)$ is justified by the excellent agreement
with the relativistic, nonzero-width approach in the region of the deuteron
data, $\wlab\lesssim100\;\MeV$; see the comparison of the energy dependence of
the polarisabilities in the two approaches in Fig.~\ref{fig:dynpolas}. The
final fits are insensitive to varying the value of $b_1$ by as much as $5\%$.
In practice, we use the \ChiEFT deuteron wave function at \NXLO{2} (cutoff
$650\;\MeV$) in the implementation of Epelbaum et al.~\cite{Epelbaum:1999dj}
and the AV18 potential~\cite{av18} for NN rescattering. As discussed in
Section~\ref{sec:dunify}, this combination provides an adequate \ChiEFT
representation of the two-nucleon system and the residual dependence on the
deuteron wave function, NN potential and numerical implementation is
minuscule.  These parameters are also used in Figs.~\ref{fig:d-rescattdep} to
\ref{fig:d-wfdep} and \ref{fig:crosssectionpionless}.

The presentation here differs from that of Hildebrandt et
al.~\cite{Hildebrandt:2005ix,Hildebrandt:2005iw} in the following: a new
parameter set $(b_1,\,g_{\pi \mathrm{N}\Delta},\,\Delta_M)$ for the
$\Delta(1232)$ from the Breit-Wigner parameters and the proton Compton data,
and not from the pole position; a more careful analysis of systematic errors
and summation of statistical and point-to-point systematic errors in
quadrature, as discussed in Section~\ref{sec:expdeutlow}; an implementation of
correlated systematic errors of the experiments by a floating normalisation as
in the proton case \eqref{eq:chisqare}; and small technical improvements to
implement the Thomson limit and increase numerical accuracy, see
Section~\ref{sec:2Nmotivation}.  However, the relatively large experimental
errors make the fit rather insensitive to the procedure used, and so none of
these changes substantially alters the conclusions of
Refs.~\cite{Hildebrandt:2005ix,Hildebrandt:2005iw}.

As in the proton case, the fractional error in the extracted polarisabilities
is $\delta^2$, namely one order past the last one for which all contributions
are consistently retained in both the one- and two-nucleon sectors.  Up to
higher-order corrections, the results should also agree with those of the
version without rescattering but with a dynamical $\Delta(1232)$, fitted in
Regime II, $\w\sim\mpi$.  Refs.~\cite{Hildebrandt:2005ix,Hildebrandt:2004hh}
performed fits of $\alphaes$ and $\betams$ to the same data without
rescattering.  However, the assumption $\w\sim\mpi$ fails at the lower end of
the data, as demonstrated in Fig.~\ref{fig:d-rescattdep}, and the residual
wave-function dependence of the extracted values is rather large, namely
$\pm1$, cf.~Fig.~\ref{fig:d-wfdep}.  Therefore, we do not use this criterion
to estimate higher-order corrections, even though we note that our final
values given below agree with theirs within error bars.  Ultimately,
higher-order effects in the chiral amplitudes at order $e^2\delta^4$ and
beyond provide by far the largest residual uncertainty, estimated at $\pm0.8$.
As in Section~\ref{sec:protonanalysis}, this estimate is justified by assuming
that the uncertainty is of order $\delta^1$ of the LO-to-NLO correction;
c.f.~also \cite{Hildebrandt:2004hh}.

\begin{figure}[!htb]
  \begin{center}
    \includegraphics*[width=0.4\linewidth]{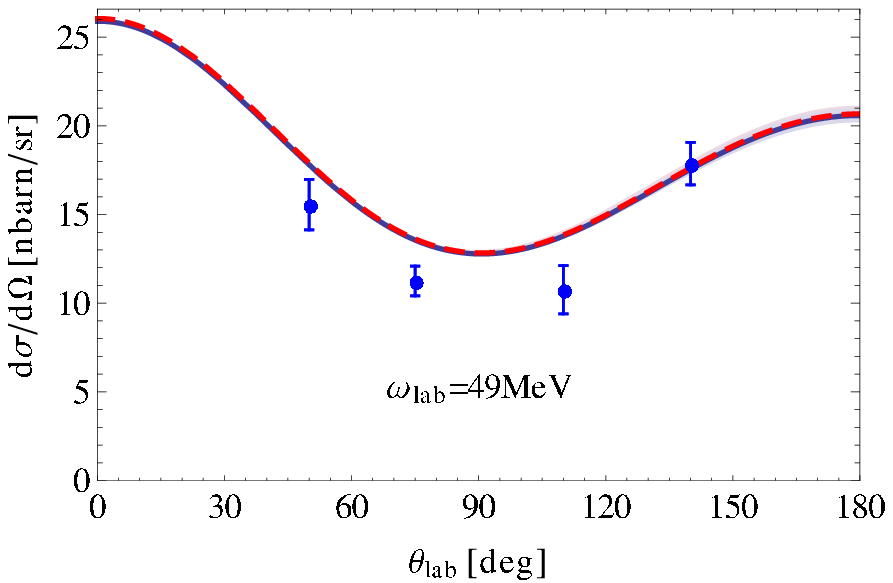}
    \hq\hq\hq\hq
    \includegraphics*[width=0.4\linewidth]{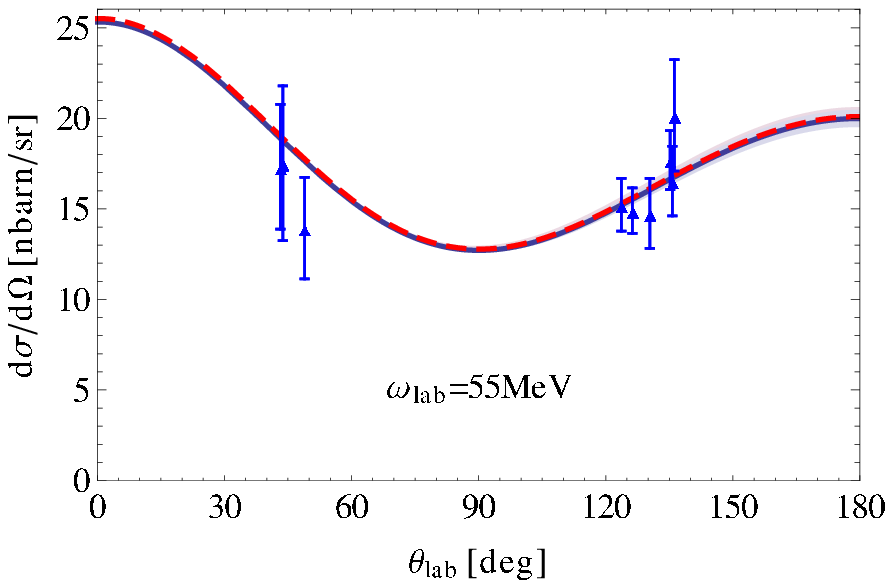}
    \\[3ex]
    \includegraphics*[width=0.4\linewidth]{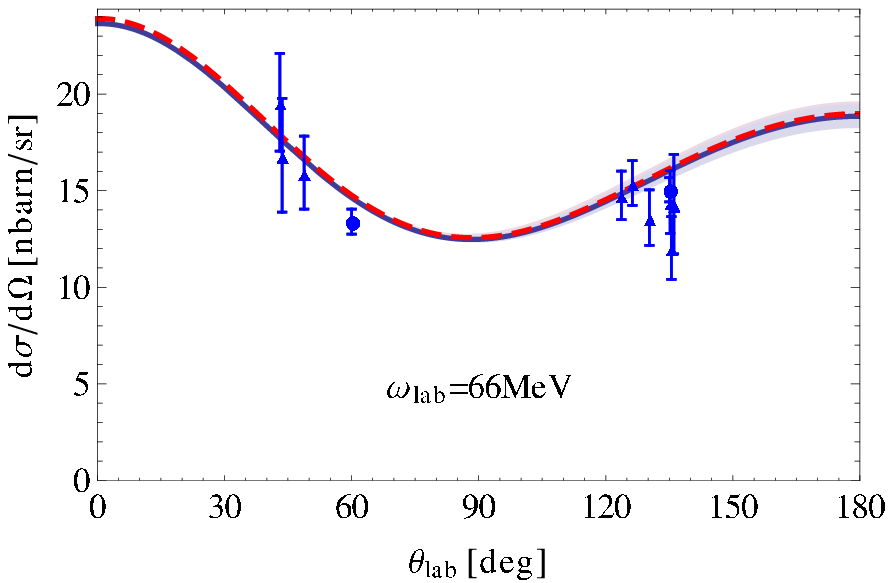}
    \hq\hq\hq\hq
    \includegraphics*[width=0.4\linewidth]{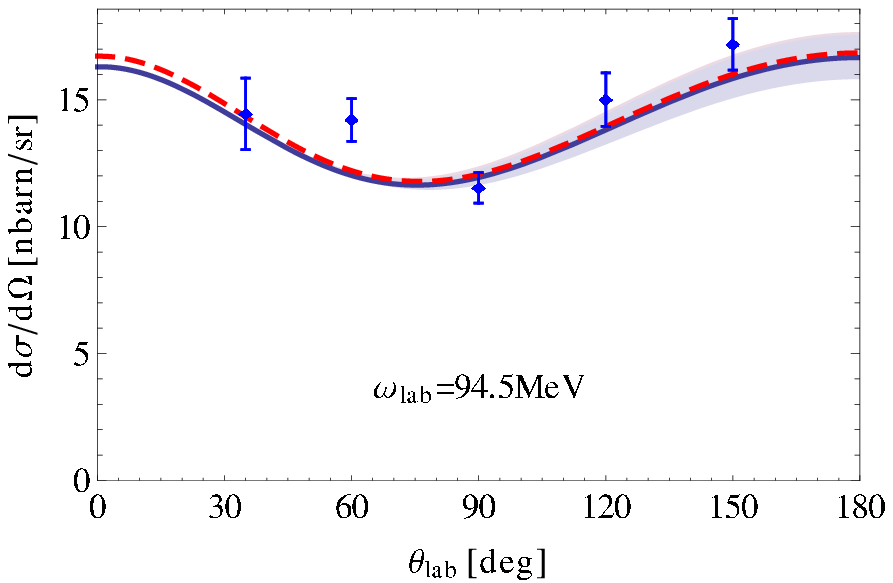}
    \caption{(Colour online) Cross sections in the two-parameter (dashed) and
      one-parameter (solid) determinations of the isoscalar spin-independent
      dipole polarisabilities. Bands: statistical error of the one-parameter
      fit. Data symbols from Table~\ref{table-deuteron-low}, statistical
      errors only. \label{fig:dfit}}
  \end{center}
\end{figure}

With the anticipated accuracy established, the isoscalar spin-independent
polarisabilities read:
\begin{equation}
  \label{eq:d2parameterfinalfit}
  \alphaes=10.5\pm 2.0(\text{stat})\pm0.8(\text{theory})\;\;,\;\;
  \betams=3.6\pm 1.0(\text{stat})\pm0.8(\text{theory})
\end{equation}
from a $\chi^2$ fit, with the corresponding cross sections shown in
Fig.~\ref{fig:dfit}.
As before, the statistical error is determined from the projection of the
$\chi^2_\mathrm{min}+1$ contours onto the $\alphaes$ and $\betams$ axes. For
$29$ data points and $5$ free parameters, one finds a total $\chi^2=24.3$, or
$1.01$ per degree of freedom. The small differences compared to the results
reported in Refs.~\cite{Hildebrandt:2005ix,Hildebrandt:2005iw},
$\alphaes=11.5\pm1.4(\text{stat})\pm1(\text{theory})$ and
$\betams=3.4\pm1.6(\text{stat})\pm1(\text{theory})$, are well within
statistical uncertainties and stem from the minor differences discussed above.
The collective data set appears consistent since each experiment contributes
about equally to the overall $\chi^2$, and the extracted parameters remain
nearly unchanged when one data set is eliminated.  This confirms the
assessment in Section~\ref{sec:dunify} that the angular dependence of
Hornidge's SAL data is described well by the theory. The floating
normalisation of that set is about $10$\% lower than the other two, which in
turn both have a floating normalisation close to $1$. All floating
normalisations are thus compatible with the correlated systematic
uncertainties ($3$\% for Lucas, $5$\% for Hornidge, $7.5$\% for Lundin) stated
in Section~\ref{sec:expdeutlow}. Only more accurate experiments, such as those
ongoing at MAX-Lab~\cite{Myers:2010,Shoniyozov:2011}, can reveal if the lower
normalisation of the Hornidge data is more than a statistical accident.

The results (\ref{eq:d2parameterfinalfit}) are in good agreement with the
proton values, confirming that isovector effects in $\alphae$ and $\betam$ are
indeed small. They also add up, within error bars, to the isoscalar Baldin sum
rule, $\alphaes+\betams=14.5\pm0.3$, Eq.~\eqref{eq:baldinisoscalar}. One can
therefore use the Baldin constraint to reduce statistical uncertainties in a
one-parameter fit for $\alphaes-\betams=7.3\pm1.8(\text{stat})
\pm0.8(\text{theory})$, thereby obtaining very similar results with about half
the statistical error:
\begin{equation}
  \label{eq:d1parameterfinalfit}
  \begin{split}
    \alphaes &=10.9\pm
    0.9(\text{stat})\pm0.2(\text{Baldin})\pm0.8(\text{theory})
    \\
    \betams &=\phantom{0}3.6\mp
    0.9(\text{stat})\pm0.2(\text{Baldin})\pm0.8(\text{theory})
  \end{split}
\end{equation}
The total $\chi^2$ is unchanged, but drops to $0.97$ per degree of freedom
since the number of degrees of freedom is increased to 25. The cross sections
of this fit, shown in Fig.~\ref{fig:dfit}, also compare well with both the
two-parameter fit and the data.  Extractions including a dynamical
$\Delta(1232)$~\cite{Hildebrandt:2005ix, Hildebrandt:2005iw} lead to
systematically higher values for $\betams$ than those without
it~\cite{Beane:2002wn,Beane:2004ra}, since the enhancement in the Hornidge
(SAL) data at backward angles must be compensated in the latter case by
decreasing $\alphaes-\betams$, cf.~Section~\ref{sec:dunify}.  In addition,
agreement with the Baldin sum rule was marginal in
Refs.~\cite{Beane:2002wn,Beane:2004ra}.

Comparing with the results of Section~\ref{sec:protonanalysis}, we see that
the static polarisabilities of the proton, \eqref{eq:p2parameterfinalfit} and
\eqref{eq:p1parameterfinalfit}, and isoscalar nucleon,
\eqref{eq:d2parameterfinalfit} and \eqref{eq:d1parameterfinalfit}, are
identical within error bars.  They may thus be combined to obtain the neutron
polarisabilities in the two-parameter analysis as
\begin{equation}
  \label{eq:n2parameterfinalfit}
  \alphaen=10.5\pm 4.0(\text{stat})\pm0.8(\text{theory})\;\;,\;\;
  \betamn=4.4\pm 2.1(\text{stat})\pm0.8(\text{theory})\;\;,
\end{equation}
and in the one-parameter fit using the Baldin sum rule as
$\alphaen-\betamn=7.0\pm 3.6(\text{stat})\pm0.8(\text{theory})$,
\begin{equation}
  \label{eq:n1parameterfinalfit}
  \begin{split}
    \alphaen &=11.1\pm
    1.8(\text{stat})\pm0.4(\text{Baldin})\pm0.8(\text{theory})
    \\
    \betamn &=\phantom{0}4.1\mp
    1.8(\text{stat})\pm0.4(\text{Baldin})\pm0.8(\text{theory})
  \end{split}
\end{equation}
with statistical errors added in quadrature.  These results are 
in good agreement with those of the quasi-free
experiments~\cite{Kossert:2002ws}, while having somewhat smaller
uncertainties. We note again that the proton results were obtained in a
slightly different \ChiEFT variant, in which $\Delta(1232)$ propagation is
treated relativistically and its width is included non-perturbatively.
However, since the omitted terms are of higher order and therefore
parametrically small, the two variants must agree within the theoretical
uncertainties; cf.~Fig.~\ref{fig:dynpolas}. In turn, comparing proton and
isoscalar results confirms that the scalar dipole polarisabilities of the
proton and neutron are largely isoscalar, i.e.~that both share essentially the
same two-photon response. It is curious to note that the proton-neutron
difference of the magnetic polarisability appears to have a statistically
insignificant tendency to be slightly negative, $\betamv\approx-1\pm2$, in
line with a recent constraint from its contribution to the electromagnetic
self-energy of the proton-neutron mass difference~\cite{WalkerLoud:2012bg}.

Finally, smaller theoretical uncertainties are expected from a fully
consistent extraction of order $e^2\delta^4$ which is under way and uses the
same variant to the same order for both the proton and deuteron data, plus
chirally-consistent interactions throughout~\cite{Griesshammer:2012}.
Higher-quality data with carefully formulated correlated and point-to-point
systematic errors would reduce the sizable statistical error.

\subsection{\it EFTs for very low energies}

\subsubsection{Compton scattering without pions}
\label{sec:pionless}

Since $\w \ll \mpi$ in Regime I, a photon cannot resolve details of pion-cloud
effects and one can formulate a more radical EFT by integrating out the pion
into contact interactions amongst nucleons and with photons. This is the
low-energy version of \ChiEFT, called ``pion-less'' EFT (\EFTNoPion). Its
typical momentum scale is $\gamma\approx50\;\MeV$; its breakdown scale
$\LambdaNoPion \sim \mpi$ is set by the mass of the pion as the lightest
particle not included as a dynamical degree of freedom; and its expansion
parameter is $\QNoPion \equiv \gamma/\LambdaNoPion$.  The lack of finite-range
forces considerably simplifies calculations, which can be carried out to high
orders with relative ease, and indeed analytic results are common.  This EFT
has been used for model-independent subtractions of ``nuclear effects'' and
predictions in a variety of very low-energy processes, also with electro-weak
probes; see
e.g.~\cite{seattle_review,Bedaque:2002mn,Braaten:2004rn,Platter:2009gz} for
reviews.  In such calculations, the expansion parameter is established as
being in the range $\QNoPion \approx 1/3$ to 1/5 for $p_\text{typ} \sim
\gamma$, and accuracy of better that 1\% has been achieved.  Since gauge
invariance and LETs such as the Thomson limit are automatically fulfilled
exactly, one can check the numerics of \ChiEFT.  Indeed, any consistent
theoretical framework, whatever its detailed treatment of meson-exchange
currents, ``off-shell effects'', cutoff dependence, etc., must agree with the
\EFTNoPion result within the mutual accuracies at energies where \EFTNoPion is
applicable.  \EFTNoPion is thus a theoretically rigorous yet numerically
simple tool to study reactions at very low energies.

For NN scattering, the theory is equivalent to Bethe's Effective Range
Expansion of the \oneS and \threeS
channels~\cite{Blatt:1949zz,Bethe:1949yr,Schwinger,ChewGoldberger,
  BarkerPeierls}, and at NLO in $\QNoPion$, the scattering lengths and
effective ranges are the only LECs. In the latter, momentum-dependent NN
couplings also give rise to photon couplings by minimal substitution.

A particularly compact NLO two-nucleon Lagrangian is given in
Refs.~\cite{Kaplan:1996nv,Bedaque:1999vb}.  At this order,
\threeS-$\wave{3}{D}{1}$ mixing is still absent, and the deuteron is a pure
S-state.  Since contributions with two isovector magnetic-moment interactions
are relatively large, they have typically been promoted by one order,
$(\kappa^{(\mathrm{v})})^2\approx 5.5\sim\QNoPion^{-1}$~\cite{Chen:1998ie,
  Chen:1998vi, Beane:1999uq}.
   
In the one-nucleon sector, the NLO terms relevant for Compton scattering are
identical to those in \ChiEFT with all pion-nucleon couplings set to zero, and
lead to the Petrun{}'kin amplitudes~\eqref{eq:low-en-amps}. The photons couple
to a point-like nucleon with spin and anomalous magnetic moment, and structure
effects enter through the photon-nucleon seagull terms of the effective
Lagrangian~\eqref{eq:polsfromints}. As discussed there, only the static
polarisabilities are relevant for $\w \ll
\LambdaNoPion$~\eqref{eq:alphaexpanded}.  At the same order, the relevant
two-nucleon diagrams are just those in which one or both photons couple to the
``effective range" NN contact terms.

\EFTNoPion is applicable in all of Regime I ($\w\lesssim20\;\MeV$). In that
range, one can again count the photon energy as either soft or hard, but now
relative to the low scale $\gamma$~\cite{Griesshammer:2000mi, Chen:1998ie,
  Chen:1998vi}.  When $\w\lesssim 3 \;\MeV \sim B_\mathrm{d} = 
{\gamma^2}{/\MN}\sim\QNoPion^2 $ is counted as soft, the photon wavelength is
larger than the deuteron size, and the photon energy may barely be sufficient
to disintegrate it.  This is the region of the very first Compton-scattering
calculation on a composite nuclear system, by Bethe and Peierls in
1935~\cite{BethePeierls:1935}. Here, one tests the stiffness of the deuteron
as a whole against deformation in the photon field, parameterised by the
electric and magnetic scalar and tensor polarisabilities of the
\emph{deuteron}. Tensor polarisabilities probe the deformation of its
quadrupole component.  Closed-form expressions exist for each, often to high
orders. Since it can be shown that the polarisabilities of the individual
nucleons only enter at the $\lesssim0.1\%$ level~\cite{Beane:2000fi,
  Griesshammer:2000mi}, we do not describe these in more detail here but refer
to the original
literature~\cite{Chen:1998vi,Chen:1999tn,Phillips:1999hh,seattle_review,Ji:2003ia}
and mention that these parameter-free predictions usually agree very well with
other theoretical and experimental determinations~\cite{Friar:1984zzb,
  Lucas:1968,Friar:1997jy,Friar:2005gv,Rodning:1982zz}.

At higher energies, the photon wavelength is comparable to the deuteron size,
and $\w \sim \gamma \sim \QNoPion$ is counted as hard. However, the expansion
is only justified if the hard scale in the intermediate-state propagator is
also small compared to the breakdown scale, $\MN \w\lesssim\mpi^2$,
i.e.~$\w\lesssim 20 \;\MeV$ as in Regime I of \ChiEFT.  The nucleon
polarisabilities then enter because of their $\w^2$ dependence, in principle
as $\QNoPion^2 \sim 10\%$ corrections, i.e.~at \NXLO{2}.  To this order, this
region was studied in \EFTNoPion, with a projected accuracy of order $\QNoPion
\sim 30\%$ for $\alphaes$ and $\betams$~\cite{Beane:2000fi, Chen:2004wv,
  Griesshammer:2000mi, Chen:1998ie}.  As the energy increases, the relative
importance of each contribution changes.  While partial-wave mixing is still
absent at \NXLO{2}, two new short-distance NN effects enter. The first one
couples the \oneS and \threeS waves by a spin-flipping magnetic photon, with
its LEC $L_1$ determined by $\mathrm{np}\to \mathrm{d}\gamma$ at thermal
energies. In addition, Chen et al.~\cite{Chen:2004wv} pointed out the
importance of a seagull term from the relativistic correction to the nucleon
magnetic-moment interaction, i.e.~from the spin-orbit coupling.  On top of the
scalar amplitudes analogous to $A_{1,2}$, they also noted that vector
amplitudes dependent on the deuteron spin and analogous to $A_{3\text{-}6}$
appear at \NXLO{2} and contribute about $20\%$ at these energies. At leading
nonvanishing order, they receive contributions from the magnetic moment,
spin-orbit coupling and $L_1$, and add about $4\;\mathrm{nb}/\mathrm{sr}$ to
the backward cross section at $\w = 50 \;\MeV$. Since all three are magnetic,
spin-flip effects, they are most prominent in backward-angle scattering and
thus can mimic a false signal of the combination $\alphaes - \betams$.

\begin{figure}[!htbp]
  \centerline{
    \includegraphics*[width=0.4\linewidth]{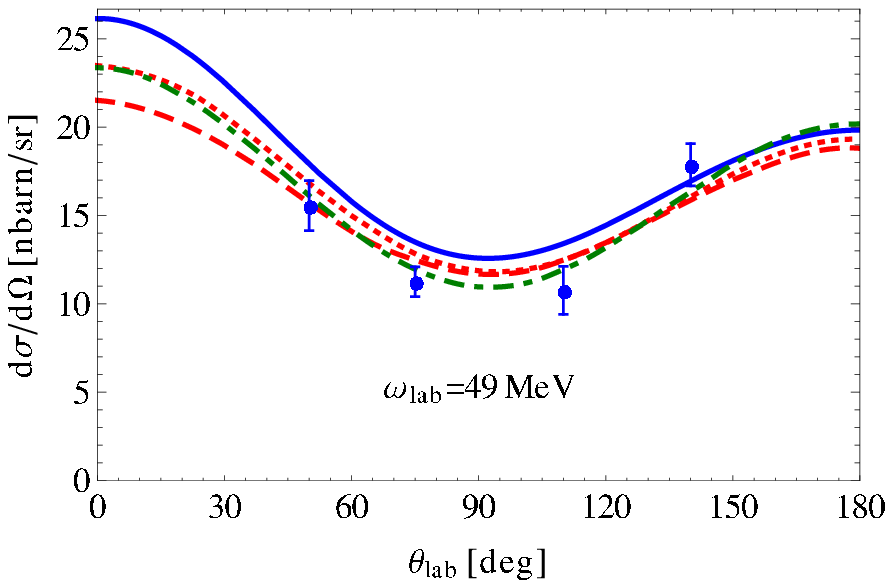}
    \hq\hq\hq\hq
    \includegraphics*[width=0.4\linewidth]{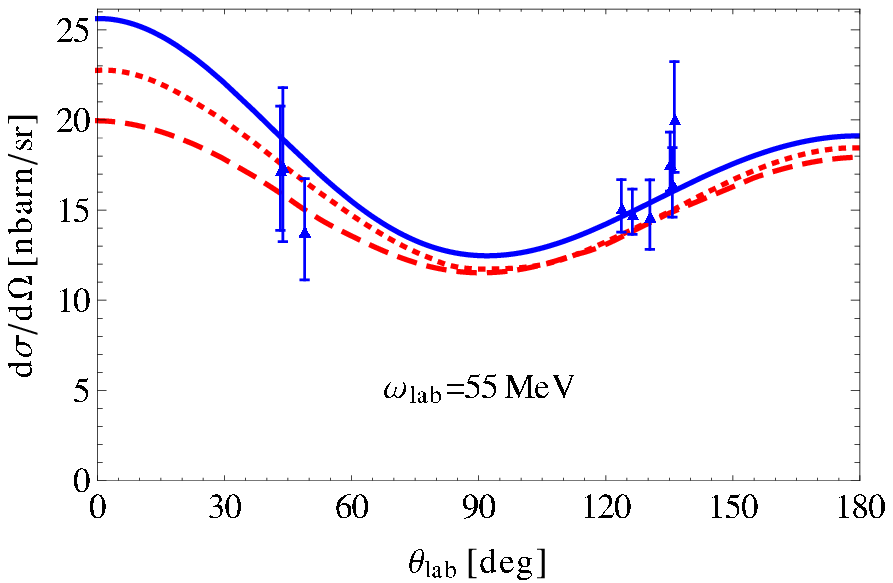}}

  \caption{\label{fig:crosssectionpionless} (Colour online) Deuteron Compton
    scattering in \EFTNoPion and \ChiEFT, compared with data at $49$ and $55
    \;\MeV$ in the lab frame (symbols from Table~\ref{table-deuteron-low},
    stat.~errors only).  Solid (blue) line: NLO \ChiEFT with dynamical
    $\Delta(1232)$ and rescattering~\cite{Griesshammer:2010pz,
      Hildebrandt:2005ix, Hildebrandt:2005iw}; dashed/dotted (red): \NXLO{2}
    \EFTNoPion with resummed/perturbative effective-range
    corrections~\cite{Chen:2004wv}; dash-dotted (green): \ChiEFT in the KSW
    scheme with perturbative pions at NLO~\cite{Chen:1998ie} (left panel
    only).  All calculations use the static scalar polarisabilities
    \eqref{eq:op3preds} and are rescaled to the flux factor~\eqref{eq:flux}. }
\end{figure}

In Fig.~\ref{fig:crosssectionpionless} the data at $\wlab = 49$ and $55
\;\MeV$ are compared to the cross section of two \EFTNoPion variants which
differ only by higher-order contributions and 
thus map out a band whose width of $\approx 5\%$ may indicate a typical
theoretical uncertainty at \NXLO{2}~\cite{Chen:2004wv}.  However, the formal
expansion parameter $\sim \sqrt{\MN \w}/\mpi \approx 1.5$ at these energies
suggests non-convergence even for the lowest-energy data set at $49 \;\MeV$.
In addition, in \EFTNoPion the deuteron is still a pure S-wave state at this
order, the tensor components of the Compton amplitude are zero, and
rescattering through higher NN partial waves is absent. Since all of these
effects add to the cross section and are well described by pion-exchange
contributions, it is no surprise that \EFTNoPion misses overall strength
compared to the \ChiEFT version. The lack of the deuteron tensor component and
of rescattering is most prominent at forward angles where the \ChiEFT result
is up to $20\%$ higher than \EFTNoPion, while both agree within errors at
backward angles.

Nonetheless, this breakdown should be gradual, not catastrophic, so that
\EFTNoPion may serve for qualitative explorations~\cite{Griesshammer:2000mi}.
Indeed, Chen et al.~fitted the scalar polarisabilities at \NXLO{2} to data at
$49$ and $55$~MeV as $\alphaes =12.3 \pm 1.4$, $\betams =5.0 \pm 1.6$, and
$\alphaes =14.2\pm2.1$, $\betams =9.3\pm2.5$ (statistical errors only) in two
variants which should differ by higher-order terms~\cite{Chen:2004wv}.  The
difference is again an indicator of systematic uncertainties, but the rather
large values and disagreement with the isoscalar Baldin sum
rule~\eqref{eq:baldinisoscalar} reflect the fact that the fit must also mock
up the physics not captured in \EFTNoPion at this order.
Extractions in \EFTNoPion are thus not reliable for the current data, but the
relative sensitivity of observables to the polarisabilities may be captured.
For example, Chen et al.~studied the double-polarisation observables
$\Sigma^\text{circ}_{x/z}$~\eqref{eq:sigmax} and the beam asymmetry
$\Sigma_3$~\eqref{eq:polasym} on a vector-polarised deuteron to
\NXLO{2}~\cite{Chen:2004wwa}. Parallel results in
\ChiEFT~\cite{Choudhury:2004yz, ShuklaThesis, Griesshammer:2009pq,
  Griesshammer:2010pz} will be discussed in Section~\ref{sec:future}.

Can nucleon polarisabilities be extracted instead from high-accuracy data
inside the formal radius of convergence of \EFTNoPion, $\w\lesssim20 \;\MeV$?
No data exist, but \HIGS can provide the high beam intensity for cross
sections at $\lesssim10\%$ (i.e.~\NXLO{2}) accuracy needed for
polarisabilities with $30\%$ errors~\cite{Weller:2009zza}. Unfortunately, this
is not competitive with the accuracy of \ChiEFT at higher energies, where
signals from nucleon polarisabilities are much stronger; see
Section~\ref{sec:2Nhigherenergies}. An attempt to decrease the error by an
\NXLO{3} calculation runs into several unknowns, including contact terms
between two photons and two nucleons which parameterise the part of the
deuteron polarisabilities not determined by long-range NN properties.

\subsubsection{Perturbative pions} 
\label{sec:pertpion}

Figure~\ref{fig:crosssectionpionless} also contains the result of a
calculation at $49 \;\MeV$ in the KSW variant of \ChiEFT, in which pion
effects are considered small enough to be included
perturbatively~\cite{Kaplan:1998we,Kaplan:1998tg}. While it still allows for
analytic expressions, with the Thomson limit manifestly fulfilled, it also
adds some aspects of the pionic tensor force. However, its radius of
convergence appears at best marginally larger than that of
\EFTNoPion~\cite{Fleming:1999ee}.  Since \EFTNoPion is recovered by setting
the pion-nucleon coupling $g_A$ to zero and adjusting the parameters, it is
not surprising that its results are close to resummed \EFTNoPion. Indeed,
Refs.~\cite{Chen:1998rz, Chen:1998ie, Chen:1998vi} also provided the first
``pion-less'' Compton results. In Fig.~\ref{fig:crosssectionpionless}, we
rescaled the NLO calculation of Ref.~\cite{Chen:1998vi} from the
nonrelativistic flux factor to the version in Eq.~\eqref{eq:flux} used for the
other curves. This 5\% change allows the dynamical content of the different
theories to be isolated, based on different kinematics
used\footnote{Incidentally, the authors of Ref.~\cite{Chen:1998ie} meant to
  include the minuscule contribution from the slope parameters
  $\alpha_{E\nu}^{(\mathrm{s})}$ and $\beta_{M\nu}^{(\mathrm{s})}$
  in~\eqref{eq:alphaexpanded} as the leading energy-dependent effects in the
  scalar polarisabilities, but made the mistake of using instead the values of
  the leading momentum dependence of the \emph{generalised polarisabilities}
  for photons of nonzero virtuality.}.
Finally, Chen's investigation of unpolarised scattering off a tensor-polarised
deuteron in the KSW approach~\cite{Chen:1998rz} led Karakowski to address the
same quantity in a potential model~\cite{Karakowski:1999pt,Karakowski:1999eb}.


\subsection{\it Deuteron Compton scattering: model calculations}
\label{sec:gammadmodels}

Several calculations of deuteron Compton scattering were performed in the
1950s and
1960s~\cite{Capps:1957A,Capps:1957B,Schult:1960,Tenore:1965,Pokorski:1967}.
The earliest were based on the impulse approximation, which is poor because of
the large size of the exchange currents depicted in
Fig.~\ref{fig:deuteronNLOThomson}(b). In the 1980s, Weyrauch and Arenh\"ovel
performed detailed calculations of deuteron Compton scattering that included
these effects~\cite{Weyrauch:1984tf}. They also performed an energy-dependent
multipole decomposition of the deuteron amplitude, akin to that for the
nucleon amplitude in Section~\ref{sec:multipoles}. From these dynamical
deuteron polarisabilities, all deuteron Compton-scattering observables can be
reconstructed, with predictions as a function of $\w$ available in
Ref.~\cite{Weyrauch:1984tf}.

Formulating the problem in terms of these quantities has the advantage that
their imaginary parts can be obtained from the optical theorem and information
on $\gamma \mathrm{d} \rightarrow \mathrm{NN}$ from dispersion relations,
parallel to the discussion in Section~\ref{sec:drs}. This accounts for the
NN-reducible piece of the Compton amplitude, see Eq.~(\ref{eq:Mrescatt}), and
Fig.~\ref{fig:deuteronLOThomson}(b) and (c), as well as
Fig.~\ref{fig:deuteronNLOThomson}(c) and (d).  The remaining effects are part
of the amplitude $T_{\gamma \mathrm{d}}$ of all diagrams that do not include
an NN intermediate state, Eq.~\eqref{eq:wein}, called the ``seagull
amplitude'' in Ref.~\cite{Weyrauch:1984tf} and elsewhere.  As already
discussed in Section~\ref{sec:2Nmotivation}, significant cancellations between
the NN-reducible and NN-irreducible diagrams lead to the correct Thomson
limit.

Based on this, Weyrauch employed current conservation to simplify the
expressions and derived an explicit representation of the NN Green's function
$G$, Eq.~\eqref{eq:Mrescatt}, using a separable NN
interaction~\cite{Weyrauch:1988zz,Weyrauch:1990zz}. This avoids DRs for the
deuteron amplitudes. Subsequently, Wilbois~\cite{Wilbois:1995}, and later
Karakowski and Miller~\cite{Karakowski:1999eb, Karakowski:1999pt}, used
increasingly sophisticated NN potentials and added contributions beyond those
mandated by current conservation, such as leading relativistic corrections and
some meson-exchange currents beyond the Siegert theorem. Both arrived at
similar cross sections.

The most sophisticated and most widely used calculations employing these
techniques in a one-boson-exchange framework were carried out by Levchuk and
L'vov~\cite{Levchuk:1995,Levchuk:1999zy}. The NN current operator ${\bf J}$ in
Eq.~\eqref{eq:Mrescatt} and the irreducible diagrams of $T_{\gamma
  \mathrm{d}}$~\eqref{eq:wein} are derived by minimal substitution from the
one-boson-exchange Hamiltonian, including form factors.  This avoids
Siegert-like theorems and makes it straightforward to compute important
dynamical effects, e.g.~nonstatic corrections to the exchange propagators and
$\Delta(1232)$ effects, both of which Ref.~\cite{Levchuk:1999zy} considered.

Turning to results, Wilbois claimed in Ref.~\cite{Wilbois:1995} that the model
dependence of the differential cross section due to the choice of NN potential
is $\approx 1$\% or less for photon energies up to 100 MeV. Levchuk and
L'vov~\cite{Levchuk:1999zy} found a variation of $\approx 5$\% for different
versions of the Bonn OBEPR potential~\cite{Machleidt:1987}, a higher cross
section than those of Refs.~\cite{Wilbois:1995,Karakowski:1999eb,
  Karakowski:1999pt}, and more angular variation at photon energies $\geq 70$
MeV. Several checks in Ref.~\cite{Levchuk:1999zy} were used to ensure that
current conservation and its consequences had been implemented at a reasonable
level of accuracy. For example, the Thomson limit is violated by 6\%.  All
implementations agreed that rescattering effects in the cross section are
$\approx 10$\% at 50 MeV and decrease further by 100 MeV, in line with the
subsequent \ChiEFT findings of Fig.~\ref{fig:d-rescattdep} in
Section~\ref{sec:dunify}.

Of all the studies discussed in this subsection, only
Refs.~\cite{Wilbois:1995,Karakowski:1999eb,
  Karakowski:1999pt,Levchuk:1995,Levchuk:1999zy} included terms in the
single-nucleon Compton amplitude beyond the nucleon Born terms. The first four
included the static values of $\alphaes$ and $\betams$, but no higher
functional dependence on $\omega$. While good agreement with the forward-angle
Lucas data~\cite{Lucas:1994} was found at low energies, it is not surprising
that the rise of the differential cross section at backward angles in the
higher-energy SAL data~\cite{Hornidge:2000} could not be reproduced, in view
of the importance of the $\Delta(1232)$ established in
Section~\ref{sec:RegimeII}.  The study of Levchuk and
L'vov~\cite{Levchuk:1999zy} implemented the most sophisticated single-nucleon
amplitude of these models. A nonrelativistic reduction of the Dirac-Pauli
Hamiltonian was performed and terms up to ${\cal O}(1/\MN^2)$ were kept in the
resulting Hamiltonian.  They then added the contributions from the static
values of $\alphaes$, $\betams$, the four spin polarisabilities and four
fourth-order polarisabilities, including the slope parameters of
Eq.~\eqref{eq:alphaexpanded}.  The eight higher-order polarisabilities were
taken from the fixed-$t$ DR calculation of Babusci et
al.~\cite{Babusci:1998ww}. As discussed in Section~\ref{sec:multipoles}, this
incorporates all dynamics of the Compton amplitude, up to this order in
$\omega$. Nevertheless, they could not describe the backward-angle Hornidge
(SAL) points, which led Hornidge et al. to infer a value of
$\betamn\approx10$---very different from $\betamp$~\cite{Hornidge:2000}.
However, this problem is now known to be the result of a coding
mistake~\cite{Levchuk:2009A,Levchuk:2009B}. Preliminary results from
calculations that rectify this error, and improve on the treatment of the
Thomson limit in Ref.~\cite{Levchuk:1999zy}, indicate that it is no longer
necessary to employ a value of $\betamn$ that differs appreciably from
$\betamp$ in order to reproduce these data~\cite{Levchuk:2009A,Levchuk:2009B}.

A similar model was used to calculate the inelastic reaction, $\gamma
\mathrm{d} \rightarrow \gamma' \mathrm{np}$ by Levchuk et
al.~\cite{Levchuk:1994,Levchuk:1995nn,Wissmann:1998ta}. The full set of
diagrams for this process was not computed, since the focus in these works was
on making predictions for the inelastic deuteron Compton reaction in neutron
quasi-free kinematics. In this Regime, the most important diagrams correspond
to photon scattering on quasi-free nucleons, NN rescattering, and the effect
of meson-exchange currents and the $\Delta(1232)$. The OBEPR potential was
again employed to compute the deuteron wave function and NN amplitude.  The
nucleon Compton amplitudes were taken from the DR calculation of
Ref.~\cite{Lvov:1996xd}. Using these ingredients, Wissmann et
al.~\cite{Wissmann:1998ta} showed that the cross section above 200 MeV in the
neutron quasi-free peak is large enough for an accurate extraction of
$\alphaen$. These computations gave strong theoretical motivation to extract
$\alphaen$ from measurements of the inelastic reaction in quasi-free
kinematics, as performed in Refs.~\cite{Kolb:2000,Kossert:2002ws}.  Different
NN models predict triple-differential cross sections in this region that vary
by 3\%~\cite{Kossert:2002ws}. From a theoretical perspective, the absence of a
full, current-conserving calculation of $\gamma \mathrm{d} \rightarrow \gamma'
\mathrm{np}$ is of some concern.
Also, as emphasised in Section~\ref{sec:drs}, significant model assumptions
exist in the DRs of Ref.~\cite{Lvov:1996xd} that were used to analyse these
data, and the unanimity in DR predictions disappears at these higher energies.
The impact of different $\gamma \mathrm{N} \rightarrow \pi \mathrm{N}$
multipole analyses in the DR construction of the single-nucleon amplitude was
assessed in Refs.~\cite{Wissmann:1998ta,Kossert:2002ws}, but this is not the
only source of uncertainty in the DR calculation. Overall, the model
uncertainty of $\pm 1.1$ quoted by Kossert et al.~is probably an
underestimate~\cite{Lvov:2011}.  Lastly, Levchuk et al.~\cite{Levchuk:1994}
emphasise that, in the kinematics relevant for this reaction, the $\pi^0$ pole
part, Eq.~\eqref{eq:pi-pole}, has a marked impact on the cross section. As
discussed in Section~\ref{sec:expdeutqf}, Kossert et al.~exploited this
sensitivity in order to perform a simultaneous extraction of $\alphaen -
\betamn$ and $\gammapin$, assuming the Baldin sum-rule value \eqref{eq:bothSR}
for $\alphaen + \betamn$~\cite{Kossert:2002ws}.

\subsection{\it Compton scattering from \textnormal{\threeHe}}
\label{sec:gammaHe3chiral}

The first computations of elastic Compton scattering on the \threeHe nucleus
were performed in the past five years by Shukla (n\'ee Choudhury) et al.,
using the \ChiEFT framework without explicit $\Delta(1232)$ degrees of freedom
at NLO ($P^3$)~\cite{Choudhury:2007bh,Shukla:2008zc,ShuklaThesis}. Since the
focus was on photon energies of 60 to 120 MeV, one can use the Regime II
approximation of instantaneous interaction kernels, Eq.~\eqref{eq:wein}.  A
variety of potentials were employed in the calculation of the \threeHe wave
function $|\Psi_{^3\mathrm{He}} \rangle$.  Effects from a three-nucleon
irreducible kernel $T_{\gamma \mathrm{NNN}}$ begin only at \NXLO{3}, and the
\ChiEFT expansion converges well.  Given an expected accuracy of $\lesssim
20$\%, the authors present exploratory studies on the kinds of information
about neutron polarisabilities that are accessible in such experiments.

\begin{figure}[!htb]
  \begin{center}
\parbox{0.86\textwidth}{
\includegraphics*[width=.48\linewidth]{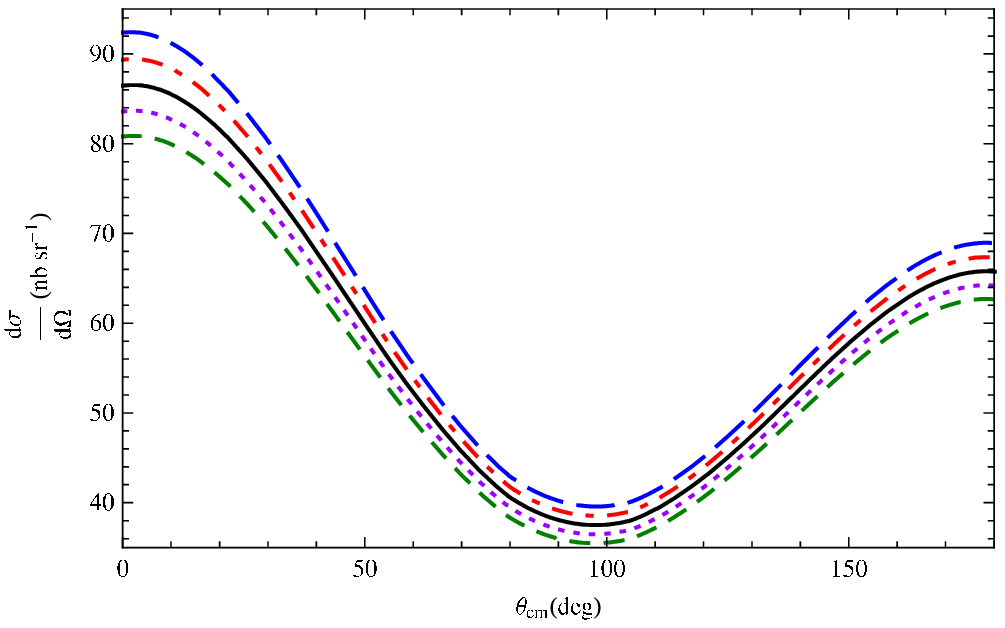}
\hfill
\includegraphics*[width=.48\linewidth]{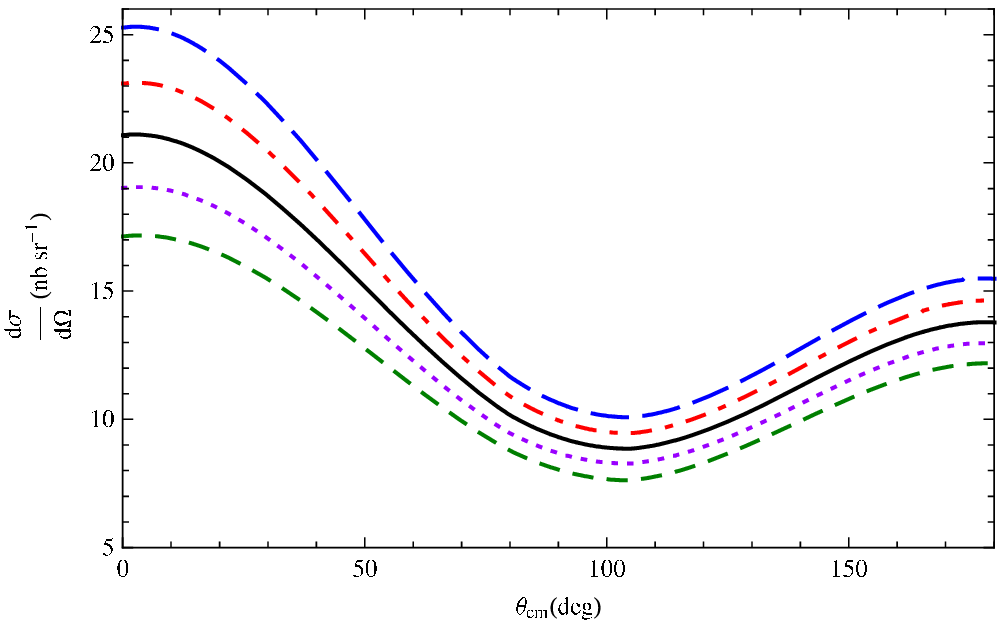}}

\caption{(Colour online) Sensitivity of the differential cross sections for
  Compton scattering on \threeHe in the centre-of-mass frame at NLO (Regime
  II, $P^3$) in \ChiEFT without explicit $\Delta(1232)$ on $\alphaen$ at $60$
  (left) and $120\,\MeV$ (right).  Solid (black) curve: central value
  $\alphaen=12.2$; long-dashed (blue): $
  \alphaen-4$, dot-dashed (red): $
  -2$, dotted (magenta): $
  +2$, dashed (green): $
  +4$.  From Ref.~\cite{Shukla:2008zc}.}
\label{fig:dcsalpha}
\end{center}
\end{figure}

Not surprisingly, \threeHe has a significantly larger cross section for
coherent Compton scattering than the deuteron, due to the presence of two
protons in the nucleus. Since the proton Thomson terms interfere with the
neutron polarisability amplitudes in Regime II, neutron polarisabilities have
a larger absolute effect on cross sections for \threeHe than for the deuteron;
see Figure~\ref{fig:dcsalpha}.
Compared to the proton and deuteron results of Figs.~\ref{fig:sect5-fig3} and
\ref{fig:d-omegadependence}, the process scales roughly with $Z^2$ and is
indeed coherent at the lower end, $\wcm\approx60\;\MeV$, while cross sections
are enhanced over the single-charge results only by $Z$ for $120\;\MeV$.

The sensitivity of the cross section to the scalar polarisabilities of the
neutron was tested by including the higher-order terms $\delta\alphaen$,
$\delta\betamn$ for variations about the $\calO(P^3)$
predictions~\eqref{eq:op3preds}. The example in Fig.~\ref{fig:dcsalpha} shows
a significant effect from the neutron electric polarisability. Varying the
magnetic polarisability produces effects of a similar magnitude. Experimental
prospects and possible improvements will be discussed in
Section~\ref{sec:future}, together with results for double-polarisation
observables.



\section{The future}
\label{sec:future}

We conclude by outlining the experimental and theoretical progress which can
be expected in this area over the next few years.

\subsection{\it The case for investigating spin
  polarisabilities} 
\label{sec:spinpols}

An exciting new frontier in low-energy Compton scattering from light nuclei is
the use of polarised beams and targets to isolate specific polarisabilities of
interest. In particular, the future holds considerable promise for gaining
access to the dipole spin polarisabilities which have thus far been rather
poorly constrained (see Section~\ref{sec:multipoles}).  The values extracted
from experiments on the proton can be compared to those of the neutron
measured in, e.g.,~\threeHe, and to different combinations of isoscalar spin
polarisabilities accessible on the deuteron. \ChiEFT predictions at order
$\epsilon^3\sim e^2\delta^3$(modified) exist for the sensitivity of the
following observables to both scalar and spin polarisabilities: circularly or
linearly polarised beams on polarised protons (and, perhaps less useful,
neutrons)~\cite{Hildebrandt:2005ix, Hildebrandt:2003md} and circularly or
linearly polarised beams on unpolarised or vector polarised
deuterons~\cite{Griesshammer:2010pz}.  A study of many deuteron observables
was first conducted in the ``$\Delta$-less'' version at NLO ($P^3$) with
strict Regime II counting in Refs.~\cite{Choudhury:2004yz, ShuklaThesis}.

For both the deuteron and the proton, the sensitivity to $\gammaee$ seems
particularly large for $\Delta_x^\text{circ}$ (\ref{eq:deltaxz}), the
difference of scattering a right-circularly polarised photon on a target
polarised in either direction perpendicular to the scattering plane.  A good
signal for $\gammamm$ is also seen in the observable $\Delta_z^\text{lin}$,
for which the target is polarised parallel to the beam and the photons are
linearly polarised either in the scattering plane or perpendicular to it.  For
all targets, the spin polarisabilities can be reliably extracted at photon
energies $\gtrsim100\;\MeV$, after measurements at $\lesssim70\;\MeV$ provide
high-accuracy determinations of the scalar polarisabilities so that they do
not contaminate the residual uncertainties of the spin
polarisabilities~\cite{Griesshammer:2010pz}. Since full information on
sensitivities and impact of constraints like the Baldin sum rule cannot
adequately be conveyed on paper, Ref.~\cite{Griesshammer:2010pz} focused only
on prominent examples and made the complete results available as an
interactive \emph{Mathematica}
notebook\footnote{Ref.~\cite{Griesshammer:2010pz} contains two coding errors.
  All ``isoscalar'' variations quoted should be interpreted as variations
  around the isoscalar values for the neutron polarisabilities \emph{only}.
  Fortunately, this leaves the conclusions qualitatively unchanged, albeit
  with all sensitivities doubled. Finally, the variation of $\gamma_{E1E1}$
  around its (correct) central value was implemented with wrong signs.}.

\begin{figure}[!htb]
  \begin{center}
\parbox{0.7\textwidth}{
\includegraphics*[width=.48\linewidth]{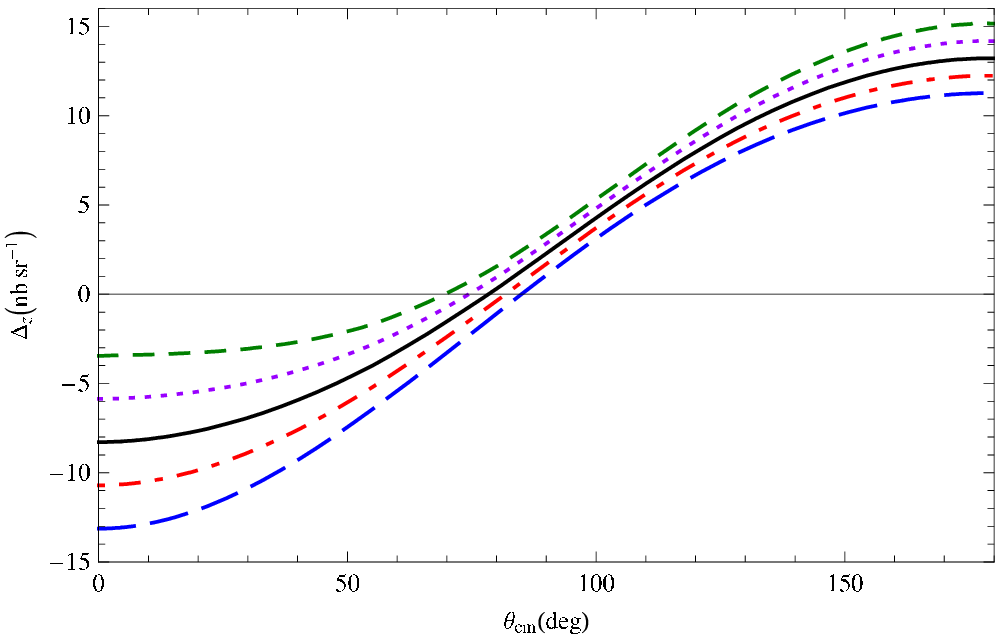} \hfill
\includegraphics*[width=.48\linewidth]{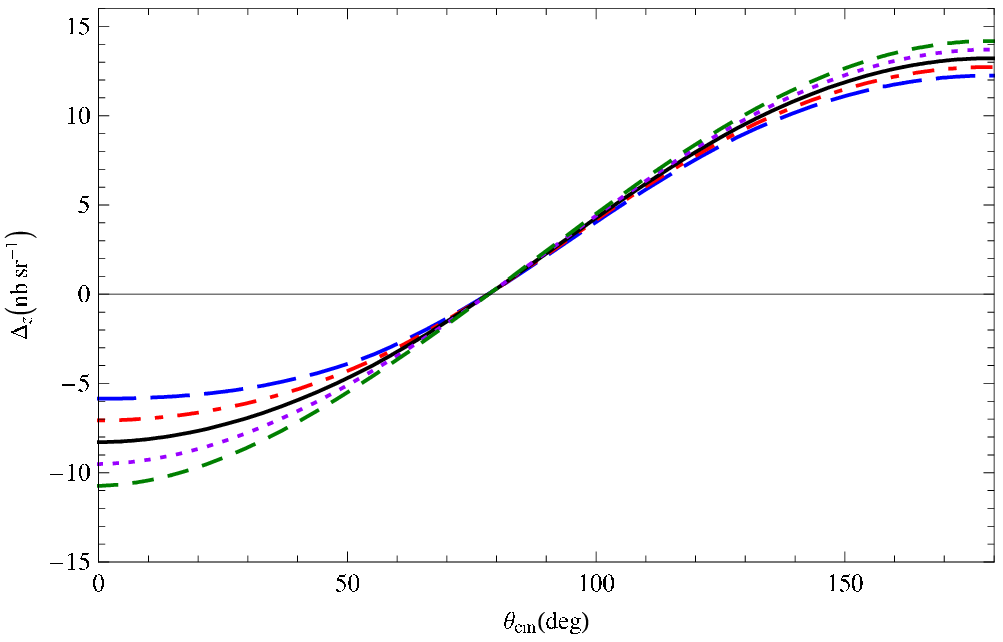}\\[2ex]
\includegraphics*[width=.48\linewidth]{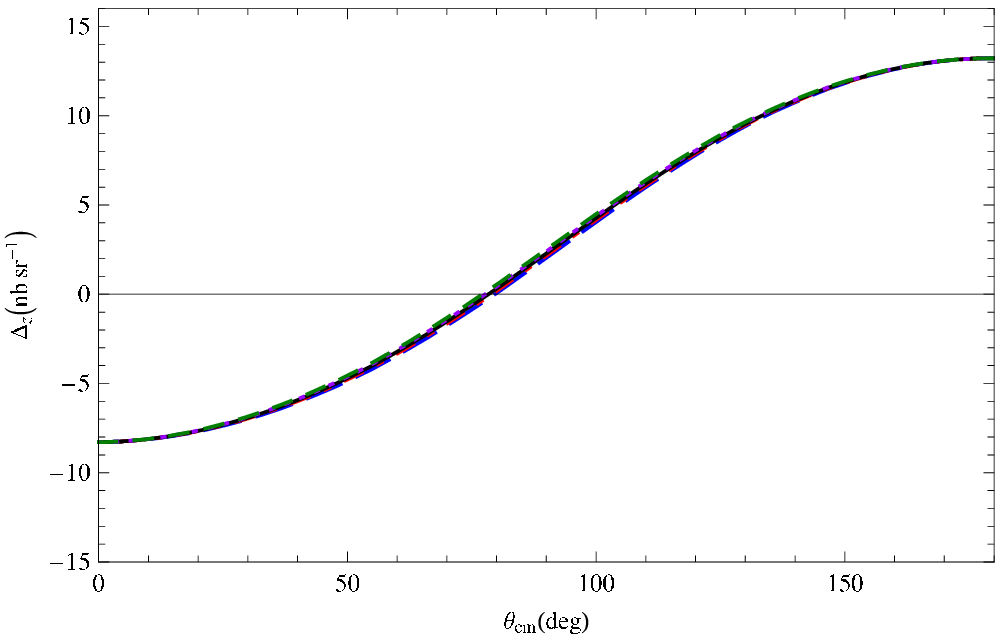} \hfill
\includegraphics*[width=.48\linewidth]{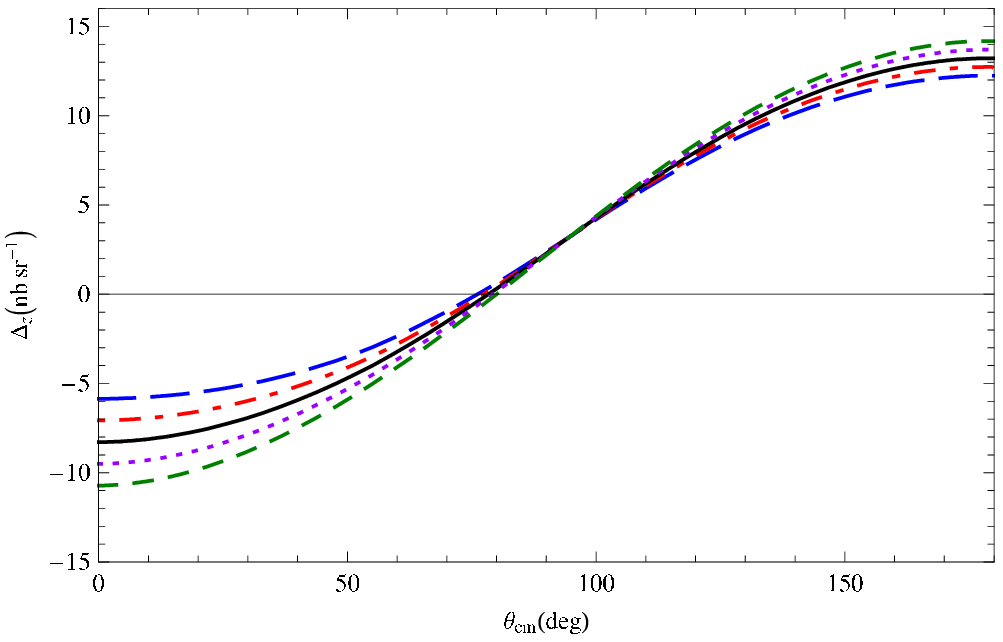}}
\caption{(Colour online) Sensitivity of the double-polarisation observable
  $\Delta_{z}^\text{circ}$ in \threeHe Compton scattering in the centre-of-mass frame at
  NLO (Regime II) in \ChiEFT without explicit $\Delta(1232)$ at $120\,\MeV$ when
  varying the neutron spin polarisabilities $\gamma_{1}^{(\text{n})}$ (top
  left), $\gamma_{2}^{(\text{n})}$ (top right), $\gamma_{3}^{(\text{n})}$
  (bottom left) and $\gamma_{4}^{(\text{n})}$ (bottom right), around the
  $\calO(P^3)$ values~\eqref{eq:op3preds}. Solid (black) curve: unperturbed
  NLO result; variations $\delta\gamma^{(\text{n})}_{i}$ of $-100$\%
  (long-dashed blue), $-50$\% (dot-dashed red), $+50$\% (dotted magenta) and
  $+100$\% (short-dashed green) of the central value. From
  Ref.~\cite{Shukla:2008zc}.
  \label{fig:ddcsz}}
\end{center}
\end{figure}

As an example of a double-polarised observable in \threeHe,
Fig.~\ref{fig:ddcsz} shows predictions at $120$~MeV for
$\Delta_z^\text{circ}$. In addition to the first \threeHe Compton-scattering
calculation discussed in Section~\ref{sec:gammaHe3chiral}, Shukla et
al.~\cite{Choudhury:2007bh,Shukla:2008zc,ShuklaThesis} demonstrated that
polarised \threeHe behaves as an ``effective polarised neutron'' in Compton
scattering. For this case, at least up to NLO, the Compton response is
dominated by the nuclear configuration in which the two protons are paired in
a relative S-wave, and the spin of the nucleus is carried by the neutron.  The
two-body currents at NLO were found to be almost completely independent of the
\threeHe target spin. These two facts lead to \threeHe double-polarisation
observables that look very similar to the neutron asymmetries in
Refs.~\cite{Bernard:1995dp,Hildebrandt:2003md,Hildebrandt:2005ix}.
In each panel, one of the four neutron spin polarisabilities in the basis
of~\eqref{eq:poltranslate} is varied. This shows that a measurement of
$\Delta_z^\text{circ}$ at this energy should enable an extraction of the
combination $\gamma_1^{(\text{n})} - (\gamma_2^{(\text{n})} + 2
\gamma_4^{(\text{n})}) \cos \thetacm=
-(\gammaeen+\gammaemn)-(\gammamen+\gammammn)\cos\thetacm$.
$\Delta_x^\text{circ}$ (\ref{eq:deltaxz}) was found to be sensitive to a
different combination of neutron spin polarisabilities.

\subsection{\it Experiment} 
\label{sec:futureexp}


The High Intensity Gamma-Ray Source (\HIGS) will soon perform its first
Compton scattering experiment that is dedicated to the extraction of nucleon
polarisabilities.  At \HIGS, the photons produced in a free-electron laser
(FEL) are backscattered by the electrons in the FEL storage ring to produce a
high-intensity beam of nearly monochromatic, 100\% polarised photons, with
either linear or circular photon polarisation~\cite{Weller:2009zza}.  The
photon energy resolution depends on the collimation of the incident beam, but
resolution of 2-3\% with photon fluxes of $\sim10^7$ Hz is readily possible.
Photon energies up to 80 MeV are already available at \HIGS, and the region
around pion threshold is anticipated to be accessible in 2 to 3 years.
Because the \HIGS photon beam does not involve bremsstrahlung, it is extremely
clean, with very low background.

Early Compton scattering at \HIGS will focus on exploiting the high intensity
of the beam to perform precision measurements on the proton and deuteron.
Unpolarised measurements on deuterium will be done at 65 and 100 MeV using a
scintillating active target~\cite{Weller:2009}. For the first time, Compton
scattering will be measured by detecting the recoil deuteron in coincidence
with the scattered photon, an advance which is expected to greatly reduce
backgrounds from inelastic scattering and other beam-induced processes. This
new data set will overlap with the existing data of
Refs.~\cite{Lucas:1994,Hornidge:2000,Lundin:2003}, but will have significantly
better statistics than these previous experiments.

Single- and double-polarised measurements are anticipated to be a high
priority at \HIGS~\cite{Weller:2009zza}.  For example, another active target
will be used for the photon-beam asymmetry $\Sigma_3$~\eqref{eq:polasym} at
$90\deg$ and $82\;\MeV$~\cite{Ahmed:2010}.  This experiment will isolate
$\alphaep$ at an angle of $90^\circ$ using linearly polarised photons, since
$\betamp$ does not contribute to $\Sigma_3$ at that angle, as discussed in
Section~\ref{sec:observables}.  The lowest energy with published
beam-asymmetry data is presently
$\wcm=213.1$~MeV~\cite{Blanpied:1996,Blanpied:1997,Blanpied:2001}, so this
experiment represents a substantial increase of the kinematic range.

Entirely new investigations of double-polarisation observables on the proton
are planned at both \HIGS and Mainz. At \HIGS, a circularly polarised beam of
energy $100$~MeV incident on a newly designed, scintillating, transversely
polarised proton target will enable a measurement of the double-spin
asymmetry, $\Sigma^{\rm circ}_{x}$~\eqref{eq:sigmax}, and provide access to
\gammaeep~\cite{Miskimen:2009}.  Meanwhile, experiments with a polarised beam
and a polarised \threeHe target will be carried out at a photon energy of
$125$~MeV~\cite{Gao:2010} to explore the neutron spin polarisabilities.  \HIGS
would also be the ideal place to measure the double-polarisation observables
on the deuteron discussed above. Such a programme requires a polarised
deuterium target, although as yet there are no specific plans at \HIGS for
such experiments.  However, we reiterate that they could provide complementary
information on neutron spin polarisabilities to the \threeHe experiments
already scheduled.

A measurement of unpolarised Compton scattering on $^6$Li at 60 MeV has
already been completed at \HIGS~\cite{Feldman:2010higs}, and data taking at
$80$~MeV is planned. With suitable theoretical interpretation, such cross
sections could provide an alternative path to the extraction of $\alphaen$ and
$\betamn$. The path towards such an extraction is further advanced for $Z=2$
nuclei, as seen for \threeHe in Section~\ref{sec:gammaHe3chiral}. Another good
candidate for complementary information is $^4$He, since it is both a scalar
and isoscalar target, with cross sections that are comparable in magnitude to
those on \threeHe. In these Helium nuclei, nuclear binding is not as complex
as in ${}^6$Li, and so measurements of their Compton cross sections would be
very worthwhile. The prospects for advances in the theory of all these nuclear
Compton reactions are discussed in Section~\ref{sec:theoryfuture}.


Another avenue to measure nucleon polarisabilities has been pursued at the
S-DALINAC at Darmstadt. A proof-of-principle experiment was performed at low
energies using a bremsstrahlung beam and a high-pressure ionisation
chamber~\cite{Yevetska:2010}. This chamber, which could be filled with
hydrogen or deuterium, functioned as both the target and as a particle
detector for the recoiling hadron.  Even with a continuous bremsstrahlung
distribution for the incident photon beam, the coincident detection of the
scattered photon in a 25.4 cm $\times$ 35.6 cm NaI detector and the
struck-particle recoil in the active target volume provided sufficient
kinematic over-determination to reduce backgrounds significantly. A test
measurement of $\gamma \mathrm{p}$ differential cross sections using
bremsstrahlung endpoints of 60 and 79 MeV obtained data of comparable
statistical quality to that of Federspiel~\cite{Federspiel:1991} for $30$ MeV
$\leq \wlab \leq 50$ MeV. Yevetska et al.\ claim that ``expected yields from
an experiment based on the technique described here are at least an order of
magnitude larger than in the experiment of [Ref.~\cite{Olmos:2001}]'' and
conclude that high-statistics experiments using this setup are feasible.
However, no full-scale production run is planned.


MAX-Lab at Lund is presently performing unpolarised measurements on the
deuteron at 145 to 170 MeV. These will provide the first high-accuracy Compton
scattering data on deuterium at higher energies~\cite{Feldman:2010lund}. Data
from previous runs below 115 MeV are being analysed in
parallel~\cite{Myers:2010,Shoniyozov:2011}. For both, the experimental setup
at Lund is the same as that described at the end of
Section~\ref{sec:expdeutlow}.  While proton data exist up to the
$\Delta(1232)$ resonance (as discussed in Sections~\ref{sec:expoverview} and
\ref{sec:protonanalysis}), there are no analogous measurements on the
deuteron, so data on the neutron amplitude in this energy range are quite
limited.  One of the primary motivations for extending the Compton experiments
at Lund above pion-production threshold is to provide benchmark data for EFT
calculations that are envisioned for the near future (see below).

A new synchrotron light facility, MAX-IV, is presently being constructed at
Lund~\cite{maxiv:2011}. Under a proposal to dedicate one of the beamlines to
nuclear physics, its $1.5$~GeV electron storage ring could open up new
opportunities.  Perhaps as soon as 2014, it will be possible to perform laser
backscattering via a technique similar to that at LEGS and produce collimated
photon beams of $>10^4$~Hz and polarisations of $> 70$\% for $130$ MeV $\leq
\wlab \leq 157$ MeV. These could be used for Compton scattering and other
photonuclear experiments~\cite{Fissum:2011}.  Even at this early stage, a
directed programme of Compton studies is being explored.


MAMI at Mainz has also been gearing up to perform polarised Compton
experiments using the Crystal Ball detector~\cite{Downie:2009}.  These
experiments cover the energy range $200$--$300$ MeV and focus on the spin
polarisabilities of the proton. Polarised photon beams are incident on either
an unpolarised liquid hydrogen target or a polarised frozen-spin butanol
target, with the target located at the centre of the Crystal Ball. Higher
photon energies are mandatory for this detection scheme because, as seen in
Section~\ref{sec:expprothigh}, Compton scattering cannot be separated from
other processes unless the recoil proton is detected in coincidence with the
scattered photon. MAMI produces linearly polarised beams using tagged coherent
bremsstrahlung on a diamond radiator, and circularly polarised beams via a
longitudinally polarised electron beam incident on an amorphous radiator.
Three separate polarisation observables are measured: $\Sigma_3$, as well as
$\Sigma^{\rm circ}_{x}$ and $\Sigma^{\rm circ}_{z}$.  These experiments have
only recently begun, and so far only preliminary results have been obtained.
Since MAMI already has a polarised \threeHe target, the possibility to perform
similar experiments for $\gamma$\threeHe scattering also
exists~\cite{Downie:2011}.  Such experiments, in concert with those at \HIGS,
will provide access to the proton and neutron spin polarisabilities.


We close our discussion of the future of experiments by reiterating a central
conclusion from the proton analysis of Section~\ref{sec:protonanalysis}.  The
unpolarised proton database is somewhat noisy, and the data that exist between
$190$ MeV and $250$ MeV are contradictory and not of particularly high
quality. The statement in Section~\ref{sec:deuteronanalysis} about improving
the deuteron database also applies to the proton: data of higher quality with
carefully formulated correlated and point-to-point systematic errors are
needed. Theorists and experimentalists should work together in order to
determine which of the sometimes-conflicting data sets are trustworthy and to
ensure that part of the future experimental Compton programme enhances the set
of measurements of unpolarised $\gamma\mathrm{p}$ scattering.

\subsection{\it Theory}
\label{sec:theoryfuture}

Indeed, it is notable that the last twenty years of the history of low-energy
Compton scattering from protons and light nuclei has been marked by
significant cooperation between experiment and theory (EFTs, DRs,\dots).  We
will now provide an outlook for the theoretical advances which can be
anticipated, bearing in mind that the future path of theoretical developments
is somewhat trickier to predict than that of experimental work.

In Section~\ref{sec:protonanalysis}, we presented results of an analysis of
proton data up to energies around $350$ MeV using \ChiEFT with a resonant
$\Delta(1232)$ pole as well as $\pi\mathrm{N}$ loops at $\calO(e^2\delta^3)$,
with a first look at extending to $\calO(e^2\delta^4)$ (\NXLO{2}) by including
higher-order $\pi$N loops.  It was noted that in order to obtain a reasonable
fit to the data at $\calO(e^2\delta^4)$, a contact term had to be added to the
calculation, thereby allowing the value of $\gammamm$ to be tuned. This
analysis is presently being refined. In particular, the impact of other spin
polarisabilities on the fit must be examined, and the effect of $\gamma
\mathrm{N} \Delta$ vertex dressing must be incorporated.  The resulting
calculation will be complete up to $\calO(e^2 \delta^4)$ in the low-energy
region and up to $\calO(e^2 \delta^0)$ in the vicinity of the resonance, and
it is expected to produce robust constraints on the scalar polarisabilities
$\alphaep$ and $\betamp$~\cite{McGovern:2012}.


In order to interpret the ongoing experiment at MAX-Lab described above, a
theory capable of describing elastic scattering on the deuteron up to $\wlab
\approx 200$ MeV is required, taking into account that pion photoproduction
channels are open. In particular, as the energy increases towards the
$\Delta(1232)$ peak, it is expected that resonant pion photoproduction will
play an increasingly important role. The $\delta$
expansion~\cite{Pascalutsa:2002pi} mandates that the $\mathrm{NN}$ and
$\mathrm{N} \Delta$ channels should be treated on an equal footing in the
vicinity of the resonance peak.  But experience from the single-nucleon
sector, combined with the power-counting arguments of
Section~\ref{sec:singleN}, implies that this is only strictly necessary within
about 50 MeV of the peak. Thus, the first step for analysing data up to $200$
MeV will be a perturbative inclusion of $\Delta$ effects, with pion
production. However, it must be noted that a reliable calculation has to
include the correct $\mathrm{d}\gamma\to \mathrm{d}\pi^0$ and $\gamma
\mathrm{d} \to \mathrm{NN}\pi$ thresholds---just as the $\pi \mathrm{N}$
threshold had to be corrected in the proton calculation. The remedy in the NN
sector is well-defined, see
Refs.~\cite{Baru:2004kw,Lensky:2006wd,Lensky:2005hb,Lensky:2006af,Lensky:2007zc,
  Hoferichter:2011zz}, but more elaborate than for the single-nucleon sector.
At higher energies, the issues of chirally consistent
currents~\cite{Kolling:2009iq, Kolling:2011mt, Kvinikhidze:2009be,
  Pastore:2009is}, wave functions and potentials~\cite{Nogga:2005hy,Birse,
  Yang:2009kx, Yang:2010, Birse:2010fj, Long:2011xw, Valderrama:2011mv} must
also be revisited. As discussed in Section~\ref{sec:dunify}, they are largely
absent at lower energies, thanks to the strict constraint imposed by the
Thomson limit.


Computing the deuteron breakup reaction $\gamma \mathrm{d} \rightarrow \gamma'
\mathrm{np}$ in \ChiEFT is a high priority for future work. The calculation of
Levchuk and L'vov was used to extract $\alphaen$ from data on this reaction in
Ref.~\cite{Kossert:2002ws}, as discussed in Section~\ref{sec:gammadmodels}. It
is important to analyse its model dependence using a theory with a consistent
description of Compton interactions in the one- and two-nucleon sectors, along
with defensible error bars. The quasi-free mechanisms are independent of the
\ChiEFT cutoff for low missing momenta, but corrections from final-state
interactions could be sensitive to details of the short-distance physics in
the NN system~\cite{Yang:2010}.  \ChiEFT therefore provides a diagnostic---the
cutoff dependence of results---which facilitates studying the reliability of
the calculation of effects in the NN sector.

The single-nucleon amplitude described in Section~\ref{sec:protonanalysis} and
presented in Ref.~\cite{McGovern:2012} will also provide an alternative to the
dispersion-relation methods used by Kossert et al.~\cite{Kossert:2002ws} for
the analysis of the quasi-free data.  In particular, one might be concerned
that the asymptotic contributions to the disperson-relation integrals
introduces uncontrolled model dependence. The \ChiEFT amplitude will be able
to demonstrate the extent to which different assumptions about the
short-distance $\gamma$N physics alter the extracted value of $\alphaen$. If
\ChiEFT confirms that the kinematics of Ref.~\cite{Kossert:2002ws} are ideal
for a reliable extraction of $\alphaen$ from inelastic deuteron Compton data,
then a by-product of such an analysis could be a \ChiEFT result for
$\gammapin$, to which these data are also sensitive (see
Section~\ref{sec:expdeutqf}).  More broadly, such a calculation could be
envisioned as a tool to search for other kinematics in order to optimise the
extraction of neutron polarisabilities.


Turning from the 2N to the 3N sector, we have already argued that ${}^3$He
provides a fine opportunity for obtaining information on the neutron Compton
amplitude (see Sections~\ref{sec:gammaHe3chiral} and \ref{sec:spinpols}). The
experiments at \HIGS and MAMI discussed in Section~\ref{sec:futureexp} aim to
use this opportunity to provide data pertinent to the spin polarisabilities of
the nucleon, but
the calculations reported in Section~\ref{sec:gammaHe3chiral} for \threeHe
were clearly exploratory in nature. They neither contained a dynamical
$\Delta(1232)$ nor assessed the importance of rescattering in the 3N
intermediate state. In combination, these effects can increase
double-polarisation observables by up to a factor of
$2$~\cite{Griesshammer:2010pz}.

In particular, the results from Compton scattering on deuterium presented in
Section~\ref{sec:dunify} suggest that rescattering significantly reduces the
wave-function dependence seen in
Refs.~\cite{Choudhury:2007bh,Shukla:2008zc,ShuklaThesis}. For the deuteron,
rescattering increases some double-polarisation observables by up to $40$\% at
$125\;\MeV$~\cite{Griesshammer:2010pz}. It is therefore imperative to include
the effect of rescattering for \threeHe in order to provide a precise \ChiEFT
extraction of neutron polarisabilities from forthcoming data.
A Bochum-J\"ulich-Bonn-Washington (DC) collaboration is embarking on such a
calculation, initially using the ``$\Delta$-less'' \ChiEFT NN and NNN
potential at \NXLO{2}.  As in the case of the deuteron, consistency between
the \threeHe wave function, the nuclear interaction and the nuclear
electromagnetic operators is a key ingredient. The correct \ChiEFT
implementation in the low-energy (Regime I) power counting can again be
checked by the Thomson and Gell-Mann-Goldberger-Low low-energy theorems for
the triton and \threeHe. 

As for the proton and deuteron, we also expect the $\Delta(1232)$ to have a
large effect on observables at $\wlab \gtrsim90$~MeV. It is straightforward to
include the single-nucleon amplitudes with an explicit $\Delta$ at orders
$\epsilon^3$ and $e^2\delta^4$, respectively, into the \threeHe calculations.
In contradistinction to the deuteron case, $\Delta$ exchange currents can
appear, and so a consistent \ChiEFT calculation requires the $\pi$-exchange
diagrams of Refs.~\cite{Choudhury:2007bh, Shukla:2008zc, ShuklaThesis} to be
complemented by currents with $\Delta$s. In the kinematic domain
$0\le\omega\sim\mpi$, the $\pi\mathrm{N}\Delta$ interactions and corrections
to the 3N$\rightarrow$3N Green's function do not have to be resummed but can
be treated in perturbation theory.  In the longer term, a calculation that
includes channels with up to one explicit $\Delta(1232)$ and a resummation of
intermediate-state interactions can be anticipated.  It would compute the
coherent reaction in a consistent manner over the entire kinematic range $0
\leq \omega \leq 350$ MeV, similar to the ongoing effort for the deuteron.
Inelastic Compton scattering on \threeHe could also be used to extract proton
or neutron polarisabilities in specific kinematics, as for the deuteron.


Compton scattering from nuclei such as ${}^4$He and ${}^6$Li is becoming
accessible to ab initio calculation. Bampa et al.~recently applied the Lorentz
Integral Transform (LIT) method to deuteron Compton scattering with a \ChiEFT
potential~\cite{Bampa:2011fq}. The currents were constructed via Siegert-like
theorems. The results up to $60$ MeV are promising. The same method is
presently being applied to ${}^6$Li~\cite{Leidemann:2011}, for comparison with
the recently completed ${}^6$Li experiment at 60 MeV at
\HIGS~\cite{Feldman:2010higs}.  This computation will then be extended to
higher energies by incorporating NN currents beyond the electric dipole and
the full single-nucleon amplitude. The ability to accurately assess the impact
of nucleon polarisabilities in Compton scattering from ${}^6$Li offers the
tantalising prospect that they could be extracted there, where the cross
section in the coherent region is about twice as large as in \threeHe---and
roughly nine times larger than for the proton or deuteron.


\subsection{\it Conclusion} 
\label{sec:conclusion}

An energy-dependent multipole analysis of Compton scattering provides
important information on the scales, symmetries and mechanisms which govern
the interactions amongst the low-energy constituents of the nucleon, and with
photons. It facilitates exploration of the chiral symmetry and isospin
dependence of the pion cloud, the properties of the $\Delta(1232)$ resonance,
and the question of which degrees of freedom dominate the response of the
nucleon spin to electromagnetic fields. It is in this context that Chiral
Effective Field Theory has emerged as a systematic and reliable tool to guide,
predict and analyse experiments.

A window of opportunity seems to be emerging between the competing demands of
theorists and experimentalists. With increasing nuclear mass, there is also an
increase in the number and viability of potential target nuclei.  Cross
sections grow quadratically with the target charge $Z$ at lower energies where
the nucleons act coherently, and still linearly with $Z$ at higher energies,
$\w\gtrsim 100\;\MeV$. This makes high-accuracy experiments more feasible for
heavier nuclear targets.  Breakup experiments on such nuclei may permit the
isolation of neutron properties in particular, but proton structure effects
may also be enhanced thanks to the interference effects with more charges in
the target. On the other hand, interpreting such data in terms of the
properties of the individual nucleons requires a model-independent analysis
which treats nuclei by ab initio methods. This poses no difficulty for
deuteron Compton scattering and is within reach for \threeHe. Given the recent
progress, one can speculate that both ${}^4$He and ${}^6$Li may be treated in
the same \ChiEFT framework.  This will also provide important benchmarks on
the accuracy with which \ChiEFT describes nuclear binding and the charged
meson-exchange currents which are directly probed by the Compton photons.

In conclusion, the future is bright for Compton scattering from protons and
light nuclei, both literally and figuratively. A number of high-luminosity
facilities around the world conduct or plan to perform experiments that will
map out the spin polarisabilities of the proton, with those of the neutron not
far behind. More sophisticated experiments, with higher fluxes and improved
target techniques, combined with ab initio computations based on \ChiEFT
Hamiltonians, current operators and photon-nucleon amplitudes, could permit
the use of nuclei up to at least $A=6$.  The resulting values for the twelve
static dipole polarisabilities of the proton and neutron will provide
important benchmarks for lattice QCD computations and models of the nucleon.
The picture that ultimately emerges of the manner in which mechanisms driven
by chiral symmetry (like pion loops) compete with other shorter-distance
effects (like the properties of the $\Delta(1232)$ resonance) will provide
fascinating and important insights into the similarities and differences of
the two-photon responses of the proton and the neutron.



\section*{Acknowledgements}
We acknowledge our many collaborators, who have shared their insights on these
topics with us. We are especially grateful to: Vadim Lensky and Vladimir
Pascalutsa for correspondence regarding Ref.~\cite{Lensky:2009uv} and for
supplying us with curves relevant to that calculation; Winfried Leidemann for
information on future plans regarding application of the Lorentz Integral
Transform to Compton scattering; Anatoly L'vov for discussion of the
theoretical uncertainties associated with his calculations; Barbara Pasquini
for comments on dispersion relations; Paul Hewson for a stunningly quick
coding of some amplitudes; Deepshikha Shukla for helping to clarify numerical
issues in deuteron Compton scattering; and the organisers and participants of
the INT workshop 08-39W: ``Soft Photons and Light Nuclei'' (all) and of the
INT programme 10-01: ``Simulations and Symmetries'' (JMcG, HWG and DRP), both
of which also provided financial support. This work has been supported in part
by UK Science and Technology Facilities Council grants ST/F012047/1,
ST/J000159/1 (JMcG) and ST/F006861/1 (DRP), by the US Department of Energy
under grants DE-FG02-06ER-41422 (GF), DE-FG02-95ER-40907 (HWG) and
DE-FG02-93ER-40756 (DRP), by the US National Science Foundation
\textsc{Career} award PHY-0645498 (HWG), and by University Facilitating Funds
of the George Washington University (HWG).
\newpage


\end{document}